# AUTOMATING THE ANALYSIS AND IMPROVEMENT OF DYNAMIC PROGRAMMING ALGORITHMS WITH APPLICATIONS TO NATURAL LANGUAGE PROCESSING

by

Tim Vieira

A dissertation submitted to The Johns Hopkins University in conformity with the requirements for the degree of Doctor of Philosophy.

Baltimore, Maryland

July, 2023

# Abstract

This thesis develops a system for automatically analyzing and improving dynamic programs, such as those that have driven progress in natural language processing and computer science, more generally, for decades. Finding a correct program with the optimal asymptotic runtime can be unintuitive, time-consuming, and error-prone. This thesis aims to automate this laborious process. To this end, we develop an approach based on

1. a high-level, domain-specific language called Dyna for concisely specifying dynamic programs

2. a general-purpose solver to efficiently execute these programs

3. a static analysis system that provides type analysis and worst-case time/space complexity analyses

4. a rich collection of meaning-preserving transformations to programs, which systematizes the repeated insights of numerous authors when speeding up algorithms in the literature

5. a search algorithm for identifying a good sequence of transformations that

reduce the runtime complexity given an initial, correct program

We show that, in practice, automated search—like the mental search performed by human programmers—can find substantial improvements to the initial program. Empirically, we show that many speed-ups described in the NLP literature could have been discovered automatically by our system. We provide a freely available prototype system at `https://github.com/timvieira/dyna-pi`.

**Primary Reader and Advisor:**

Jason Eisner (Johns Hopkins University)

**Secondary Readers:**

David Chiang (University of Notre Dame)

Scott Smith (Johns Hopkins University)

# Acknowledgments

## Academic

I am incredibly grateful to my advisor, Jason Eisner, for taking me on as a student and patiently working with me all these years. I have learned *so much* about how to teach, present, write, think, get the details right—and have fun doing it! My PhD has been a truly transformational experience thanks to Jason.

I want to thank Ryan Cotterell, who acted as an unofficial secondary advisor in my last two years. Without Ryan's support, knowledge, patience, cheerleading, and insights, I would not have been able to write this dissertation.

I want to thank Wes Filardo, who taught me an incredible amount of what underlies this thesis. Sitting next to such a knowledgeable and generous human made my PhD much more valuable and enjoyable.

I want to thank my committee members—Jason Eisner, David Chiang, and Scott Smith—for patiently reading this *long* manuscript and providing useful and insightful feedback. Their feedback has greatly improved this manuscript. Any mistakes are my own.

I want to thank Andrew McCallum and Dan Roth for employing me as a research programmer, which set me on track to do a PhD. Andrew even provided the crucial introduction to Jason Eisner!

I want to thank Gerald DeJong, Jeff Erickson, Viraj Kumar, and Cinda Heeren for their support during my undergraduate years. Viraj is the reason why I first felt empowered to go to graduate school—I am very grateful to him.

My PhD experience would not have been the same without the extraordinary academic community I have been fortunate to be part of.

- **Collaborators:** Afra Amini, Alexandra Butoi, Andreas Opedal, Anej Svete, Benjamin Dayan, Brian DuSell, Clara Meister, Clemente Pasti, He He, Jiarong Jiang, Jiawei Huang, Matthew Francis-Landau, Qi He, Ran Zmigrod, Tiago Pimentel, and Wes Filardo.

- **Labmates (JHU):** Adam Teichert, Brian Lu, Chu-Cheng Lin, Darcey Riley, Dingquan Wang, Hongyuan Mei, Juneki Hong, Leo Du, Matthew Francis-Landau, Matt Gormley, Michael Paul, Nick Andrews, Ryan Cotterell, Sabrina Mielke, and Wes Filardo.

- **Labmates (UMass):** Greg Druck, Sameer Singh, Sebastian Riedel, Michael Wick, Kedar Bellare, Adam Saunders, Rachel Shorey, David Mimno, Pallika Kanani, Anton Bakalov, Karl Schultz, Limin Yao, Meagan Day, Laura Dietz, Hanna Wallach, and Jason Naradowsky.

- **Labmates (UIUC):** Alex Klementiev, Dan Goldwasser, Jeff Pasternack, Kevin Small, Lev Ratinov, Mark Sammons, Michael Connor, Ming-Wei

Chang, Nick Rizzolo, Quang Do, Rajhans Samdani, Subhro Roy, V.G.Vinod Vydiswaran, Vivek Srikumar, and Yuancheng Tu.

- **Broader JHU:** Ben Van Durme, Colin Lea, Elizabeth Salesky, Frank Ferraro, Hainan Xu, Huda Khayrallah, Juri Ganitkevitch, Kim Franklin, Keith Levin, Sanjeev Khudanpur, Tongfei Chen, Mark Dredze, Naomi Saphra, Olivia Buzek, Pushpendre Rastogi, Rachel Rudinger, Ruth Scally, Sheng Zhang, Travis Wolfe, and Yanif Ahmad.

- **Broader Community:** Aaron Schein, David Bellanger, David Smith, Hal Daumé, Kianté Brantley, Mathieu Blondel, Nikos Karampatziakis, Phil Thomas, Pierre-Luc Bacon, Sasha Rush, and Vlad Niculae.

# Non-Academic

I want to thank my parents, Marcia and Terry Vieira, for their unconditional love and support, for always prioritizing my education, and for their sacrifices to send me to private school and out-of-state college. I could not ask for better parents. Thank you, Mom and Dad!

I want to thank my incredible friends and family for all their love, support, and entertainment: Hanna Wallach, Maia Papaya, Marcia Vieira, Terry Vieira, Tamara Vieira, Roseli Fabbri, Rachael Wallach von Portheim, Luna Petunia, Robin Wallach, Rob Wallach, Parth Naik, Ryan Cotterell, Boris Umanov, Alex O'Neill, Keanu, Dan Muriello, Jason Naradowsky, Ian Baran, Dan Schreiber,

Abbie Jacobs, Aaron Schein, Clint Adams, Hal Daumé, and Amber Boydstun.

The biggest and most important thank you is to my partner, Hanna Wallach. She has been with me through my entire PhD journey—she even helped me with my application essay! I would not be here at the finish line without her love, encouragement, generosity, patience, and support. Thank you, Hanna.

# Previous Publications

Portions of this thesis have been published in the following:

- Tim Vieira, Matthew Francis-Landau, et al. (2017). "Dyna: Toward a Self-Optimizing Declarative Language for Machine Learning Applications". In: *Proceedings of the ACM SIGPLAN Workshop on Machine Learning and Programming Languages)*, pp. 8–17. ISBN: 978-1-4503-5071-6. URL: http://cs.jhu.edu/~jason/papers/#vieira-et-al-2017
- Tim Vieira, Ryan Cotterell, et al. (2021). "Searching for More Efficient Dynamic Programs". In: *Findings of the Association for Computational Linguistics*. URL: https://arxiv.org/abs/2109.06966

# Contents

# List of Tables

# List of Figures

# List of Algorithms

# Chapter 1

# Introduction

Algorithmic research in natural language processing (NLP) has focused—in large part—on developing dynamic programming solutions to combinatorial problems that arise in the field. Such algorithms have been introduced over the years for numerous linguistic formalisms, such as

- finite-state transduction (e.g., Mohri (1997), Eisner (2002)),
- context-free parsing (e.g., Stolcke (1995), Goodman (1999)),
- dependency parsing (e.g., Eisner (1996), Koo and Collins (2010))
- mildly context-sensitive parsing (e.g., Vijay-Shanker and Weir (1989), Vijay-Shanker and Weir (1990), Kuhlmann, Satta, and Jonsson (2018)).

In recent years, the same algorithms have often been used for deep structured prediction, using a neural scoring function that decomposes over the structure.[1]

[1]E.g., Durrett and Klein (2015), Rastogi et al. (2016), Lee et al. (2016), Dozat and Manning (2017), Stern et al. (2017), Kim et al. (2017), Hong and Huang (2018), Wu, Shapiro, et al. (2018), Wu and Cotterell (2019), Qi et al. (2020), and Rush (2020).

**Designing efficient algorithms is challenging.** Designing and implementing dynamic programs which are as efficient as possible is challenging and error-prone. When a dynamic programming algorithm for a new problem is first introduced in the literature, its runtime is sometimes suboptimal—faster versions are often published over time (see Example 1). Indeed, introducing the first algorithm and subsequently finding improvements is common throughout computer science. We seek to facilitate this process through automation.

**Example 1.** *Consider the following instances*[2] *of published dynamic programs whose runtime bounds were later improved. In the examples below, $n$ denotes the length of the input sentence.*

- *Projective dependency parsing: Collins (1996) gave an $\mathcal{O}(n^5)$ algorithm that was sped up to $\mathcal{O}(n^4)$ by Eisner and Satta (1999a).*

- *Split-head-factored dependency parsing: implemented naïvely runs in $\mathcal{O}(n^5)$; with some effort, an $\mathcal{O}(n^3)$ algorithm can be derived (Eisner, 1996; Johnson, 2007; Eisner and Blatz, 2007).*

- *Linear index-grammar parsing: $\mathcal{O}(n^7)$ in Vijay-Shanker and Weir (1989), sped up to $\mathcal{O}(n^6)$ by Vijay-Shanker and Weir (1993).*

- *Lexicalized tree adjoining grammar parsing: $\mathcal{O}(n^8)$ in Vijay-Shankar and Joshi (1985), sped up to $\mathcal{O}(n^7)$ by Eisner and Satta (2000).*

[2] Many of these examples were brought to our attention in the works of Eisner and Blatz (2007) and Gildea (2011); further discussion can be found therein.

- *Inversion transduction grammar: $\mathcal{O}(n^7)$ in Wu (1996), sped up to $\mathcal{O}(n^6)$ by Huang, Zhang, et al. (2005).*

- *CKY parsing (Cocke and Schwartz, 1970; Younger, 1967; Kasami, 1965) is typically presented in a suboptimal $\mathcal{O}(k^3 n^3)$ form, but can be sped up to $\mathcal{O}(k^2 n^3 + k^3 n^2)$ (Lange and Leiß, 2009; Eisner and Blatz, 2007) where $k$ is the number of nonterminal symbols in the grammar.*

- *Tomita's context-free parsing algorithm (1985) runs in $\mathcal{O}(n^{\rho+1})$ where $\rho$ is the length of the longest right-hand side of a context-free production in the grammar (Johnson, 1989). However, it can be made to run in $\mathcal{O}(n^3)$ by binarizing the production rules.*

> **Central question.** This dissertation explores the question:
>
> Can we *automatically* discover these faster algorithms?

**A unified notation for dynamic programming.** Our approach starts with an observation made by Goodman (1999) that a dynamic programming algorithm can be regarded as performing inference in a semiring-weighted deduction system. Their work builds upon a rich history of (unweighted) deduction systems that we review in §9.1. Eisner, Goldlust, et al. (2005) provided a programming language, Dyna, for expressing such deduction systems, along with a compiler that produced fast inference code. We will use Dyna in this thesis as our algorithm-specification language. We introduce the language in §2 and give formal semantics in §4.

Below is an example Dyna program that computes the total weight of all paths between pairs of nodes in a weighted directed graph.

```
path(I,I) += 1.
path(I,K) += edge(I,J) * path(J,K).
```

Here the weight of an edge comes from the relation `edge(I,J)`. For example, to declare that the weight of the edge from a node labeled `"a"` to a node labeled `"b"` is 13, we write `edge("a","b") += 13`. From there, the `path(I,K)` items encode a weighted relation between all the nodes in the graph. Specifically, for all pairs of nodes `I,K`:

$$\mathsf{path}(\mathtt{I},\mathtt{K}) = \underbrace{1[\mathtt{I}{=}\mathtt{K}]}_{\#1} + \sum_{\mathtt{J}} \underbrace{\mathsf{edge}(\mathtt{I},\mathtt{J}) \cdot \mathsf{path}(\mathtt{J},\mathtt{K})}_{\#2} \tag{1.1}$$

Here the variable `J` ranges over the set of nodes. The reason why the variable `J` is summed in (1.1), while `I` and `K` are not, is because it is local to the right-hand side of #2. The reader may recognize (1.1) as the mathematical recurrence for the total weight of all paths between pairs of nodes in a graph. Turning this mathematical specification into an efficient computational procedure requires several implementation details to be specified. In this example, we need to select data structures to store and index the edge and path relations—but, more importantly, we need to *solve* the recursive equation to derive the total weight. In § 5, we describe general-purpose procedures for taking the mathematical recurrences (written in Dyna) and turning them into efficient algorithms.

**Automated analysis.** A key ingredient to our system is static analysis: the ability to reason about the program without running it on specific inputs.

Such analyses include type analysis, runtime complexity analysis, and space complexity analysis (Cousot and Cousot, 1977; McAllester, 2002). These analyses are important for understanding the space of possible program behaviors on unknown input data.[3] *Automated* program analysis aids programmers because *manual* program analysis may be tedious and error-prone. In the context of NLP, automated program analysis of dynamic programs (for instance, as specified in Dyna) has the potential to help NLP algorithms researchers correctly and speedily analyze their creations. Indeed, the NLP literature provides examples where such automated analysis could have been of use.

**Example 2.** *Consider the following example from the NLP literature:*

- *Kuhlmann, Gómez-Rodríguez, et al. (2011, §5.4) discovered a surprising property of their tabular dependency parsing algorithm. Their initial naive analysis gave a time complexity of $\mathcal{O}(n^5)$ and space complexity of $\mathcal{O}(n^3)$. Still, they were able to tighten this analysis to a time complexity of $\mathcal{O}(n^3)$ and space complexity of $\mathcal{O}(n^2)$ after noticing that the objects built by their transition system always had two equal variables (which could be proved inductively). They then revised their program formulation to eliminate the redundant variable.*

- *Huang and Sagae (2010) derive a projective dependency parsing algorithm with a stated complexity of $\mathcal{O}(n^7)$, which was later discovered to actually be $\mathcal{O}(n^6)$ according to Shi et al. (2017).*[4]

[3]Although we do not explore it in this thesis, we also mention that static program analysis is essential for compiling efficient code. Type inference, for example, allows the compiler to synthesize efficient data structures for representing internal objects and efficient code for pattern-matching against them.

[4]This is stated in Shi et al. (2017) without proof. The only explanation we could find in the

Our automated analysis system yields big-$\mathcal{O}$ bounds on runtime and space complexity (§6.3). These bounds are derived by inferring types for each variable in a given rule and using each type to bound the number of ways the rule will fire during execution (§6.3). At a high level, to apply this technique to the path program, we first see that the variables `I`, `J`, and `K` are nodes. Supposing that the set of nodes has size $n$. The number of ways that the rule `path(I,K) += edge(I,J) * path(J,K)` will fire during inference is $\mathcal{O}(n^3)$. The foundations of this technique are given by McAllester (2002)'s *meta theorem*. Many authors[5] manually apply the meta theorem to analyze the runtime of their recursive algorithms. Our runtime analyzer (§6.3) automatically applies the meta theorem to derive tight bounds on the worst-case space and runtime complexity of executing a given program (using the execution methods in § 5).

**Systematizing algorithmic speed-ups.** All of the runtime improvements mentioned in Example 1 can be cast as source-to-source program transformations on Dyna programs (Eisner and Blatz, 2007); we develop these transformations in § 7.[6] In dynamic programming, program transformations may be exploited to literature is in Hong and Huang (2018). In that work, the authors explain that one of the variables in the reduction rule (of a closely related model), which appears to take on $\mathcal{O}(n)$ values, is functionally dependent on the other variables. This means that it may take on at most one value given the values of other variables in scope; thus, one of the factors of $n$ in the runtime becomes a constant. This observation appears to also hold in the model of Huang and Sagae (2010).

[5] A list of NLP papers that explicitly use this analysis technique includes the following citations: Gildea (2011), Nederhof and Satta (2011), Gilroy et al. (2017), Melamed (2003), Kuhlmann (2013), Nederhof and Sánchez-Sáez (2011), Büchse et al. (2011), Lopez (2009), Eisner and Blatz (2007), Huang and Sagae (2010), and Kuhlmann, Gómez-Rodríguez, et al. (2011).

[6] In a seminal paper, Burstall and Darlington (1977) introduced the idea of synthesizing faster/-better (functional) programs by source-to-source transformation: given an initial simple, lucid, and—most importantly—*correct* program, we may repeatedly transform it by altering its recursive structure to eventually reach an improved version of the program. Later, Tamaki and Sato (1984) extended the technique to logic programming.

derive algorithms with a faster runtime. These transformations map a program to another with the same meaning (i.e., given the same inputs, it will produce the same outputs) but with a possibly different running time. An example of such a speed-up is given below.

Suppose that we only want to compute the total weight of paths from one node set `start(I)` to another node set `stop(K)`. The following program does this by first computing the total weights for all pairs and then summing over a subset:

```
3 goal += start(I) * path(I,K) * stop(K).
4 path(I,I) += 1.
5 path(I,K) += edge(I,J) * path(J,K).
```

However, this program is not as efficient as it can be. By exploiting the distributive property of addition and multiplication, we can sum over the possible `stop` nodes sooner in the recurrence:

```
6 goal += start(I) * pathstop(I).
7 pathstop(I) += stop(I).
8 pathstop(I) += edge(I,J) * pathstop(J).
```

Here `pathstop(I)` computes the total weight of all paths from node `I` to a node in `stop`. The running time of this program is $\mathcal{O}(n^2)$. In §7, we provide a collection of program-to-program transformations capable of recovering this speedup.

**Program improvement through automated transformation search.** This work shows how to search over program transformation sequences to automatically discover faster algorithms, automating the work of the NLP algorithmist. Our work poses program optimization as search over transformed versions of the initial program. We describe how to use search to rapidly rediscover many of the known optimizations listed in Example 1. The program optimizer ties together many aspects of the thesis: Our space of possible programs (§ 2), a cost function that serves as a proxy for program runtime (§ 6), and a set of directed edges that connect semantically equivalent programs (§ 7). Our search starts at the initial program and seeks a low-cost equivalent program that can be reached by traversing directed edges. In § 8, we review the search techniques that we will use for this purpose. We provide an experimental demonstration in §8.3.

**Summary.** We motivated our approach with many examples from NLP illustrating that designing and analyzing efficient dynamic programs is challenging. We will develop a framework that decomposes the problem of systematically analyzing and improving dynamic programs into the following components:

1. a domain-specific language to logically specify an algorithm (§ 2 and § 4) and general machinery to execute them (§3.1 and § 5)
2. a system for the automated analysis of the space/runtime complexity of dynamic programs as well as type analysis (§ 6)
3. a rich collection of meaning-preserving transformations to programs (§ 7)
4. a search algorithm for identifying a good sequence of transformations that reduce the running time complexity (§ 8).

We deliver a powerful system that can recover the kinds of speedups in Example 1.[7] We present an experimental demonstration of our system in §8.3. A prototype implementation is freely available at `https://github.com/timvieira/dyna-pi`.

[7]Our experiments include all of the speedup scenarios of Example 1 *except* for the linear index-grammar and lexicalized tree adjoining grammar parsers. We include many other interesting speedup scenarios in our experiments that we have not mentioned here.

# Chapter 2

# A Domain-Specific Language for Dynamic Programming

Dyna is a high-level, domain-specific language for concisely expressing dynamic programs as recurrence relations. It was first introduced by Eisner, Goldlust, et al. (2005), where it was demonstrated to be a useful notation and software library for rapidly developing efficient algorithms for a variety of popular models of linguistic structure (such as tagging, parsing, and translation) as well as many combinatorial optimization problems that can be solved by dynamic programming (such as shortest path, knapsack, and sequence alignment).

Our exposition of the Dyna language will unfold as a sequence of simple and hopefully familiar computations—introducing concepts and terminology when necessary or convenient.

**Example 3.** *The total weight of length-4 paths in a weighted directed graph:*

```
9  total += w(Y₁,Y₂) * w(Y₂,Y₃) * w(Y₃,Y₄) * w(Y₄,Y₅).
```

This program defines the **value** of a derived **item** `total` in terms of input items `w(...)`. This program assumes that we have encoded the set of edges in the given graph by declaring values for `w(...)` items. For example, to declare that there is an edge from a node labeled `"a"` to a node labeled `"b"` with weight 13, we write `w("a","b") += 13`. The value of `total` is $\sum_{Y_1} \cdots \sum_{Y_5} w(Y_1,Y_2) \cdot w(Y_2,Y_3) \cdot w(Y_3,Y_4) \cdot w(Y_4,Y_5)$.

Dyna rules may be recursive:

**Example 4.** *Minimum-cost path in a graph.*

```
10  goal min= start(I) + β(I).
11  β(J) min= stop(J).
12  β(I) min= w(I,J) + β(J).
```

This program defines `β(I)` as the cost of the minimum-cost path in a graph from the node `I` to any terminal node `J`. Here `stop(J)` must first be defined to be the cost of stopping at `J` when `J` is a terminal node (and should be left undefined otherwise). Also, `w(I,J)` must first be defined to be the cost of the directed edge from `I` to `J` (and should be left undefined if there is no such edge).

**Separation of logic and control.** Example 4 specifies the problem of finding the single-target shortest path in a weighted directed graph. However, the details of the implementation are abstracted away. Thus, many well-known single-target shortest path algorithms, such as the Bellman–Ford–Moore algorithm (Bellman, 1958; Ford, 1956; Moore, 1959) and Dijkstra's algorithm (Dijkstra, 1959) are represented as the same Dyna program (Example 4) as they are based on the same recursive factoring of paths. The differences between these algorithms are lower-level than the basic Dyna notation allows as they are essentially only differences in evaluation order (§5.1.3). One of the benefits of the Dyna notation is that it separates the mathematical specification of the algorithm from the details of the execution of that algorithm (e.g., data structures, control flow, execution order). This idea is often referred to as *the separation of logic and control* (Kowalski, 1979).

Notice that if we change the equations from $\langle$`min,+`$\rangle$ to $\langle$`max,*`$\rangle$, we get the Viterbi algorithm (Viterbi, 1967) for finding the most probable path in a graph from nodes in a set `start` to nodes in a set `stop` where the probability of a path is given by the product of edge probabilities (or weights) `w(...)`:

**Example 5.** *Finding the most probable path in a graph.*

```
13  goal max= start(J) * β(J).
14  β(J) max= stop(J).
15  β(J) max= β(I) * w(I,J).
```

What is interesting about the connection between Example 4 and Example 5 is that we kept the same recurrence, but we have changed the aggregate quantity that we seek to compute. Both programs define the same set of derivations, which we can interpret as paths in the graph. How these paths are weighted and interpreted differs. In Example 4, paths have costs; the costs are extended with addition and aggregated with min. In Example 5, paths have probabilities; the probabilities are extended with multiplication and aggregated with addition. We will return to the extension and aggregation operations momentarily.

Next, we give a fundamental algorithm to NLP.

**Example 6.** *Weighted context-free parsing with CKY (Cocke and Schwartz, 1970; Younger, 1967; Kasami, 1965), or, more precisely, the inside algorithm (Baker, 1979; Jelinek, 1985):*[8]

```
16 β(X,I,K) += γ(X,Y,Z) * β(Y,I,J) * β(Z,J,K).
17 β(X,I,K) += γ(X,Y) * β(Y,I,K).
18 β(X,I,K) += γ(X,Y) * word(Y,I,K).
19 goal += β(s,0,N) * len(N).
```

The values of the γ items should be defined to be the weights of the corresponding context-free grammar rules: for example, the item `γ(s,np,vp)` $= 0.7$ encodes the production $\mathrm{s} \xrightarrow{0.7} \mathrm{np}\,\mathrm{vp}$. Also, the item `word(X,I,K)` should be 1 if the input word `X` appears at position `K` of the input sentence and `I` $=$ `K` $-1$, and should be left undefined otherwise. Then for any nonterminal symbol `X` and any substring spanning positions (`I`,`K`] of the input sentence, the item `β(X,I,K)` represents the total weight of all grammatical derivations of that substring from `X`.

**Syntax and Terminology.** A Dyna **program** $\mathfrak{p}$ serves to define **values** for **items**. Thus, the program defines a function from items to values; we denote the value of an item $x$ as $[\![x]\!]$. We call $[\![\cdot]\!]$ the **valuation function**; we defer its formal definition to Definition 22. Each item is named by an element of the **Herbrand universe** $\mathbb{H}$—that is, a (possibly nested) labeled tuple such as `f(g(z,h(3)))`, which is a 1-tuple labeled `f` whose single element is a 2-tuple labeled `g`. The elements of $\mathbb{H}$ are known as **ground terms** because they contain no variables. The program $\mathfrak{p}$ itself is an unordered collection of **rules**[9] of the form $x \oplus\!= y_1 \circ \cdots \circ y_K$. We call $x$ the **head**, and $y_1, \ldots, y_K$ the **body** of the rule. Each $y_k$ in the body is called a

[8] If the reader is not familiar with context-free parsing, we recommend Jurafsky and Martin (2020, chapters 12–13).

[9] Note that the collection may contain duplicates (§3.1.1).

**subgoal**.

The operations $\oplus$ and $\circ$ manipulate[10] values from a set $\mathbb{V}$ and may be any pair of operations that form a **semiring** (§3.3). Let $\underline{0}$ and $\underline{1}$ denote the semiring's additive and multiplicative identity elements, respectively. If an item's value is left undefined or has no $\oplus$-summands, this is equivalent to saying it has value $\underline{0}$, which means it cannot affect the values of any other items; in effect, it does not exist. Common semirings include total weight $\langle \mathbb{R}_{\geq 0}, \texttt{+}, \texttt{*} \rangle$, boolean $\langle \{\texttt{false}, \texttt{true}\}, \vee, \wedge \rangle$, minimum cost $\langle \mathbb{R} \cup \{\infty\}, \texttt{min}, \texttt{+} \rangle$, and Viterbi $\langle \mathbb{R}_{\geq 0}, \texttt{max}, \texttt{*} \rangle$.[11] The distributive property of the semiring gives rise to an important interpretation: The value of a given item is a certain sum over all of its **derivations** (i.e., proof trees) under the program's rules.[12] The value of the derivation is the product of the values of its axioms.[13] In this paper, we assume that all rules in the program use the same semiring.[14]

Rules do not have to name specific items. The terms in a rule may include **variables**—denoted by capitalized letters, such as $\texttt{X}$—that are universally quantified over $\mathbb{H}$. This allows them to pattern-match against many items. Let $\texttt{X}_1, \ldots, \texttt{X}_N$ denote the distinct variables in a rule R. A **grounding** of R is a variable-free rule $\{X_1 \mapsto x_1, \ldots, X_N \mapsto x_N\}(\mathbf{R})$ (that is, the result of substituting $x_i$ for all copies of $X_i$ in R, for each $i$) for some $x_1, \ldots, x_N \in \mathbb{H}$. $\Gamma(\mathbf{R})$ denotes the **set of all groundings** of R; including a non-ground rule in $\mathfrak{p}$ is equivalent to including all of its groundings.

---

[10]The notation used to denote the semiring operations varies from author to author. Many authors (Lehmann, 1977; Kuich, 1997; Green, Karvounarakis, et al., 2007; Esparza et al., 2007) use $\langle +, \cdot, 0, 1 \rangle$, which leads to confusion since these operations are often used to denote their ordinary meanings in the same text. In the NLP literature, it is most common to $\langle \oplus, \otimes, ⓪, ① \rangle$ to avoid excessive overloading of $\langle +, \cdot, 0, 1 \rangle$. We have chosen to use $\langle \oplus, \circ, \underline{0}, \underline{1} \rangle$ since they are easier to distinguish visually.

[11]Semirings provide an elegant abstraction for computing many common quantities in NLP models, such as expected values, covariance, gradients, entropies, and $K$-best (Goodman, 1999; Huang, 2008; Eisner, Goldlust, et al., 2005; Huang and Chiang, 2005; Li and Eisner, 2009).

[12]The precise semantics, including the definition of the set of derivations, is given in §4.

[13]We formalize the sum-over-derivations interpretation in §4.

[14]Dyna $\underline{2}$ (Eisner and Filardo, 2011), relaxes this restriction. We discuss future work on Dyna $\underline{2}$ in §D.

The Dyna language allows logical **side conditions** on a rule, such as `goal += f(X) for X < 9.` This is syntactic sugar for `goal += f(X) * lessthan(X,9)`, where the value of the `lessthan` constraint is given by

$$[\![\texttt{lessthan(X,Y)}]\!] \stackrel{\text{def}}{=} \underline{1} \textbf{ if } \texttt{X} < \texttt{Y} \textbf{ else } \underline{0}$$

The constraint `X < 9` makes use of the **built-in relation** <; Dyna's standard library supports many other common arithmetic relations, such as

- $[\![\texttt{X is Y+Z}]\!] \stackrel{\text{def}}{=} \underline{1} \textbf{ if } \texttt{X} = \texttt{Y} + \texttt{Z} \textbf{ else } \underline{0}$
- $[\![\texttt{X is Y}^K]\!] \stackrel{\text{def}}{=} \underline{1} \textbf{ if } \texttt{X} = \texttt{Y}^{\texttt{K}} \textbf{ else } \underline{0}$

Unlike the items defined by the program's rules, built-in relations are opaque—they are not defined by rules but rather by executable (non-Dyna) code.

**Input data** is represented as program $\mathfrak{D}$ consisting of **axioms**—i.e., rules of the form $(\boldsymbol{x} \oplus\!= v)$ where $v \in \mathbb{V}$. For example, consider the shortest-path program as $\mathfrak{p}$ and the specific graph to run it on as $\mathfrak{D}$.

$\mathfrak{p} =$

```
% definition of shortest
% path from S to stop
β(S) min= w(S,S') + β(S').
β(S) min= stop(S).
```

$\mathfrak{D} =$

```
% define graph (w, stop)
w(bal,phl) min= 3.
w(phl,nyc) min= 1.
...
stop(nyc) min= 0.
```

Adding the $\mathfrak{D}$ rules to $\mathfrak{p}$ results in a new program $\mathfrak{p}'$ that can subsequently be queried, as we will show below.

**Querying a Program.** The way that a user interacts with a Dyna program is through queries. A **query** against program $\mathfrak{p}$ is specified as a (possibly nonground) term $x$, and the **result** of a query is a **specialized program** that encodes the restriction of the valuation function $[\![\cdot]\!]$ (Definition 22) to the possible groundings of $x$. Below is an example of a program, a query, and the query's results (going left to right).

```
29 % definition of shortest
30 % path from S to stop
31 β(S) min= w(S,S') + β(S').
32 β(S) min= stop(S).
33 % define graph (w, stop)
34 w(bal,phl) min= 3.
35 w(phl,nyc) min= 1.
36 ...
37 stop(nyc) min= 0.
```

query β(S) ⟶

```
38 β(bal) min= 4.
39 β(phl) min= 1.
40 β(nyc) min= 0.
41 ...
```

The query `β(S)` returns a map: each item of the form `β(S)` maps `S` to the cost of its shortest path to any of the `stop` nodes. As the example output above shows, Dyna represents the result of a query as specialized programs, i.e., programs that contain only the answers to the specific query. The specialized programs that are answers to queries are restricted to have the property that all their right-hand sides are constants.[15]

Notice that the results for a query may be an infinite-sized valuation. To compactly represent certain infinite valuations, we allow the heads of the rules in the results to contain free variables. For example,

[15]It is possible to relax this condition to allow the right-hand side of a rule to contain built-in relations. For example, `f(X) += 3 for (1 is X mod 2)` says that `f(X)` items have a contribution of `3` if `X` is an odd number.

```
42  f(X) += 7.     % free variable
43  f(3) += 11.
```

$$[\![ \mathtt{f(X)} ]\!] = \begin{cases} 18 & \textbf{if } \mathtt{X} = 3 \\ 7 & \textbf{otherwise} \end{cases}$$

For efficiency, we apply **constant folding** to simplify the contribution to each head wherever possible. For example, the following results would be replaced by #42–43.

```
44  f(X) += 6.
45  f(Y) += 1.
46  f(3) += 9.
47  f(3) += 2.
```

# Summary

This chapter has provided a high-level overview of the Dyna syntax. We introduced the main concepts and terminology Dyna users should be familiar with. We will describe the concepts pertaining to the internal representation of programs in §3, the precise meanings of a derivation and the valuation mapping in §4, and how to execute programs in §5.

# Chapter 3

# Notation, Background, and Utilities

This chapter includes background material that we will use throughout this thesis. This background material includes:

# 3.1 Internal Representation of Programs

This section describes important notation and our internal program representation. We will describe an efficient data structure for representing the results of a query as a *result stream* (§3.1.2.1). We will develop a representational abstraction that allows us to regard rules as variable-free when manipulating them (e.g., during inference (§5)).

## 3.1.1 Notation and Representation

We will now describe how programs are represented internally. When the user writes a program such as

```
48 f(X) += g(X,Y) * h(Y).
49 f(X) += 1.
```

it is represented as a bag of rules. Each rule R is of the form $x \oplus\!= y_1 \circ \cdots \circ y_K$. It is represented internally as a pair of a head and a body. The body is a list of subgoals. Let $\mathsf{head}(\mathrm{R})$ and $\mathsf{body}(\mathrm{R})$ denote the head and body of the rule R. The head $x$ and each subgoal $y_k$ are represented as nonground terms. A **nonground term**[16] is a labeled tuple that may contain named variables (denoted by capital letters, such as `X` and `Y` in the example). Let $\mathsf{vars}(\cdot)$ be a function that returns the set of free variables contained in a nonground term, e.g., $\mathsf{vars}(\texttt{f(g(X),4,X)}) \mapsto \{\texttt{X}\}$. Variables can be shared between nonground terms, e.g., `X` is shared between the head `f(X)` and the subgoal `g(X,Y)` of #48 and `Y` is shared between its two subgoals.

[16]When we write "nonground term," we mean a *possibly* nonground term, i.e., a term that is allowed to be nonground but does not necessarily contain free variables.

**Substitution** is a function $\boldsymbol{\theta}(\cdot)$ from nonground terms to nonground terms; it is parameterized by a **substitution mapping** $\boldsymbol{\theta}$ that maps variables to nonground terms, such as $\boldsymbol{\theta} = \{\texttt{Y} \mapsto \texttt{g(X)}, \texttt{Z} \mapsto \texttt{4}\}$. When we apply the substitution map, $\boldsymbol{\theta}(x)$, it will replace all instances of the variable on the left-hand side of an arrow ($\mapsto$) with whatever is on the right-hand side of the arrow.

**Unification.** The primary operation performed on nonground terms is **unification**; it is performed internally by our solvers (§ 5) to bind variables. Unification solves equations of the form $\boldsymbol{\theta}(x) = \boldsymbol{\theta}(x')$ where $x$ and $x'$ are nonground terms. Such constraints can be solved efficiently using a unification algorithm, $\boldsymbol{\theta} \leftarrow \mathsf{unify}(x, x')$ (e.g., Robinson (1965) and Martelli and Montanari (1982)).[17] Here unify returns a substitution mapping $\boldsymbol{\theta}$ or FAIL if no solution exists. For example, $\boldsymbol{\theta} \leftarrow \mathsf{unify}(\texttt{f(Y,4)}, \texttt{f(g(X),Z)})$ returns the solution $\boldsymbol{\theta} = \{\texttt{Y} \mapsto \texttt{g(X)}, \texttt{Z} \mapsto \texttt{4}\}$.[18] When we apply this substitution map to both sides, the terms become equal, $\boldsymbol{\theta}(\texttt{f(Y,4)}) = \boldsymbol{\theta}(\texttt{f(g(X),Z)})$. On the other hand, $\boldsymbol{\theta}(\texttt{f(X,g(X))}) = \boldsymbol{\theta}(\texttt{f(3,g(4))})$ is unsatisfiable; thus, $\boldsymbol{\theta} = \text{FAIL}$. We also mention equations of the form $\boldsymbol{\theta}(x) = x'$. Such constraints occur in program transformations §7, and they can be solved in a similar fashion using $\boldsymbol{\theta} \leftarrow \mathsf{cover}(x, x')$; pseudocode is given in Algorithm 35.

Since variable names are global, we often need to rename them (the X in #48 is supposed to be different from the X in #49. For this purpose, we use a function fresh, which renames all variables within a term, rule, or entire program to novel names in order to avoid unwanted name conflicts. For example,

[17] We give pseudocode and implementation details for our unification algorithm in §C.1.

[18] We allow unify to take a substitution mapping as an optional third argument, which enables algorithms to chain together the results of unification calls.

$\mathsf{fresh}(\texttt{f(g(X),X)}) = \texttt{f(g(X}_*\texttt{),X}_*\texttt{)}$ where $\texttt{X}_*$ is a novel variable name.

**Destructuring.** We will use destructuring, such as $(x \mathrel{\oplus}= y_1 \circ \cdots \circ y_K) \leftarrow \mathrm{R}$, to break up the rule R into its head and variable-length body. It is equivalent to $x \leftarrow \mathsf{head}(\mathbf{R})$, $K \leftarrow |\,\mathsf{body}(\mathbf{R})\,|$, $\langle y_1, \ldots, y_K \rangle \leftarrow \mathsf{body}(\mathbf{R})$. If $K = 0$, we regard the body as $\underline{1}$. We may write $f(x_1, \ldots, x_N) \leftarrow x$ to destructure a term into its arguments and bind $f$ to its label. For example, $f(x_1, \ldots, x_N) \leftarrow \texttt{p(q(X),r(X))}$ sets $N \leftarrow 2$, $f \leftarrow \texttt{p}$, $x_1 \leftarrow \texttt{q(X)}$, $x_2 \leftarrow \texttt{r(X)}$. In a destructuring expression, such as $(_ \mathrel{\oplus}= y_1 \circ \cdots \circ y_K) \leftarrow \mathrm{R}$, we may use underscore (_) to match a subexpression that we do not wish to name.

**Bags.** We will use bags (aka multisets) to represent a collection of rules (programs). Literal bags are denoted by the paired delimiters $\wr \cdot \int$. Let $\alpha$ and $\beta$ be bags. We write operations $\alpha \cup \beta$ for bag union and $\setminus$ for bag difference. When used as a condition, $x \in \beta$ is true if and only if $x$ appears at least once in $\beta$. However, when used in a loop, $x \in \beta$ will repeat duplicate instances of $x$. We will access elements of $\beta_i$ of $\beta$ by their "position" $i$—even though the position is arbitrary—these positions are simply a canonical linear order of the (possibly duplicated) elements.

**Programs.** The notation $\mathfrak{p}[x] \mathrel{\oplus}= y_1 \circ \cdots \circ y_K$ adds the rule $x \mathrel{\oplus}= y_1 \circ \cdots \circ y_K$ to the program $\mathfrak{p}$. We use the notation $\mathfrak{p}[x]$ to retrieve the collection containing the bodies of the (specialized) rules in $\mathfrak{p}$ whose heads have been unified with the subgoal $x$.[19] To take the product of two subgoals $x$ and $x'$, we may write $\mathfrak{p}[x \circ x']$ or $\mathfrak{p}[x] \circ \mathfrak{p}[x']$. The notation can be used to read (rule bodies) and write (new rules) across programs with expressions like $\mathfrak{p}'[x] \oplus\!= \mathfrak{p}[y_1] \circ \cdots \circ \mathfrak{p}[y_K]$. We write $\mathfrak{p} \oplus \mathfrak{p}'$ to concatenate the rules of two programs: $\mathfrak{p}$ and $\mathfrak{p}'$. To create an empty program $\mathfrak{p}$, we write $\mathfrak{p} \leftarrow \wr\!\int$. We explain all of this notation in more detail in §3.1.2.

[19]We may optionally pair the rule body with the head-unificiation's substitution mapping.

**Shortcuts for defining integer intervals.** We find the following notation to be useful in defining lists of contiguous integers, $i, j \in \mathbb{Z}$:

$$
\begin{aligned}
[i{:}j) &\stackrel{\text{def}}{=} [\, i, i{+}1, \ldots, j{-}1 \quad ] \\
[i{:}j] &\stackrel{\text{def}}{=} [\, i, i{+}1, \ldots, j{-}1, j \,] \\
(i{:}j) &\stackrel{\text{def}}{=} [\quad i{+}1, \ldots, j{-}1 \quad ] \\
(i{:}j] &\stackrel{\text{def}}{=} [\quad i{+}1, \ldots, j{-}1, j \,]
\end{aligned}
$$

When an endpoint is not specified, we take it to be an extreme value, e.g., $[-3{:}) = [-3, -2, \ldots]$ and $[{:}10) = [\ldots, 9, 10]$.[20] We use this notation primarily to index rule bodies.

**Rule bodies.** We make use of the following notational shortcuts for manipulating rule bodies. Let $\boldsymbol{\mu} = y_1 \circ \cdots \circ y_K$ be a list of subgoals. Let $\boldsymbol{\beta}$ be a list of indices, then we define $\boldsymbol{\mu}_{\boldsymbol{\beta}} \stackrel{\text{def}}{=} [y_i \mid i \in \boldsymbol{\beta} \cap [1{:}K]]$. We frequently use the integer-range syntax for $\boldsymbol{\beta}$, e.g., $\boldsymbol{\mu}_{[i:j)}$ denotes the contiguous range of subgoals in $[i{:}j)$. We may also write $y_{[i:j]} \circ y_{(j:k]} = y_{[i:k]}$. Similarly, $\boldsymbol{\mu}_{[i:j)} \circ \boldsymbol{\mu}_j \circ \boldsymbol{\mu}_{(j:k]} = \boldsymbol{\mu}_{[i:k]}$, and $\boldsymbol{\mu}_{[:j)} \circ \boldsymbol{\mu}_j \circ \boldsymbol{\mu}_{(j:]} = \boldsymbol{\mu}$.

[20] In these extreme cases, the difference between an open ("(" or ")") and closed ("[" or "]") endpoint is irrelevant.

### 3.1.2 Lookup and Update Represention

In this section, we will develop a notation that substantially reduces the clutter that results from explicitly passing around substitution mappings and from iteration over the results of queries. On the left is the notation that we will develop, and on the right is the code that it represents.

$$\mathfrak{p}[\texttt{a(I,L)}] \oplus\!= \mathfrak{m}[\texttt{b(I,J)}] \circ \mathfrak{m}[\texttt{c(J,K)}] \circ \mathfrak{m}[\texttt{d(K,L)}] \quad \overset{\text{translation}}{\Longrightarrow} \quad \begin{array}{l} \textbf{for } \langle v_1, \boldsymbol{\theta}_1 \rangle \in \mathfrak{m}[\texttt{b(I,J)}]: \\ \quad \textbf{for } \langle v_2, \boldsymbol{\theta}_2 \rangle \in \mathfrak{m}[\boldsymbol{\theta}_1(\texttt{c(J,K)})]: \\ \quad\quad v' \leftarrow v_1 \circ v_2 \\ \quad\quad \textbf{for } \langle v_3, \boldsymbol{\theta}_3 \rangle \in \mathfrak{m}[\boldsymbol{\theta}_2(\texttt{d(K,L)})]: \\ \quad\quad\quad \mathfrak{p} \cup\!= \boldsymbol{\theta}_3((\texttt{a(I,L)} \oplus\!= v' \circ v_3)) \end{array}$$

In the example, there are two programs: $\mathfrak{p}$ and $\mathfrak{m}$. We are looking up values in $\mathfrak{m}$, and we are updating $\mathfrak{p}$. The bracket notation $\mathfrak{m}[x]$ returns a stream of pairs of values and bindings to the variables in $x$.[21] The notation $\mathfrak{p}[x] \oplus\!= \beta$ will add rules to $\mathfrak{p}$ that result from expanding the queries in $\beta$.[22] We will formalize the lookup and update notation shortly.

An **explicit result stream** for that query $x$ represents the results of a query against a given program as a stream: $[\boldsymbol{\theta}^{(t)}((x \oplus\!= v^{(t)}))]_{t=1\ldots}$ where each $\boldsymbol{\theta}^{(t)}$ is a substitution map (over $x$'s variables) and $v^{(t)} \in \mathbb{V}$ is a contribution to the value of the item $\boldsymbol{\theta}^{(t)}(x)$.[23] We will employ one more trick: we will *implicitly* update the bindings of variables in $x$ as our index $t$ in the stream progresses. We do so by representing variables as pointers and the substitution mapping by what each variable (pointer) points to. We call this an **implicit result stream**. This has the benefit that if a variable in $x$ occurs outside of $x$, its bindings are seamlessly and efficiently propagated.[24] For example, if we create a result stream for `f(X)`, then the bindings to `X` would be propagated to `g(X)` because they share the variable `X`. This implementation strategy has the benefit that $\boldsymbol{\theta}^{(t)}$—and substitutions by it—can be abstracted away in most algorithms—allowing for concise pseudocode that does not need to directly manipulate or manage $\boldsymbol{\theta}^{(t)}$.

[21] The notation $\mathfrak{m}[x]$ should not be confused with $\mathfrak{m}[\![x]\!]$, which computes the valuation mapping for items matching $x$.

[22] Note that the result of expanding these queries could be an arbitrary rule body.

[23] The result streams have the added benefit that laziness supports infinitely long results. Such flexibility can be useful if the stream's consumer will not consume it entirely (i.e., the infinite stream is *short-circuited*).

#### 3.1.2.1 Implicit Result Stream Data Structure

Our implicit result system implements substitution by directly binding the variables in $x$ through in-place unification (Algorithm 1). Each variable is represented as a pointer (or reference) in this system.Whenever distinct variables `X` and `Y` are unified, we can simply bind the `X` pointer to `Y`. Later, if we want to test the equality of two variables, we follow each of their pointers to their respective roots and compare literal equality.[25] We extend $\mathsf{vars}(\cdot)$ to return the set of variables under the control of a given result stream.

Since our iterators are in-place, in order to proceed to iteration $t+1$, the iterator needs to *undo* the unifications from iteration $t$ before it can safely apply any new unifications. The in-place implementation avoids unnecessary copying operations (e.g., calls to $\mathsf{fresh}$, copying $\boldsymbol{\theta}$). To facilitate undoing a unification, we have implemented unify_implicitly as a Python-style iterator, which has localized *do* and *undo* code. If unification fails, this iterator loops zero times. If unification succeeds, the iterator modifies the variable bindings as a side effect and yields a value, then suspends while this result is processed. Once this iterator resumes, it will undo its bindings—restoring its arguments to their prior state from before unify_implicitly was called.

[24] A similar data structure underlies Warren's Abstract Machine, which is the basis of most Prolog implementations (Warren, 1983; Aït-Kaci, 1991).

[25] Our data structure is inspired by the well-known union–find data structure (Galler and Fischer, 1964) for representing equivalence relations. The use of union–find for unification was explored by Mannila and Ukkonen (1986).

Pseudocode for implicit unification is provided in Algorithm 1; it uses syntax for creating iterators based on Python's generators.[26] The logic in the procedure is the same as unify (Algorithm 34). The pseudocode makes use of the following helper functions: isbound$(\cdot)$, which returns true if and only if its argument is not a null pointer (i.e., not a free variable), and deref$(\cdot)$ follows the pointer.

Lastly, we write $[x = y]$ as a shorthand for unify_implicitly$(x, y)$,[27] and $v[x = y]$ as shorthand for $v \circ [x = y]$ where $v \in \mathbb{V}$ is constant value.

#### 3.1.2.2 Lookup ($\mathfrak{p}[x]$)

In this section, we show how to create (implicit) result streams by intersecting a query with the rules of a program and joining result streams so that they correspond to product queries. Pseudocode for the intersection of a query $x$ against a program $\mathfrak{p}$ is given in Algorithm 2. In this procedure, the query $x$ is

[26]Python's generator syntax (`https://peps.python.org/pep-0255/`) includes the **yield** keyword, similar to **return** in regular functions. However, the key difference is that **return** provides a *single* value, whereas **yield** allows for the production of a *sequence* of values. Moreover, generators can be paused and resumed: a **yield** statement outputs a value and temporarily suspends execution. When the generator is next called, it resumes from where it left off, continuing until it encounters another **yield** or it reaches the end of the function. Additionally, the expression "**yield from** xs" is a shorthand for "**for** x ∈ xs: **yield** x".

[27]This notation was inspired by Iverson bracket notation.

**Algorithm 1** unify_implicitly: Implicit unification algorithm.

```
1.  def unify_implicitly(x, y):
2.    if x = y: yield 1                                              ▷ Success: suspend
3.    else if isvar(x):
4.      if isbound(x): yield from unify_implicitly(deref(x), y)
5.      else if isvar(y) and isbound(y): yield from unify_implicitly(x, deref(y))
6.      else if x does not occur in y:
7.        try:
8.          bind x to y                                     ▷ x is free; bind it to y here
9.          yield 1                                                  ▷ Success: suspend
10.       finally:
11.         unbind x                                                 ▷ Undo binding
12.   else if isvar(y): yield from unify_implicitly(y, x)        ▷ Handled by symmetry
13.   else if isterm(x) and isterm(y) and their labels are equal:
14.     yield from unify_implicitly(x.args, y.args)
15.   else if isseq(x) and isseq(y) and |x| = |y|:             ▷ Unify argument sequences
16.     if |x| = 0: yield 1                                          ▷ Success: suspend
17.     else:
18.       for _ ∈ unify_implicitly(x_1, y_1):
19.         yield from unify_implicitly(x_(1:], y_(1:])
```

nondeterministically matched against different cases, including values, builtins, and program rules. In the case of program rules, if the query $x$ unifies with the head $x'$ of a rule $x' \oplus= y_1 \circ \cdots \circ y_K$, the dispatch procedure *suspends*—allowing the caller of the method to peer at the state of the variables in $x$ and the value $y_1 \circ \cdots \circ y_K$ emitted by **yield**. Once the caller is ready to move to the next match, the process continues to the following rules in $\mathfrak{p}$. Notice that this procedure does *not* compute $\mathfrak{p}[\![x]\!]$; it only expands the query $x$ by a single step.

**Built-in relations.** On line 2.6, we indicate that built-in relations should insert their logic into the lookup procedure. To give a sense of what that logic

[28]In practice, we speed-up the loop line 2.8 by indexing the rules of $\mathfrak{p}$ by a "hash code" that ensures that for any $q$, $\text{unify}(\text{head}(s), q) \neq \text{FAIL} \Rightarrow \text{hash}(\text{head}(s)) = \text{hash}(q)$. Additionally, the cost of fresh can be amortized somewhat by keeping a pool of fresh copies for each rule and reusing a copy from the pool if there are no active references to it.

**Algorithm 2** $\mathfrak{p}[x]$: Intersect a query $x$ with the rules of a program $\mathfrak{p}$. Returns an implicit result stream (i.e., an iterator that mutates $x$'s variable bindings and emits rule bodies).

```
1.  def p[x]:
2.    ▷ Implicitly unify x with the heads of the rules in p; return a stream of their bodies
        subject to implicit unification. If x is a value, return it. If x is a built-in relation,
        dispatch to the built-in result stream method.
3.    if x ∈ 𝕍: yield x
4.    else if x is a builtin:
5.      ▷ Dispatch to builtin-specific code; example code is given in Algorithm 3.
6.      (...)
7.    else
8.      for R ∈ p:                                          ▷ Use rules in the program[28]
9.        ▷ Below, we freshen to avoid variable-name collisions in recursion
10.       (x′ ⊕= y_1 ∘ ⋯ ∘ y_K) ← fresh(R)     ▷ Often, the body of a rule is just a constant.
11.       for _ ∈ [x′ = x]:
12.         yield y_1 ∘ ⋯ ∘ y_K
```

might look like, consider the case where the query $x$ matches the built-constraint `A is B + C`. The dispatching procedure looks like the pseudocode in Algorithm 3.

Notice that the pseudocode has several special cases that depend on whether or not `A`, `B`, and `C` are bound to actual numbers. In the final "else" case, we see an issue regarding built-in relations: they must be *sufficiently instantiated* to be evaluated. For example, the addition builtin `A is B+1` is not sufficiently instantiated: it refuses to enumerate the infinitely many pairs ⟨`A`, `B`⟩ that satisfy the relation. Determining when a relation is sufficiently instantiated is left to the built-in library designer's discretion. Thus, by modifying Algorithm 3, the library *could* (lazily) enumerate all infinitely many pairs of floating-point numbers as a proxy for the uncountable set of real solutions. Generally, built-in relations will throw instantiation fault exceptions when the number of solutions is infinite because it is more likely that the user made a mistake than they wanted an

**Algorithm 3** Built-in relation dispatching.

```
1.  match x:
2.      ⋮
3.    case A is B + C:
4.      if isnum(B) and B = 0:                      ▷ Special cases for identity element (i.e., 0)
5.        yield from [A=C]
6.      else if isnum(C) and C = 0:
7.        yield from [A=B]
8.      else if isnum(B) and isnum(C):                                   ▷ Solve for A given B and C
9.        a ← B + C                           ▷ The + operation here is part of the host language.
10.       yield from [A=a]
11.     else if isnum(A) and isnum(C):                                   ▷ Solve for B given A and C
12.       b ← A − C
13.       yield from [B=b]
14.     else if isnum(A) and isnum(B):                                   ▷ Solve for C given A and B
15.       c ← A − B
16.       yield from [C=c]
17.     else
18.       if optional:                                                   ▷ if delaying is allowed
19.         yield x                                               ▷ returns a “delayed constraint”
20.       else
21.         raise InstantiationFault()
22.     ⋮
23.   case A is B * C:
24.       ⋮
25.   case A is B / C:
26.       ⋮
```

infinite stream of solutions. An alternative strategy (labeled as “optional” in the code) is to *delay* the processing of the constraint in hopes that it becomes more instantiated later. We note that the implementation of certain builtins may throw runtime exceptions, and other implementations may encode errors as special values (akin to NaN in floating-point numbers). This choice is up to the designer of the specific built-in relation.

**Join.** Next, we describe how to *lift* the $\circ$ operator to join streams of values and their associated variable bindings. Suppose $\beta_1 \leftarrow \mathfrak{p}[y_1]$ and $\beta_2 \leftarrow \mathfrak{p}[y_2]$ for queries $y_1$ and $y_2$. As suggested by the choice of notation $\beta_1 \circ \beta_2$, the meaning of joining result streams is the same as the meaning of the right-hand side of a rule with $y_1 \circ y_2$. An algorithm for creating a result stream for $\beta_1 \circ \beta_2$ is given in Algorithm 4. The algorithm is shockingly simple: it is just a pair of nested loops. It is so simple due to the implicit unification strategy: variables are seamlessly and efficiently propagated between subgoals.[29] In particular, there is no need to pass around or return explicit substitution mappings ($\boldsymbol{\theta}$). These nested loops correspond to a join operation because the bindings from the outer loop over $\beta_1$ are passed to the inner loop over $\beta_2$. The inner loop iterator is re-initialized on each iteration of the outer loop: $\beta_2$ starts from the beginning, given the initial bindings, which are (generally) different from the previous iteration of the outer loop. In other words, to ensure the correct join behavior, it is important that the iterators for $\mathfrak{p}[y_1]$ and $\mathfrak{p}[y_2]$ are only *initialized* when they are put in an iteration context (e.g., such as line 4.3 or line 4.4). This ensures that the inner loop $\mathfrak{p}[y_2]$ corresponds to a new iterator for each round of the outer loop.

**Algorithm 4** $\beta_1 \circ \beta_2$: Algorithm for joining result streams $\beta_1$ and $\beta_2$.

1. **def** $\beta_1 \circ \beta_2$:
2.     ▷ *Each iteration of the outer iterator $\beta_1$ binds variables through implicit unification; those bindings are propagated to the inner iterator $\beta_2$.*
3.     **for** $v_1 \in \beta_1$ :
4.         **for** $v_2 \in \beta_2$ :
5.             **yield** $v_1 \circ v_2$

The result of the join operation is another implicit result stream. We write $\mathfrak{p}[y_1 \circ \cdots \circ y_K]$ as a shorthand for $\mathfrak{p}[y_1] \circ \cdots \circ \mathfrak{p}[y_K]$. This shorthand can be formalized as adding a case to Algorithm 2 that will dispatch to Algorithm 4.[30] Note that the for-loops in the algorithm can be performed in any order; however, the choice of order may affect the running time.[31] Since the join operation is associative, a multi-way join need not be parenthesized; it is well-defined as the $\circ$-product of multiple result streams. In §6.3, we consider an automated method for determining which ordering of a multi-way product is most efficient to use.

[29] Such propagation is often called a *sideways information passing strategy* in the logic programming literature.

#### 3.1.2.3 Update ($\mathfrak{p}[x] \oplus\!= \ldots$)

This section discusses methods for aggregating result streams as updates to a program. Let be $\mathfrak{p}$ a program. Since $\mathfrak{p}$ is just a specification of a valuation mapping, it is a glorified map data structure (like a Python dictionary or Java hash map). Thus, it seems reasonable that a user or algorithm should be able to update a given item's value by a specific value. Therefore, we allow $\mathfrak{p}$ to be updated using the same notation for updating maps:[32]

$$\mathfrak{p}[\texttt{f(1)}] \oplus\!= 3 \quad \overset{\text{translates to}}{\Longrightarrow} \quad \mathfrak{p} \leftarrow \mathfrak{p} \cup \wr \texttt{f(1)} \oplus\!= \texttt{3} \wr$$

Similarly,

$$\mathfrak{p}[\texttt{f(X)}] \oplus\!= 3 \quad \overset{\text{translates to}}{\Longrightarrow} \quad \mathfrak{p} \leftarrow \mathfrak{p} \cup \wr \texttt{f(X)} \oplus\!= \texttt{3} \wr$$

We can also add a new rule with subgoals

$$\mathfrak{p}[\texttt{f(X)}] \oplus\!= \texttt{g(X,Y)} \circ \texttt{h(Y)} \quad \overset{\text{translates to}}{\Longrightarrow} \quad \mathfrak{p} \leftarrow \mathfrak{p} \cup \wr \texttt{f(X)} \oplus\!= \texttt{g(X,Y)} \circ \texttt{h(Y)} \wr$$

[30] Another choice is to make $\circ$ as builtin.

[31] Some care is needed to ensure that the multiplicands of noncommutative $\circ$ operations appear in the correct order.

[32] Note that the updates here are additive (as suggested by $\oplus\!=$); they are not replacements or overrides.

Now, what happens with new rules containing subgoals that are result streams (possibly from other programs)?

$$\mathfrak{p}[\texttt{f(X)}] \oplus\!= \texttt{g(X,Y)} \circ \mathfrak{m}[\texttt{h(Y)}] \overset{\text{translates to}}{\Longrightarrow} \quad \textbf{for } v \in \mathfrak{m}[\texttt{h(X)}] : \quad \mathfrak{p}[\texttt{f(X)}] \oplus\!= \texttt{g(X,Y)} \circ v$$

Notice that the elements of the stream are not required to be values in $\mathbb{V}$; they may contain delayed subgoals, for example, if $\mathfrak{m}$ has the rule `h(X)` $\oplus$= `p(X,Z)` in the example above. Using this notation requires some care to ensure that the meaning of the `p(X,Z)` items is consistent across $\mathfrak{m}$ and $\mathfrak{p}$. A safe setting is one where the delayed subgoals are guaranteed to have the same meaning across programs; builtins are a good example of relations that have the same meaning across programs. This flexibility allows us to include `h(X) += 3 for (1 is X mod 2)` as a rule in $\mathfrak{m}$—broadening the set of valuations that we can compactly represent.

Lastly, we mention an important special case we implicitly handle in this method. Replacing a subgoal with a set of values (or rule bodies) makes it possible to "lose" a variable. Consider the following,

**Example 7.** *Suppose we have an empty program* $\mathfrak{p}$ *and*

$\mathfrak{m} = \Big\{$ `f(X) += 4.`

*We wish to apply the rule* `goal += f(X)` *to the values in* $\mathfrak{m}$ *using the following:*

$\mathfrak{p}[\texttt{goal}] \oplus\!= \mathfrak{m}[\texttt{f(X)}]$

*If we are* not *careful, the rule added to* $\mathfrak{p}$ *will be* `goal += 4`*, but that would be incorrect, as the value* `4` *needs to be summed infinitely many times. We fix this edge case by including a* `4` with an infinite multiplicity *factor into the update: the* *case of the real semiring* $\infty \cdot 4 = \infty$. *Thus, the correct update is* `goal += ∞`.[33]

The dropped variable check is encapsulated in the process of adding a rule. In our implementation, we do the following. Let $\mathfrak{p}[\boldsymbol{x}] \oplus\!= \beta$ be an update operation where $\beta$ is a stream. Let $v$ be the current value in $\beta$'s stream. We use the following **multiplicity correction**:

$$m = \begin{cases} 1 & \text{if } \mathsf{vars}(\boldsymbol{x} \oplus\!= \beta) \subseteq \mathsf{vars}(\boldsymbol{x} \oplus\!= v) \\ \infty & \text{otherwise} \end{cases} \tag{3.1}$$

where $\mathsf{vars}(\cdot)$ returns the set of free variables in its argument. Here is what the condition says: if replacing the rule $\boldsymbol{x} \oplus\!= \beta$ by $\boldsymbol{x} \oplus\!= v$ loses a variable, we need to sum infinite copies of $v$. For each iterate of the stream $\beta$, we use $v \cdot m$ where $m$ is the result of (3.1) applied to the iterate (in its current state). This correction is applied whenever we add rules using the update notation ($\mathfrak{p}[\boldsymbol{x}] \oplus\!= \beta$).

## 3.2 General Mathematics

For completeness, this section includes precise definitions of the key mathematical concepts used throughout this work.

**Definition 1.** *A* ***binary relation*** $\mathcal{R}$ *on a set* $\mathcal{A}$ *is a subset of the Cartesian product:* $\mathcal{R} \subseteq \mathcal{A} \times \mathcal{A}$*. We typically denote inclusion tests using infix operators, e.g.,* $a\,\mathcal{R}\,b \iff \langle a, b\rangle \in \mathcal{R}$.

Adjectives used to define binary relations include:

[33] In the case of an idempotent semiring (e.g., boolean, min-plus), the multiplicity $\geq 1$ are all the same. This is why the added complexity of checking update multiplicity does not appear in the ordinary logic programming setting: the boolean semiring is idempotent.

- **Reflexive**: $\forall\, a \in \mathcal{A}:\; a\,\mathcal{R}\,a$
- **Symmetric**: $\forall\, a, b \in \mathcal{A}:\; a\,\mathcal{R}\,b \Longrightarrow b\,\mathcal{R}\,a$.
- **Asymmetric**: $\forall\, a, b \in \mathcal{A}:\; a\,\mathcal{R}\,b \Longrightarrow \neg(b\,\mathcal{R}\,a)$
- **Antisymmetric**: $\forall\, a, b \in \mathcal{A}:\; a\,\mathcal{R}\,b,\, b\,\mathcal{R}\,a \Longrightarrow a = b$
- **Transitive**: $\forall\, a, b, c \in \mathcal{A} : a\,\mathcal{R}\,b,\, b\,\mathcal{R}\,c \Longrightarrow a\,\mathcal{R}\,c$

**Definition 2.** *A* ***partition*** $\sigma$ *of a set* $\mathcal{A}$ *is a set of (non-empty) subsets* $\{\sigma_1, \ldots, \sigma_K\}$ *of* $\mathcal{A}$ *such that each element* $a \in \mathcal{A}$ *is a member of precisely one of these subsets. Put differently,* $\sigma$ *must be*

*(i) mutually exclusive:* $\forall\, i \neq j : \sigma_i \cap \sigma_j = \emptyset$

*(ii) exhaustive:* $\mathcal{A} = \bigcup_{k=1}^{K} \sigma_k$

### 3.2.1 Ordering Relations

**Definition 3.** *A binary relation* $\equiv$ *on a set* $\mathcal{A}$ *defines an* ***equivalence relation*** *on the elements of* $\mathcal{A}$ *if and only if the following conditions are satisfied for all* $a, b, c \in \mathcal{A}$*:*

*(i) Reflexive:* $(a \equiv a)$.

*(ii) Symmetric:* $(a \equiv b) \Longrightarrow (b \equiv a)$.

*(iii) Transitive:* $(a \equiv b), (b \equiv c) \Longrightarrow (a \equiv c)$.

*For a given element* $a \in \mathcal{A}$ *its* ***equivalence class*** $[a]^{(\equiv)} \stackrel{\text{def}}{=} \{a' \mid a \equiv a', a' \in \mathcal{A}\}$*. The set of all equivalence classes* $\left\{[a]^{(\equiv)} \,\middle|\, a \in \mathcal{A}\right\}$ *forms a partition of* $\mathcal{A}$*.*

**Definition 4.** *A binary relation* $\preccurlyeq$ *on a set* $\mathcal{A}$ *defines a* ***partial order*** *on the elements of* $\mathcal{A}$ *if and only if the following conditions are satisfied for all* $a, b, c \in \mathcal{A}$*:*

(i) *Reflexive:* $a \preccurlyeq a$

(ii) *Antisymmetric:* $a \preccurlyeq b$, $b \preccurlyeq a \implies a = b$

(iii) *Transitive:* $a \preccurlyeq b$, $b \preccurlyeq c \implies a \preccurlyeq c$

**Definition 5.** *A binary relation* $\prec$ *on a set* $\mathcal{A}$ *defines a* ***strict partial order*** *on the elements of* $\mathcal{A}$ *if and only if the following conditions are satisfied for all* $a, b, c \in \mathcal{A}$*:*

(i) *Asymmetric:* $a \prec b \implies \neg(b \prec a)$

(ii) *Transitive:* $a \prec b$, $b \prec c \implies a \preccurlyeq c$

**Definition 6.** *A binary relation* $\preccurlyeq$ *on a set* $\mathcal{A}$ *defines a* ***total order*** *if and only if it is both a partial order, and it is total:* $a \preccurlyeq b$ or $b \preccurlyeq a$ *for all* $a, b, c \in \mathcal{A}$.

**Definition 7.** *A binary relation* $\lesssim$ *on a set* $\mathcal{A}$ *defines a* ***preorder*** *on the elements of* $\mathcal{A}$ *if and only if the following conditions are satisfied for all* $a, b, c \in \mathcal{A}$*:*

(i) *Reflexive:* $a \lesssim a$

(ii) *Transitive:* $a \lesssim b$, $b \lesssim c \implies a \lesssim c$

**Definition 8.** *A binary relation* $\prec$ *over a set* $\mathcal{A}$ *is a* ***well-founded ordering*** *if and only if it does not contain any* ***infinitely descending chains****: there is no infinite sequence* $a_0, a_1, a_2, \ldots \in \mathcal{A}$ *such that* $a_{n+1} \prec a_n$ *for all* $n \in \mathbb{N}$.

| name | reflexive | transitive | antisymmetric | symmetric | asymmetric | total |
|---|---|---|---|---|---|---|
| preorder | ✔ | ✔ | | | | |
| partial order | ✔ | ✔ | ✔ | | | |
| total order | ✔ | ✔ | ✔ | | | ✔ |
| strict partial order | | ✔ | | | ✔ | |
| equivalence relation | ✔ | ✔ | | ✔ | | |

**Figure 3.1:** Taxonomy of order relations.

### 3.2.2 Functions

**Definition 9.** *A **function** $\phi$ from a set $\mathcal{A}$ to a set $\mathcal{B}$ is a binary relation that assigns exactly one element of $\mathcal{B}$ to each element of $\mathcal{A}$. We write $\phi : \mathcal{A} \to \mathcal{B}$ for conciseness. We often refer to $\mathcal{A} \to \mathcal{B}$ as $\phi$'s type.*

**Definition 10.** *The mapping $\phi : \mathcal{A} \to \mathcal{B}$ is a **bijection** if for all $a \in \mathcal{A}$:*

- ***surjective**: $\forall\, b \in \mathcal{B} : \exists\, a \in \mathcal{A} : \phi(a) = b$.*
- ***injective**: $\forall\, a, a' \in \mathcal{A} : (\phi(a) = \phi(a')) \implies (a = a')$.*

An equivalent definition is that $\phi$ is a bijection from $\mathcal{A} \to \mathcal{B}$ if and only if both of the following conditions hold:

- $\forall\, a \in \mathcal{A} : \exists!\, b : \phi(a) = b$
- $\forall\, b \in \mathcal{B} : \exists!\, a : \phi(a) = b$

Furthermore, we say that the function $\phi^{-1}$ is the **inverse** of $\phi$ if and only if $\forall\, b : \phi^{-1}(b) = a$ where $a$ is the unique solution to $\phi(a) = b$.

### 3.2.3 Graphs

**Definition 11.** *A directed subgraph is **strongly connected** if and only if for each pair of nodes $i$ and $j$ in the subgraph, there exist paths within the subgraph from $i \rightsquigarrow j$ and $j \rightsquigarrow i$.*

**Definition 12.** *The **strongly connected components** (SCCs) of a graph $G$ partition it into strongly connected subgraphs. The SCCs are the equivalence classes given by co-accessibility equivalence relation $(i \leftrightsquigarrow j) \iff (i \rightsquigarrow j)$ and $(j \rightsquigarrow i)$. The graph relating SCCs is called the **condensation** of the graph $G$.*

**Example 8.** *Below is a diagram[34] of a directed graph where its strongly connected components have been annotated with dashed ellipses.*

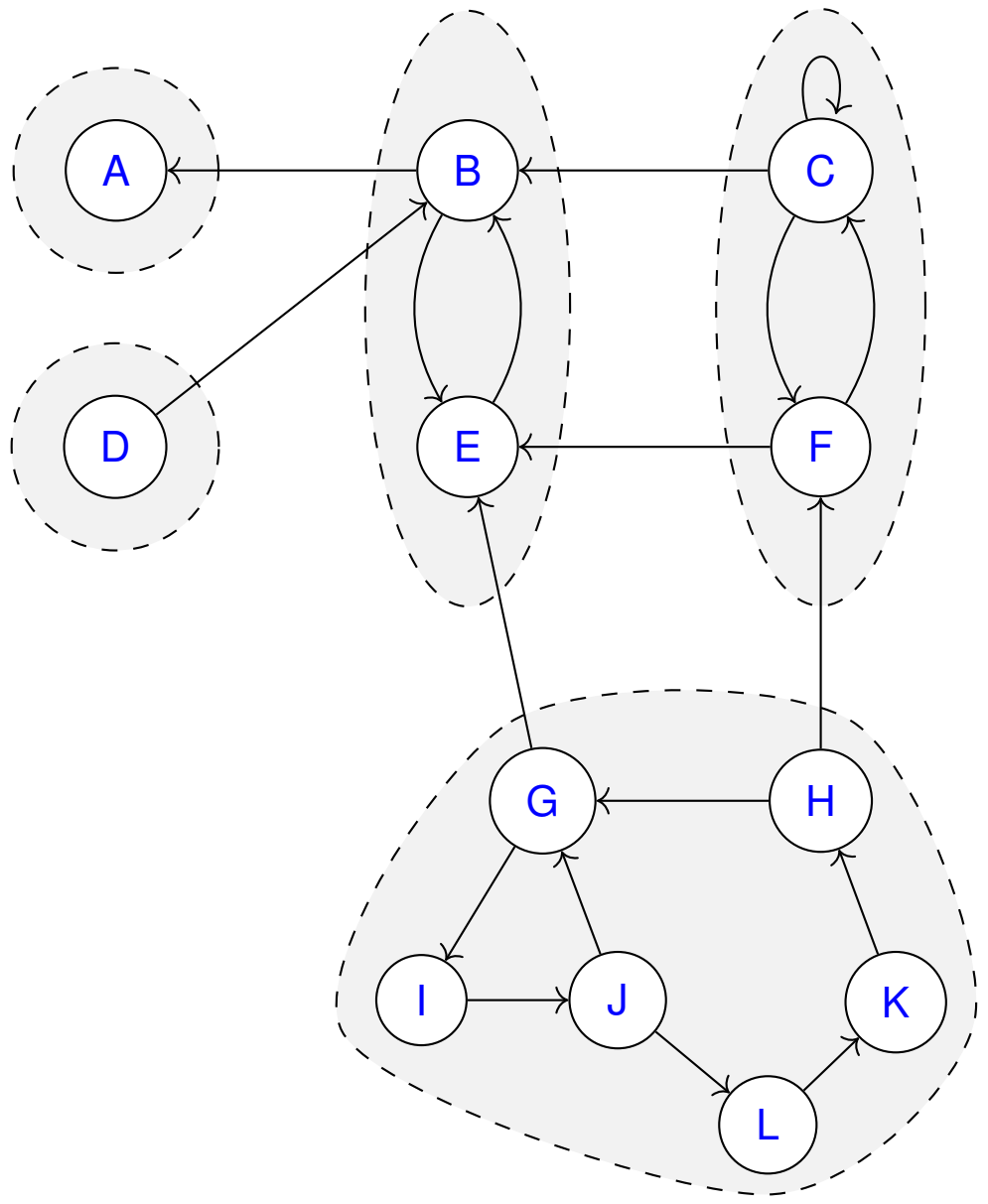

**Definition 13.** *A **topological sort** (**toposort**, for short) of a directed graph is a total ordering (Definition 6) of its nodes such that for every edge $i \to j$, $i$ comes before $j$ in the ordering. A toposort exists if and only if the graph is acyclic.*

[34] This diagram is based on `https://tex.stackexchange.com/questions/592539/how-to-create-figures-with-strongly-connected-components`.

We often talk about the toposort of the condensation of a cyclic graph; the condensed graph is always acyclic. For completeness, we provide pseudocode for Tarjan (1972)'s algorithm for finding a topologically sorted list of the strongly connected components of a graph in Algorithm 5.

**Algorithm 5** sccs(incoming, targets): Enumerate the strongly connected components (SCC; Definition 12) of a directed graph in topological order.

▷ *Here the graph $\langle V, E\rangle$ is represented implicitly using a function* incoming$(v)$ *that enumerates the nodes $u$ such that $(u \to v) \in E$, and a subset of target nodes* targets $\subseteq V$ *to begin traversal from; if none of the nodes in* targets *depend (transitively) on a given SCCs, it will be omitted from the list of SCCs returned. The algorithm is due to Tarjan (1972). It runs in $\mathcal{O}(|V| + |E|)$ time and uses $\mathcal{O}(|V|)$ additional space. Note that* sccs(outgoing, $V$)*, where* outgoing$(u)$ *enumerates the nodes $v$ such that $(u \to v) \in E$, will return the same set of SCCs* sccs(incoming, $V$)*, but in a* reverse *topological order.*

```
def sccs(incoming, targets):
  stack ← []                         ▷ We maintain a stack of distinct items
  pos ← map()                        ▷ We record the position of each node on the stack
  visited ← {}                       ▷ We maintain a set of nodes that have been visited
  sccs_list ← []                     ▷ Will be the return value
  def traverse(v):
    low ← |stack|                    ▷ Will be the return value
    if v ∉ visited:                  ▷ first visit
      visited.add(v)
      pos[v] ← |stack|
      stack.push(v)                  ▷ we may undo this
      for u ∈ incoming(v):
        if u ∈ pos:                  ▷ cycle detected
          low min= pos[u]
        else
          low min= traverse(u)
      if low = pos[v]:               ▷ If low is unchanged; nothing beneath v is in its SCC
        σ ← {}
        while |stack| > low:
          u ← stack.pop()
          σ.add(u)
          pos.remove(u)
        sccs_list.append(σ)          ▷ v's SCC is ready
    return low
  for v ∈ targets: traverse(v)
  return sccs_list
```

### 3.2.4 Abstract Algebra

**Definition 14.** *A* ***binary operation*** *on a set* $\mathbb{V}$ *is any function* $\odot : \mathbb{V} \times \mathbb{V} \to \mathbb{V}$. *Binary operators are often written using infix notation* $a \odot b$ *rather than* $\odot(a, b)$, *which is typical for function applications.*

**Definition 15.** *A* ***monoid*** $\langle \mathbb{V}, \odot, \underline{i} \rangle$ *is a set* $\mathbb{V}$ *equipped with a binary operation* $\odot$ *and an identity element* $\underline{i}$ *such that the following properties hold:*

*(i)* $\odot$ *is associative:* $\forall\, a, b, c \in \mathbb{V} : (a \odot b) \odot c = a \odot (b \odot c)$

*(ii)* $\underline{i}$ *is the left and right identity element:* $\forall\, a \in \mathbb{V} : \underline{i} \odot a = a \odot \underline{i} = a$

Some adjectives used to describe monoids include:

- **commutative** $\iff \forall\, a, b \in \mathbb{V} : \; a \odot b = b \odot a$
- **idempotent** $\iff \forall\, a \in \mathbb{V} : a \odot a = a$

**Definition 16.** *A* ***group*** *is a monoid* $\langle \mathbb{V}, \odot, \underline{i} \rangle$ *where additionally, every element has an inverse element:* $\forall\, a \in \mathbb{V}, \exists\, b \in \mathbb{V} : b \odot a = a \odot b = \underline{i}$. *The element* $b$ *may be denoted as* $a^{-1}$ *if the group is considered product-like or* $-a$ *if the group is considered sum-like.*

**Definition 17.** *A* ***semiring*** $\langle \mathbb{V}, \oplus, \circ, \underline{0}, \underline{1} \rangle$ *is a set* $\mathbb{V}$ *equipped with two monoid operations* $\langle \mathbb{V}, \oplus, \underline{0} \rangle$ *and* $\langle \mathbb{V}, \circ, \underline{1} \rangle$ *such that the following properties hold:*

*(i)* $\langle \mathbb{V}, \oplus, \underline{0} \rangle$ *is a commutative monoid*

*(ii)* $\langle \mathbb{V}, \circ, \underline{1} \rangle$ *is a monoid*

*(iii)* Distributivity*:* $\forall\, a, b, c \in \mathbb{V}$*:*

$$a \circ (b \oplus c) = (a \circ b) \oplus (a \circ c) \quad \textit{and} \quad (b \circ a) \oplus (c \circ a) = (b \oplus c) \circ a$$

*(iv)* Multiplicative annihilation *by* $\underline{0}$*:* $\forall\, a \in \mathbb{V}$*:*

$$\underline{0} \circ a = a \circ \underline{0} = \underline{0}$$

Additionally, we adopt the following notation for the $n$**-fold sum**

$$b \cdot n \overset{\text{def}}{=} n \cdot b \overset{\text{def}}{=} \bigoplus_{i=1}^{n} b \quad \text{for } n \in \mathbb{N} \cup \{\infty\}, b \in \mathbb{V} \tag{3.2}$$

and the $n$**-fold product**

$$b^n \overset{\text{def}}{=} \bigotimes_{i=1}^{n} b \quad \text{for } n \in \mathbb{N} \cup \{\infty\}, b \in \mathbb{V} \tag{3.3}$$

Some adjectives used to describe semirings include:

- **commutative** $\iff$ $\circ$ is commutative
- **idempotent** $\iff$ $\oplus$ is idempotent

**Definition 18.** *A **closed semiring** has an additional operator, called the (Kleene) **star operator** $(\cdot)^\star$. This operation must satisfy the following axioms for all $a \in \mathbb{V}$:*

$$a^\star = \underline{1} \oplus a \circ a^\star \quad \textit{and} \quad a^\star = \underline{1} \oplus a^\star \circ a$$

We often interpret $a^\star$ as the solution to the infinite series that arises from unrolling either of its recursive identities:

$$a^\star = \bigoplus_{n=0}^{\infty} a^n \quad \text{for } a \in \mathbb{V}$$

The star operation is useful for solving certain linear equations. For example, $x = b \oplus a \circ x \oplus c \circ y$ can be rearranged to "solve" for $x$ *without subtraction or division*: $x = a^\star \circ (b \oplus c \circ y)$.

A special family of semirings that always have a star operation is those that are $K$-closed.

**Definition 19.** *For $K \in \mathbb{N}$, a semiring $\langle \mathbb{V}, \oplus, \circ, \underline{0}, \underline{1} \rangle$ is $K$-**closed** if and only if*[35]

$$\forall\, a \in \mathbb{V}, \forall\, K' > K \colon \bigoplus_{k=0}^{K'} a^k = \bigoplus_{k=0}^{K} a^k.$$

Intuitively, $K$-closed semirings imply that we only have to sum contributions of a cycle by iterating around the cycle at most $K$ times; further iterations do not affect the value of the summation. When $K = 0$, the condition simplifies to $\forall\, a \in \mathbb{V} \colon \underline{1} \oplus a = \underline{1}$. We briefly mention some examples of $K$-closed semirings; these examples are defined in §3.3. Examples of $0$-closed semirings include the boolean, min-plus, and Viterbi semirings. An example of a general $K$-closed semiring include the $(K+1)$-best semirings (Mohri, 2002, §6) and the $K$-collapsed natural numbers. Non-examples of $K$-closed semirings include the arithmetic semirings.

Each $K$-closed semiring has simple expressions for computing the star operation because they are no longer infinite sums.

[35] It is possible to replace $\forall K'$ with $K' = K + 1$—the conditions are equivalent (Mohri, 2002, Lemma 4).

$$\begin{cases} \text{0-closed} : a^* = \underline{1} \\ \text{1-closed} : a^* = \underline{1} \oplus a \\ \text{2-closed} : a^* = \underline{1} \oplus a \oplus a^2 \\ \quad \vdots \\ K\text{-closed} : a^* = \bigoplus_{k=0}^{K} a^k \end{cases} \tag{3.4}$$

**Definition 20.** *A semiring is **monotonic** under the partial ordering relation* $\preccurlyeq$ *if*

$$\forall\, a \preccurlyeq b, \forall\, c \in \mathbb{V} : \begin{cases} a \oplus c \preccurlyeq b \oplus c \\ a \circ c \preccurlyeq b \circ c \\ c \circ a \preccurlyeq c \circ b \end{cases}$$

**Where to now?** In §3.3, we will describe a number of examples of semirings. In §4.4, we will describe some special families of semirings that are useful in providing convergence guarantees for Dyna programs.

## 3.3 Semirings: A Powerful Abstraction for Dynamic Programming

Semirings are a powerful abstraction for dynamic programming. They provide a useful and elegant abstraction level for

- algorithms for NLP (e.g., Teitelbaum (1973), Mohri (1997), Goodman (1999), and Huang (2008))
- probabilistic inference (e.g., Aji and McEliece (2006))
- database provenance (e.g., Green, Karvounarakis, et al. (2007), Green and Tannen (2017), and Cheney et al. (2009))
- network analysis (e.g., Gondran and Minoux (2008) and Baras and Theodorakopoulos (2010))
- program analysis (e.g., Cousot and Cousot (1977), Esparza et al. (2010), and Dolan (2013))

Semirings enable the following separation of concerns:

(1) devising a compact encoding of a set of derivations (e.g., paths in a graph, parses of a sentence)

(2) how to evaluate and aggregate the derivations (e.g., the highest-weight derivation, the total weight of all derivations)

In Dyna, part 1 is specified by the program rules, and part 2 is given by choice of semiring. Modularizing by semiring gives different semantics for the same algorithm with the same time/space complexity (assuming semiring operations have unit time), and we can even reuse the implementation.[36]

The key property of a semiring is distributivity, as it allows us to factor out common subexpressions: we can safely rewrite $(a \circ b) \oplus (a \circ c)$ (three operations) into an equivalent smaller expression $a \circ (b \oplus c)$ (two operations). The factored expression can be significantly smaller.

**Example 9.** *The example below may be regarded as encoding a weighted directed graph with two states* x *and* y*. The weights in the graph have been left abstract as symbols* a*,* b*,* c*, and* d*.*

```
x += a * x.
x += b * y.
y += c * x.
y += d * y.
y += 1.  % final state
```

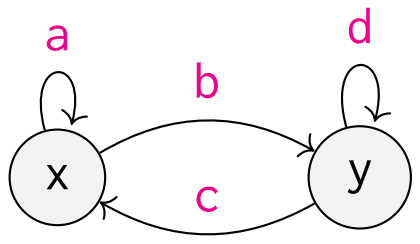

*We can translate the Dyna program into a system of equations:*

$$\mathtt{x} = \mathtt{a} \circ \mathtt{x} \oplus \mathtt{b} \circ \mathtt{y}$$
$$\mathtt{y} = \mathtt{c} \circ \mathtt{x} \oplus \mathtt{d} \circ \mathtt{y} \oplus \underline{1}$$

*The solution to the system of equations is a map where the value associated with* x *is the set of all paths starting in state* x *and ending in state* y*, and the value associated with* y *is the set of all paths starting in state* y *and ending in state* y*. In this graph, there are infinitely many paths, and yet, we can express them as a small set of equations. Below we consider replacing the weights (*a*,* b*,* c*,* d*) with*

[36] Although certain semirings may allow for additional efficiency shortcuts that others do not, certain implementations may assume properties beyond the basic semiring axioms.

*values in a variety of semirings. In each case, we describe the interpretation of the semiring values associated with* x *and* y*.*

- $\langle\{\texttt{false}, \texttt{true}\}, \vee, \wedge, \texttt{false}, \texttt{true}\rangle$ *where* true *is placed on each edge* $\Longrightarrow$ y *is reachable from* x*, and* y *is reachable from* y*.*

- $\langle\mathbb{R}\cup\{\infty\}, +, \cdot, \infty, 0\rangle \Longrightarrow$ *the total weight of all paths from* x *to* y *and* y *to* y*.*

- $\langle\mathbb{R}, \min, +, 0, 1\rangle \Longrightarrow$ *the least cost path from* x *to* y *and* y *to* y*.*

- $\langle 2^{\Sigma^*}, \cup, \cdot, \emptyset, \varepsilon\rangle$ *where* $\Sigma$ *is the alphabet* $\{\texttt{a}, \texttt{b}, \texttt{c}, \texttt{d}\} \Longrightarrow$ *the (regular) formal language that describes the set of paths from* x *to* y *and from* y *to* y*.*

*This is just a small example of what we can do by swapping the semiring and keeping the Dyna program essentially the same.*

## Semiring Compendium

Below we list several commonly used semirings and some applications where they appear. Many of these semirings are available in the Dyna library https://github.com/timvieira/dyna-pi. We do not have a particular organizing principle for this list, but we hope that it may serve as a useful reference for some readers and that it illustrates that semirings can do some surprising things.

### Logic

**Boolean semiring** $\langle\{\texttt{false}, \texttt{true}\}, \vee, \wedge, \texttt{false}, \texttt{true}\rangle$ are used to test reachability in graphs and, more generally, provability of items under the rules of the program. Its star operation is $(\texttt{false})^* = (\texttt{true})^* = \texttt{true}$

**Multi-valued logic.**

- **Three-valued logic** $\langle\{\texttt{true}, \texttt{false}, \texttt{unk}\}, \max, \min, \texttt{false}, \texttt{true}\rangle$ where $\texttt{false} < \texttt{unk} < \texttt{true}$.[37]

- **Fuzzy sets** $\langle[0,1], \max, \min, 0, 1\rangle$ is related to fuzzy sets (Zadeh, 1965).[38]

- **Łukasiewicz semiring** $\langle[0,1], \max, \max(0, x+y-1), 0, 1\rangle$ (Łukasiewicz, 1920)

These semirings are each 0-closed; thus, their star operations can be obtained using the $K$-closed shortcut (3.4).

## Arithmetic semirings

- **Counting semiring**: $\langle\mathbb{N}, +, \cdot, 0, 1\rangle$. As the name suggests, this semiring is used to count things, such as the number of paths between two nodes in a graph. If $\mathbb{N}$ is augmented with $\infty$, then $a^* = 1$ **if** $a = 0$ **else** $\infty$.

- **Integer semiring**: $\langle\mathbb{Z}, +, \cdot, 0, 1\rangle$. Its star operation is $0^* = 1$, $a^* = \infty$ **if** $a > 0$, and undefined if $a < 0$.

- **Real semiring**: $\langle\mathbb{R}, +, \cdot, 0, 1\rangle$. If we extend $\mathbb{R}$ with $\infty$, we have $a^* = \frac{1}{1-a}$ if $|\,a\,| < 1$, $\infty$ if $a \geq 1$, and it is undefined if $a \leq -1$. So, unfortunately, this semiring does not have complete support for solving the infinite series.

[37] Often $\langle\texttt{false}, \texttt{unk}, \texttt{true}\rangle$ are encoded as $\langle -1, 0, +1\rangle$.

[38] A fuzzy set extends an ordinary set to support graded membership that indicates the extent to which a given element belongs to a given set (0 meaning definitely not in the set, and 1 meaning definitely in the set).

- **Nonnegative-real semiring**: $\langle \mathbb{R}_{\geq 0} \cup \{\infty\}, +, \cdot, 0, 1\rangle$. This semiring is used for computing the normalization constant of a probabilistic model (Goodman, 1999). Its star operation is $a^* = \frac{1}{1-a}$ for $a < 1$ and $\infty$ otherwise. Thus, to be closed, we require $\infty$ to be added to the set $\mathbb{R}$.

- **Log-nonnegative-real semiring**: $\left\langle \mathbb{R} \cup \{-\infty, \infty\}, \underset{\log}{+}, +, -\infty, 0 \right\rangle$ where $x \underset{\log}{+} y \overset{\text{def}}{=} \log(\exp(x)+\exp(y))$.[39] This semiring is isomorphic to the nonnegative-real semiring; however, it often preferred as it can avoid certain numerical issues.[40] Its star operation is $a^* = -\log(1-\exp(a))$ if $a \leq 0$ and $\infty$ otherwise.

- **First-order expectation semirings** (Eisner, 2001) can be used to compute first-order statistics, such as the expected error rate, the entropy, or the expected feature counts.

- **Second-order expectation semirings** (Li and Eisner, 2009) can be used to compute second-order statistics, such as the variance of the error rate, the gradient of entropy, and the covariance matrix of feature counts.

- $K$**-collapsed natural numbers** $\langle\{0, \cdots, K\}, \min(K, x+y), \min(K, x\cdot y), 0, 1\rangle$ can be used as guard rails that prevent infinite running times due to counting off to infinity. It is isomorphic to the boolean semiring when $K=2$. Its star operation is $a^* = 1$ **if** $a=0$ **else** $K$.

[39] For speed and numerical precision, the log-space addition operation should not be computed as written, it should be computed using a library function (Mächler, 2012). The numpy library for Python provides the function `numpy.logaddexp` for this purpose.

[40] We also mention the log-real semiring (Li and Eisner, 2009, Table 3), which extends the log-nonnegative-real semiring with an additional sign bit so that negative numbers can be represented.

## Optimization

These semirings are often used when computing the minimum/maximum values over some set of values, for example, the weighted set of paths in a graph.

- The **tropical semiring** (aka **min-plus semiring**) $\langle \mathbb{R} \cup \{\infty\}, \min, +, \infty, 0\rangle$ is used to find the shortest path or cheapest derivation. Its star operation is $a^* = 0$

- The **arctic semiring** (aka **max-plus semiring**) $\langle \mathbb{R} \cup \{-\infty\}, \max, +, -\infty, 0\rangle$ is used to find the longest path or highest-weight derivation. Its star operation is $a^* = \underline{1} \oplus a$ if $a \leq 0$ else $\infty$

- The **Viterbi semiring** (aka **max-times semiring**) $\langle [0, 1], \max, \cdot, 0, 1\rangle$ is used to find the most probable explanation (MPE) in a probabilistic model. MPE corresponds to maximum a posteriori (MAP) decoding when the model does not have latent variables. Its star operation is $a^* = 1$. If we broaden the carrier set with numbers $> 1$, we have to extend the definition to $a^* = \infty$ for $a > 1$.

- The **min-times semiring** $\langle \mathbb{R}_{\geq 0} \cup \{\infty\}, \min, \cdot, \infty, 1\rangle$ can be used to find the least probable derivation. Its star operation is $a^* = \underline{1} \oplus a$

These semirings can be augmented with additional bookkeeping to return an argmin/argmax (or even the set of argmins/argmaxes) in addition to the min/max. It is also possible to make the elements being compared in these optimization semirings tuples and to compare them under a lexicographic ordering. This is one way to keep track of back pointers and to add tie-breaking operations.

The **maximum-capacity semiring** $\langle \mathbb{R} \cup \{-\infty, \infty\}, \max, \min, -\infty, \infty \rangle$ can be used to find the maximum capacity path in a graph. It is 0-closed so $a^* = \underline{1}$.

The $K$**-best semiring** (top-$K$ semiring or bottom-$K$ semiring) extends the optimization semirings to provide the $K$-best alternatives where $K \geq 0$ is a fixed constant.[41] The $K$-best semirings work with any monotonic product operations, which include multiplication (by nonnegative numbers), Cartesian products, and addition. Let $\langle \mathcal{A}, \circ, \varepsilon \rangle$ be a monoid. Suppose that it is partially ordered under $\preccurlyeq$ and the $\circ$ operation is monotonic.

- $\mathbb{V} \overset{\text{def}}{=} \left\{ \mathsf{sort}^K_{\preccurlyeq}(a_1 \cdots a_K) \mid a_1 \ldots a_K \in \mathcal{A}^K \right\}$: each element of this set is a length-$K$ sorted list of elements from $\mathcal{A}$. Here the function $\mathsf{sort}^K_{\preccurlyeq}$ takes a sequence, sorts it under $\preccurlyeq$, and truncates only to include the $K$ smallest elements.

- $A \oplus B \overset{\text{def}}{=} \mathsf{sort}^K_{\preccurlyeq}(A \cup B)$

- $A \circ B \overset{\text{def}}{=} \mathsf{sort}^K_{\preccurlyeq}([a \circ b \mid a \in A, b \in B])$

- $\underline{0} \overset{\text{def}}{=} []$

- $\underline{1} \overset{\text{def}}{=} [\varepsilon]$

[41] Goodman (1998) gave the $K$-best semiring, which eagerly combines the top-$K$ values for some fixed $K$.[42] Later, this approach was generalized (Huang and Chiang, 2005) to allow the $K$-best list to be computed lazily (and without knowledge of $K$), which can be significantly faster as many of those $K$-best lists are not even used (e.g., a $K$-best list of a span that does not appear in any complete parse). The implementation details for lazily combining streams efficiently are given in Huang and Chiang (2005).

Since this semiring is $(K-1)$-closed. We can use the truncated sum to implement the star operation $a^* = \sum_{k=0}^{K-1} a^k$.

## Matrices

The semiring of matrices (Lehmann, 1977) is important as we will use it in §5.3 to solve certain linear families of infinite sums.

**Definition 21** (Matrices over a semiring)**.** *We can "lift" any base semiring* $\langle \mathbb{V}, \oplus, \circ, \underline{0}, \underline{1} \rangle$ *into a semiring over square matrices* $\left\langle \mathbb{V}^{D\times D}, +, \llcorner\lrcorner, \mathbf{0}, \mathbf{1} \right\rangle$ *where* $D$ *is a set.*[43] *The matrix addition and multiplication operations (given below) are a straightforward generalization of the usual matrix multiplication on* $\mathbb{R}$*. Let* $\mathbf{A}, \mathbf{B} \in \mathbb{V}^{D\times D}$.

- $[\mathbf{A} + \mathbf{B}]_{ij} \overset{\text{def}}{=} \mathbf{A}_{ij} \oplus \mathbf{B}_{ij}$
- $[\mathbf{A}\,\mathbf{B}]_{ik} \overset{\text{def}}{=} \bigoplus_{j\in D} \mathbf{A}_{ij} \circ \mathbf{B}_{jk}$
- $[\mathbf{0}]_{ij} \overset{\text{def}}{=} \underline{0}$
- $[\mathbf{1}]_{ij} \overset{\text{def}}{=} \underline{1}$ ***if*** $i = j$ ***else*** $\underline{0}$

*The star operation* $\mathbf{A}^*$ *can be computed as follows. If* $|\,D\,| = 1$*, then* $\mathbf{A}^* = [a]^* = [a^*]$*. Otherwise,* $|\,D\,| > 1$ *and we can decompose* $\mathbf{A}$ *into nontrivial blocks* $\langle \mathbf{B}, \mathbf{C}, \mathbf{D}, \mathbf{E} \rangle$ *and apply the following blockwise recursion.*

$$\mathbf{A}^* = \begin{bmatrix} \mathbf{B} & \mathbf{C} \\ \mathbf{D} & \mathbf{E} \end{bmatrix}^{\star} = \begin{bmatrix} \mathbf{B}^* + \mathbf{B}^*\,\mathbf{C}\,\boldsymbol{\Delta}^*\,\mathbf{D}\,\mathbf{B}^* & \mathbf{B}^*\,\mathbf{C}\,\boldsymbol{\Delta}^* \\ \boldsymbol{\Delta}^*\,\mathbf{D}\,\mathbf{B}^* & \boldsymbol{\Delta}^* \end{bmatrix} \tag{3.5}$$

[43] We have denoted the juxtaposition operation (e.g., $\mathbf{A}\,\mathbf{B}$) as $\llcorner\lrcorner$, the so-called "visible space."

*where* $\boldsymbol{\Delta} = \mathbf{E} + \mathbf{D}\,\mathbf{B}^*\mathbf{C}$*. It is straightforward to verify that each recursive call to* $(\cdot)^*$ *is decreasing in size; thus, the recursive definition is well-founded. Furthermore, if the blocks are chosen to be equal in size, then the runtime is* $\mathcal{O}(|\,D\,|^3)$*. We refer the reader to Lehmann (1977) for proof of correctness and further discussion.*

## Provenance semirings

**Provenance semirings** (Green, Karvounarakis, et al., 2007; Green and Tannen, 2017; Cheney et al., 2009; Schlund, 2016) are used in database systems to validate or explain query results. Similarly, "derivation semirings" (Goodman, 1998) can be used to encode the set of "derivations" for a given item.[44] Some examples of provenance semirings given in Green, Karvounarakis, et al. (2007) include the following:

- The **set semiring** $\langle 2^{\Sigma}, \cup, \cap, \emptyset, \Sigma \rangle$ where $\Sigma$ is a base set can be used to track the possibility of events in a system.

- The **why semiring** $\left\langle 2^{2^{\Sigma}}, \cup, \{x \cup y\}_{x \in X, y \in Y}, \emptyset, \{\emptyset\} \right\rangle$ where $\Sigma$ is a base set. The base set $\Sigma$ is a set of annotations; the values computed with this semiring are the combinations of annotations required for a given tuple to appear in answer to a query.

- The **lineage semiring** $\langle 2^{\Sigma} \cup \{\bot\}, \cup, \cup, \bot, \emptyset \rangle$ where $\Sigma$ is a base set and $\bot$ is a special element satisfying the semiring properties of $\underline{0}$. In the context of a graph, the lineage semiring determines the subset of reachable arc labels.

[44]Note: Goodman's notation of a "derivation" is different from our notion (which is formalized later in § 4). In particular, his notion is less general than ours because ours is not built from semiring operations: our notion of a derivation is given by nested rules, which are not multiplicatively associative.

## Formal Languages

**Formal languages**: $\left\langle 2^{\Sigma^\star}, \cup, \cdot, \emptyset, \varepsilon \right\rangle$ where $\Sigma$ is a set of symbols and $\Sigma^\star$ denotes the set of strings composed of those symbols, $\varepsilon$ is the empty string, and $\cdot$ denotes string concatenation. The semiring of formal languages can be used to enumerate the set of strings accepted by a context-free grammar, a finite-state automaton, and many other formalisms. There are, unfortunately, no shortcuts for computing the star operation. Finite-state automata can compactly represent certain infinite sets of strings; they are closed on the union, concatenation, and star operations.[45]

**Weighted formal languages**: The semiring of formal languages can be extended to semiring-weighted languages, where a semiring weight is assigned to each string. Let $\langle \mathbb{V}, \oplus, \circ, \underline{0}, \underline{1} \rangle$ be a base semiring. Let $\Sigma$ be an alphabet as in the semiring of formal languages. The semiring of weighted formal languages $\left\langle \widehat{\mathbb{V}}, \widehat{\oplus}, \widehat{\circ}, \widehat{0}, \widehat{1} \right\rangle$ is defined as follows. Let $\widehat{\mathbb{V}} = \mathbb{V}^{\Sigma^*} = \Sigma^* \to \mathbb{V}$ be the set of all functions from strings in $\Sigma^*$ to values in the base semiring. Let $a, b \in \widehat{\mathbb{V}}$ and $x \in \Sigma^*$:

- $[a \mathbin{\widehat{\oplus}} b](x) \stackrel{\text{def}}{=} a(x) \oplus b(x)$

[45] Although context-free languages (CFLs) are also closed under the semiring operations, the representations of CFLs (i.e., context-free grammars and pushdown automata) do not have (complete) algorithms to test the equality of the context-free languages that they represent, severely limiting their utility as a computational representation.

- $[a \widehat{\circ} b](x) \stackrel{\text{def}}{=} \bigoplus_{x:\ x=y\cdot z} a(y) \circ b(z)$

- $[\widehat{0}](x) \stackrel{\text{def}}{=} \underline{0}$

- $[\widehat{1}](x) \stackrel{\text{def}}{=} \underline{1}$ **if** $x = \varepsilon$ **else** $\underline{0}$

There are, unfortunately, no shortcuts for computing the star operation.

## Sets of Vectors

- The **semiring of semilinear sets** is $\langle 2^{\mathbb{N}^{\Sigma}}, \cup, +, \emptyset, \{\mathbf{0}\}\rangle$ where $X + Y \stackrel{\text{def}}{=} \{x + y\}_{x \in X, y \in Y}$. This semiring can be used to compute the Parikh image of a formal language with alphabet $\Sigma$.[46] We refer the reader to Schlund (2016, Ch. 5) for further discussion and efficient implementation.

- The **Pareto frontier semiring** is $\langle$Pareto efficient sets of vectors, $\mathsf{pareto}(X \cup Y), \mathsf{pareto}(X + Y), \emptyset, \{\mathbf{0}\}\rangle$. The function pareto filters the set of vectors to those that are Pareto efficient. A set of vectors is **Pareto efficient** if and only if each of its vectors is non-dominated. A vector is said to dominate another vector if it is $\leq$ all the other vectors in all dimensions and $<$ in at least one dimension. The set of Pareto efficient vectors represents the optimal trade-offs between different objectives or criteria (i.e., where improving one objective comes at the cost of worsening another). The Pareto frontier semiring on $\mathbb{R}^D_{\geq 0}$ is 0-closed, so its star operation is $X^* = \underline{1}$. This semiring is discussed further in Gondran and Minoux (2008).

[46]The **Parikh image** of a language $L$ is defined as $\mathsf{Parikh}(L) \stackrel{\text{def}}{=} \{\mathsf{bag}(w) \mid w \in L\}$ where $\mathsf{bag}(w) \stackrel{\text{def}}{=} \wr a \mid a \in w \wr$. Placing the symbols of each word into a bag removes their order.

- The **semiring of convex sets in a vector space** is $\langle$convex sets over a vector space, $\text{convexhull}(X \cup Y)$, $\text{convexhull}(X + Y)$, $\emptyset$, $\{\mathbf{0}\}\rangle$. For a finite-dimensional vector space, we may restrict from convex sets to polytopes and still have a semiring—the advantage being that each polytope can be computationally represented as a finite set of vertices, yielding polytime operations.[47] Dyer (2013) showed how to use this semiring to perform efficient minimum error rate training (Och, 2003), a common technique used in machine translation. An earlier citation for this semiring is Iwano and Steiglitz (1990).

## Other

- Polynomials, and formal power series (Kuich, 1997; Green, Karvounarakis, et al., 2007) over semirings form a semiring in each case.

- Baras and Theodorakopoulos (2010) and Gondran and Minoux (2008) give many interesting semirings for network analysis, including semirings to enumerate graph cuts and bridges.

- Many works use semirings to perform dataflow analyses on programs (Dolan, 2013; Esparza et al., 2010; Cousot and Cousot, 1977).

[47] The semiring of convex sets is similar to the Pareto frontier semiring. The difference is that convex sets find the set of optimal derivations with respect to linear combinations of the base objectives.

# Chapter 4

# Semantics of Dyna Programs

In § 2, we introduced the Dyna language through a number of examples. In this chapter, we will formalize the meaning of a Dyna program—i.e., what it actually computes—as a certain sum over potentially infinitely many semiring-weighted derivations (§4.1).[48] We illustrate the challenges with several simple examples. Quite surprisingly, we can often compute—or, at least accurately approximate—these infinite sums. We describe fixpoint iteration as the mathematical algorithm framework for computing these semantics. We later (§ 5) reify the framework in executable algorithms.

[48]The prior formalization of Dyna (Eisner, Goldlust, et al., 2005) did not describe Dyna's semantics as a sum over weighted derivations, which meant that it was mismatched with other work on semiring-weighted deduction, e.g., Goodman (1999). This mismatch, however, enables the generalization of Dyna 1 to the Dyna 2 language (Eisner and Filardo, 2011).

## 4.1 Total Value of All Derivations

**Example 10.** *Consider the following program, which encodes a geometric series:*

$\mathfrak{p} = \begin{cases} \texttt{x += b.} \\ \texttt{x += a * x.} \end{cases}$

*The meaning of the item* x *in this program is a semiring value, which we denote as* $\mathfrak{p}[\![\mathrm{x}]\!]$*. The value of* x *as a function of* a *and* b *is* $\mathfrak{p}[\![\mathrm{x}]\!] = \mathtt{b} + \mathtt{a} * \mathtt{b} + \mathtt{a} * \mathtt{a} * \mathtt{b} + \mathtt{a} * \mathtt{a} * \mathtt{a} * \mathtt{b} + \mathtt{a} * \mathtt{a} * \mathtt{a} * \mathtt{a} * \mathtt{b} + \mathtt{a} * \mathtt{a} * \mathtt{a} * \mathtt{a} * \mathtt{a} * \mathtt{b} + \mathtt{a} * \mathtt{a} * \mathtt{a} * \mathtt{a} * \mathtt{a} * \mathtt{a} * \mathtt{b} + \cdots$*. We arrive at this (infinite) sequence by enumerating the set of derivations of* x *under the rules in* $\mathfrak{p}$*. The derivations of an item are a set denoted* $\mathfrak{p}\{\mathrm{x}\}$*. Below, we show 7 (of the infinitely many) derivations of the item* x*. We see that the values at the leaves of each derivation correspond to the terms in the summation above.*

$\mathfrak{p}\{\mathrm{x}\} =$

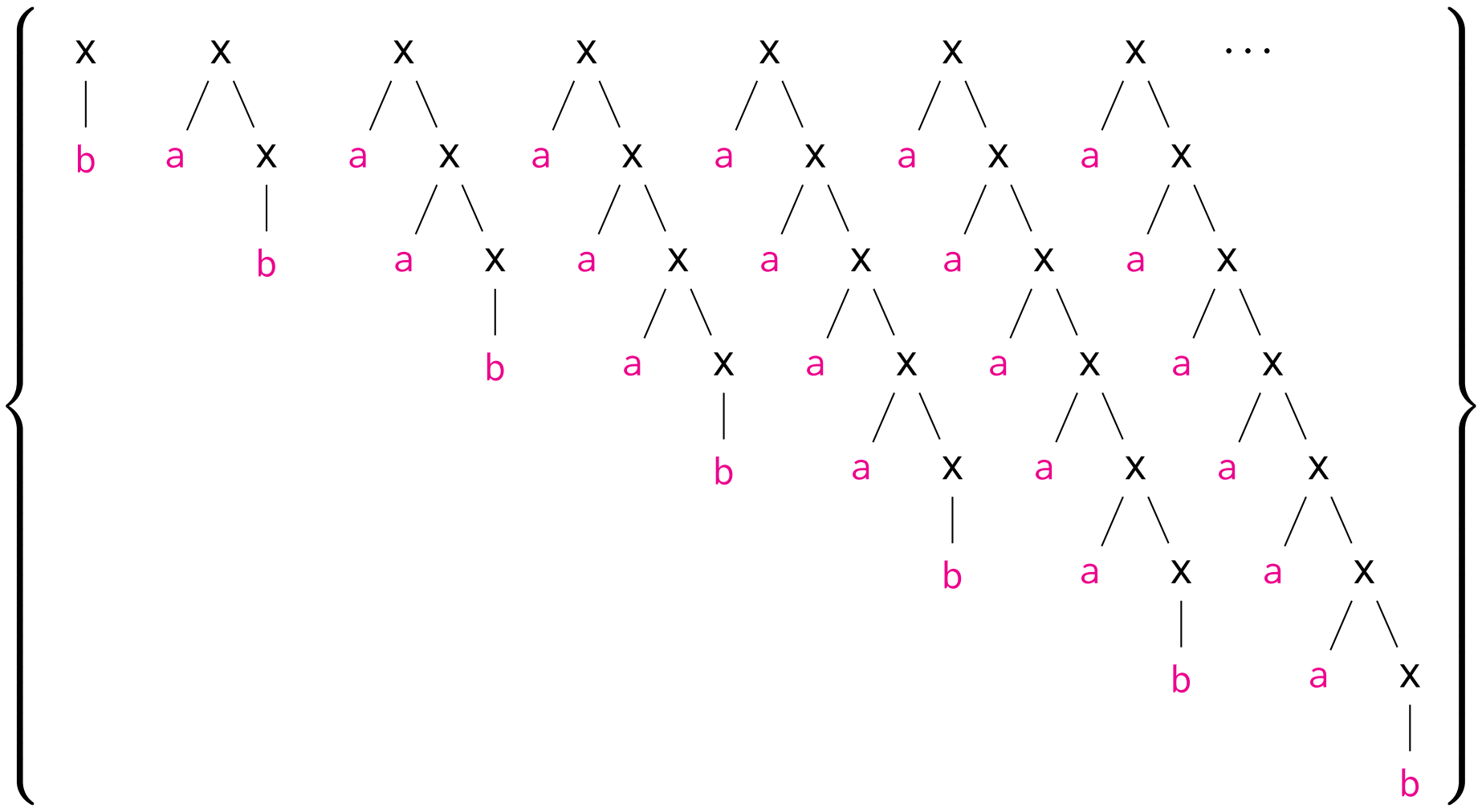

*At a high level, we generate these derivations by repeated substitution in the rules matching a subgoal.*[49] *Below, we show these rule substitutions by nesting rules within other rules. Each of these nested rules encodes a derivation tree.*

[49]We will make this formal in Definition 23.

```
58  x += b.
59  x += a * (x += b).
60  x += a * (x += a * (x += b)).
61  x += a * (x += a * (x += a * (x += b))).
62  x += a * (x += a * (x += a * (x += a * (x += b)))).
63  x += a * (x += a * (x += a * (x += a * (x += a * (x += b))))).
64  x += a * (x += a * (x += a * (x += a * (x += a * (x += a * (x += b)))))).
65  ···
```

*Next, we can* summarize *each derivation by its value—i.e., the product of the values of its leaves.*[50] *By doing so, we recover the terms of the infinite series:*

```
66  x += b.
67  x += a * b.
68  x += a * a * b.
69  x += a * a * a * b.
70  x += a * a * a * a * b.
71  x += a * a * a * a * a * b.
72  x += a * a * a * a * a * a * b.
73  ···
```

*Notice that our ability to expand* `x` *into a set of derivations whose values can be summed requires distributivity.*

We now look at an example program containing variables.

**Example 11.** *Consider CKY (Example 6; repeated below) where we have augmented the original program with some input data:*

```
74  β(X,I,K) += γ(X,Y,Z) * β(Y,I,J) * β(Z,J,K).     78  % context-free grammar rules
75  β(X,I,K) += γ(X,Y) * β(Y,I,K).                  79  γ(s,a) += v₁.
76  β(X,I,K) += γ(X,Y) * word(Y,I,K).               80  γ(s,s,s) += v₂.
77  goal += β(s,0,N) * len(N).

81  % encoding of the input sentence "a a a a"
82  word(a,0,1) += w₁. word(a,1,2) += w₂. word(a,2,3) += w₃. word(a,3,4) += w₄.
83  len(4) += 1.
```

*where* $v_1, v_2, w_1, w_2, w_3, w_4 \in \mathbb{V}$. *Now, consider the* `goal` *item. It has five derivations under the CKY program's rules:*

---

[50]The product uses an in-order traversal of the tree given by the nested terms to preserve the order of possibly noncommutative product operations.

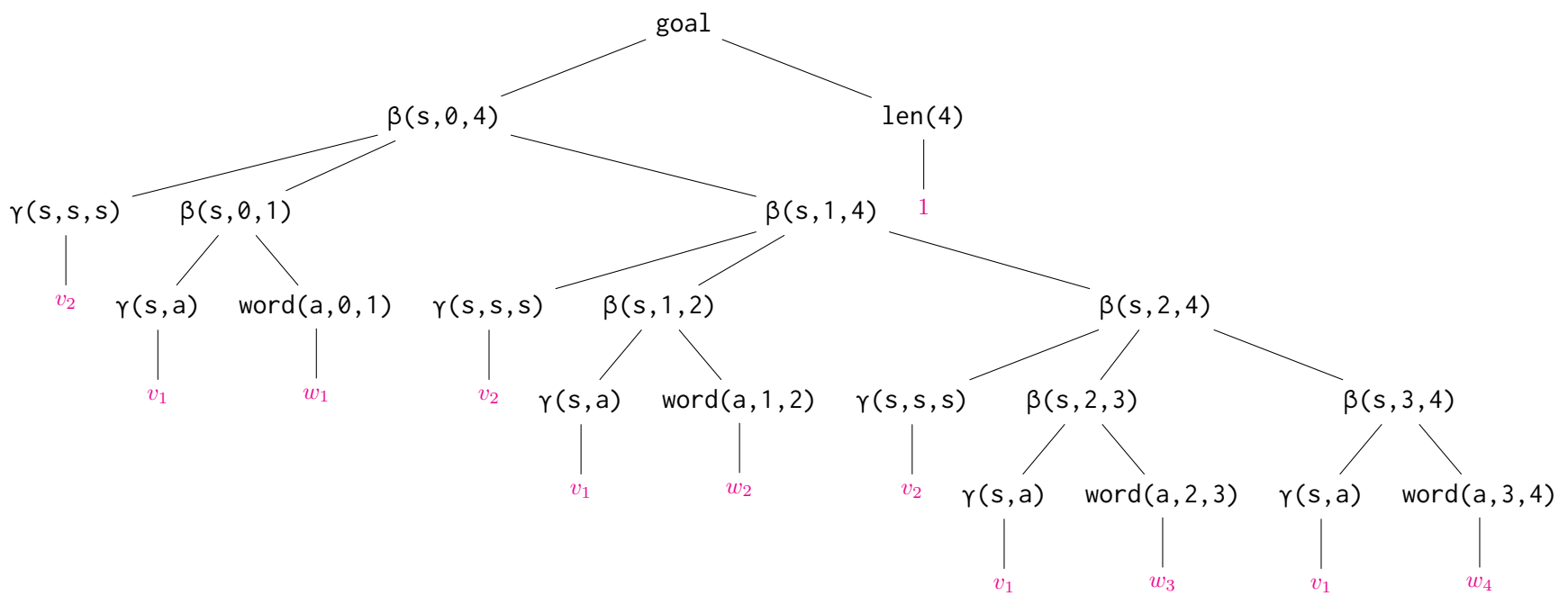
goal
β(s,0,4)
len(4)
γ(s,s,s)
β(s,0,1)
β(s,1,4)
1
$v_2$
γ(s,a)
word(a,0,1)
γ(s,s,s)
β(s,1,2)
β(s,2,4)
$v_1$
$w_1$
$v_2$
γ(s,a)
word(a,1,2)
γ(s,s,s)
β(s,2,3)
β(s,3,4)
$v_1$
$w_2$
$v_2$
γ(s,a)
word(a,2,3)
γ(s,a)
word(a,3,4)
$v_1$
$w_3$
$v_1$
$w_4$

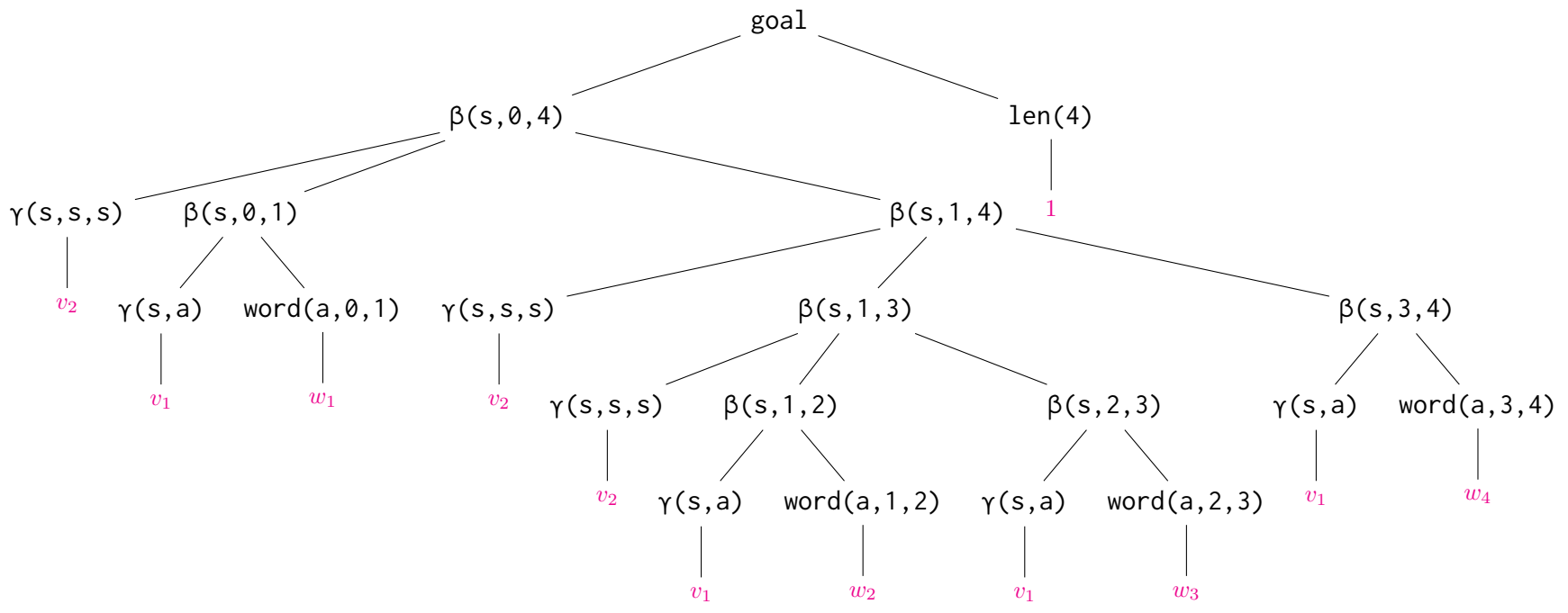
goal
β(s,0,4)
len(4)
γ(s,s,s)
β(s,0,1)
β(s,1,4)
1
$v_2$
γ(s,a)
word(a,0,1)
γ(s,s,s)
β(s,1,3)
β(s,3,4)
$v_1$
$w_1$
$v_2$
γ(s,s,s)
β(s,1,2)
β(s,2,3)
γ(s,a)
word(a,3,4)
$v_2$
γ(s,a)
word(a,1,2)
γ(s,a)
word(a,2,3)
$v_1$
$w_4$
$v_1$
$w_2$
$v_1$
$w_3$

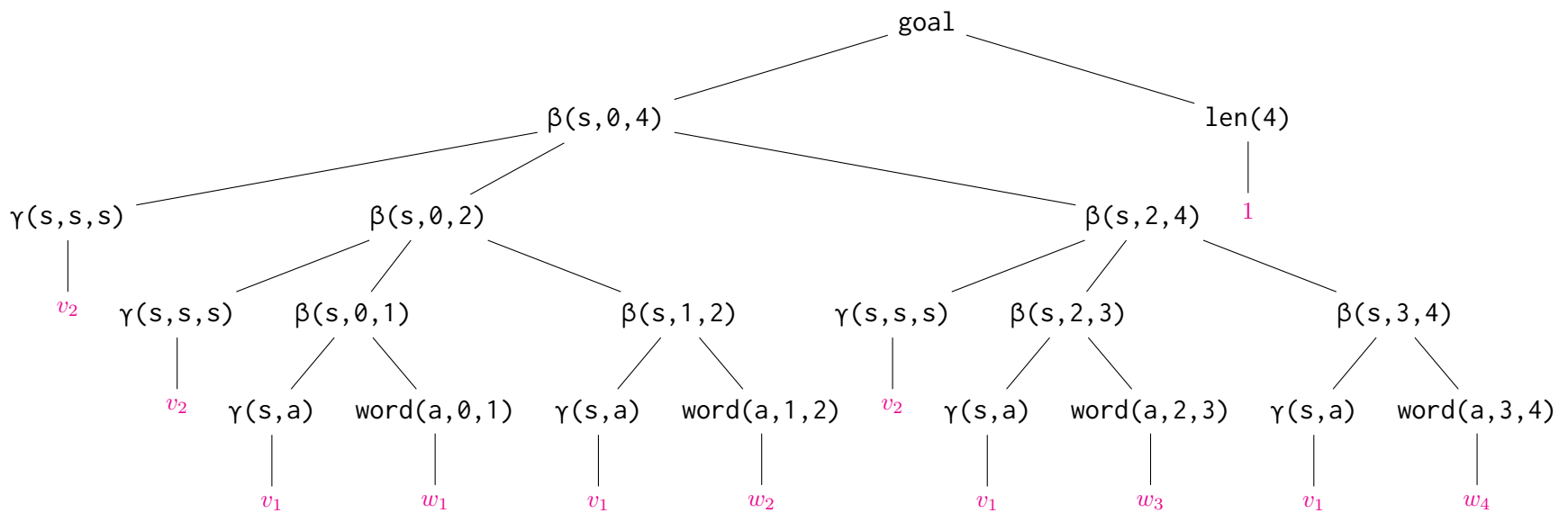
goal
β(s,0,4)
len(4)
γ(s,s,s)
β(s,0,2)
β(s,2,4)
1
$v_2$
γ(s,s,s)
β(s,0,1)
β(s,1,2)
γ(s,s,s)
β(s,2,3)
β(s,3,4)
$v_2$
γ(s,a)
word(a,0,1)
γ(s,a)
word(a,1,2)
$v_2$
γ(s,a)
word(a,2,3)
γ(s,a)
word(a,3,4)
$v_1$
$w_1$
$v_1$
$w_2$
$v_1$
$w_3$
$v_1$
$w_4$

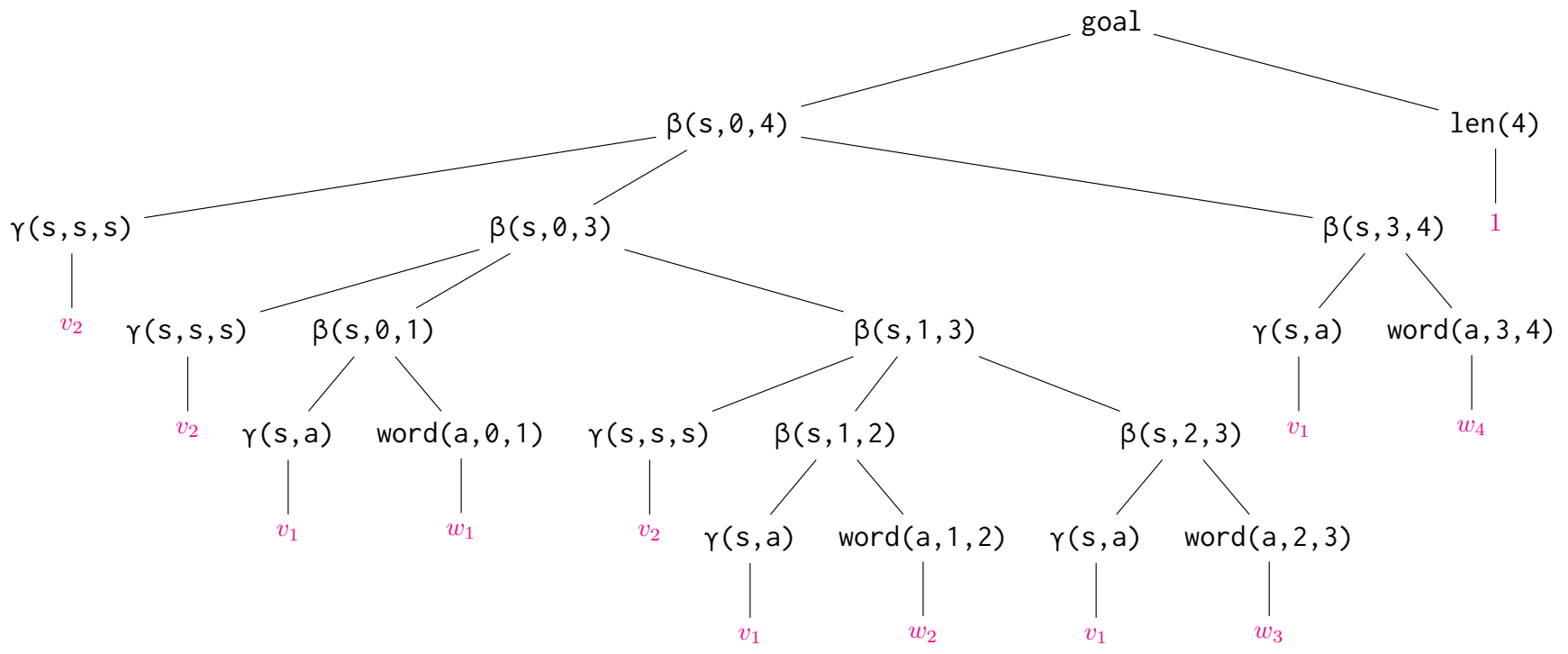
goal
β(s,0,4)
len(4)
γ(s,s,s)
β(s,0,3)
β(s,3,4)
1
$v_2$
γ(s,s,s)
β(s,0,1)
β(s,1,3)
γ(s,a)
word(a,3,4)
$v_2$
γ(s,a)
word(a,0,1)
γ(s,s,s)
β(s,1,2)
β(s,2,3)
$v_1$
$w_4$
$v_1$
$w_1$
$v_2$
γ(s,a)
word(a,1,2)
γ(s,a)
word(a,2,3)
$v_1$
$w_2$
$v_1$
$w_3$

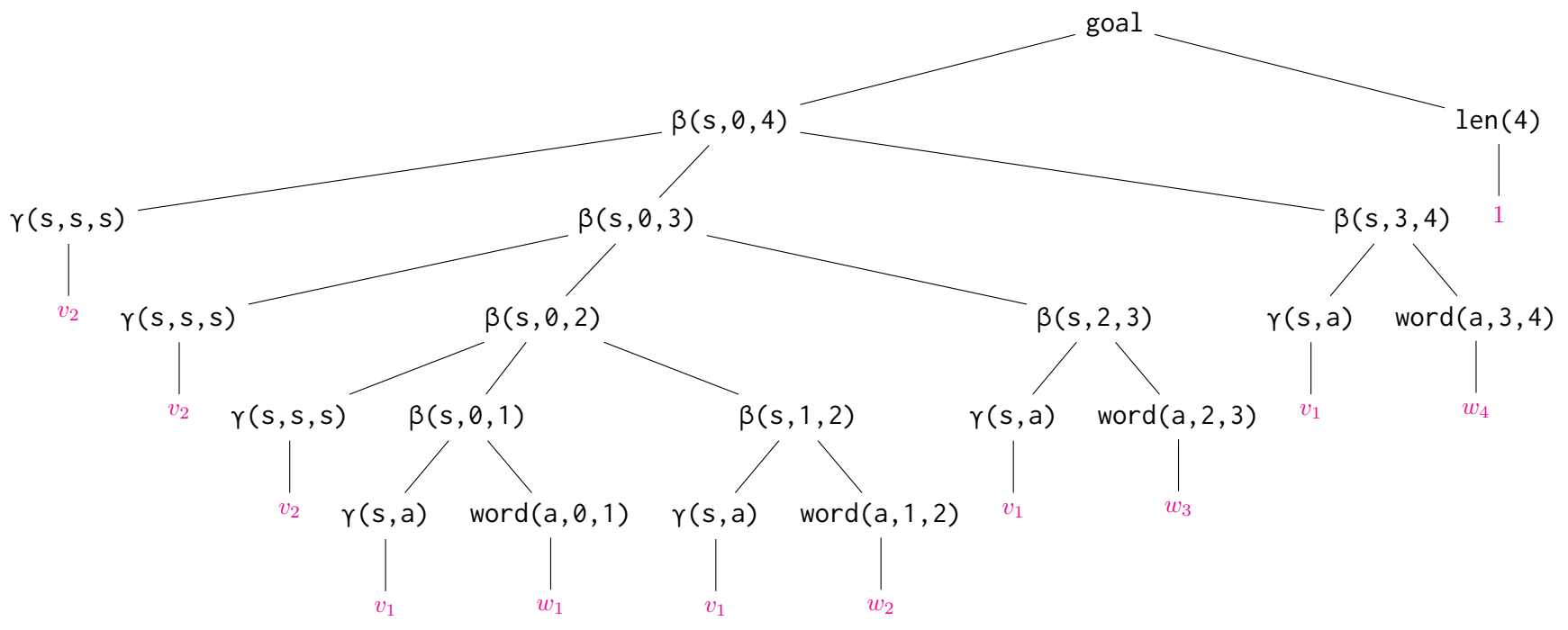
goal
β(s,0,4)
len(4)
γ(s,s,s)
β(s,0,3)
β(s,3,4)
1
$v_2$
γ(s,s,s)
β(s,0,2)
β(s,2,3)
γ(s,a)
word(a,3,4)
$v_2$
γ(s,s,s)
β(s,0,1)
β(s,1,2)
γ(s,a)
word(a,2,3)
$v_1$
$w_4$
$v_2$
γ(s,a)
word(a,0,1)
γ(s,a)
word(a,1,2)
$v_1$
$w_3$
$v_1$
$w_1$
$v_1$
$w_2$

*Again, we see that each derivation's internal nodes are labeled with items, and its leaves are semirings values. Each derivation's value is the product of its leaves. To determine the value of the* `goal` *we sum the values of each of its derivations:*

$$
\begin{aligned}
[\![\texttt{goal}]\!] = \big(& \texttt{v}_2 * \texttt{v}_1 * \texttt{w}_1 * \texttt{v}_2 * \texttt{v}_1 * \texttt{w}_2 * \texttt{v}_2 * \texttt{v}_1 * \texttt{w}_3 * \texttt{v}_1 * \texttt{w}_4 * \underline{1} \\
&+ \texttt{v}_2 * \texttt{v}_1 * \texttt{w}_1 * \texttt{v}_2 * \texttt{v}_2 * \texttt{v}_1 * \texttt{w}_2 * \texttt{v}_1 * \texttt{w}_3 * \texttt{v}_1 * \texttt{w}_4 * \underline{1} \\
&+ \texttt{v}_2 * \texttt{v}_2 * \texttt{v}_1 * \texttt{w}_1 * \texttt{v}_1 * \texttt{w}_2 * \texttt{v}_2 * \texttt{v}_1 * \texttt{w}_3 * \texttt{v}_1 * \texttt{w}_4 * \underline{1} \\
&+ \texttt{v}_2 * \texttt{v}_2 * \texttt{v}_1 * \texttt{w}_1 * \texttt{v}_2 * \texttt{v}_1 * \texttt{w}_2 * \texttt{v}_1 * \texttt{w}_3 * \texttt{v}_1 * \texttt{w}_4 * \underline{1} \\
&+ \texttt{v}_2 * \texttt{v}_2 * \texttt{v}_2 * \texttt{v}_1 * \texttt{w}_1 * \texttt{v}_1 * \texttt{w}_2 * \texttt{v}_1 * \texttt{w}_3 * \texttt{v}_1 * \texttt{w}_4 * \underline{1}\big)
\end{aligned}
$$

*The other items in the program also have derivations, and values derived from those derivations, but we will not show those to keep the example brief.*

Next, we formalize the concepts introduced in the example above.

**Definition 22.** *The meaning of a Dyna program*[51] $\mathfrak{p}$ *is a* ***valuation*** $\mathfrak{p}[\![\cdot]\!] : \mathbb{H} \to \mathbb{V}$,[52] *which computes the sum over derivation values.*[53]

$$
\mathfrak{p}[\![x]\!] \stackrel{\text{def}}{=} \begin{cases} x & \textit{if } x \in \mathbb{V} \\ \mathfrak{B}[\![x]\!] & \textit{if } x \in \mathfrak{B} \\ \bigoplus\limits_{\delta \in \mathfrak{p}\{x\}} \mathsf{yield}(\delta) & \textit{otherwise} \end{cases} \tag{4.1}
$$

*where* $\mathfrak{B}[\![x]\!]$ *is the valuation for built-in relations (items in* $\mathfrak{B}$*),*[54] *and* $\mathfrak{p}\{x\}$ *(Definition 23, below) is the (possibly infinite) set of all derivations of the item* $x$ *and* $\mathsf{yield}(\delta) \in \mathbb{V}$ *is the product of the values in the leaves of the derivation* $\delta$ *(in left-to-right order). We note that when* $\mathfrak{p}\{x\}$ *is infinite,* (4.1) *might be undefined*

[51]Note that the Dyna $\underline{2}$ semantics (discussed in §D) are *not* based on such a sum over derivations.
[52]We may sometimes denote the set of valuations as $\mathbb{V}^{\mathbb{H}}$ rather than $\mathbb{H} \to \mathbb{V}$.
[53]Typically, $\mathfrak{p}[\![\cdot]\!]$ assigns non-$\underline{0}$ values to only a finite subset of $\mathbb{H}$ (that are not builtins).
[54]Unlike program items, built-in relations are defined directly by their valuation; In other words, the valuation for builtins is opaque: we do not have a collection of rules that define the builtins—they are essentially just foreign procedure calls. For simplicity, we assume that the set of builtins (i.e., $x \in \mathfrak{B}$) is disjoint from the items defined in the program.

*(e.g., Example 12, below). When the program* $\mathfrak{p}$ *is clear from context, we may write* $[\![\cdot]\!]$ *as a shorthand for* $\mathfrak{p}[\![\cdot]\!]$.

**Definition 23.** *We define a* ***derivation*** $\delta$ *in program* $\mathfrak{p}$ *as a tree structure that satisfies one of the following conditions:*

*(i)* $\boldsymbol{\delta} \in \mathbb{V}$

*(ii)* $\boldsymbol{\delta}$ *is of the form* $x \mathrel{\oplus=} v$ *where* $x$ *is a builtin and* $v = \mathfrak{B}[\![x]\!]$

*(iii)* $\boldsymbol{\delta}$ *is a product of derivations* $\boldsymbol{\delta}_1 \circ \cdots \circ \boldsymbol{\delta}_K$

*(iv)* $\boldsymbol{\delta}$ *is of the form* $(x \mathrel{\oplus=} \boldsymbol{\delta}_1 \circ \cdots \circ \boldsymbol{\delta}_K)$ *where*

*(a)* $\boldsymbol{\delta}_k$ *are valid derivations for each* $k$

*(b)* $x \mathrel{\oplus=} \mathsf{head}(\boldsymbol{\delta}_1) \circ \cdots \circ \mathsf{head}(\boldsymbol{\delta}_K)$ *is a grounding of a rule in* $\mathfrak{p}$ *where* $\mathsf{head}(\boldsymbol{\delta}_k)$ *gives* $\boldsymbol{\delta}_k$*'s root label for each* $k$.[55]

*These cases define a labeled tree with its internal nodes labeled by rules (or built-in relations) and its leaves labeled by values. Notice that valid derivations may contain variables—meaning that any grounding of its variables is also a valid derivation. The* ***value*** *of a derivation* $\boldsymbol{\delta}$ *is* $\mathsf{yield}(\boldsymbol{\delta})$*; we sometimes write* $[\![\boldsymbol{\delta}]\!]$ *intead of* $\mathsf{yield}(\boldsymbol{\delta})$. *Finally, let* $\mathfrak{p}\{x\}$ *denote the* ***set of all derivations*** *of the item* $x$*. Technically, we want the* smallest *set of objects satisfying the description of a derivation, as this disallows derivations of infinite height.*[56]

[55]Recall that we also use $\mathsf{head}(\cdot)$ to access the head of a rule; the reuse of this notation is not coincidental, as each derivation is built from nested rules (with ground variables).

[56]The set of derivations can often be represented compactly using a *packed forest* data structure, which is also known as a *B-hypergraph* (Gallo et al., 1993). By analogy, a *set of derivations* is to a *packed forest* as a *set of strings* is to a *finite-state automaton*. In both cases, we can represent certain possibly infinite sets of the respective structures efficiently by exploiting structure sharing. Packed forests are commonly used in NLP to represent parse forests and possible translation outputs (e.g., Klein and Manning (2001), Li and Eisner (2009), Huang (2008), and Eisner, Kornbluh, et al. (2006)).

**Definition 24.** *The **height** of a derivation is defined as follows*

$$\mathsf{height}(\boldsymbol{\delta}) \stackrel{\text{def}}{=} \begin{cases} 0 & \textit{if } \delta \textit{ is a constant or builtin} \\ \max(\mathsf{height}(\boldsymbol{\delta}_1), \ldots, \mathsf{height}(\boldsymbol{\delta}_K)) & \textit{if } \delta \textit{ is a product } \boldsymbol{\delta}_1 \circ \cdots \circ \boldsymbol{\delta}_K \\ 1 + \mathsf{height}(\mathsf{body}(\boldsymbol{\delta})) & \textit{if } \delta \textit{ is a derivation by a rule} \end{cases}$$

Not all infinite sums converge to a well-defined value. If a given item in the program does not have a well-defined sum over derivations, then its semantics are also not well-defined. In §4.4, we describe the family of $\omega$-continuous semirings (Kuich, 1997), which are guaranteed to have a well-defined value. However, we do not require the user to use such semirings.

## 4.2 Fixpoint Approximation

**Definition 25.** *We can rearrange* (4.1) *into a system of equations, known as the **fixpoint equations**[57] for a given program* $\mathfrak{p}$*:*[58]

$$[\![x]\!] = \bigoplus_{\substack{\mathbf{R} \in \mathfrak{p}, \\ (x \oplus= y_1 \circ \cdots \circ y_K) \in \Gamma(\mathbf{R})}} [\![y_1]\!] \circ \cdots \circ [\![y_K]\!] \quad \textit{for } x \in \mathbb{H} \tag{4.2}$$

*Recall that* $\Gamma(\mathrm{R})$ *is the set of all groundings of the rule* R *(defined in §2).*

The fixpoint equations suggest a recursive algorithm for evaluating $[\![\cdot]\!]$. That algorithm would compute $[\![x]\!]$ by recursively computing the values of all $[\![y_k]\!]$ items that $[\![x]\!]$ depends on and plugging them into (4.2). However, this algorithm will recurse forever anytime it encounters a cycle.

[57] A **fixpoint equation** has the form $x = f(x)$ for some vector-valued function $f$.

[58] Dyna 2 does not have the sum-over-derivations semantics—its semantics are any fixpoint of a generalized version of (4.2). For further discussion, we refer the reader to §D.

Another important point about cyclic fixpoint equations is that they may admit multiple solutions, including non-solutions to Definition 22.[59] Thus, the fixpoint equations are *underspecified*: they provide necessary, but not sufficient, conditions for a solution to equal the potentially infinite sum in Definition 22. Consider the following example.

**Example 12.** *The following Dyna program encodes the geometric series:* $[\![\texttt{x}]\!] = 1 + a + a^2 + a^3 + \cdots$ *where* $a = [\![\texttt{a}]\!]$.

```
x += 1.
x += a * x.
```

*Each term in the geometric series corresponds to the value of a derivation of* x*. There are infinitely many derivations for* x *under the program. If we plug this program into the fixpoint equations* (4.2)*, we see that* x *depends on itself:*

$$[\![\texttt{x}]\!] = [\![1]\!] + [\![\texttt{a}]\!] \cdot [\![\texttt{x}]\!] = 1 + a \cdot [\![\texttt{x}]\!]$$

*Since we are in the real semiring, we can use subtraction and division as a shortcut to solve this equation:*

$$[\![\texttt{x}]\!] = \frac{1}{1 - a}$$

*Notice, however, the solution only corresponds to the sum over derivations (i.e., the value of the geometric series) when* $|\, a \,| < 1$*. The sum-over-derivations semantics for* $[\![\texttt{x}]\!]$ *when* $a \geq 1$ *ought to be* $\infty$*. Reassuringly,* $\infty$ *satisfies the necessary conditions given by the fixpoint equations:* $\infty = 1 + a \cdot \infty$ *when* $a > 0$*. Notice,*

[59] In the acyclic case, however, the fixpoint equations define a unique value for each item $x$ that equals the sum over (finitely many) derivations $[\![x]\!]$ (Definition 22).

*also, that desired semantics of* $\infty$ *does* not *correspond to the minimal fixpoint since* $\frac{1}{1-a} < \infty$*. When* $a < -1$*, the series does not have a well-defined solution as it oscillates between* $\pm\infty$*; however, the fixpoint equations have the solution* $\frac{1}{1-a}$*.*

Below, we give another example of underspecification that arises from having multiple solutions to a quadratic system:

**Example 13.** *The following program encodes the Catalan series:* $[\![\mathtt{x}]\!] = c + 2\,c^2 + 5\,c^3 + 14\,c^4 + 42\,c^5 + 132\,c^6 + \cdots$ *where* $c \stackrel{\text{def}}{=} [\![\mathtt{c}]\!]$.

```
x += c.
x += x * x.
```

*Now, we consider the fixpoint equations for this program's value. We see that the fixpoint equations admit* two *solutions:*

$$
\begin{aligned}
[\![\mathtt{x}]\!] &= [\![\mathtt{c}]\!] + [\![\mathtt{x}]\!]^2 && \textit{[def. of fixpoint]} \\
[\![\mathtt{x}]\!]^2 - [\![\mathtt{x}]\!] + c &= 0 \\
[\![\mathtt{x}]\!] &\in \frac{1 \pm \sqrt{1 - 4\,c}}{2} && \textit{[by quadratic formula}^{60}\textit{]}
\end{aligned}
$$

*Assuming* $0 \le c \le \frac{1}{4}$*, the smaller solution* $\frac{1-\sqrt{1-4\,c}}{2}$ *corresponds to the sum over derivations. The larger solution* $\frac{1+\sqrt{1-4\,c}}{2}$ *is a fixpoint, but it is not equal to the sum over derivations.*

[60] Quadratic formula: $a\,x^2 + b\,x + c = 0 \wedge a \neq 0 \iff x = \frac{-b \pm \sqrt{b^2 - 4\,a\,c}}{2a}$.

## 4.3 Fixpoint Iteration

**Fixpoint iteration**[61] is a general scheme for approximating $[\![\cdot]\!]$. This scheme defines a sequence of approximate valuations $\nu^{(0)}, \nu^{(1)}, \ldots, \nu^{(T)}$, which approach $[\![\cdot]\!]$ as $T \to \infty$. The sequence is defined as

$$\begin{cases} \nu^{(0)} = \bot \\ \nu^{(t+1)} = \mathbf{T}_{\mathfrak{p}}\, \nu^{(t)} \end{cases} \tag{4.3}$$

where $\bot$ is the valuation for a program with no rules,[62] and $\mathbf{T}_{\mathfrak{p}}$ is the **step operator** for a program $\mathfrak{p}$:

$$(\mathbf{T}_{\mathfrak{p}}\, \nu)[x] \overset{\text{def}}{=} \bigoplus_{\substack{\mathbf{R} \in \mathfrak{p}, \\ (x \oplus\!= y_1 \circ \cdots \circ y_K) \in \Gamma(\mathbf{R})}} \nu[y_1] \circ \cdots \circ \nu[y_K] \qquad \text{for } x \in \mathbb{H} \tag{4.4}$$

Equation (4.5) is simply the abstraction of the right-hand side of (4.2) from Definition 25 into a function of the valuation $\nu$. We can rephrase the fixpoint equations (Definition 25) as a solution $\nu^*$ such that

$$\nu^* = \mathbf{T}_{\mathfrak{p}}\, \nu^* \tag{4.5}$$

Thus, like (4.2), (4.5) may admit multiple fixpoints or none at all. In practice, a suitable fixpoint $\nu^*$ is often computed by some variant of fixpoint iteration, which initializes $\nu$ to $\bot$, and repeatedly updates $\nu$ to $\mathbf{T}_{\mathfrak{p}}\, \nu$ until it converges to some $\nu^* = \mathbf{T}_{\mathfrak{p}}\, \mathbf{T}_{\mathfrak{p}} \cdots \mathbf{T}_{\mathfrak{p}}\, \bot$.[63]

[61] Fixpoint iteration—in the setting of logic programming—is known as forward chaining or bottom-up evaluation (Ullman, 1988; Goodman, 1999; McAllester, 2002; Eisner, Goldlust, et al., 2005).

[62] Note: according to Definition 22, the valuation $\wr\!\wr[\![\cdot]\!]$ for a program with no rules $\wr\!\wr$ is not a completely empty valuation, which would map all of its arguments to the value $\underline{0}$; rather, constants and built-in relations may have non-$\underline{0}$ values under $\bot = \wr\!\wr[\![\cdot]\!]$.

[63] For some programs, this procedure may never converge or may take infinite time to converge.

**Proposition 1.** *The valuation estimate at time $t$ given by fixpoint iteration $\nu^{(t)} = \mathbf{T}_{\mathsf{p}}^{t} \perp$ equals the total value of all derivations of height $\leq t$.*

*Proof (Sketch).* We start with $\nu^{(0)} = \perp$. From there, we build height $\leq 1$ derivations by extending the height $\leq 0$ derivations (i.e., constants and builtins appearing on the right-hand sides of the rules in the program). From there, each application of the step operation combines the valuations of height $\leq t$ derivations by applying the rules of the program to derive the values of all height $\leq t{+}1$ derivations.[64] ■

Fixpoint iteration is the mathematical basis for the algorithms we develop in § 5. Assuming that the valuation has a convergent sum, fixpoint iteration will diligently compute the sum (up to each height) until every height has been covered; thus, it converges in the limit to the value of $[\![\cdot]\!]$. In that later section, we will discuss and address computational inefficiencies and provide implementation details. We will also describe an accelerated fixpoint iteration variant based on Newton's method for solving equations.

In the boolean semiring, however, there is always a unique minimal fixpoint $\nu^*$, and for each item, $x$, $\nu[x]$ reaches $\nu^*[x]$ in finite time.

[64]This is a form of dynamic programming: $\nu^{(t)}$ is a memo table for a version of the computation, which makes the height $t$ an argument to break cycles.

# 4.4 Convergence Theory

This section briefly overviews the basic results found in the literature on semiring-weighted systems. Our goal is primarily to review; no new results are presented in this section.

An easy case where convergence is guaranteed is when the fixpoint equations are acyclic (i.e., the set of derivations is finite); here, there is no issue computing its value by fixpoint iteration or—better yet—by dynamic programming. The tricky cases are cyclic equations, as they denote sums over of infinitely many derivations, which may not be well-defined.

We have seen that *not all* fixpoints (Definition 25) are solutions to (4.1). However, fixpoint iteration (§4.3) is a procedure that sums the total value of derivations by their height (Proposition 1); thus, the fixpoint iterates approach $[\![\cdot]\!]$ assuming it is well-defined. To formalize what is meant by *approaching* a value, we require a notion of topology; we will describe a commonly used $\omega$-continuous topology in this section.

Kuich (1997) provides sufficient conditions for when fixpoint iteration—starting from a valuation that assigns all items the value $\underline{0}$—converges in the limit to $[\![\cdot]\!]$ (Definition 22).[65] The theory assumes that the semiring is $\omega$-continuous (Definition 28), which fortunately includes most semirings of interest in NLP. The class of $\omega$-continuous semirings has useful closure properties to ensure that the solution to the infinite sum over derivations equals the supre-mum of the partial sums (under an appropriate ordering relation). We note that some semirings—including $\langle \mathbb{R}, +, \cdot, 0, 1 \rangle$ (unless it is restricted to $\mathbb{R}_{\geq 0}$)—are not $\omega$-continuous—essentially because they include negative numbers, which means they can lead to undefined limits.

[65]We found out about this work thanks to Goodman (1998, §2.3.1–2.3.2).

Before defining $\omega$-continuous semirings, we define an ordering relation on semirings as it provides a topology on the semiring with which we can employ the usual mathematical concepts relating to convergence, such as the supremum.

**Definition 26.** *The **natural order** (denoted by $\sqsubseteq$) on the members of a semiring $\langle \mathbb{V}, \oplus, \circ, \underline{0}, \underline{1} \rangle$ is defined follows: $\forall\, a, b \in \mathbb{V} :\ (a \sqsubseteq b) \iff (\exists\, c \in \mathbb{V} : a \oplus c = b)$. The natural order is a preorder on $\mathbb{V}$. However, when $\sqsubseteq$ defines a partial order, we say that the semiring is **naturally ordered**.*

**Example 14** (Examples and Non-examples of Naturally Ordered Semirings)**.**

- *The natural order on real (and integer) semirings does not define a partial order: $\forall\, a, b$ there always exists $c = (b - a)$ that satisfies $a \oplus c = b$. Thus, $\sqsubseteq$ does not satisfy the antisymmetry requirement of partial order.*

- *The natural order on the counting semiring (or the nonnegative-real semiring) coincides with $\leq$: $\forall\, a, b$ there exists $c$ such that $a \oplus c = b$ if and only if $b - a \in \mathbb{N}$. In other words, if $a \sqsubseteq b \iff a \leq b$.*

- *The natural order on formal languages coincides with the $\subseteq$ relation on sets: For $L_1, L_2 \in 2^{\Sigma^\star}$, $L_1 \sqsubseteq L_2 \iff L_1 \subseteq L_2$.*

**Proposition 2.** *(Kuich, 1997, Theorem 2.1): Every naturally ordered semiring is monotonic (under $\sqsubseteq$), and $\underline{0}$ is its smallest element ($\forall\, a \in \mathbb{V} : \underline{0} \sqsubseteq a$).*

Thus, for naturally ordered semirings, each semiring operation is nondecreasing on the $\sqsubseteq$-lattice on $\mathbb{V}$. The natural order can be extended to valuations ($\mathbb{V}^{\mathbb{H}}$): let $\nu, \nu' \in \mathbb{V}^{\mathbb{H}} : \nu \sqsubseteq \nu' \iff \forall\, \boldsymbol{x} \in \mathbb{H} : \nu[\boldsymbol{x}] \sqsubseteq \nu'[\boldsymbol{x}]$. Next, we observe that the step operator $\mathbf{T}_{\mathfrak{p}}$ for a given program $\mathfrak{p}$ is monotone: $\nu \sqsubseteq \nu' \implies \mathbf{T}_{\mathfrak{p}}\, \nu \sqsubseteq \mathbf{T}_{\mathfrak{p}}\, \nu'$. Thus, the fixpoint iteration iterates increase $\sqsubseteq$ on each iteration until they eventually saturate; no oscillation can cause the sum to be ill-defined. However, additional technical details are required to formally push this idea into a theorem; specifically, it is sufficient for the semiring to be $\omega$-continuous (defined below).

**Definition 27.** *(Kuich, 1997, pages 611–612): A semiring is **complete** if it has an infinite sum operation ($\bigoplus$) that satisfies the following properties for all countably infinite sets $I$:*

*(i) Base cases:*

$\bigoplus_{i\in\emptyset} a_i = \underline{0}$, $\bigoplus_{i\in\{j\}} a_i = a_j$, *and* $\bigoplus_{i\in\{j,k\}} a_i = a_j \oplus a_k$ *for* $j \neq k$

*(ii) Infinite commutativity and associativity:*

$\bigoplus_{J\in\mathcal{J}} \bigoplus_{j\in J} a_j = \bigoplus_{i\in I} a_i$ *where $\mathcal{J}$ is any partitioning of $I$*

*(iii) Infinite distributivity:*

$c \circ \big(\bigoplus_{i\in I} a_i\big) = \bigoplus_{i\in I} (c \circ a_i)$ *and* $\big(\bigoplus_{i\in I} a_i\big) \circ c = \bigoplus_{i\in I} (a_i \circ c)$ *for all* $c \in \mathbb{V}$

**Definition 28.** *Suppose that $\langle \mathbb{V}, \oplus, \circ, \underline{0}, \underline{1} \rangle$ is a naturally ordered semiring, and its infinite sum operator is satisfies*

$$\bigoplus_{i\in\mathbb{N}} a_i = \sup\{a_1 \oplus \cdots \oplus a_i \mid i \in \mathbb{N}\} \quad \textit{for all infinite sequences } \langle a_i \rangle_{i\in\mathbb{N}} \tag{4.6}$$

*where the* $\sup$[66] *is taken with respect to the ordering* $\sqsubseteq$*. If this semiring is complete, then we say that the semiring is* $\omega$***-continuous***.

The definitive reference for the study of these assumptions is Kuich (1997). These are the primary theoretical assumptions studied in work on semirings (e.g., Goodman (1998), Goodman (1999), Esparza et al. (2007), Green, Karvounarakis, et al. (2007), and Mohri (2009)).

**Convergence-rate analysis.** It is possible to study fixpoint iteration's *rate* of convergence when $\mathfrak{p}$ is a *contraction mapping*. The limit analysis above makes weaker assumptions (i.e., $\omega$-continuity), which are sufficient to ensure convergence in the limit, but it does not characterize the *rate* of convergence.

To use the contraction mapping analysis in our setting, we require that the set of valuation mappings $\mathbb{V}^{\mathbb{H}}$ is a **complete**[67] **metric space**—i.e., it must be equipped with a **distance metric** $\rho\colon \mathbb{V}^{\mathbb{H}} \times \mathbb{V}^{\mathbb{H}} \to \mathbb{R}_{\geq 0}$. For example, in the case of the real semiring, we can use the Chebyshev distance $\rho(\nu, \nu') = \max_{x\in\mathbb{H}} |\, \nu[x] - \nu'[x] \,|$.[68] A *complete* metric space requires that every Cauchy sequence $\nu^{(0)}, \nu^{(1)}, \ldots$ has a limit also in $\mathbb{V}^{\mathbb{H}}$. We say $\nu^{(0)}, \nu^{(1)}, \ldots$ is a **Cauchy sequence** if, for any arbitrarily small distance $r > 0$, $\exists N > 0$ such that

[66] The **supremum** $\sup S$ of a subset $S \subseteq P$ from a partially ordered set $\langle P, \sqsubseteq\rangle$, is the least element in $P$ that is greater than or equal to each element of $S$ if such an element exists. The supremum is sometimes called the **least upper bound**.

[67] Note that the meaning of *complete* here is different from a *complete* semiring (Definition 27).

[68] Plenty of other distance metrics are possible for the real semiring, such as the Euclidean and Manhattan distance metrics.

$\forall m, n > N$, $\rho\big(\nu^{(n)}, \nu^{(m)}\big) < r$. We say the step operation $\mathbf{T}_{\mathfrak{p}}$ for a given program $\mathfrak{p}$ is a **contraction mapping** if

$$\forall \nu, \nu' \in \mathbb{V}^{\mathbb{H}} \colon \rho(\mathbf{T}_{\mathfrak{p}}\, \nu, \mathbf{T}_{\mathfrak{p}}\, \nu') \leq \lambda \cdot \rho(\nu, \nu')$$

with a **contraction rate** $\lambda < 1$.[69] Intuitively, a contraction mapping is a function that, when applied to any pair of points $\langle \nu, \nu' \rangle$, brings them closer together by at least a factor of $\lambda$. Note that it is *not* guaranteed that $\mathbf{T}_{\mathfrak{p}}$ is a contraction mapping for all programs $\mathfrak{p}$.

If $\mathbf{T}_{\mathfrak{p}}$ is a contraction mapping, then the **Banach fixpoint theorem** guarantees that the fixpoint iteration sequence $\nu^{(0)}, \nu^{(1)}, \ldots$ is Cauchy, and converges at a linear rate (as a function of $n$) to a unique fixpoint $\nu^*$:

$$\rho\big(\nu^{(n)}, \nu^*\big) \leq \lambda^n \cdot \rho\big(\nu^{(0)}, \nu^*\big) \tag{4.7}$$

Note that the quantity $\rho\big(\nu^{(0)}, \nu^*\big)$ in the right-hand side of (4.7) cannot be computed; fortunately, we can relax the bound so that it is computable by replacing this quantity with one that can be computed: the size of the first update $R = \rho\big(\nu^{(0)}, \nu^{(1)}\big)$. This gives the so-called **prior bound**:

$$\rho\big(\nu^{(n)}, \nu^*\big) \leq \frac{\lambda^n}{1-\lambda} \cdot R \tag{4.8}$$

This bound can be rearranged to bound the number of iterations ($n$) that are needed to ensure that the distance to $\nu^*$ is $\leq \epsilon$:

[69] Contraction mappings are a special case of Lipschitz-continuous functions where the function maps onto itself (i.e., the functions domain and range the same). Thus, the constant $\lambda$ is sometimes called a Lipschitz constant.

$$n \geq \frac{\log\left(\frac{\epsilon\,(1-\lambda)}{R}\right)}{\log \lambda} \quad \Longrightarrow \quad \rho\big(\nu^{(n)}, \nu^{*}\big) \leq \epsilon \tag{4.9}$$

Another useful bound is the **posterior bound**:

$$\rho\big(\nu^{(n)}, \nu^{*}\big) \leq \frac{\lambda}{1-\lambda} \cdot \rho\big(\nu^{(n)}, \nu^{(n-1)}\big) \tag{4.10}$$

The posterior bound relates the update's size at iteration $n$ to the remaining distance to $\nu^{*}$. The posterior bound shows that if the most recent update (at iteration $n$) is small, we are close to the solution. Thus, the posterior bound gives an approximate termination condition.

$$\frac{\lambda}{1-\lambda} \cdot \rho\big(\nu^{(n)}, \nu^{(n-1)}\big) \leq \epsilon \quad \Longrightarrow \quad \rho\big(\nu^{(n)}, \nu^{*}\big) \leq \epsilon \tag{4.11}$$

In practice, however, we do not know $\lambda$, so the threshold is just on $\rho\big(\nu^{(n)}, \nu^{(n-1)}\big)$.

# Summary

This section formalized the semantics of Dyna programs as a sum over semiring-weighted derivations. We described fixpoint iteration, the mathematical basis of the execution algorithms we will develop in § 5. We provided an overview of the sufficient conditions for which fixpoint iteration converges to $[\![\cdot]\!]$.

# Chapter 5

# Inference Algorithms

This section develops general algorithms for performing inference in Dyna programs (i.e., evaluating $[\![\cdot]\!]$). These general algorithms are why Dyna programs do not require the specification of implementation details: programmers provide a specification of the *logic*, and the Dyna system takes care of the *control flow*. We describe several algorithms:

- In §5.1, we describe three implementations of fixpoint iteration of increasing sophistication: naïve (§5.1.1), semi-naïve (§5.1.2), and prioritized (§5.1.3).
- In §5.2, we then describe an *accelerated* fixpoint algorithm, which is based on a version of Newton's method for solving fixpoint equations over a semiring (Esparza et al., 2007).
- In §5.3, we describe a specialized solver for certain linearly recursive programs. This specialized solver is leveraged as a subroutine by the accelerated algorithm.

- In §5.4, we describe a unified, hybrid algorithm that solves large programs by breaking them up into smaller subprograms that can be solved separately with the best applicable method.

**Solution representations.** Recall from §2 that we represent the answer to a query as a program of a restricted form where each of its rules is simple. A rule is simple if its body contains only constant values and builtins. We call a program consisting of only simple rules a **simple program**. In this chapter, we use simple programs to represent the valuation function $[\![\cdot]\!]$. This allows us to compactly represent the solutions to certain programs with infinite valuations, such as[70]

```
{ x(I) += (I > 0).
{ x(I) += 0.5 * x(I).
```

$\xrightarrow{\text{solve}}$

```
{ x(I) += 2 * (I > 0).
```

This generalizes prior work by Eisner, Goldlust, et al. (2005), which required solutions to be represented as a hash table mapping ground terms to their values. In our notation, that translates to a restriction that each rule in the solution has a ground head, and its body is a constant value. The consequence of their assumption is that they can only represent the solutions to programs with finitely many non-$\underline{0}$ items.

We will use simple programs in our algorithms to represent intermediate valuation functions. We will use the result-stream manipulation operations (developed in §3.1) in the algorithms of this section, as they make working it possible to manipulate these program representations seamlessly.

[70]We provide additional examples in §A.

# 5.1 Algorithms for Fixpoint Iteration

In §4.3, we discussed the mathematical foundation for fixpoint iteration. Here, we will describe the algorithm procedurally and discuss implementation details. In the logic programming literature, fixpoint iteration is known under the names *bottom-up evaluation* or *forward chaining*.[71,72] We will give three algorithms of increasing sophistication: naïve (§5.1.1), semi-naïve (§5.1.2), and prioritized (§5.1.3). We do not recommend using the algorithms naïve (§5.1.1) and semi-naïve (§5.1.2); we have included them for pedagogical purposes to motivate the prioritized algorithm (§5.1.3). Our version of prioritized fixpoint iteration generalizes the forward chaining algorithm of Eisner, Goldlust, et al. (2005) to support non-range-restricted rules, built-in relations, and delayed constraints.

## 5.1.1 Naïve Fixpoint Iteration

This section describes our simplest implementation of fixpoint iteration. In the pseudocode provided in Algorithm 6, we can see that the algorithm is a remarkably straightforward translation of the mathematical definition of (4.4) thanks to our abstractions for manipulating streams (§3.1).[73]

In the code, we initialize the **chart** $\mathfrak{m}$ to the empty program $\Lbag\Rbag$ (i.e., the program with no rules) so its valuation equals $\bot$, as in (4.3). We then repeatedly

[71] E.g., Ullman (1988), Goodman (1999), McAllester (2002), and Eisner, Goldlust, et al. (2005).

[72] Our version is most like Eisner, Goldlust, et al. (2005), as it has been generalized to support semiring values.

[73] These abstractions allow us to specify each of our inference algorithms *as if* the program's rules do not contain variables.

**Algorithm 6** naive: Solve program $\mathfrak{p}$ by (naïve) fixpoint iteration

```
1.  def naive(𝔭):
2.    ▷ This procedure implements fixpoint iteration (4.3) on the function step_𝔭.
3.    𝔪 ← ⦃⦄                ▷ ⦃⦄ is a program with no rules, its valuation equals ⊥ as in (4.3).
4.    while true:
5.      𝔪′ ← step_𝔭(𝔪)                                   ▷ Apply the step operation
6.      if 𝔪 = 𝔪′: return 𝔪
7.      𝔪 ← 𝔪′

8.  ▷ The procedure step_𝔭(𝔪) implements T_𝔭(ν) from (4.4); here, the valuation ν is represented as
    a program 𝔪.
9.  def step_𝔭(𝔪):
10.   𝔪′ ← ⦃⦄
11.   for (x ⊕= y_1 ∘ ⋯ ∘ y_K) ∈ 𝔭:
12.     𝔪′[x] ⊕= 𝔪[y_1] ∘ ⋯ ∘ 𝔪[y_K]
13.   return 𝔪′
```

apply the step function $\mathsf{step}_{\mathfrak{p}}$, which implements $\mathbf{T}_{\mathfrak{p}}$ from (4.4), until a fixpoint is reached (i.e., the condition on line 6.6 is satisfied).

We see that $\mathsf{step}_{\mathfrak{p}}(\mathfrak{m})$ implements $\mathbf{T}_{\mathfrak{p}}(\mathfrak{m}[\![\cdot]\!])$ as it sums the contributions of each rule in the program $(x \mathrel{\oplus=} y_1 \circ \cdots \circ y_K) \in \mathfrak{p}$ by instantiating its body against the current chart $\mathfrak{m}$ with the help of the result-stream abstractions: $\mathfrak{m}[y_1] \circ \cdots \circ \mathfrak{m}[y_K]$. Line 6.12 adds the contributions of the current rule to the next iterate $\mathfrak{m}'$ using $\mathfrak{m}'[x] \mathrel{\oplus=} \mathfrak{m}[y_1] \circ \cdots \circ \mathfrak{m}[y_K]$. It is straightforward to verify that $\mathfrak{m}$ and $\mathfrak{m}'$ are always simple programs.

We note that the runtime of Algorithm 6 is not as good as it could be. In the following sections, we will improve upon it in several ways.

**Termination condition.** Our algorithm's termination condition (line 6.6) is $\mathfrak{m} = \mathfrak{m}'$. Termination can be guaranteed for the following cases:

- Finite acyclic programs (regardless of the semiring).

- Finite programs over a semiring that is guaranteed to saturate after a finite number of iterations. The family of $K$-closed semirings (Definition 19) is a setting where this is always true; this family includes many important semirings such as the boolean semiring and the optimization semirings (§3.3). However, it does not include many important semirings, such as the real semiring.

**Approximate termination condition.** Defining an approximate termination condition to replace line 6.6 depends on the specific semiring.[74] A natural choice for the real semiring is $\forall x\colon\ |\,\mathfrak{m}[x] - \mathfrak{m}'[x]\,| \le \epsilon$ for some small threshold $\epsilon \ge 0$. For other semirings, $|\,v - v'\,|$ should be replaced with an appropriate distance metric $\rho(v, v')$. This choice is inspired by the posterior bound (4.10) in the Banach fixpoint theorem (discussed in §4.4). The number of fixpoint iterations required to reach the threshold $\epsilon$ may be difficult to bound unless the step function is a contraction mapping (§4.4).

### 5.1.2 Semi-Naïve Fixpoint Iteration

The seminaive algorithm is a more efficient implementation of Algorithm 6. The improvement is based on the observation that on iteration $h+1$, many of the values for items in $\mathfrak{m}^{(h)}$ are unchanged; however, the step operator *rederives* them for $\mathfrak{m}^{(h+1)}$. The seminaive algorithm refines the induction on height by precisely deriving the set of derivations with height $=h$ rather than $\le h$ on each step. Doing so improves efficiency, as each step does not need to rederive derivations of lower heights.

[74]Our implementation has hooks for defining approximate equality for a given semiring.

Pseudocode is given in Algorithm 7. The algorithm's structure is similar to Algorithm 6, except that in addition to $\mathfrak{m}$ (and $\mathfrak{m}'$), we maintain the values of new heights $\partial$ (and $\partial'$). It is straightforward to verify the invariant that $\mathfrak{m}$, $\partial$, $\mathfrak{m}'$, and $\partial'$ are simple programs. We argue for the correctness of the algorithm in Proposition 3.

**Algorithm 7** seminaive: Solve program $\mathfrak{p}$ by semi-naïve fixpoint iteration

```
1.  def seminaive(𝔭):
2.    𝔪 ← ⦃⦄
3.    ∂ ← step_𝔭(𝔪)                        ▷ Initialize with step from Algorithm 6
4.    while true:
5.      ⟨𝔪′, ∂′⟩ ← step′_𝔭(𝔪, ∂)
6.      if 𝔪 = 𝔪′: return 𝔪
7.      ⟨𝔪, ∂⟩ ← ⟨𝔪′, ∂′⟩

8.  ▷ Incremental step function; applies change ∂ to chart 𝔪 and derives a new change ∂′.
9.  def step′_𝔭(𝔪, ∂):          Should omit constants and builtins;
10.   ∂′ ← ⦃⦄  ←                 these are added by line 7.3.
11.   𝔪′ ← 𝔪 ⊕ ∂                           ▷ Accumulate changes
12.   for (x ⊕= y_1 ∘ ··· ∘ y_K) ∈ 𝔭:
13.     for k ∈ [1:K]:[75]
14.       ∂′[x] ⊕= 𝔪′[y_[:k)] ∘ ∂[y_k] ∘ 𝔪[y_(k:]]
15.   return ⟨𝔪′, ∂′⟩
```

**Proposition 3.** *On iteration $h$ of Algorithm 7 (line 7.7), the value of $\mathfrak{m}$ is the valuation function for heights $<h$, and the value of $\partial$ is the valuation function restricted height $=h$.*

*Proof (Sketch).* According to Definition 24, we obtain a $=h+1$ derivation when we apply one rule to a collection of *at least one* $=h$ derivation and any number of

[75] For efficiency, the join algorithm should use $\partial[y_k]$ as its outermost loop.

$<h$ derivations. To illustrate, consider the case of a derivation with 2 children. A rather direct way to partition the events that create height $=h+1$ derivations is the following three cases, which identify all positions where the $=h$ subderivation can appear:

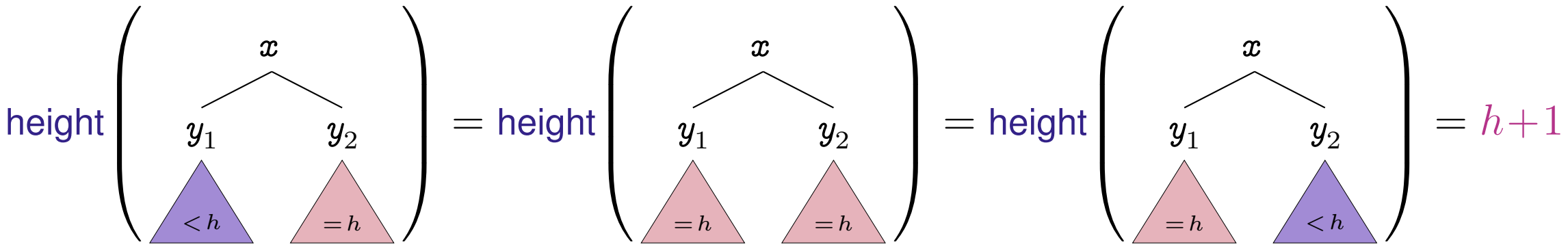

However, a more efficient—albeit less obvious—partitioning of the event above merges the first two cases into one case ($\leq h$)—giving two cases instead of three:

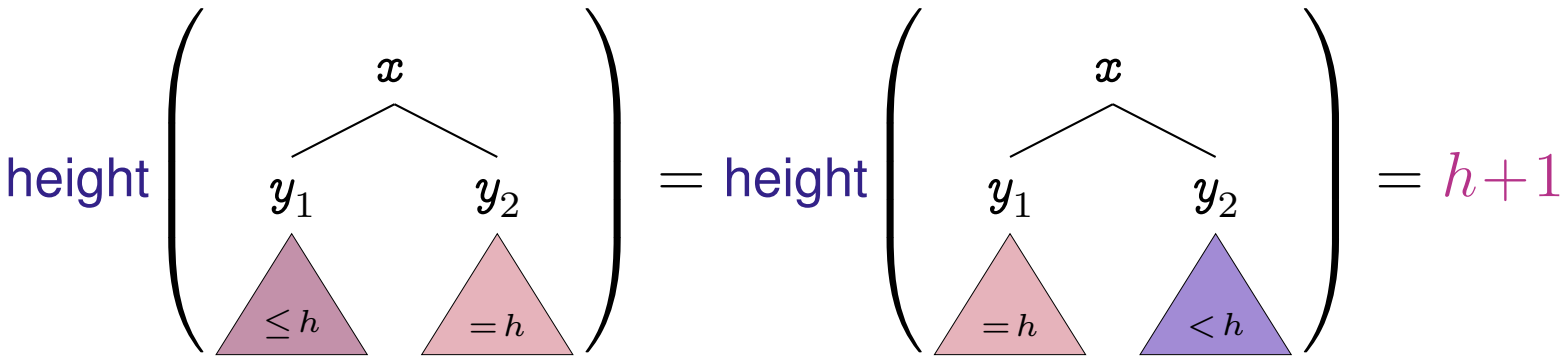

What we have done here is to identify events with their rightmost $=h$ subderivation.[76] We depict the general case below (for $k = 1 \ldots K$):

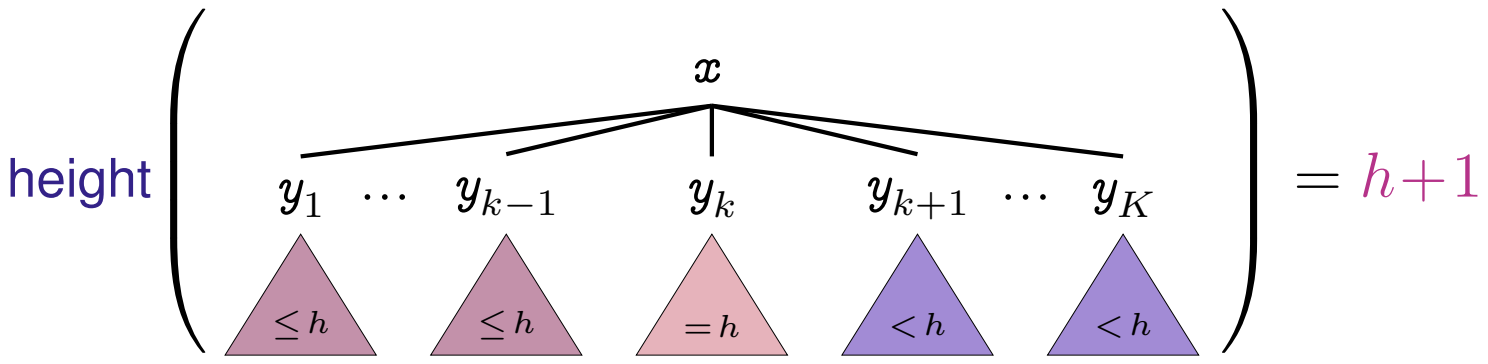

(5.1)

We refer to this strategy for grouping events as the **cursor trick**. We refer to the position of the rightmost $=h$ subtree as the *cursor* as it moves from left to right in a given rule body to control which version of each subgoal's value ($\mathfrak{m}$, $\mathfrak{d}$, or $\mathfrak{m}'$) is used in the update.

[76]When we consider derivations with more than two children, the benefit of the latter decomposition becomes more apparent. In the latter decomposition, there are $K$ cases to consider, whereas the former has $\mathcal{O}(2^K)$ because it considers each subset of positions where the maximal height can appear.

Algorithm 7 operationalizes this construction as follows. The function $\mathsf{step}'$ derives the values for the set of derivations of the next height ($=h+1$) by applying each rule in the program once: for a given rule $x \oplus= y_1 \circ \cdots \circ y_K \in \mathfrak{p}$, we consider plugging in each item from $\mathfrak{d}$ as the rightmost $=h$ child. We do so by looping over values of the cursor $k$ from 1 to $K$ (line 7.13), and applying the update on line 7.14 (repeated below):

$$\mathfrak{d}'[x] \oplus= \mathfrak{m}'[y_1] \circ \cdots \circ \mathfrak{m}'[y_{k-1}] \circ \mathfrak{d}[y_k] \circ \mathfrak{m}[y_{k+1}] \circ \cdots \circ \circ \mathfrak{m}[y_K]$$

We see that the update has the same pattern as (5.1), which is why $\mathfrak{d}'$ corresponds to values contributed by the set of the derivation of height $=h+1$. ■

**Example 15.** *We give a simple example of the cursor trick in action.*

```
91 x += y * y * y.
92 y += 2.
93 y += z.
94 z += 1.
```

*We begin by initializing* $\mathfrak{m} = \{\!\!\{\}\!\!\}$ *and* $\mathfrak{d} = \mathsf{step}_{\mathfrak{p}}(\mathfrak{m})$. *Then, on the first iteration:*

$$\mathfrak{m} = \{\!\!\{\}\!\!\} \quad \mathfrak{d} = \begin{cases} \texttt{y += 2.} \\ \texttt{z += 1.} \end{cases} \quad \overset{\mathsf{step}'}{\Longrightarrow} \quad \mathfrak{m}' = \begin{cases} \texttt{y += 2.} \\ \texttt{z += 1.} \end{cases} \quad \mathfrak{d}' = \begin{cases} \texttt{x += 2 * 2 * 2.} \\ \texttt{y += 1.} \end{cases}$$

*Next, we see the cursor trick in action in* $\mathfrak{d}'$*:*

$$\mathfrak{m} = \begin{cases} \texttt{y += 2.} \\ \texttt{z += 1.} \end{cases} \quad \mathfrak{d} = \begin{cases} \texttt{x += 8.} \\ \texttt{y += 1.} \end{cases} \quad \overset{\mathsf{step}'}{\Longrightarrow} \quad \mathfrak{m}' = \begin{cases} \texttt{x += 8.} \\ \texttt{y += 3.} \\ \texttt{z += 1.} \end{cases} \quad \mathfrak{d}' = \begin{cases} \texttt{x += 1 * 2 * 2.} \\ \texttt{x += 3 * 1 * 2.} \\ \texttt{x += 3 * 3 * 1.} \end{cases}$$

*Next, we apply the update and see that* x *has the correct value of* 27 *in* $\mathfrak{m}'$*:*

$$\mathfrak{m} = \begin{cases} \texttt{x += 8.} \\ \texttt{y += 3.} \\ \texttt{z += 1.} \end{cases} \quad \mathfrak{d} = \begin{cases} \texttt{x += 19.} \end{cases} \quad \overset{\mathsf{step}'}{\Longrightarrow} \quad \mathfrak{m}' = \begin{cases} \texttt{x += 27.} \\ \texttt{y += 3.} \\ \texttt{z += 1.} \end{cases} \quad \mathfrak{d}' = \{\!\!\{\}\!\!\}$$

*The* seminaive *halts on the next iteration with the final chart:*

$$\mathfrak{m} = \begin{cases} \texttt{x += 27.} \\ \texttt{y += 3.} \\ \texttt{z += 1.} \end{cases}$$

**Efficiency.** In logic programming, values are always boolean; thus, once a *single* derivation for a given item is found, we no longer need to consider expanding. In other words, once an item has a single derivation, its value has *converged*—i.e., it has reached its fixpoint. Thus, the seminaive strategy is particularly efficient in the boolean semiring because it only does work proportional to the change $\mathfrak{m}$, and it does not perform any unnecessary rule application.[77] However, when considering general semiring values, we see that seminaive algorithms can be inefficient. In particular, seminaive will combine values for items before they have converged, and it will repeat the combination each time a taller derivation of an item is discovered.

**Example 16.** *Below is an example family of acyclic programs with a linear size in a parameter $n$. On this family of programs, semi-naive runs in $O(n^2)$ time due to unnecessary repropagation.*[78]

```
x(I+1) += x(I), I < n.
x(1) += y(I).
y(I+1) += y(I), I < n.
y(1) += 1.
```

*We can visualize the direct dependencies among the items as a graph*

[77] This requires that the loop over rules in line 7.12 is modified so that only rules with a non-$\underline{0}$ grounding are enumerated.

[78] We are using a short-hand that allows certain built-in relations, e.g., `I+I`, to be written in place: `x(I+1) += x(I), I < n` translates to `x(I') += x(I), I' is I+1, I < n`.

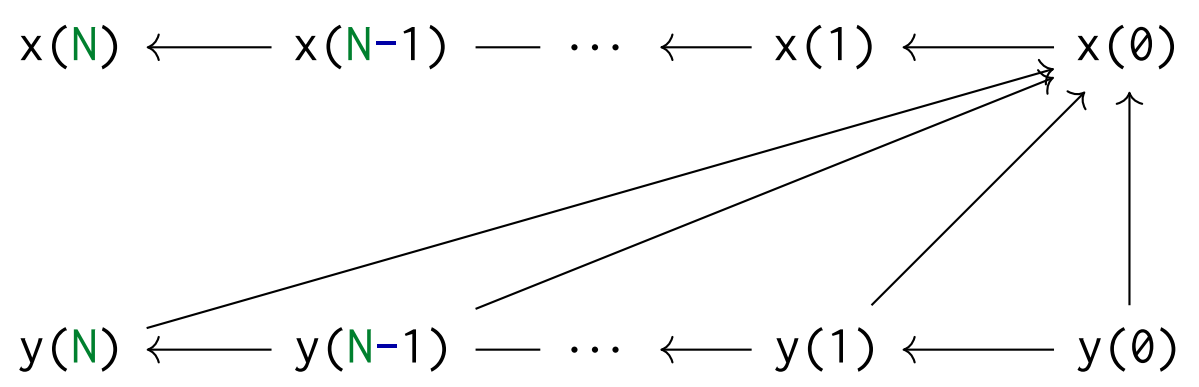

*At a high level, the reason why* seminaive *runs in quadratic time is because* x(0)*'s value will update $\mathcal{O}(n)$ times and, thus, items that depend on* x(0) *will update $\mathcal{O}(n)$ times as well. After a warm-up period, $\partial$ will have $\mathcal{O}(n)$ values propagated per iteration. These big changes happen $\mathcal{O}(n)$ times before converging, which gives the quadratic time.*

*In the next section, we will develop another algorithm that allows us to compute the value of each item in this program in topologically sorted order—achieving a much-improved runtime of $\mathcal{O}(n)$ on this example.*

### 5.1.3 Prioritized Fixpoint Iteration

In this section, we will give a more flexible strategy for semi-naive evaluation, which uses an agenda (work list) to safely eliminate height as a synchronization barrier. The benefit is that we can avoid *repropagation*: rather than advancing all items from height $\leq h$ to height $\leq h{+}1$ in lockstep, we can allow some items to advance way beyond the others. Ideally, we would defer propagation until each item reaches its fixpoint value. Once a set of items converges, we do not need to continue combining with other converged items (we only need to combine a set of converged items once). In cyclic graphs, blocking updates until a fixpoint is reached is impossible (deadlock), but we can focus updates on propagating

**Algorithm 8** agenda: Solve program $\mathfrak{p}$ by agenda-based fixpoint iteration

1. **def** agenda($\mathfrak{p}$):
2. $\quad \mathfrak{m} \leftarrow \{\!\!\{\}\!\!\}$; $\mathfrak{m}' \leftarrow \{\!\!\{\}\!\!\}$
3. $\quad \partial \leftarrow \text{step}_{\mathfrak{p}}(\mathfrak{m})$ ▷ *Initialize with* step *from Algorithm 6*
4. $\quad$ **while** $\partial \neq \{\!\!\{\}\!\!\}$: — *May contain delayed constraints*
5. $\quad\quad (y' \mathrel{\oplus}= v) \leftarrow \partial.\text{pop}()$ ▷ *Pick any update from* $\partial$ *to apply*
6. $\quad\quad \mathfrak{m}'[y'] \mathrel{\oplus}= v$
7. $\quad\quad$ **if** $\mathfrak{m}'[y'] = \mathfrak{m}[y']$: **continue** ▷ *Update has no effect; no need to propagate*[80,81]
8. $\quad\quad$ **for** $(x \mathrel{\oplus}= y_1 \circ \cdots \circ y_K) \in \mathfrak{p}$:
9. $\quad\quad\quad$ **for** $k \in [1{:}K]$:
10. $\quad\quad\quad\quad \partial[x] \mathrel{\oplus}= \mathfrak{m}'[y_{[:k)}] \circ v[y_k = y'] \circ \mathfrak{m}[y_{(k:]}]$
11. $\quad\quad \mathfrak{m}[y'] \mathrel{\oplus}= v$ ▷ *Now,* $\mathfrak{m} = \mathfrak{m}'$
12. $\quad$ **return** $\mathfrak{m}$

within a cyclic subgraph before moving on. In other words, we can prioritize updates within a strongly connected component (Definition 12), allowing it to reach a fixpoint before advancing to the next strongly connected component.[79]

The key to allowing this generalization is to generalize $\partial$ from Algorithm 7 to buffer changes before they are committed to $\mathfrak{m}$. This allows some items to advance their valuations to larger heights while others are deferred. The values accumulated in $\partial$ correspond to the sums over heights that have yet to be accumulated into $\mathfrak{m}$—it is no longer limited to a single next height. Assuming that no updates are deferred indefinitely, the agenda strategy's valuation estimate should converge to $[\![\cdot]\!]$ (under the same conditions that seminaive would converge).

The policy for determining which **update** to pop from the **agenda** $\partial$ (line 8.5) is called the **prioritization policy**. It is typically implemented using a priority queue. We provide guidance on prioritization function design later in §5.1.4.

[79]Goodman (1998) calls SCCs "looping buckets."

Pseudocode for our algorithm is provided in Algorithm 8.[82] We note that Algorithm 8 is based closely on the agenda-based algorithm given in Eisner, Goldlust, et al. (2005). We have made subtle but necessary modifications to their algorithm to support nonground updates.[83] Much like Algorithm 7, the agenda-based algorithm uses the cursor trick,[84] which—at least conceptually—requires multiple versions of the value of the **driver** item $y'$ that is popped from the agenda ($\mathfrak{d}$): the new version $\mathfrak{m}'[y']$, the change $\mathfrak{d}[y']$, and the old version $\mathfrak{m}[y']$. We call the queries $\mathfrak{m}'[y_{[:k)}]$ and $\mathfrak{m}[y_{(k:]}]$ on line 8.10 **passenger** queries. If the $y'$ does not match multiple subgoals, then the passenger queries will see the values with the most up-to-date value that is not pending on the change buffer. The update $y' \oplus\!= v$ is only *committed* to $\mathfrak{m}$ on line 8.11 after the update has been carefully pushed through the rules in the program (using the cursor trick to avoid double counting the update); at this point $\mathfrak{m} = \mathfrak{m}'$. The strategy used by Eisner, Goldlust, et al. (2005) is essentially the same. However, their strategy is only correct for ground updates: it does not extend to the case of nonground

[80]Note that it is safe to allow false negatives on this equality test on line 8.7, but not false positives. However, false negatives may result in more iterations for $\mathfrak{d}$ to become empty (line 8.4).

[81]The equality test on line 8.7 is between the results of the query specified by $y'$ in each of $\mathfrak{m}'$ and $\mathfrak{m}$. When $y'$ is nonground, this test may no longer run in constant time as we are comparing the equality of result streams.

[82]Much like Algorithm 7 is straightforward to verify the invariant that $\mathfrak{m}$, $\mathfrak{d}$, $\mathfrak{m}'$, and $\mathfrak{d}'$ are simple programs.

[83]The key difference is that we represent the chart and agenda as programs, whereas Eisner, Goldlust, et al. (2005) require them to each be a hash table mapping ground items to semiring values. This representation difference allows us to represent certain infinite valuation functions efficiently. A more superficial difference is that we maintain two versions of the chart $\mathfrak{m}$ and $\mathfrak{m}'$, whereas Eisner, Goldlust, et al. (2005) represent the old value of the driver (i.e., that in $\mathfrak{m}$) using a local variable. A nonground update would cause Eisner, Goldlust, et al. (2005)'s pseudocode to loop over the infinite Herbrand universe matching the update; their implementation, on the other hand, would throw an exception in this case.

[84]A version of the cursor trick appears in Eisner, Goldlust, et al. (2005).

updates that may contain delayed constraints.[85]

### 5.1.4 Prioritization Functions

We use the prioritization policies in each iteration of Algorithm 8 (line 8.5) to determine which update to propagate next. In principle, any prioritization policy that does not *starve* any item from popping results in convergence (in the limit) to a fixpoint—assuming that fixpoint iteration converges at all. A simple default policy with this property is first-in, first-out (FIFO); however, FIFO is essentially just the seminaive algorithm, which we saw in Example 16 can be unnecessarily slow to converge.

This section describes a particular form of prioritization policy called a prioritization function. The way it works is that the prioritization function assigns each update in $\partial$ a numeric priority at the time the update is *pushed* (i.e., when it is added to $\partial$ on either line 8.3 or line 8.10).[86] When we call $\partial$.pop() on line 8.5, the highest-priority update (with ties broken arbitrarily) is returned, and it is removed from $\partial$. The priority values of each update can be stored in a priority queue data structure that indexes the updates in $\partial$ to make these push and pop operations efficient.

An approach to prioritization proposed by Goodman (1998) first runs the program in the boolean semiring to estimate the set of items $\overline{\mathfrak{m}}$ that have non-$\underline{0}$

[85]Please refer to §A for examples involving nonground updates and delayed constraints.
[86]It is also possible to assign items non-numeric scores; the only requirement is that the priority values are partially ordered.

**Algorithm 9** $\overline{\mathsf{support}}$: Estimate an upper bound on the support set of $\mathfrak{p}$.

1. **def** $\overline{\mathsf{support}}(\mathfrak{p})$:
2. $\quad\overline{\mathfrak{p}} \leftarrow \mathsf{booleanize}(\mathfrak{p})$ ▷ *Convert $\mathfrak{p}$ into a program $\overline{\mathfrak{p}}$ in the boolean semiring by replacing each non-$\underline{0}$ constant in $\mathfrak{p}$ with* `true`.[88]
3. $\quad\overline{\mathfrak{m}} \leftarrow \mathsf{solve}(\overline{\mathfrak{p}})$ ▷ *Warning: May not terminate when $\overline{\mathfrak{p}}$ has infinitely many* `true` *items (see text).*
4. $\quad$**return** $\mathsf{disjoin}(\overline{\mathfrak{m}})$ ▷ *Make sure $\overline{\mathfrak{m}}$'s rules denote disjoint sets.*
5. **def** $\mathsf{disjoin}(\overline{\mathfrak{m}})$:
6. $\quad$▷ *Transform $\overline{\mathfrak{m}}$ into a program whose rules denote nonoverlapping sets*
7. $\quad$▷ *This procedure can be sped up by partitioning $\overline{\mathfrak{m}}$ into hash buckets where items in the same bucket might overlap, and those in different buckets cannot.*
8. $\quad$**while** `true`:
9. $\quad\quad$find $r, s \in \overline{\mathfrak{m}}$ such that $r \cap s$ and $r \neq s$; if no such pair exists, **break**
10. $\quad\quad$▷ *Replace the overlapping items with their tightest upper bound (*$\mathsf{merge}$*).*[89,90]
11. $\quad\quad\overline{\mathfrak{m}} \leftarrow (\overline{\mathfrak{m}} \smallsetminus \{r, s\}) \cup \{\mathsf{merge}(r, s)\}$
12. $\quad$**return** $\overline{\mathfrak{m}}$

value.[87] Algorithm 9 provides pseudocode this approach; it has been generalized beyond Goodman (1998)'s setting to allow nonground items.

Here any solver ($\mathsf{solve}$) can be used to solve $\overline{\mathfrak{p}}$, such as $\mathsf{seminaive}$, or $\mathsf{agenda}$ (with any nonstarving prioritization function). However, for this method to be effective, we need $\mathsf{solve}$ to terminate.[91] Fortunately, we only require an *upper bound* on the support to move forward. In §6.1, we describe a "relaxed" forward chaining method that can be used to estimate a upper bound $\overline{\mathfrak{m}}$ in finite time.

[87] In §6.1, we further discuss the use of "booleanization" to estimate the set of non-$\underline{0}$ items.

[88] Here $\mathsf{booleanize}$ guarantees that $\forall \boldsymbol{x} \in \mathbb{H}\colon \mathfrak{p}[\![\boldsymbol{x}]\!] \neq \underline{0} \Longrightarrow \overline{\mathfrak{p}}[\![\boldsymbol{x}]\!] = \mathtt{true}$.

[89] Here $\mathsf{merge}$ is the anti-unification algorithm (Reynolds, 1970; Plotkin, 1970). For example, the pair $\boldsymbol{x} =$ `f(3,Y)`, $\boldsymbol{y} =$ `f(X,4)` is rounded up to $\boldsymbol{z} =$ `f(X,Y)`.

[90] Using $\mathsf{merge}$ loses precision as it finds an *upper bound* on the original $\overline{\mathfrak{m}}$ that is disjoint. In principle, we could completely remove the imprecision introduced by $\mathsf{merge}$. By replacing line 9.11 with $\overline{\mathfrak{m}} \leftarrow (\overline{\mathfrak{m}} \smallsetminus \{r, s\}) \cup \{r \cap s, r \smallsetminus s, s \smallsetminus r\}$. However, this choice has the computational challenge of efficiently representing differences of sets of nonground terms (Filardo, 2017, chapter 4); we use $\mathsf{merge}$ in practice.

[91] In some cases, a booleanization may drastically overestimate the set of items that forward chaining needs to care about. For example, a program like `goal += p(N). p(z). p(s(N)) += 0.5 * p(N).` has infinitely many nonzero items, but the value of `p(s`$^k$`(z))` is very small as a function of $k$: $[\![$`p(s`$^k$`(z))`$]\!] = 0.5^k$ so the values go to $0$ quickly. So, we obtain a good estimate of `goal` with just a few iterations of fixpoint iteration. However, booleanization enumerates an infinite number of items since $\overline{\mathfrak{p}}[\![$`p(s`$^k$`(N))`$]\!] = \mathtt{true}$ for all $k \in \mathbb{N}$.

**Algorithm 10** incoming and outgoing: Lazily enumerate incoming and outgoing arcs in program $\mathfrak{p}$'s dependency graph given its estimated support $\overline{\mathfrak{m}}$.

```
▷ Lazily enumerate head-to-subgoal arcs
def incoming(x):
  for (y_1 ∘ ⋯ ∘ y_K) ∈ p[x]:
    for _ ∈ m̄[y_1 ∘ ⋯ ∘ y_K]:
      for k ∈ [1:K]:
        yield y_k
```

```
▷ Lazily enumerate subgoal-to-head arcs
def outgoing(y′):
  for (x ⊕= y_1 ∘ ⋯ ∘ y_K) ∈ p:
    for k ∈ [1:K]:
      for _ ∈ m̄[y_[:k) ∘ (y_k=y′) ∘ y_(k:]]:
        yield x
```

From the estimated set of non-$\underline{0}$-valued items $\overline{\mathfrak{m}}$, we can find a topologically ordered list of strongly connected components of the item–item dependency graph, $G = \langle V, E \rangle$ where $V$ is the set of items in $\overline{\mathfrak{m}}$, and $E$ is the set of edges given by the head-to-subgoal relationship in the instantiated program. Algorithm 10 describes how to enumerate the incoming (and outgoing) arcs lazily.[92,93]

The prioritization function is then the mapping from each item to the (integer) position of its SCC in the reverse toposort. Determining the toposorted SCCs with Algorithm 5 runs in linear time in the instantiated program's size $\mathcal{O}(|V| + |E|)$ using linear space in the number of items $\mathcal{O}(V)$.[94] This same SCC analysis will also be used in our hybrid algorithm (§5.4).

[92]For completeness, we have also included outgoing.

[93]Note that the pseudocode implicitly relaxes any delayed constraints coming from $\overline{\mathfrak{m}}$ because it ignores them in the "**for** _ $\in \overline{\mathfrak{m}}[\ldots]$" loops. This may overestimate the set of arcs in $G$, and consequently, the SCC analysis will be coarse. We also assume that builtins *delay* rather than throw instantiation faults (§3.1.2.2).

[94]Goodman (1998) missed a trick when they first proposed this scheme: the graph edges do not need to be materialized to run Algorithm 5. Specifically, we can use incoming (or outgoing) to enumerate the dependents of a node lazily. This procedure only requires constant space because the queries and unification operations performed therein only require constant space using our result-stream implementation (§3.1.2.1). Simultaneously with this dissertation, Eisner (2023) published a detailed overview of algorithms for efficient inference in weighted deduction systems. Both our approach and theirs leverage the observation above to keep space complexity linear in $|V|$. Their "weighted cyclic forward-chaining algorithm" is closely related to our agenda with SCC-based prioritization. The difference is that our agenda algorithm coordinates all of its control flow through an agenda; they use a recursive algorithm for recursive solving of each SCCs (like our hybrid algorithm (§5.4)). Their approach considers weighted deduction systems that are not limited to semiring weights.

We note that if the values of only a subset of items $\text{targets} \subseteq \mathbb{H}$ are required, then we may wish to use the toposorted SCCs of $\text{sccs}(\text{incoming}, \text{targets})$, as it will only enumerate SCCs that targets depend on. Skipping SCCs that are not needed by the targets nodes can provide large savings in cases where the omitted SCCs are slow to converge, or the semiring operations are expensive.

To avoid additional factors in the runtime, we can use a bucket queue (Dial, 1969) to implement the pop operation in this case efficiently. With this prioritization scheme, we can solve Example 16 using Algorithm 8 in linear time, which is considerably better than the quadratic runtime of Algorithm 7. Notice that if $G$ contains cycles, each cycle is "solved" before moving to the next. If convergence is not guaranteed (e.g., in the real semiring), this can create a barrier to progress on other program components (i.e., starvation of updates).

**Discussion.** Below are some alternative techniques for specifying a prioritization function:

- User-customizable prioritization functions (Eisner and Filardo, 2011).
- Synthesize a prioritization function using static analysis and satisfiability modulo theories (Colón and Sipma, 2001; Podelski and Rybalchenko, 2004; Bagnara et al., 2010).
- Lastly, we mention that there may be benefits to using carefully designed prioritization *within* an SCC. In other related settings, prioritization has been shown to improve convergence rates.[95] Future work may want to investigate

[95]For example, solvers for Markov decision processes (see Wingate and Seppi (2005) for a survey),

methods based on minimum feedback arc sets, which may be regarded as an approximate topological ordering when cycles are present (e.g., Baharev et al. (2016)), and machine learning of prioritization heuristics (e.g., Jiang et al. (2012) and Vieira, Francis-Landau, et al. (2017)).

## 5.2 Newton's Method

Our newton method is an accelerated solver for fixpoint problems, based on Esparza et al. (2007)'s generalization of Newton's algorithm for solving fixpoint equations over a semiring. What is interesting about newton relative to the fixpoint iterations methods is that it can sum infinitely many derivations in a single iteration. More specifically, each iteration sums the set of derivations by increasing *dimension* (defined below) rather than by its height as in fixpoint iteration (Proposition 1).[96]

Etessami and Yannakakis (2009) showed that fixpoint iteration could converge exponentially slower than Newton's method.

loopy belief propagation (see Elidan et al. (2006) for a representative method), nonlinear systems of equations (see Baharev et al. (2016) for a survey), and for inference in superior semirings (e.g., Felzenszwalb and McAllester (2007)).

[96]In our setting, we use newton to sum over semiring-weighted derivations; thus, our presentation will differ from that of Esparza et al. (2007)'s which takes Newton's method as their starting point. Newton's method for solving a fixpoint equation $\mathbf{x} = \mathbf{f}(\mathbf{x})$ *over the reals* where $\mathbf{x} \in \mathbb{R}^D$ and $\mathbf{f} : \mathbb{R}^D \to \mathbb{R}^D$ is an iterative scheme with the update rule: $\mathbf{x}^{(t+1)} \leftarrow \mathbf{x}^{(t)} + (\mathbf{I} - \mathbf{J}^{(t)})^{-1}\big(\mathbf{f}(\mathbf{x}^{(t)}) - \mathbf{x}^{(t)}\big)$ where $\mathbf{J}^{(t)}$ is the Jacobian matrix of partial derivatives of $\mathbf{f}$ evaluated at $\mathbf{x}^{(t)}$. Generalizing to the semiring setting requires retrofitting Newton's method to accommodate semirings that do not have subtraction, division, or differentiation. For our purposes, such retrofitting would be a distraction; we refer the reader to any of the following works, which discuss the connection: Esparza et al. (2007), Reps et al. (2017), and Luttenberger (2009). We found the presentation in Schlund (2016) the clearest.

**Example 17.** *Consider the following program based on Etessami and Yannakakis (2009, Theorem 3.2). This program has a single fixpoint* $[\![\mathsf{x}]\!] = 1$.[97]

```
x += 0.5.
x += 0.5 * x * x.
```

*Etessami and Yannakakis (2009) show that fixpoint iteration is exponentially slower than Newton's method in this example. Specifically, fixpoint iteration requires roughly* $2^k$ *iterations to gain* $k$ *bits of precision, whereas Newton's method gains one bit of precision per iteration.*

**Definition 29.** *The* ***dimension*** $\mathsf{dim}(\boldsymbol{\delta})$ *of a derivation* $\boldsymbol{\delta}$ *is defined inductively as*

*(i) Base case:* $\mathsf{dim}(\boldsymbol{\delta}) \stackrel{\text{def}}{=} 0$ *if* $\boldsymbol{\delta}$ *is a value or builtin*

*(ii) If* $\boldsymbol{\delta}$ *is of the form* $\boldsymbol{x} \oplus\!= \boldsymbol{\delta}_1 \circ \cdots \circ \boldsymbol{\delta}_K$ *where no* $\boldsymbol{\delta}_k$ *is a product*

- *If* $K = 0$*:* $\mathsf{dim}(\boldsymbol{\delta}) = 0$.
- *If* $K \geq 1$*: Let* $d = \max(\mathsf{dim}(\boldsymbol{\delta}_1), \ldots, \mathsf{dim}(\boldsymbol{\delta}_K))$*. If the argmax is unique (i.e., no ties for the max),* $\mathsf{dim}(\boldsymbol{\delta}) \stackrel{\text{def}}{=} d$*. Otherwise,* $\mathsf{dim}(\boldsymbol{\delta}) \stackrel{\text{def}}{=} d + 1$.

*(iii) If* $\boldsymbol{\delta}$ *a product of the form* $\boldsymbol{\delta}_1 \circ \cdots \circ \boldsymbol{\delta}_K$*:* $\mathsf{dim}(\boldsymbol{\delta})$ *is not defined for products because the tie-checking definition is not associative.*

*Please refer to Figure 5.1 for an illustration.*

The dimension of a derivation is always $\leq$ its height, but it is often $\ll$. The set of derivations with dimension $d$ can be infinite. In contrast, the set of derivations with fixed height is finite.[98] For example, each derivation in the geometric series program (Example 10) has dimension $\leq 1$. What is remarkable about the set of derivations with dimension $d{+}1$ is that there is an efficient *linear* structure relating it to the sets of derivations with dimensions $\leq d$, $= d$, and $< d$. This allows us to grow the set of derivations we have summed over much more quickly using a specialized linear solver (§5.3).

[97] $\frac{1}{2}x^2 - x + \frac{1}{2} = 0$ which, by the quadratic formula ($a\,x^2 + b\,x + c = 0 \land a \neq 0 \iff x = \frac{-b \pm \sqrt{b^2 - 4\,a\,c}}{2a}$ with $\{a \mapsto \frac{1}{2}, b \mapsto -1, c \mapsto \frac{1}{2}\}$) has solutions $\{1\}$.

[98] The statement about fixed height derivations assumes ignores the possibility that a built-in relation creates an infinite number of derivations with height $= h+1$.

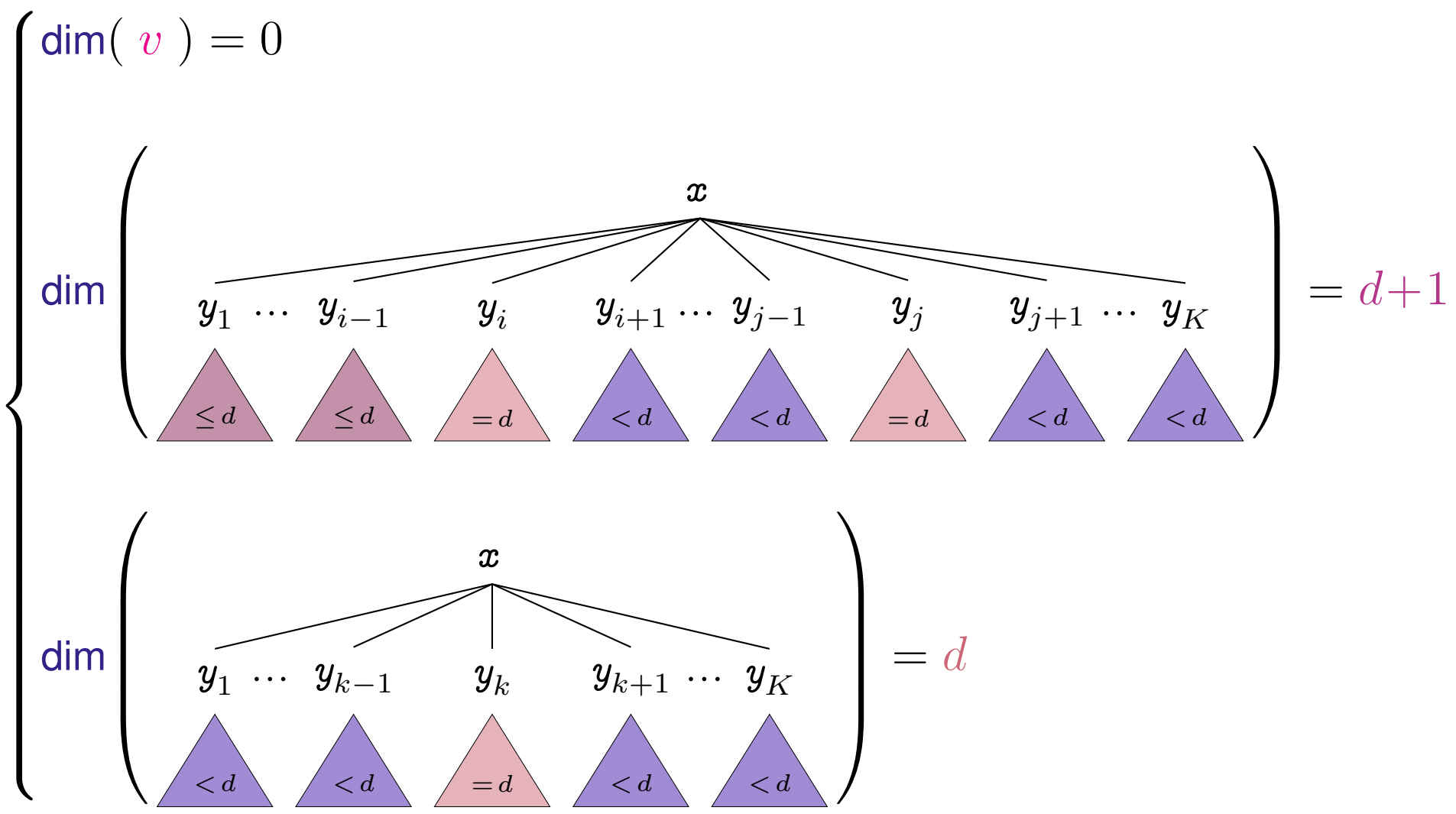

**Figure 5.1:** Illustration of the dimension of a derivation (Definition 29).

We provide pseudocode for newton in Algorithm 11.[99] The high-level picture of newton is much like seminaive forward chaining: newton tracks its analogs of $\mathfrak{m}$ and $\partial$, except here $\mathfrak{m}$ is the set of dimension $< d$ derivations, and $\partial$ is the set of dimension $= d$ derivations. To efficiently determine the values for all $= d{+}1$ derivations, newton solves a *linearized* program.

We discuss solving the linear subproblems in §5.3. Unfortunately, not all subproblems $\partial'$ can be solved exactly *and* efficiently for all semirings; important

[99]Our pseudocode generalizes Schlund (2016, Definition 3.9), which is for the special case of context-free grammars, to programs. The generalization is relatively seamless thanks to the building blocks from §3.1.

**Algorithm 11** newton: Solve program 𝔭 by Newton's method

```
1.  def newton(𝔭):
2.    ⟨𝔪, 𝔡⟩ ← newton_init(𝔭)
3.    while true:
4.      ⟨𝔪′, 𝔡′⟩ ← newton_step(𝔭, 𝔪, 𝔡)
5.      if 𝔪′ = 𝔪: break
6.      ⟨𝔪, 𝔡⟩ ← ⟨𝔪′, 𝔡′⟩
7.    return 𝔪

8.  def newton_step(𝔭, 𝔪, 𝔡):
9.    𝔪′ ← 𝔪 ⊕ 𝔡
10.   𝔡′ ← ⦃⦄        ← Should omit constants and builtins; these are added by line 11.2.
11.   for (x ⊕= y_1 ∘ ··· ∘ y_K) ∈ 𝔭:
12.     ▷ Base case: combine at least two derivations with dimension =d
13.     for 1 ≤ i < j ≤ K:
14.       𝔡′[x] ⊕= 𝔪′[y_[:i)] ∘ 𝔡[y_i] ∘ 𝔪[y_(i:j)] ∘ 𝔡[y_j] ∘ 𝔪[y_(j:]]
15.     ▷ Recursive case: derive =d+1 derivations by combining one =d+1 derivation
          with ≤d derivations
16.     for k ∈ [1:K] where y_k ∉ 𝕍 ∪ 𝔅:
17.       𝔡′[x] ⊕= 𝔪′[y_[:k)] ∘ y_k ∘ 𝔪′[y_(k:]]
18.   return ⟨𝔪′, solve_linear(𝔡′)⟩          ▷ Solve using specialized linear solver (§5.3)

19. def newton_init(𝔭):
20.   ▷ Initial set of new derivations
21.   𝔪 ← ⦃⦄; 𝔡′ ← ⦃⦄
22.   for (x ⊕= y_1 ∘ ··· ∘ y_K) ∈ 𝔭 where at most 1 of y_1, ..., y_K ∉ 𝕍 ∪ 𝔅:
23.     𝔡′[x] ⊕= y_1 ∘ ··· ∘ y_K
24.   return ⟨𝔪, solve_linear(𝔡′)⟩           ▷ Solve using specialized linear solver (§5.3)
```

tractable special cases include commutative semirings and cases where the subproblems are either left linear or right linear. Other cases are discussed in the literature (e.g., Esparza et al. (2007), Schlund (2016), and Reps et al. (2017)). We note that fixpoint iteration methods are always available for (approximately) solving $\mathfrak{d}'$; however, fixpoint iteration may not converge quickly (or at all).

As noted in other work on Newton's method (e.g., Schlund (2016)), decomposing the program into strongly connected components—as we did for prioritization (§5.1.4)—can speed the algorithm up considerably. Our hybrid algorithm (§5.4) speeds up Newton's method in this way.

We argue for the correctness of newton in the following proposition.

**Proposition 4.** *On iteration $d$ of Algorithm 11 (line 11.6), the value of $\mathfrak{m}$ is the valuation function restricted to dimensions $<d$, and the value of $\mathfrak{d}$ is the valuation function restricted to dimension $=d$.*

*Proof (Sketch).* We first argue that the value $\mathfrak{d}'$ returned by the call to newton_step on line 11.4 is precisely the sum of dimension $=d{+}1$ derivations. Below, we illustrate how to build dimension $=d{+}1$ derivations inductively from the set of dimension $=d$ derivations, the set of dimension $\leq d$ derivations, and the set of $<d$ derivations.

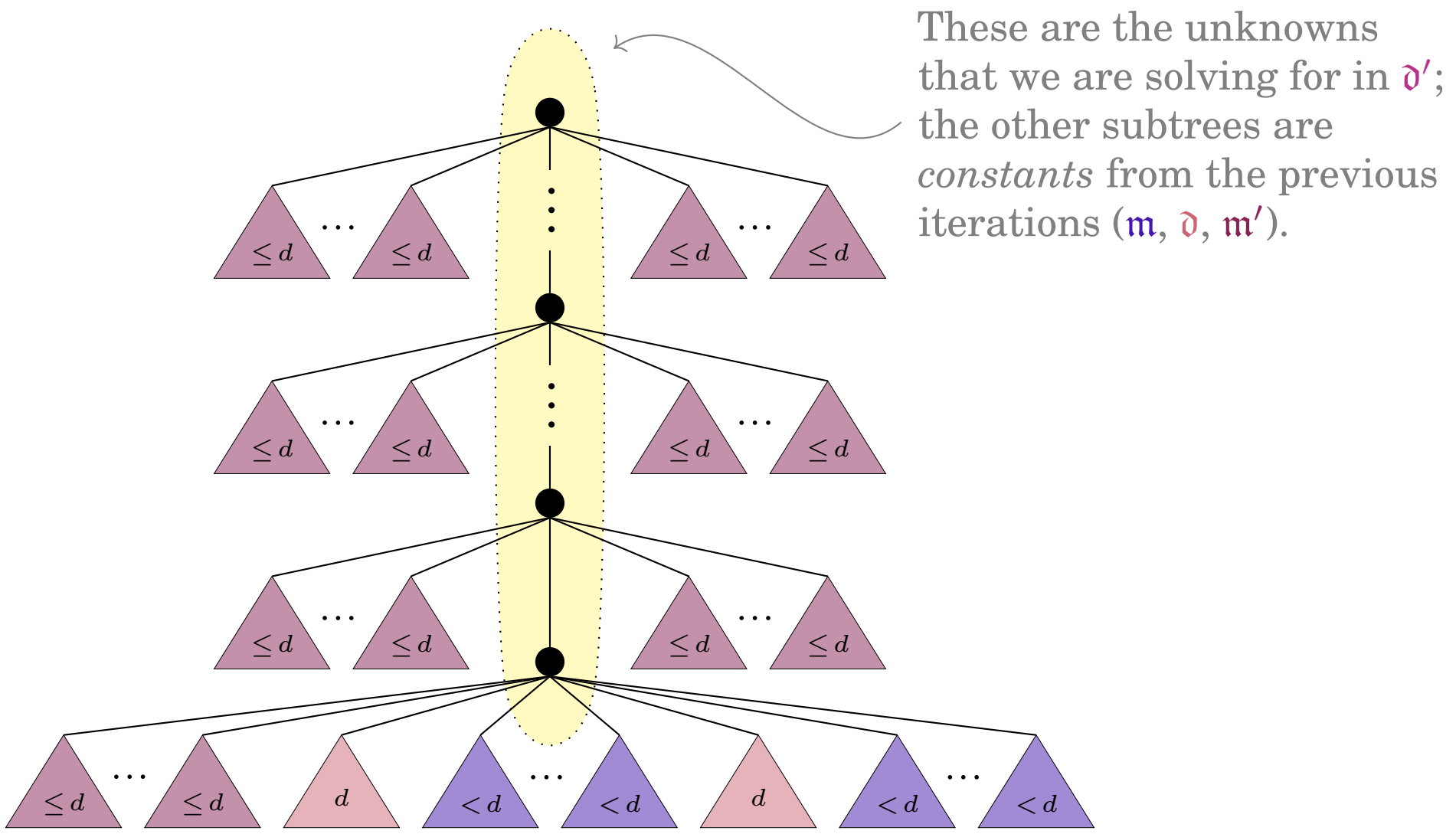

We argue that our scheme for creating new derivations generates derivations of dimension $=d{+}1$. According to Definition 29, to create a dimension $\geq d{+}1$ derivation, we need to combine at least two $=d$ derivations. That is why the base case of newton_step (lines 11.13–11.14) combines at least two $=d$ derivations;

this is depicted as the rule connecting at least two adjacent $=d$ derivations; this step creates a subderivation with dimension equal to $=d+1$. Now, because all subderivations that can combine along the spine (the path up to the root of the derivations) of the new derivation have dimension $\leq d$, the new derivation has dimension $=d+1$; we see this in the recursive case of newton_step (lines 11.15–11.17). Thus, the dimension of the new derivation is precisely dimension $=d+1$. To ensure that the enumeration of $=d+1$ is exhaustive, the program $\mathfrak{d}'$ includes every rule that can combine the $\leq d$ subderivations and the base case of at least two $=d$ subderivations. The program $\mathfrak{d}'$ is a compact encoding of the value of this potentially infinite set of $=d+1$ derivations. Its solution is found by calling a specialized solver (described in §5.3) for programs with linear rules.

On line 11.9, we see that $\mathfrak{m}'$ is the sum of $\mathfrak{d}$ and $\mathfrak{d}$. On line 11.6, $\mathfrak{m}'$ overwrites $\mathfrak{m}$. Thus, at iteration $d$, $\mathfrak{m}$ the total value of all derivations with $<d$. ■

## 5.3 Specialized Right-Linear Solver

This section gives an algorithm that solves the special family of right-linear programs, such as those generated by Newton's method (§5.2).[100] Unlike fixpoint iteration and Newton's algorithm, this algorithm is exact: it can sum over the infinite set of right-linear derivations in closed form. Until now, our only exact algorithms were for acyclic programs (i.e., those defining only finitely many derivations). We can perform inference efficiently because the solution can be

[100] We note that all algorithms we derive in this section for the right-linear case work equally well for the left-linear case with straightforward modifications.

expressed in terms of finitely many regular operations. A simple example of a right-linear program is

```
127 x += 1.
128 x += 0.5 * x.
129 a += 0.5.
```

Although this program has infinitely many derivations of x, we can solve for the total value of x using grade school algebra (see earlier discussion Example 10). Right-linearity can be checked syntactically: each rule has one of two possible forms (1) a completely constant body side like #127, or (2) the body contains a single subgoal as its rightmost subgoal like #128 (that subgoal cannot be a builtin). A program can often be put into a right-linear form using simple program transformations. The example below is not (immediately) right-linear.

```
130 x += 1.
131 x += x * a.
132 a += 0.5.
```

However, if we unfold the a subgoal in #131 and exploit the commutativity of multiplication (if permitted by the program's semiring) to reorder the right-hand side, then we arrive at the right-linear program from before. Unfortunately, not all semirings are commutative; thus, permuting the right-hand side is not allowed. The following example, using the semiring of formal languages (§3.3), is linear but not right-linear.[101]

```
133 x += 1.
134 x += a * x * b.
```

Unlike the previous example, we cannot transform this program into a right-linear program because its semiring is not commutative. Unfortunately, the right-linear algorithm cannot solve this program efficiently; we would need a general linear algorithm.

[101] The solution to this program is $\mathrm{x} = \{a^n b^n \mid n \in \mathbb{N}\}$, which is not a regular language. Thus, it cannot be expressed as a (finite) expression involving the operations $\circ$, $\oplus$, and $(\cdot)^*$.

Algorithm 12 leverages a general-purpose algorithm for solving right-linear recurrences, i.e., equations of the form $\mathrm{x} = \mathrm{b} \oplus \mathbf{A}\,\mathrm{x}$ where $\mathbf{A} \in \mathbb{V}^{N \times N}, \mathrm{b} \in \mathbb{V}^{N}$. The solution is found by the matrix star operation $\mathrm{x} = \mathbf{A}^{*}\,\mathrm{b}$, which we defined in Definition 21. Special cases of this algorithm for specific applications are well-known, e.g.,

- matrix inversion (Gauss–Jordan Elimination)[102]
- all-pairs shortest path (Floyd, 1962; Warshall, 1962; Roy, 1959)
- finite-state automaton to regular expressions conversion (Kleene, 1956)
- transitive closure (Warshall, 1962)

These problems were unified into what is often called **algebraic path problems** (Lehmann, 1977; Tarjan, 1981). Unifying these problems was just the (closed) semiring abstraction (§3.3) at work again! A choice of semiring—with the same underlying algorithm that sums over all paths in a graph—solves all the above problems and many others.

A useful perspective on the matrix star operation is that it computes the total weight of paths in a weighted directed graph. Paths may be regarded as a restricted family of right-recursive derivations (Definition 23).[103] Suppose the

[102] Althoen and McLaughlin (1987) provide an interesting historical overview of the Gauss–Jordan Elimination algorithm.

[103] That is the interpretation used in Algorithm 12. Similarly, Algorithm 11 uses this interpretation to sum over linearly recursive derivations. Notice, however, that Algorithm 11 requires solutions to be linearly recursive. To use the efficient right-recursive solver, we need the subgoals to be permutable (e.g., a commutative semiring).

**Algorithm 12** solve_linear: Solve a right-linear program $\mathfrak{p}$ by reduction to the matrix star operation (Definition 21).

```
def solve_linear(𝔭):
  m̄ ← support̄(𝔭)                                  ▷ Estimate 𝔭's support (Algorithm 9)
  N ← {x | m̄[x] = true}
  A ← 0^{N×N}; b ← 0^N
  ▷ The following two loops instantiate 𝔭's rules against the estimated support. In
    the loop's body, we assemble the matrix A and vector b that will be passed to the
    right-linear solver.
  for (x ⊕= y_1 ∘ ⋯ ∘ y_K) ∈ 𝔭:
    for _ ∈ m̄[y_1 ∘ ⋯ ∘ y_K]:
      if x ⊕= y_1 ∘ ⋯ ∘ y_K is not ground:
        raise Error("nonground reasoning not supported")
      else if y_1 ∘ ⋯ ∘ y_K = v where v ∈ 𝕍:
        b_x ⊕= v
      else if y_1 ∘ ⋯ ∘ y_K = v ∘ y where v ∈ 𝕍, y ∉ 𝔅[104]:
        A_{x,y} ⊕= v
      else
        raise Error("program is not right linear")
  ▷ Solve the right-linear system with the matrix star operation (Definition 21).
  s ← A* b
  ▷ Convert solution vector s into a program in the obvious way
  𝔪 ← ⦃⦄
  for x ∈ N: 𝔪[x] ⊕= s_x
  return 𝔪
```

matrix $\mathbf{A}$ is the weighted adjacency matrix of a graph. Then $\mathbf{A}_{ik}$ is the total weight of all paths of length $=1$. Similarly, the total weight of all paths between $i$ and $k$ of length $=2$ is $(\mathbf{A}\,\mathbf{A})_{ik}$. Thus, the star operation on matrices $\mathbf{A}^* \overset{\text{def}}{=} \sum_{i=0}^{\infty} \mathbf{A}^i$ computes the total weight of paths of length $0, 1, \ldots$ and adds them up.

Definition 21 provides a simple recursive method for computing $\mathbf{A}^*$. It runs in $\mathcal{O}(|\,N\,|^3)$ time and can be made to run in $\mathcal{O}(|\,N\,|^2)$ space.[105] The runtime can be improved by decomposition into strongly connected components (Definition 12).[106] Our hybrid algorithm (§5.4) performs this optimization, which is not in Algorithm 12. Note that there is a speed–accuracy tradeoff when using the right-linear solver. Often fixpoint iteration can reach a tolerable level of accuracy faster than the exact algorithm.[107] Lastly, we mention that solve_linear requires the program's rules to come to ground (see line 12.8).[108]

[104] We may take $v = \underline{1}$ to match the pattern.
[105] Sub-cubic algorithms exist when $\mathbb{V}$ is a *ring*, e.g., Strassen (1969).
[106] The SCC-based algorithms for transitive closure date back to Purdom (1970).

## 5.4 Hybrid Algorithm

This chapter has provided several algorithms that each work best under different conditions. This section provides a hybrid algorithm that tries to use the best algorithms available to solve a given program $\mathfrak{p}$. We saw that if $\mathfrak{p}$ is acyclic, it can be solved efficiently by agenda with toposorted prioritization (§5.1.4). However, if $\mathfrak{p}$ is cyclical, it is best to solve $\mathfrak{p}$ by breaking it up into cyclical subprograms $\mathfrak{p}^{(1)}, \ldots, \mathfrak{p}^{(M)}$ by strongly connected component (SCC) decomposition.[109] This decomposition allows each subprogram to be solved separately, provided that the solutions to the subprograms it depends on have been computed beforehand; this can be achieved by solving the subprograms in a topological order. By analyzing the structure of $\mathfrak{p}^{(m)}$, we can determine what the best method for

[107] The convergence rate of fixpoint iteration depends on the properties of the $\mathbf{A}$ matrix. In the case of the real semiring, we can bound the number of iterations to reach a given error tolerance in terms of the spectral radius of $\mathbf{A}$.

[108] It is often possible to work around this restriction. For example, the abbreviation transformation (§6.2) can remove certain nonground rules that cause the assertion to fail. For example, the program {`x(I) += 1. x(I) += 0.5 * x(I).`} is linear, but unfortunately not ground because of the free variable `I`. Abbreviation transforms it to {`x' += 1. x' += 0.5 * x'.`}, which can be solved by solve_linear. It also provides a recovery rule `x(I) += x'` to rebuild the original nonground items.

[109] SCC decomposition is a common optimization for solving cyclical equations (e.g., Goodman (1998), Nederhof and Satta (2008), Esparza et al. (2007), Schlund et al. (2013), Mohri (2002), Eisner (2023), Baharev et al. (2016), Chiang, McDonald, et al. (2023), and Purdom (1970)).

solving it is likely to be. Below is our ranking:

1. solve_linear: Because it is exact. However, it requires a closed semiring (§5.3), at least one of {right-linearity, left-linearity[100], linearity+commutativity}, and that the program can be grounded (see line 12.8).

2. newton: Because it converges more quickly than fixpoint iteration methods. However, it requires that the induced linear subproblems can be solved efficiently.

3. agenda: When all else fails, the subprogram can be solved with agenda, as it runs under the most general conditions.

Algorithm 13 provides pseudocode for our hybrid algorithm, which solves each subprogram in a topological order where the solutions to earlier subprograms are taken as inputs to later subprograms. It makes use of a subroutine best_solve that uses the ranking of solvers above to select the best strategy for solving the given subprogram.[110]

**Discussion.**

- The discussion related to the termination of $\overline{\text{support}}$ applies here (§5.1.4).

- When solving an SCC with either newton or a fixpoint algorithm, it may be necessary to use a numerical convergence threshold (or iteration limit), as

[110] A simple implementation of best_solve would use exception handling to move down the list of preferred solvers. This can be sped up when the necessary properties can be checked ahead of any expensive computations. For example, linearity can be checked without instantiating the program's rules in Algorithm 12.

**Algorithm 13** solve: Solve program $\mathfrak{p}$ using the hybrid algorithm.

```
1.  def solve(𝔭):
2.      m̄ ← support(𝔭)                          ▷ Estimate 𝔭's support (Algorithm 9)
3.      𝔪 ← ⟅⟆
4.      for σ ∈ sccs(incoming, m̄):               ▷ Topologically ordered list of SCCs (Algorithm 5)
5.          ▷ Restrict 𝔭 to compute the items in the current component, σ.
6.          𝔭′ ← ⟅⟆
7.          for x ∈ σ:
8.              for (y_1 ∘ ··· ∘ y_K) ∈ 𝔭[x]:    ▷ Add the specialized rules to the subprogram;
9.                  𝔭′[x] ⊕= y_1 ∘ ··· ∘ y_K         Note: 𝔭's rules may participate in multiple SCCs.
10.         ▷ Solve the subprogram 𝔭′ with the best applicable method, taking the solutions to
              earlier subprograms as input through 𝔪.[111]
11.         𝔪 ← best_solve(𝔭′ ⊕ 𝔪)
12.     return 𝔪
```

discussed earlier in this chapter. Otherwise, the Algorithm 13 could run forever on a single subprogram with making progress on the others.

- Looking forward to other chapters, we note that it is often best to *optimize* the program before executing. We note, in particular, the program specialization techniques of (§6.2) and the program optimizer (§8). These methods produce a semantically equivalent program that is more efficient to execute.

# Summary

In this chapter, we described many algorithms for solving Dyna programs. We gave three implementations of fixpoint iteration. We extended Eisner, Goldlust, et al. (2005)'s forward chaining algorithm to allow nonground updates and delayed constraints. We adapted Esparza et al. (2007)'s semiring-weighted Newton algorithm to Dyna programs, which provides a powerful new solver to the Dyna toolbox. We also gave a specialized solver for programs with a certain linear structure. Lastly, we gave a unified, hybrid algorithm that breaks up a large program into smaller subprograms that can be solved separately with the best applicable method.

[111] The $\oplus$-addition below should be implemented carefully to avoid the cost of copying.

# Chapter 6

# Program Analysis

This chapter is about static program analysis: proving that the program satisfies certain properties without actually running it on *all possible* input data sets. Analyzing program behavior without actually running the program on concrete data is known in the programming languages community as *abstract interpretation* (Cousot and Cousot, 1977); it can be viewed as the partial evaluation of a program to acquire constraints on its *possible* semantics.

- Abstract analyses are useful for code generation because code generation can take advantage of any guarantee that an item will have a particular type or property.[112] In §6.2, we leverage the output of abstract analysis to generate specializations of the program (i.e., Dyna-to-Dyna code generation) that can compute a subset of the items more efficiently by removing unnecessary intermediate items.

[112]In the context to code generation into an object-oriented host language, these guarantees can be used to represent the item efficiently as an instance of a class specialized to objects of that type and to speed up tests (e.g., pattern-matching).

- Abstract analyses are useful for complexity analysis: If the program stores and iterates over a collection of distinct objects of a particular type, then knowing the type's cardinality on the concrete data can yield bounds on the program's space and runtime requirements (§6.3); our program optimizer (§8) uses such bounds as cost function.

Below is a listing of this chapter's contents.

## 6.1 Type Analysis

### 6.1.1 Formulation

This section describes a system that infers the set of items that may have a non-$\mathbb{0}$ value in a program $\mathfrak{p}$ for all input data $\mathfrak{D}$ that satisfy user-provided input constraints $\overline{\mathfrak{D}}$. We call the set of non-$\mathbb{0}$-valued items the **support** of the program. Let $\mathsf{nz}(\mathfrak{D})$ and $\mathsf{nz}(\mathfrak{p}(\mathfrak{D}))$ denote the support of their respective arguments. Since we do not know the *concrete* data $\mathfrak{D}$ to execute $\mathfrak{p}(\mathfrak{D})$,[113] we compute an upper bound $\overline{\mathfrak{m}}$ on its support through *abstract* interpretation such that[114]

$$\underbrace{\mathsf{nz}(\mathfrak{D}) \subseteq \overline{\mathfrak{D}}}_{\textit{input satisfies constraints}} \quad\Longrightarrow\quad \underbrace{\mathsf{nz}(\mathfrak{p}(\mathfrak{D})) \subseteq \overline{\mathfrak{m}}}_{\textit{valid upper bound on support}}$$

In other words, $\overline{\mathfrak{D}}$ characterizes the subset of $\mathbb{H}$ for which the input items $\mathfrak{D}$ is non-$\mathbb{0}$, and $\overline{\mathfrak{m}}$ characterizes the subset of $\mathbb{H}$ for which the derived items $\mathfrak{p}(\mathfrak{D})$ is non-$\mathbb{0}$. Therefore, we call $\overline{\mathfrak{D}}$ the **input type** and $\overline{\mathfrak{m}}$ the **derived type**.

We now describe our type specification language and how it is used to reason about possible items. Continuing with the CKY example (Example 6), we provide the specification of the input data. However, as we are interested in the *possibility* that an item may have a non-$\mathbb{0}$ value, rather than finding its specific value, we first transform the program as illustrated below.

**Example 18.** *Recall, the program to analyze is*

[113]Recall that $\mathfrak{p}(\mathfrak{D})$ is shorthand for $\mathfrak{p}'[\![\cdot]\!]$ and $\mathfrak{p}' = \mathfrak{p} \oplus \mathfrak{D}$.

[114]For naturally ordered semirings (Definition 26), it should be possible to refine our technique to produce an upper bound on the *value* of each derived item as a function of upper bounds on the values of its input items.

```
135 params: word; len; γ.
136 β(X,I,K) += γ(X,Y,Z) * β(Y,I,J) * β(Z,J,K).
137 β(X,I,K) += γ(X,Y) * β(Y,I,K).
138 β(X,I,K) += γ(X,W) * word(W,I,K).
139 goal += β(s,0,N) * len(N).
```

Because no rules have been specified to define the `word`, `len`, and `γ` items, one might think that these items have value $\underline{0}$.[115] However, the `params` declaration says that the program's input will define these items.

The CKY program $\mathfrak{p}$ above can be automatically "booleanized" to yield the **type program** $\overline{\mathfrak{p}}$ below.[116] The `:-` and `,` symbols traditionally denote the $\oplus$= and $\circ$ operations in the Boolean semiring.

```
140 params: w̅o̅r̅d̅; l̅e̅n̅; γ̅.
141 β̅(X,I,K) :- γ̅(X,Y,Z), β̅(Y,I,J), β̅(Z,J,K).
142 β̅(X,I,K) :- γ̅(X,Y), β̅(Y,I,K).
143 β̅(X,I,K) :- γ̅(X,W), w̅o̅r̅d̅(W,I,K).
144 g̅o̅a̅l̅ :- β̅(s,0,N), l̅e̅n̅(N).
```

The intention is that `β̅(X,I,K)` will have value `true` in the type program for all `X,I,K` such that `β(X,I,K)` has a non-$\underline{0}$ value in the CKY program. Of course, each program needs input: if $\mathfrak{p}$ is run on data $\mathfrak{D}$, then $\overline{\mathfrak{p}}$ must be run on a booleanized version $\overline{\mathfrak{D}}$. The valuation $\overline{\mathfrak{p}}(\overline{\mathfrak{D}})$ now effectively specifies an inferred type of items that *might* be derived by the program.[117]

To abstract away from the specific input data $\mathfrak{D}$, users may *directly* specify the booleanized input data $\overline{\mathfrak{D}}$ (mentioned above). This $\overline{\mathfrak{D}}$ may be any upper bound on its support. $\overline{\mathfrak{D}}$ serves as a type declaration for the unknown input $\mathfrak{D}$.

---

[115] In other words, a program that is expecting inputs is only *partially specified*. When input items are queried, they are returned as delayed constraints (rather than incorrectly assigning them a value of $\underline{0}$).

[116] The translation additionally maps any constant values to `false` or `true`.

[117] For example, $\overline{\mathfrak{p}}(\overline{\mathfrak{D}})[\![$`β̅(np,3,5)`$]\!]$ = `true` if $\mathfrak{p}(\mathfrak{D})[\![$`β(np,3,5)`$]\!] \neq \underline{0}$—and possibly even if $\mathfrak{p}(\mathfrak{D})[\![$`β(np,3,5)`$]\!] = \underline{0}$, for example, if $\underline{0}$ can result from adding non-$\underline{0}$ values under $\oplus$. Such overestimates can arise because the boolean program does not track actual values.

The type defined by $\overline{\mathfrak{p}}(\overline{\mathfrak{D}})$ then continues to include all items in $\mathtt{nz}(\mathfrak{p}(\mathfrak{D}))$, for any and every dataset $\mathfrak{D}$ for which this upper bound holds. For example, $\overline{\mathfrak{p}}(\overline{\mathfrak{D}})$ still gives value `true` to $\bar{\beta}$`(np,3,5)` if $\beta$`(np,3,5)` could possibly have a non-$\underline{0}$ value.

An example of an input type declaration for CKY is the following boolean program $\overline{\mathfrak{D}}$:

```
145 params: k; w; n.
146 word̅(W:w,I:n,K:n) :- I < K.
147 len̅(N:n).
148 γ̄(X:k,Y:k,Z:k).
149 γ̄(X:k,Y:k).
150 γ̄(X:k,W:w).
```

The lines #146–150 specify 5 sets whose union is a type that contains all input items. We use the notation $\overline{\mathtt{word}}$`(W:w,I:n,K:n) :- I < K` in #146 as a shorthand for $\overline{\mathtt{word}}$`(W,I,K) :- w(W), n(I), n(K), I < K`. This type indicates that items in the set $\{\mathtt{word(W,I,K)} \mid \mathtt{w(W)}, \mathtt{n(I)}, \mathtt{n(K)}, \mathtt{I} < \mathtt{K}\}$ may have non-$\underline{0}$ values in the input data $\mathfrak{D}$.[118,119] The predicates such as `n` could be defined by additional lines like `n(I) :- 0` $\leq$ `I` $\leq$ `20`, which defines the possible sentence positions for sentences of length 20. However, to abstract away from any particular sentence length, we omit such a definition, leaving `n` as a **parameter** (#145). Thus, the above program $\overline{\mathfrak{D}}$ actually defines a *parametric* type, just as `List[`$\alpha$`]` in many programming languages is a parametric type that depends on a concrete choice such as

[118]Note that this input type is compatible with *lattice parsing* (e.g., Tomita (1986) and Hall (2005)); in that setting, the nodes of the lattice (i.e., weighted acyclic finite-state automaton) would be labeled by their position in a topological ordering of the input, and `word(W,I,K)` encodes the weight of the arc $\mathtt{I} \xrightarrow{\mathtt{W}} \mathtt{K}$.

[119]A natural alternative to #146 is $\overline{\mathtt{word}}$`(W:w,I:n,K:n) :- (K is I + 1)`, which uses the built-in addition constraint. This type denotes the set $\{\mathtt{word(W,I,I+1)} \mid \mathtt{w(W)}, \mathtt{n(I)}, \mathtt{n(I+1)}\}$ and it directly encodes the description of the CKY input data that we gave in Example 6. Although this input characterization may be more precise for some applications, the conclusions that we can derive from it appear to be the same (i.e., we are unable to infer a tighter derived type).

$\alpha =$ `string`. It depends on three other relations, which specify the allowed terminals, nonterminals, and sentence positions. The concrete input type would have the form $\overline{\mathfrak{D}}(\mathtt{k}, \mathtt{w}, \mathtt{n})$. Our program analyzer reasons about these user parameters symbolically; thus, it performs *symbolic evaluation* (e.g., Baldoni et al. (2018)).

Our abstract forward chaining algorithm (§6.1.2) will derive the following rules that specify a valid type for the derived items of $\overline{\mathfrak{p}}(\overline{\mathfrak{D}}(\mathtt{k}, \mathtt{w}, \mathtt{n}))$:

```
β̄(X:k,I:n,K:n) :- I < K.
goal.
```

Adding these rules and $\overline{\mathfrak{D}}$ yields the desired description of a type that covers all non-$\underline{0}$ items. Call the resulting program $\overline{\mathfrak{m}}$. The program $\overline{\mathfrak{m}}$ uses the same parameters as $\overline{\mathfrak{D}}$ and provides an upper bound on $\overline{\mathfrak{p}}(\overline{\mathfrak{D}}(\mathtt{k}, \mathtt{w}, \mathtt{n}))$. The bound implies that we will only build $\bar{\beta}$ items where `X` is a nonterminal (has type `k`), and the `I` and `K` variables are sentence positions (have type `n`) with `I < K`. Additionally, the bound says (reassuringly) that it may be possible to build a `goal` item. If `goal` were impossible, we would have detected a mistake in the code. Similarly, if the bound had included `β̄(X:k,I:k,K:n)`, it would be a suspiciously loose upper bound, since `I` should never be a nonterminal.

Notice that $\overline{\mathfrak{p}}' = \overline{\mathfrak{p}} \oplus \overline{\mathfrak{D}}$ would *also* be a valid type for $\overline{\mathfrak{p}}(\overline{\mathfrak{D}}(\mathtt{k}, \mathtt{w}, \mathtt{n}))$. However, $\overline{\mathfrak{p}}'$ is not easily consumed by downstream applications.[120] Our analyzer (§6.1.2) is guaranteed to output a derived type, which is a program (like $\overline{\mathfrak{m}}$ above) where each rule is *simple*. We define a rule to be **simple** if its body contains only constants, user parameters, and built-in constraints. Simple rules may be regarded as the *symbolic* generalization of *axioms* (§ 2).

[120] E.g., space and time complexity analysis (§6.3) and program specialization (§6.2).

In summary, our type analyzer is given the program $\mathfrak{p}$ together with a parametric input type $\overline{\mathfrak{D}}$, and computes a parametric derived type such as #151–152 for the items derived by $\mathfrak{p}$. In effect, it runs the booleanized program $\overline{\mathfrak{p}}$ on the input upper bound $\overline{\mathfrak{D}}$. However, as the derived type is only required to be an upper bound, we will permit the analyzer to "round up" as it goes, as needed to keep the type simple.[121]

### 6.1.2 Abstract Forward Chaining

We now develop our type inference algorithm, which is based on fixpoint iteration (§4.3) in the boolean semiring.[122] Here, we generalize fixpoint iteration's valuation estimate $\nu$ to an *abstract* valuation where the user's type parameters (such as $\texttt{n}$) are used to define $\nu$ symbolically, as we saw in the derived type inferred for the CKY example. This generalization is necessary because we no longer have a concrete upper bound on the input data with which to initialize forward chaining: the $\texttt{n}$, $\texttt{k}$, and $\texttt{w}$ parameters are unknown relations. Instead, we specify parametric rules that upper-bound the unknown input data. Forward chaining derives new simple rules, which may also refer to the relations like $\texttt{n}$ and $\texttt{<}$. Indeed, these relations may emerge in the derived type $\overline{\mathfrak{m}}$, as we saw in #151.

We provide the abstract forward chaining algorithm in §6.1.2.1. We address the challenges arising from constraint accumulation in §§6.1.2.2–6.1.2.5.

[121]The abstract interpretation literature often calls rounding up *widening*.

[122]Recall that fixpoint iteration is also the basis for ordinary forward chaining (Ullman, 1988; Eisner, Goldlust, et al., 2005) and many of the algorithms that we developed in §5.

#### 6.1.2.1 Type Inference Algorithm

To generalize forward chaining in the boolean semiring, we will represent the valuation $\nu$ from §4 as a *program* $\overline{\mathfrak{m}}$—just like we did in §5. We will reason in the boolean semiring. This representation is helpful because we can use programs to represent unknown and infinite sets. We require each rule in $\overline{\mathfrak{m}}$ to have a specific syntactic form: each must be a *simple type*. A **simple type** is a rule of the form $\bar{x}$ :- $c_1, \ldots, c_M$ where each $c_m$ is externally defined as either a built-in constraint or a parametric constraint and may contain variables. Note that $\bar{x}$ cannot appear in the body of any other simple type. We saw the derived type for the CKY example #146–152 is the collection of simple types.

In some sense, program rules #141–150 already describe the derived type, but they do so indirectly and not as a collection of simple types. This form ensures that each simple type is self-contained and can be implemented as a (nonrecursive) object-oriented class.

Type inference is performed by fixpoint iteration on a symbolic `step` operation. This should be viewed as a "lifted" version of the forward chaining algorithm from §5: it derives simple types from other simple types rather than deriving ground items from other ground items. The constraints that appear in these simple types are passed around during execution. Such constraints are known in the constraint logic programming literature as "delayed constraints," in contrast to constraints like `2 < 3` or `3 < 2` that can be immediately evaluated and replaced with `true` or `false` (respectively).

**Algorithm 14** abstract_fc: Abstract forward chaining algorithm for type program $\overline{\mathfrak{p}}$. This procedure assumes that the input type has been included in $\overline{\mathfrak{p}}$.

**def** abstract_fc($\overline{\mathfrak{p}}$):
  ▷ *This algorithm a straightforward modification of fixpoint iteration (Algorithm 6) to find the fixpoint of the* relaxed *step operation* relax(propagate($\text{step}_{\overline{\mathfrak{p}}}(\cdot)$)).
  $\overline{\mathfrak{m}} \leftarrow \{\!\!\{\}\!\!\}$[124]
  **while** `true`:
    $\overline{\mathfrak{m}}' \leftarrow$ relax(propagate($\text{step}_{\overline{\mathfrak{p}}}(\overline{\mathfrak{m}})$))   ▷ *Apply the relaxed step operation*
    **if** $\overline{\mathfrak{m}}' = \overline{\mathfrak{m}}$: **return** $\overline{\mathfrak{m}}$
    $\overline{\mathfrak{m}} \leftarrow \overline{\mathfrak{m}}'$

*defined in Algorithm 6*[123]

However, the challenge that arises from the repeated application of step is that constraints may accumulate.[123] They can accumulate indefinitely in the case of recursive rules since they effectively unroll an unbounded chain of recursive calls, as will be illustrated in §6.1.2.2. To prevent divergence, we propose the following strategies: constraint relaxation (relax; §6.1.2.2), and constraint propagation (propagate; §6.1.2.3). We discuss termination in §6.1.2.5. We give pseudocode for the overall approach in Algorithm 14.

#### 6.1.2.2 Relaxation

Consider the following input type $\overline{\mathfrak{D}}$ for Example 4.

```
params: n.
stop(S) :- n(S).
edge(S,S') :- n(S), n(S').
```

Applying the step operation once, which renames variables to avoid conflicts, we

[123] Note that when user parameters are queried in $\overline{\mathfrak{p}}$, they are returned as delayed constraints by Algorithm 2. For example, let $\overline{\mathfrak{p}}$ be the booleanized CKY program together with its input type. Then, if we perform the lookup $\overline{\mathfrak{p}}$[n(K)], we would get the result stream containing n(K) as its only element.

[124] We represent the $\overline{\mathfrak{m}}$ as a *set* of rules. In our implementation, this set replaces rules with their most general version, i.e., if there exists a pair of simple types $r, r'$ such that $r' \subseteq r$, we remove $r'$ and keep $r$. This ensures, among other things, that rules that are equivalent under variable renaming do not both appear in the set. It can also help to achieve convergence. See §6.1.2.4 for further discussion.

derive the following simple type for β items (other types not shown):

```
156 β̄(S1) :- n(S1).
```

Applying step a second time, we derive

```
157 β̄(S1) :- n(S1).
158 β̄(S1) :- n(S1), n(S2).
```

Notice in the second rule that n(S$_2$) is lingering as a constraint from the recursive call. In keeping with the (Prolog-like) semantics of boolean programs, the variable S$_2$ is existentially quantified since it is a "local" variable that appears only in the body of the (:-) rule. After applying step $k$ times, we derive

```
159 β̄(S1) :- n(S1).
160 β̄(S1) :- n(S1), n(S2).
161    ⋮
162 β̄(S1) :- n(S1), n(S2), n(S3), …, n(Sk).
```

This procedure is diverging by elaborating the type to describe paths of every length, rather than converging to a fixpoint (i.e., a type that describes all nodes that can reach a terminal state on paths of *any* length). Our strategy for eliminating constraint accumulation from previous levels of recursion is to drop (i.e., relax) any constraint that refers to local variables:

**Definition 30** (relax)**.** *Let* $\bar{x}$ :- $c_1, \ldots, c_M$ *be a simple type. The operation* relax *drops each delayed constraint* $c_m$ *that refers to a variable that does not appear in* $\bar{x}$*. In other words, we obtain a new simple type with the same head, but with the following subset of constraints* $\{c_m \mid \mathsf{vars}(c_m) \subseteq \mathsf{vars}(\bar{x}), m \in [1{:}M]\}$*. To* relax *a set of simple types, we simply map* relax *over the set.*

Applying relax to the second application of step to #157–158, we see that it would drop the second constraint of the second simple type because it depends on the non-head variable $S_2$, yielding:

```
163 β̄(S1) :- n(S1).
164 β̄(S1) :- n(S1).
```

We may merge these rules because they are duplicates. We now have the following type

```
165 β̄(S1) :- n(S1).
```

We now have a fixpoint: this iteration's type (after relaxation) is equal to the previous iteration's type (#156). Notice that some precision is lost because β̄(S) is `true` even for nodes `S` for which there does not exist a path ending in $\overline{\texttt{stop}}$.

Why is relax valid? In general, dropping constraints from a simple type simply makes the type larger, raising the upper bound. Dropping constraints that depend on non-head variables is simply a convenient choice.

We now consider an example where relax leads to an overly loose type. Recall CKY (Example 18). The simple types from $\overline{\mathfrak{D}}$ (#145–150) are added to $\overline{\mathfrak{m}}$ on the first step, such as

```
166 γ̄(X,W) :- k(X), w(W).
167 word̄(W,I,K) :- w(W), n(I), n(K), I < K.
```

On the next step, #143 from $\overline{\mathfrak{p}}$

```
168 β̄(X,I,K) :- γ̄(X,W), word̄(W,I,K).
```

gives us an initial type for β:

```
169 β̄(X,I,K) :- k(X), w(W), n(I), n(K), I < K.
```

After relax, we have

```
170 β̄(X,I,K) :- k(X), n(I), n(K), I < K.
```

However, on the next step, #141 from $\overline{\mathfrak{p}}$

```
171 β̄(X,I,K) :- γ̄(X,Y,Z), β̄(Y,I,J), β̄(Z,J,K).
```

will expand $\overline{\mathrm{m}}$ to the following type:

```
172 β̄(X,I,K) :- k(X), k(Y), k(Z), n(I), n(J), n(K), I < J, J < K.
```

relax will drop all the constraints on local variables:

```
173 β̄(X,I,K) :- k(X), n(I), n(K).
```

Unfortunately, relaxation resulted in the ordering information (<) being lost. In the next section, we will see how to combine `I < J` and `J < K` (before relaxing them) to deduce that `I < K`. Since `I < K` only constrains head variables, it will survive the relaxation step—allowing the analyzer to deduce that `β̄(X,I,K)` items *all* have property `I < K` (rather than the first generation having `I < K`, the second having $(\exists \mathtt{J})$ `I < J, J < K`, the third having $(\exists \mathtt{J}_1, \mathtt{J}_2, \mathtt{J}_3)$ $\mathtt{I} < \mathtt{J}_1,\ \mathtt{J}_1 < \mathtt{J}_2,\ \mathtt{J}_2 < \mathtt{J}_3,\ \mathtt{J}_3 < \mathtt{K}$, and so on.).

### 6.1.2.3 Constraint Propagation

**Constraint propagation rules** (Frühwirth, 1998), are commonly used in constraint logic programming, such as ECLiPSe (Apt and Wallace, 2007), computer algebra systems, such as Mathematica (Wolfram, 2003), and theorem provers, such as Z3 (Moura and Bjørner, 2008). In our setting, they provide a mechanism for users to supply domain-specific knowledge about type parameters and built-in constraints. They are used to derive new constraints during inference, which may serve to counteract excessive relaxation (§6.1.2.2).

Recall the loose type inferred in the CKY example (#172; repeated below):

```
174 β̄(X,I,K) :- k(X), k(Y), k(Z), n(I), n(J), n(K), I < J, J < K.
```

We saw that relax causes the ordering relationship `I < K` to be lost in #173. Con-straint propagation rules combat this loss by inferring the additional constraints, such as `I < K` by the transitive property of `<`:[125]

```
175 β̄(X,I,K) :- k(X), k(Y), k(Z), n(I), n(J), n(K), I < J, J < K, I < K.
```

After applying relax, we arrive at our desired type

```
176 β̄(X,I,K) :- k(X), n(I), n(K), I < K.
```

Below, we give motivating examples of constraint propagation rules and how they are useful for our running examples. Constraint propagation rules are specified using the `<==` syntax seen in the examples below.[126] They may be specified along with the input type.

- In CKY (Example 6), the (constant) grammar start symbol `s` is an element of the parametric nonterminal type `k`, and `0` is an element of the parametric sentence position type `n`:

```
177 k(s) <== true.
178 n(0) <== true.
```

- To assert that the `k` and `n` types are disjoint:

```
179 fail <== k(X), n(X).
```

  Here `fail` is a special constraint that is never satisfied (a contradiction). If `k(X)` and `n(X)` ever both appear in the body of a simple type, the simple type is empty and can be deleted. Notice that #177–179 are sufficient to derive a contradiction from either `n(s)` or `k(0)` (i.e., both are false).

- To assert that a type `d` is a subset of a type `c`:

```
180 c(X) <== d(X).
```

[125]The specific syntax for specifying the transitive property is on #181.
[126]We use `<==` to distinguish propagation rules from program rules and input types.

- To assert that < is a strict partial order:

```
181  (I < K) <== (I < J), (J < K).
182  fail <== (I < I).
```

- To reason about a recursive type, such as a list[127] of nonterminals (i.e., type k):

```
183  ks([]) <== true.
184  k(X) <== ks([X|Xs]).
185  ks(Xs) <== ks([X|Xs]).
```

Propagation rules for builtins (like those for < above) are provided in a standard library, whereas the user must provide rules for user parameters. If the user does not provide certain rules, our system will still produce a valid upper bound, but it may be looser than necessary. Thus, writing propagation rules may be an interactive experience. However, if the user provides unsound propagation rules, it will likely lead our system to produce an incorrect upper bound.

The propagate method used in abstract forward chaining (abstract_fc) runs a standard constraint propagation algorithm (Frühwirth, 1998)—which is just another forward chaining algorithm—to deduce all transitive consequences of the constraints in the body of a simple type.

**Checking constraints.** In the propagation step, we check if each of its built-in constraints is satisfied or unsatisfiable. Any satisfied constraint can be safely dropped from its simple type, and it does not loosen the type. If an unsatisfiable constraint exists in a given simple type, then that simple type can be safely dropped from $\overline{\mathrm{m}}$.

[127] The bracket-pipe notation is a shorthand (borrowed from Prolog (Colmerauer and Roussel, 1996)) for working with lists. A list such as `[1,2,3]` desugars to `cons(1,cons(2,cons(3,nil)))`. The notation `[X,Y|Rest]` would match as `X=1,Y=2,Rest=[3]`.

### 6.1.2.4 Removing Redundant Simple Types

In Algorithm 14, we may remove a simple type $s$ from $\overline{\mathfrak{m}}$ if there exists a supertype $t \in \overline{\mathfrak{m}}$ that is $s \subseteq t$ (see also footnote 124). Below is a method for testing such subtype relationships that leverages propagate. We also include a method for merging simple that we use later in the chapter.

**Algorithm 15** Subtype, intersection, and merge algorithms for simple types.

```
▷ Returns true if s ⊆ t; may have false negatives.
def subtype(s, t):
    ▷ (s ⊆ t) ⟺ (s ∩ t) = s
    r ← intersect(s, t)
    R ← propagate(r)
    S ← propagate(s)
    return R = S                          ▷ equal up to variable renaming
▷ Return a simple type representing the intersection of s and t.
def intersect(s, t):
    θ ← unify(head(s), head(t))
    s′ ← θ(s)
    t′ ← θ(t)
    return (head(s′) :- body(s′), body(t′))
▷ Given simple types r and s, return a simple type t such that r ⊆ t and s ⊆ t. Note that t may
  be a loose bound on r and s.
def merge(r, s):
    (r :- ℛ) ← r
    (s :- 𝒮) ← s
    X ← fresh()                ▷ New variable; used to represent the common head shape
    ℛ′ ← propagate(ℛ ∘ (X=r))
    𝒮′ ← propagate(𝒮 ∘ (X=s))
    𝒯 ← ℛ′ ∩ 𝒮′                          ▷ Only keep common constraints
    return (X :- 𝒯)
```

Note that this test may fail to detect the subset relation if the rules for propagate are incomplete. Future systems might use more powerful theorem provers, such as Z3 (Moura and Bjørner, 2008), to test the subset relation.

#### 6.1.2.5 Termination

This section describes a technique for preventing abstract forward chaining from running forever. It ensures finite-time termination with a valid upper bound but may come at the expense of precision. We require a valid upper bound to proceed with downstream analyses, such as the space and time complexity analyzer (§6.3), the program specialization system (§6.2), and the program optimizer (§8).[128]

The challenge in ensuring termination is that abstract forward chaining must provide a valid upper bound, so we cannot simply run it for a finite number of iterations and return the chart because it may underestimate the type. Our approach will be to round up the type (as we did in relax). As an extreme, observe that we can round $\overline{\mathtt{m}}$ all the way up to the unconstrained type `X :- true`, as it is always a valid upper bound. Unfortunately, this type is completely informative because it is loose for *all* subtypes; having better bounds for at least *some* of the subtypes and looser types for the others would be preferable.

Our termination strategy involves **truncation** according to three user-controllable parameters $\langle D_{\text{depth}}, D_{\text{builtin}}, D_{\text{propagation}} \rangle$ that work as follows:

- **terms**: We can build infinitely many simple types corresponding to ever-larger terms. To avoid this, we truncate them to an (optional) user-specified **depth limit** $D_{\text{depth}}$. This depth limit will rewrite a simple type in the following way: any subterm that appears in the head at depth $D$ is replaced with a fresh,

[128] They may also be used to ensure that Algorithm 9, which performs support estimation on concrete data, returns a valid upper bound.

unconstrained variable.[129]

- **builtins**: The challenge of built-in constraints is that they may introduce an unbounded number of new symbols that can be used in simple types. To prevent this, we keep track of the number of distinct symbols that have been introduced by the builtin relations (across all iterations); once more than $D_{\text{builtin}}$ have been introduced, we prevent any of the built-in constraints from evaluating when they attempt to introduce a new symbol.

- **propagation**: We also must ensure that constraint-propagation rules are not infinitely productive (i.e., they cannot produce an unbounded number of new constraints). To do so, we limit the maximum number of distinct constraints that the constraint propagation rules can produce in a given call to propagate. Once more than $D_{\text{propagation}}$ constraints have been deduced (in a given simple type), the iterative constraint-propagation procedure stops with whatever it has concluded so for.

These termination heuristics (along with relax) are sufficient to ensure that Algorithm 14 terminates with a valid upper bound.[130] However, when the heuristic is enacted, the resulting analysis will be loose when the heuristic is enacted. Thus, a warning message is provided to the user. If better results are required, the programmer can interact with the system (like an interactive theorem prover) to improve the analysis by refining the input type or providing missing propagation rules.

[129] We apply the depth limit modulo structural equality constraints, so that `f([1,2,3,4,5])` and `f([1,2|Xs]) :- Xs=[3,4,5]` are handled the same, as the latter is transformed into the former.

[130] Given finite values for the parameters, only finitely many simple types can be generated. Since the relaxed step operation is monotone nondecreasing, the abstract forward chaining cannot continue to increase $\overline{\mathfrak{m}}$ forever.

In general, the truncation parameter limits should be set conservatively (to large values); however, there is an efficiency benefit to selecting small values as the type found can be exponentially large in $D_{\text{depth}}$ (see Example 19). We cannot simply set the value of $D_{\text{depth}}$ to be small in all cases. For example, if $D_{\text{depth}} = 4$, then the simple type $\texttt{f}_1\texttt{(f}_2\texttt{(f}_3\texttt{(f}_4\texttt{(f}_5\texttt{(X))))) :- n(X)}$ would be approximated by $\texttt{f}_1\texttt{(f}_2\texttt{(f}_3\texttt{(f}_4\texttt{(Trunc)))) :- true}$ which is a much looser type. If the $D_{\text{depth}}$ had been $\geq 5$, the analysis would not have lost any precision.

**Example 19.** *Abstract forward chaining produces an exponentially large type (in the $D_{depth}$ parameter) for the program below.*

```
bitstring([]).
bitstring([X|Xs]) += bit(X) * bitstring(Xs)
bit(0). bit(1).
```

*because it will enumerate exponentially many lists (in $D_{depth}$) before truncation kicks in.*[131] *The example below has $D_{depth} = 5$:*

| | | |
|---|---|---|
| $\overline{\texttt{bit}}$(0). $\overline{\texttt{bit}}$(1). | $\overline{\texttt{bitstring}}$([0,0,0]). | $\overline{\texttt{bitstring}}$([0,0,0,Trunc$_1$\|Trunc$_2$]). |
| $\overline{\texttt{bitstring}}$([]). | $\overline{\texttt{bitstring}}$([0,0,1]). | $\overline{\texttt{bitstring}}$([0,0,1,Trunc$_1$\|Trunc$_2$]). |
| $\overline{\texttt{bitstring}}$([1]). | $\overline{\texttt{bitstring}}$([0,1,0]). | $\overline{\texttt{bitstring}}$([0,1,0,Trunc$_1$\|Trunc$_2$]). |
| $\overline{\texttt{bitstring}}$([0]). | $\overline{\texttt{bitstring}}$([0,1,1]). | $\overline{\texttt{bitstring}}$([0,1,1,Trunc$_1$\|Trunc$_2$]). |
| $\overline{\texttt{bitstring}}$([0,0]). | $\overline{\texttt{bitstring}}$([1,0,0]). | $\overline{\texttt{bitstring}}$([1,0,0,Trunc$_1$\|Trunc$_2$]). |
| $\overline{\texttt{bitstring}}$([0,1]). | $\overline{\texttt{bitstring}}$([1,0,1]). | $\overline{\texttt{bitstring}}$([1,0,1,Trunc$_1$\|Trunc$_2$]). |
| $\overline{\texttt{bitstring}}$([1,0]). | $\overline{\texttt{bitstring}}$([1,1,0]). | $\overline{\texttt{bitstring}}$([1,1,0,Trunc$_1$\|Trunc$_2$]). |
| $\overline{\texttt{bitstring}}$([1,1]). | $\overline{\texttt{bitstring}}$([1,1,1]). | $\overline{\texttt{bitstring}}$([1,1,1,Trunc$_1$\|Trunc$_2$]). |

*In the case of* `bitstring` *example, no finite value $D_{depth}$ that will cover the type without rounding up. While it is true that making $D_{depth}$ larger produces a tighter analysis, it is unlikely to improve any downstream analyses. For example, the*

[131] Note that this example does not require any input data.

*space complexity analysis that we will see in §6.3.1 will still (correctly) predict that the amount of space required by fixpoint iteration is* $\infty$ *regardless of whether* $D_{depth}$ *is 5 for 5 billion.*[132]

*We can obtain a more precise analysis that constrains the list argument to be well-formed. We use the following:*

```
197  % Declare user parameters
198  params: b; bs.
199  % type suggestion for bitstring
200  bitstring(Xs) :- bs(Xs).
201  % propagation rules
202  b(0) <== true.
203  b(1) <== true.
204  bs([]) <== true.
205  b(X) <== bs([X|Xs]).
206  bs(Xs) <== bs([X|Xs]).
```

*Notice that #200 initializes the type for* `bitstring(Xs)`*. From there, abstract forward chaining proves that it is a valid upper bound. The output is*

```
207  bit(0).
208  bit(1).
209  bitstring(Xs) :- bs(Xs).
```

It is no surprise that our symbolic evaluation method (abstract forward chaining) requires these kinds of termination strategies, as symbolic evaluation techniques, more broadly, are plagued by *path explosion*, and they employ truncation techniques (like ours) to make their methods practical (Baldoni et al., 2018, §5). State merging (Baldoni et al., 2018, §5) is another practical technique where two or more analyzer states are merged, i.e., replacing a single state representing their union, possibly a lossy union. Future work may wish to explore state-merging methods as a way to mitigate the aggressive truncation

[132] In our experience as users of the type analyzer, we have seen that there are advantages to re-running the type analyzer with smaller values of the truncation parameters since they result in a smaller output that illustrates the same divergence issue.

that we do here.[133]

In summary, we have described a truncation technique for ensuring that abstract forward chaining produces a valid upper bound in finite time. The bound is sound but may be quite loose. The hope is that a loose bound is more useful than a “type analysis did not converge” error. Users can inspect the system’s output and provide additional constraint-propagation rules to avoid truncation. Thus, the type analysis system can be used as a theorem-proving assistant.[134] Lastly, truncation allows downstream analyses to proceed, albeit with the loose type.

[133] We gave a simple-type merging method in Algorithm 15 that may be useful for this purpose.

[134] It would be interesting to explore the use of dynamic analysis as a means to inform static analysis. For example, a future system could synthesize plausible invariants that happen to be true of the program items when running on concrete data. From there, the system could recommend some assertions to the user that would tighten the static program analysis of the program.

# 6.2 Type-based Program Specialization

This section describes a system that specializes a given program to a specific input–output API. The system exploits the type analyzer of §6.1 to perform a program transformation (i.e., Dyna-to-Dyna code generation) that produces the specialized program. Consider the following example.

```
210 goal += a(I,I).
211 a(I,K) += b(I,J) * c(J,K).
212 goal += dead(X).
```

with the input type[134]

```
213 b̄(_,_).
214 c̄(_,_).
```

and output type[134]

```
215 ĉ(_,_).
```

We can see that `dead(X)` is not defined; thus, the bottom rule (#212) provably never fires given the declared input type. Now, suppose we work backward from the outputs (i.e., `goal` in the example). In that case, we see that we only require a subset of `a(I,K)`—specifically, just its diagonal `a(I,I)`. We call `a(I,I)` *useful*, whereas the off-diagonal elements are *useless*.[135] Useful items should also be non-0. Usefulness analysis is a form of *sensitivity analysis*: an item is **useful** if it is non-0 and setting its value to 0 would change the value of any output item, and **useless** otherwise. Computing the precise set of useful items requires concrete input data, so much like §6.1, we will estimate a (symbolic)

[134]Recall that the underscore (_) creates a fresh, unnamed variable.
[134]Here the marking $\hat{\cdot}$ is like the $\bar{\cdot}$ marking from §6.1, except for *useful* types (see below).
[135]In the parlance of finite-state machines and context-free grammars, a useful item is *both* accessible and co-accessible.

upper bound on the set of useful items for a specific input–output type.

Below, we exploit our knowledge of the useful items to manually *specialize* each rule of the example program so it computes goal more efficiently:

```
216 goal += a(I,I).
217 a(I,I) += b(I,J) * c(J,I).  % only builds a's diagonal
```

Notice that this program is not fully equivalent to the previous one because the off-diagonal a(I,K) items are now 0-valued.[136] However, the declared input–output API is preserved: the output values (namely, goal) are guaranteed to be equal to those of the original program.

We can improve this program further by *abbreviating* the a(I,I) items to a'(I) because they always carry a redundant variable:

```
218 goal += a'(I).
219 a'(I) += b(I,J) * c(J,I).
```

We can optionally include rules, such as a(I,I) += a'(I), that recover the original items from their abbreviated versions (assuming they are useful). Later, we will show examples of abbreviations that flatten nested terms, remove redundant variables, and drop variables that are never instantiated.

This section will develop an automated system for performing specialization transformations like these.

- First (§6.2.1), we describe extension to our type analysis method (§6.1) to estimate the set of useful items.
- Next (§6.2.2), we give a specialization method that transforms each rule—

[136] In some parts of the thesis, this kind of semantics-altering transformation would require us to rename the specialized version a as it has a different meaning. To keep the exposition simple, we will allow it.

**Algorithm 16** useful: For a given program $\mathfrak{p}$, input type $\overline{\mathfrak{D}}$, and output type $\widehat{\mathfrak{D}}$, define a new set of items $\widehat{\cdot}$ where $\widehat{x}$ is true if $x$ might be useful.

▷ *Note: The function* $\widehat{\cdot}$ *wraps a term in a novel function symbol.*
**def** useful$(\mathfrak{p}, \overline{\mathfrak{D}}, \widehat{\mathfrak{D}})$:
  $\widehat{\mathfrak{p}} \leftarrow \{\}$ ▷ *New program*
  $\widehat{\mathfrak{p}} \oplus\!= \overline{\mathfrak{p}} \oplus \overline{\mathfrak{D}}$ ▷ *Defines the set of the possible items* $\overline{\cdot}$*; same as §6.1*
  $\widehat{\mathfrak{p}} \oplus\!= \widehat{\mathfrak{D}}$ ▷ *Base case: output types are useful*
  **for** $(x \oplus\!= y_1 \circ \cdots \circ y_K) \in \mathfrak{p}$:
    **for** $k \in [1{:}K]$:[137]
      $\widehat{\mathfrak{p}}[\widehat{y}_k]$ :- $\widehat{x}, \overline{y_1}, \ldots, \overline{y_K}$ ▷ *The subgoal* $y_k$ *is useful if it appears in an active rule with a useful head*
  **return** $\widehat{\mathfrak{p}}$

by expanding it into multiple versions per possible type—with the aim of reducing the total number of useless items and useless rule firings.

- Lastly (§6.2.3), we refine the specialization method to perform abbreviation, as in the example above.

### 6.2.1 Inferring a Type for Useful Items

To reason about the set of useful items for a specific API (i.e., a pair of an input type program $\overline{\mathfrak{D}}$ and an output type program $\widehat{\mathfrak{D}}$, we transform the program $\mathfrak{p}$ using Algorithm 16. The output will be a new program $\widehat{\mathfrak{p}}$ that defines new items marked with $\widehat{\cdot}$ such that $\widehat{x}$ is `true` if $x$ might be useful. Please review the annotated pseudocode below for a further explanation.

Performing type analysis on $\widehat{\mathfrak{p}} =$ useful$(\mathfrak{p}, \overline{\mathfrak{D}}, \widehat{\mathfrak{D}})$ using abstract forward chaining (§6.1) will infer a type for $\widehat{\texttt{X}}$. Below is $\widehat{\mathfrak{p}}$

[137] It is safe to skip $k$ when $y_k$ is a constant or builtin.

```
220 % base case: output types are useful
221 goal̂.
222 % goal += a(I,I).
223 â(I,I) :- goal̂, ā(I,I).
224 % goal += dead(X).
225 dead̂(X) :- goal̂, dead̄(X).
226 % a(I,K) += b(I,J) * c(J,K).
227 b̂(I,J) :- â(I,K), b̄(I,J), c̄(J,K).
228 ĉ(J,K) :- â(I,K), b̄(I,J), c̄(J,K).
```

```
229 % booleanized program
230 goal̄ :- ā(I,I).
231 goal̄ :- dead̄(X).
232 ā(I,K) :- b̄(I,J), c̄(J,K).
233 % input type
234 b̄(I,J).
235 c̄(J,I).
```

The inferred type $\widehat{\mathfrak{m}} = \textsf{abstract_fc}(\widehat{\mathfrak{p}})$ is

```
236 goal̂.      goal̄.
237 â(I,I).    ā(I,K).
238 b̂(I,J).    b̄(I,J).
239 ĉ(J,K).    c̄(J,K).
```

We see on #237 that only the diagonal `a(I,I)` is useful and the whole of `a(I,K)` is possible, as we observed in our manual analysis earlier.

Lastly, we note that $\widehat{\mathfrak{m}} \subseteq \overline{\mathfrak{m}}$ because we have added more constraints to the inference problem. We can see on line 16.8 that $\widehat{y_k}$ is only true if $\overline{y_k}$.

### 6.2.2 Specialization Transformation

Given a particular type $\widehat{\mathfrak{m}}$,[138] we can specialize our program $\mathfrak{p}$ by instantiating its rules against this type to see which rules will never (usefully) fire. Better yet, we can specialize each rule to its useful instantiations. We provide pseudocode in Algorithm 17.

**Correctness.**

↪ **Disjoint $\widehat{\mathfrak{m}}$.** We require the branches of $\widehat{\mathfrak{m}}$ to be pairwise disjoint. Disjointness guarantees that we can replace a rule $r$ by nonoverlapping instanti-

[138]Note that the type $\overline{\mathfrak{m}}$ from §6.1 would work too (swapping $\widehat{\cdot}$ with $\overline{\cdot}$ where it is needed); however, but the resulting specialized program would be less efficient since $\widehat{\mathfrak{m}} \subseteq \overline{\mathfrak{m}}$.

**Algorithm 17** specialize: Specialize program $\mathfrak{p}$ to the type $\widehat{\mathfrak{m}}$.

**def** specialize($\mathfrak{p}, \widehat{\mathfrak{m}}$):
  **assert** the simple types in $\widehat{\mathfrak{m}}$ are pairwise disjoint
  $\mathfrak{p}' \leftarrow \{\!\!\{\}\!\!\}$ ▷ *New program*
  **for** $(x \mathrel{\oplus=} y_1 \circ \cdots \circ y_K) \in \mathfrak{p}$:
    **for** $\langle \mathcal{C}_0, \mathcal{C}_1, \ldots, \mathcal{C}_K \rangle \in \widehat{\mathfrak{m}}[\widehat{x}] \times \widehat{\mathfrak{m}}[\widehat{y}_1] \times \cdots \times \widehat{\mathfrak{m}}[\widehat{y}_K]$: ▷ *Expand against the type*
      $\mathcal{C} \leftarrow$ propagate$(\mathcal{C}_0 \circ \mathcal{C}_1 \circ \cdots \circ \mathcal{C}_K)$
      **if** `fail` $\in \mathcal{C}$: **continue** ▷ *Safe to drop since it cannot fire*
      $\mathfrak{p}'[x] \mathrel{\oplus=} y_1 \circ \cdots \circ y_K \circ \mathcal{C}$ ▷ *The constraints $\mathcal{C}$ become boolean side conditions*
  **return** $\mathfrak{p}'$

ations. This is needed to avoid double-counting. Given a disjoint useful type $\widehat{\mathfrak{m}}$, we can specialize the rules of the program and its different cases. The way that the specialization transformation works is to use the abstract instantiation of the rule to gather constraints of the types that it can take on. Using those types, we can add (concrete) constraints to the rule that partition it into its possible cases. If there are overlapping types, we can take an inferred type $\widehat{\mathfrak{m}}$ and make it disjoint by applying disjoin($\widehat{\mathfrak{m}}$) (Algorithm 9) as a preprocessing step.

↪ **User parameters required.** To leverage the user's parametric types, they must provide them in the input. Consider the two possible types form the `γ(X,Y)` items in CKY (Example 18):

```
γ̄(X,Y) :- k(X), k(Y)
γ̄(X,Y) :- k(X), w(Y)
```

When the user provides a collection `γ(X,Y)` we do not know which have `k(Y)` and which have `w(Y)`. Thus, we require the user to provide these sets for our system to make their type distinctions.[139]

We require the user to provide their parameters in addition to the input data. For instance, the CKY example requires the user to provide methods that implement the `n`, `k`, and `w` constraints. This is done by defining them as items along with the input data. Since these items are constraints, they are $\{\underline{0}, \underline{1}\}$-valued; only the $\underline{1}$-valued items need to be enumerated in the input data $\mathfrak{D}$.[140] Note that the `<` constraint that appears in CKY's input type is a built-in relation provided by the system.

[139] In some cases, it may be possible to automatically determine the user's parameter sets from the input data. A warning is printed if the user forgets to provide at least one instance for each parameter.

### 6.2.3 Abbreviation Transformation

We extend the specialization transformation (§6.2.2) to *abbreviate* the names of items as we did in the matrix-diagonal example earlier. Abbreviation is a scheme that partitions each item by its possible simple types. It defines new items that are named by their simple type and any (necessary) arguments. Pseudocode is provided in Algorithm 18.

**Correctness.** Abbreviation and specialization (§6.2.2) are correct under the same conditions.

**Abbreviation vs. specialization.** Unlike the specialization transformation, the abbreviation transformation can omit certain constraints from the abbreviated program because they are statically guaranteed to hold given the types in the body of the rule (see line 18.27). The specialization transformation

[140]They can be defined using rules, e.g., if sentence positions (`n`) are encoded as the integers $\{1, \ldots, 10\}$, we can write `n(X) :- 1` $\leq$ `X` $\leq$ `10`.

[141] The order of the tuple $\langle \mathtt{X}_1, \ldots, \mathtt{X}_M \rangle$ is arbitrary, but it must be consistent for each $t$. Completely free variables may be omitted from the tuple.

[142]When adding the rule on line 18.30, we must apply the multiplicity correction (§3.1.2.3) if any free variables are summed in the abbreviation. We compute the set of local free variables in the original rule to those in the abbreviated rule. If these sets of variables are unequal, we $\circ$-multiply the body of the new rule by $(\infty \cdot \underline{1})$.

**Algorithm 18** abbreviate: Abbreviate program $\mathfrak{p}$ according to the type $\widehat{\mathfrak{m}}$.

```
def abbreviate(𝔭, 𝔪̂):
  assert the simple types in 𝔪̂ are pairwise disjoint
  𝔭′ ← ⦃⦄                                                    ▷ Will be the output program
  ▷ Map the item x to the abbreviated name for the type identified by t
  def abbrev_t(x):
    if x is a builtins and constants: return x
    (x′ :- 𝒞′) ← fresh(𝔪̂_t)
    ⟨x_1, … , x_M⟩ ← the ordered[141] set of variables in (x′ :- 𝒞′)
    θ ← unify(x, x′)
    return θ(gen_t(x_1, … , x_M))                ▷ Here gen_t is a novel symbol for simple type t
  ▷ API preservation rules
  for (x :- 𝒞) ∈ 𝔪̂:
    Assign a unique identifier t to the simple type (x :- 𝒞)
    if t is an input:
      ▷ Input items should not be abbreviated to preserve 𝔭's API. The rule below feeds
        the input items to the abbreviated items for its type t.
      𝔭′[abbrev_t(x)] ⊕= x ∘ 𝒞     ↰ Here, the constraints 𝒞 are required to ensure that the correct set
                                     of input items are passed to this abbreviated item.
    else
      ▷ The rule below recovers the original item from the abbreviated item; Note: the
        rule added below is useless in 𝔭′ unless x is an output item.
      𝔭′[x] ⊕= abbrev_t(x)
  for (x ⊕= y_1 ∘ ⋯ ∘ y_K) ∈ 𝔭:
    for ⟨𝒞_0, 𝒞_1, … , 𝒞_K⟩ ∈ 𝔪̂[x̂] × 𝔪̂[ŷ_1] × ⋯ × 𝔪̂[ŷ_K]:
      𝒞 ← propagate(𝒞_1 ∘ ⋯ ∘ 𝒞_K)
      if fail ∈ 𝒞: continue
      𝒞′ ← propagate(𝒞_0 ∘ 𝒞)
      if fail ∈ 𝒞′: continue
      ▷ The head's type constraints must be checked, but the constraints in 𝒞 can be
        removed, as they are guaranteed to be satisfied.
      𝒞″ ← 𝒞_0 ∖ 𝒞
      ⟨t_0, t_1, … , t_K⟩ ← identifiers for the current simple types
      ▷ Replace x and each y_k with its type-specific abbreviation.[142]
      𝔭′[abbrev_{t_0}(x)] ⊕= abbrev_{t_1}(y_1) ∘ ⋯ ∘ abbrev_{t_K}(y_K) ∘ 𝒞″
  return 𝔭′
```

could not do that because the subgoals are not split into their different types. So a subgoal with more than one simple type could satisfy the constraints in only *some* of the cases. This is not an issue in the case of abbreviation because the items are split into their disjoint simple types. Consider, again, CKY. The subgoal `rewrite(X,Y)` has two possible simple types: `γ̄(X,Y) :- k(X), k(Y)` and `γ̄(X,Y) :- k(X), w(Y)`. So specialization cannot guarantee much about the constraint on `Y`, but abbreviation can because `rewrite(X,Y)` split into its two types.[143]

We provide an NLP example of abbreviate in Example 21.

**Example 20.** *Consider the following program for the total weight of all paths between pairs of nodes in a weighted graph:*

```
242 path(I,I).
243 path(I,K) += edge(I,J) * path(J,K).
```

*Suppose the input and output types are*

$\overline{\mathfrak{D}} = \{$`params: n. ` $\overline{\texttt{edge}}$`(I,J) :- n(I), n(J).`$\}$ *and* $\widehat{\mathfrak{D}} = \{$`path(I,K) :- n(I), n(J).`$\}$

*Abbreviation produces the following program:*

```
244 % map each input type to its abbreviation
245 edge₀(I,J) += edge(I,J) * n(I) * n(J).
246 % recovery rules for useful output items
247 path(I,K) += path₁(I,K).
248 % abbreviated rules
249 path₁(I,I) += n(I).
250 path₁(I,K) += edge₀(I,J) * path₁(J,K).
```

*Notice that the nonground* `path(I,I)` *items are gone because the variable* `I` *is now constrained to* `n(I)`.

*Now, suppose we introduce a type (*`a`*) for the* useful *starting states:*

[143]That said, we at least know that `k(X)` holds in both cases so that we could drop that constraint. Our specialization transformation does not do this optimization, but a subsequent analysis could do it.

$\overline{\mathfrak{D}} = \{$`params: a; n.` $\overline{\texttt{edge}}$`(I,J) :- n(I), n(J).`$\}$ *and* $\widehat{\mathfrak{D}} = \{$`path(I,K) :- a(I), n(J).`$\}$

*Under this API,* abbreviate *produces:*

```
251 % map each input type to its abbreviation
252 edge₀(I,J) += edge(I,J) * n(I) * n(J).
253 % recovery rules for useful items
254 path(I,K) += path₁(I,K).
255 % abbreviated rules
256 path₁(I,I) += a(I) * n(I).
257 path₁(I,K) += edge₀(I,J) * path₁(J,K).
```

*This version further constrains the* `path(I,K)` *items, as they must also satisfy* `a(I)`.

# 6.3 Automatic Space and Time Analysis

This section builds on the type analysis provided by the methodology of §6.1 to develop an automated method for analyzing the space and time complexity of a Dyna program $\mathfrak{p}$ given data $\mathfrak{D}$. Given the types $\overline{\mathfrak{m}}$ associated with $\mathfrak{p}$, the key is to find parametric upper bounds on their *cardinality*. We will then use these to predict a parametric $\mathcal{O}$-bound on the running time of forward chaining execution of $\mathfrak{p}(\mathfrak{D})$. The bounds of this section assume that the program is executed with the agenda-based fixpoint iteration algorithm of §5.

## 6.3.1 Space Analysis

To bound the space complexity of our fixpoint iteration algorithms for evaluating $\mathfrak{m} = \mathfrak{p}(\mathfrak{D})$, it suffices to bound the size of $\mathsf{nz}(\mathfrak{m})$ because these algorithms only have to store $\mathfrak{m}$.[144] We will build on §6.1's upper bound $\overline{\mathfrak{m}} \supseteq \mathsf{nz}(\mathfrak{m})$. We will use $|\, s\,|$ to denote the true cardinality of a set $s$ and $\lceil s \rceil \geq |\, s\,|$ to denote an upper bound; we extend the notation to programs by implicitly coercing them to sets through $\mathsf{nz}(\cdot)$. Our high-level approach is $\lceil \overline{\mathfrak{m}} \rceil \stackrel{\text{def}}{=} \sum_{s \in \overline{\mathfrak{m}}} \lceil s \rceil \geq \sum_{s \in \overline{\mathfrak{m}}} |\, s\,| \geq |\, \overline{\mathfrak{m}}\,| \geq |\, \mathfrak{m}\,|$,[145] where the size $|\, s\,|$ of a simple type $s$ counts the number of satisfying assignments

[144] We assume that any additional data structures, such as those used to speed up the lookup operations, increase the space complexity by only a constant factor. Additionally, we assume that each value in $\mathbb{V}$ is constant-sized.

[145] If simple types in $\overline{\mathfrak{m}}$ overlap considerably, then the last inequality will be loose because summation double-counts the overlapping regions. Removing subtypes (§6.1.2.4) addresses the most egregious case, but future work may consider transforming $\overline{\mathfrak{m}}$ into a *partition* of simple types as follows. Replace any overlapping pair of simple types $s$ and $t$ with $s \cap t$, $s \smallsetminus t$, and $t \smallsetminus s$. Repeat until no pairs $s$ and $t$ overlap. Set difference $\smallsetminus$ can be implemented using negated versions of the elementary constraints—including equality (i.e., unification), which is usually represented implicitly.

to its variables, and $\lceil s \rceil$ is an upper bound on this.

It is well-known that counting the variable assignments that satisfy a set of constraints can be computationally expensive. However, the more important issue in our case is that we do not even know how the parametric constraints are defined. The remainder of this section considers how to construct a bound $\lceil s \rceil$ from some user-supplied information about the parametric constraints.

**Input Size Specification.** Consider the simple type `γ̄(X,W) :- k(X), w(W)`. A straightforward upper bound can be found:

$$\left| \texttt{\=γ(X,W) :- k(X), w(W)} \right| \leq \lceil \texttt{k(X)} \rceil \cdot \lceil \texttt{w(W)} \rceil = k \cdot w$$

where $k$ and $w$ are user-specified size parameters. (We will explain shortly how size parameters are specified.) However, matters become more complicated when there are multiple constraints on the same variable. Consider the simple type

```
β̄(X,I,K) :- k(X), n(I), n(K), I < K.
```

It is clear that at most $k \cdot n \cdot n$ assignments can satisfy the first 3 constraints. Additionally, imposing the fourth constraint `I < K` can only reduce that number—even though the fourth constraint *in isolation* would have infinitely many satisfying assignments ($|\texttt{I < K}| = \infty$).

We will describe how to derive an upper bound from the *sequence* of constraints $c_1, \ldots, c_M$ in the body of the simple type $s$ (including any constraints added by propagation, as they may improve the bound). Reordering the constraints does not affect the number of satisfying assignments; however, each of the $M!$ orders may give a different upper bound, so we select the tightest of these.

The idea is to use **conditional cardinalities**. Let $\mathcal{V}$ be any set of variables. While $c$ may be satisfied by many assignments to its variables, only some are compatible with a *given* assignment to $\mathcal{V}$. Suppose the user supplies a number $\lceil c \mid \mathcal{V} \rceil$ such that no assignment to $\mathcal{V}$ can be extended into more than $\lceil c \mid \mathcal{V} \rceil$ assignments to $\mathsf{vars}(c)$. The case $\lceil c \mid \mathcal{V} \rceil = 1$ is of particular interest: this is necessarily true if $\mathsf{vars}(c) \subseteq \mathcal{V}$ and asserts a functional dependency otherwise.

For a simple type $s$ of the form $\bar{x}$ :- $c_1, \ldots, c_M$, we can obtain an upper bound on the cardinality of $s$: $|\, s \,| \leq \prod_{m=1}^{M} \lceil c_m \mid \mathcal{V}_{<m} \rceil$, where $\mathcal{V}_{<m} = \bigcup_{i=1}^{m-1} \mathsf{vars}(c_i)$.[146] More generally, an upper bound on the *conditional* cardinality of $s$, given an assignment to some set of variables $\mathcal{V}$, is $\prod_{m=1}^{M} \lceil c_m \mid \mathcal{V} \cup \mathcal{V}_{<m} \rceil$. Minimizing this over all orderings gives our final upper bound $\lceil s \mid \mathcal{V} \rceil$ (with $\lceil s \rceil \stackrel{\text{def}}{=} \lceil s \mid \emptyset \rceil$).

But where do the bounds $\lceil c \mid \mathcal{V} \rceil$ come from? Consider the built-in constraint `times`, which is an infinitely large relation on three numbers. Because `times` has functional dependencies, we can take $\lceil \texttt{times(X,Y,Z)} \mid \mathcal{V} \rceil = 1$ whenever $\mathcal{V}$ contains at least two of $\{\texttt{X}, \texttt{Y}, \texttt{Z}\}$. We allow a user to declare this conditional cardinality with the following notation:

```
259  |times(+X, +Y, Z)| ≤ 1.
260  |times(X, +Y, +Z)| ≤ 1.
261  |times(+X, Y, +Z)| ≤ 1.
```

For example, #261 declares that given values for `X` and `Z`, the `times(X,Y,Z)` constraint will be satisfied by at most one value of `Y`.

Given all such declarations, we obtain $\lceil c \mid \mathcal{V} \rceil$ as the min of the upper bounds provided by conditional cardinality declarations that unify with $c$ and whose variables marked with + are all in $\mathcal{V}$. This includes an implicit declaration that $\lceil \texttt{c} \mid \texttt{vars}(\texttt{c}) \rceil \leq 1$. (Another implicit declaration is $\lceil \texttt{fail} \rceil \leq 0$.)

[146] Provided that all variables in the head $\bar{x}$ are constrained somewhere in the body. If not, $|\, s \,| = \infty$.

The same notation is useful for bounding input relations, such as the $\overline{\texttt{word}}$ relation in Example 6.

```
|word(W, +I, K)| ≤ 1.
|word(W, I, +K)| ≤ 1.
|word(W, I, K)| ≤ n.
```

The first declaration says that there is at most one word `W`, paired with a single end position `K`, starting at each position `I`. The second declaration is the mirror image of this. The third says that there are at most $n$ words overall in the input sentence—where the symbol $n$ is a **size parameter**.

We may also use size parameters to bound the cardinality of type parameters:

```
|k(X)| ≤ k.    |n(I)| ≤ n.    |w(W)| ≤ w.
```

which bounds the number of nonterminals by $k$, sentence positions by $n$ and terminals by $w$. Here $k$, $n$, and $w$ are interpreted as symbols. We will combine these symbols to form a **parametric expression** that upper bounds the cardinality of the items in the program. For example, as we saw earlier, $\lceil \overline{\beta}(\texttt{X,I,K}) \rceil = kn^2$. After adding up the sizes of simple types, our system can compute an asymptotic upper bound on the resulting expression, e.g., $kn^2 + kn + k^2n = \mathcal{O}(kn\max(n, k))$.

A final trick is to specify the cardinality or conditional cardinality of a composite constraint, which can be matched against a subsequence of $\texttt{c}_1, \ldots, \texttt{c}_M$ when constructing an upper bound. For example, `|even(X),prime(X)| ≤ 1` says that there is only one even prime, while `|γ̄(X,Y), k(Y)|` $\leq g$ says that no nonterminal symbol `X` can rewrite as more than $g$ different nonterminal symbols `Y`. (It does

not, however, constrain how many ways `X` can rewrite as a *terminal* symbol `Y` because these do not satisfy the `k(Y)` constraint.)

**Limitations.**

- Certain infinite relations, such as those that can be expressed with free variables and delayed constraints, have an estimated size of $\infty$ under this approach (cf. footnote 146). However, certain infinite relations can be represented efficiently by the solvers in § 5 as finite-size programs. However, because our approach is based on bounding the number of non-$\underline{0}$ items rather than the size of their representation, our size estimates end up extremely loose.

- We assume that all built-in relations ground out all variables that appear in them. Future work should integrate a more sophisticated mode system (e.g., Overton (2003) and Filardo (2017)) for reasoning about the instantiation states of variables. Mode analysis can also be used to reason about the set of delayed constraints, so it should help address the limitation mentioned earlier.

- Our space bounds are always (piecewise) polynomials in the input-size parameters; they cannot express exponential size bound, for example.

### 6.3.2 Time Analysis

Suppose that the Dyna program is executed by the agenda-based algorithm (Algorithm 8). This algorithm maintains two programs: the **chart** $\mathfrak{m}$ and the **agenda** $\partial$. The chart represents the current estimate of the fixpoint (§4.3). The agenda is a queue of updates to be applied to the values in $\mathfrak{m}$. When an update to $y'$ is popped, its value is updated in the chart. This change is then propagated to other items $x$. The updates to these $x$ items are placed on the agenda and carried out only later. Crucially, the affected $x$ items are found by matching $y'$ (the **driver**) to any subgoal of any rule $x \oplus\!= \cdots$, and then querying the chart for items (**passengers**) that match that rule's other subgoals.

Propagating an update from a given driver $y'$ through a rule $r$ requires us to find all the satisfying assignments to $\mathsf{body}(r)$ that are consistent with binding the driver subgoal to $y'$. Although $r$ is not a type, a subgoal in its body such as $\mathtt{\gamma(X,Y)}$ does impose a constraint on $\mathtt{X}$ and $\mathtt{Y}$—namely the constraint $\overline{\mathtt{\gamma}}\mathtt{(X,Y)}$ from our type analysis (§6.1), which is true only if $\mathtt{\gamma(X,Y)}$ could have a value in the chart.

Suppose we have matched a driver $y'$ against a subgoal of $r$ and have already matched 0 or more other subgoals to passengers retrieved from the chart. These matches have instantiated some variables $\mathcal{V}$ in the rule. For each subgoal match, we also know from our static type analysis (§6.1.2) that the variables involved in that match must have satisfied certain constraints $\mathcal{C}$. If $\mathtt{\gamma(X,Y)}$ is the next subgoal that we will match against the chart, we will get at most $N \overset{\text{def}}{=} \lceil \mathcal{C}, \overline{\mathtt{\gamma}}\mathtt{(X,Y)} \mid \mathcal{V} \rceil$ answers. Our methods from the previous section can compute this upper bound even though $\overline{\mathtt{\gamma}}$ is not a built-in or parametric constraint: simply refine the item type $\overline{\mathfrak{m}}$ by unifying the heads of all simple types $s \in \overline{\mathfrak{m}}$ with $\overline{\mathtt{\gamma}}\mathtt{(X,Y)}$, and then upper-bound the conditional cardinality of this refined type $\overline{\mathfrak{m}}'$.

Thanks to the indexing of the chart, the runtime of getting and processing $N$ answers is proportional to $1 + N$ in the worst case (with the $1$ representing the overhead of attempting the query even if there are no answers). For each answer, we must then proceed to match the remaining passengers in all possible ways.

The pseudocode in Algorithm 19 makes this precise. Two subtleties are worth pointing out. First, $\overline{\mathrm{m}}$ is disjunctive. The constraints $\mathcal{C}$ that we accumulate from previously matched passengers depend on the particular simple types that cover those passengers, so we must iterate through all possibilities. Second, while the driver subgoal is chosen by the agenda, the order in which to instantiate the remaining subgoals (passengers) is up to us. The pseudocode always chooses the next subgoal that will minimize the upper bound on runtime, given the simple types of the previously matched passengers. In other words, instead of following the fixed subgoal order stated in rule $r$, we are using our static analysis to optimize the subgoal ordering to have better running time.[147]

**Correctness and Limitations.** The running time bound produced by Algorithm 19 assumes that the number of times an item drives is *constant*. If the program is acyclic, then the topologically ordered agenda (§5.1.4) guarantees that each item is popped at most once, so the bound is correct. However, if the program is cyclic, items may drive more than a constant number of times. In the cyclic case, we can interpret our bound as the worst-case complexity of one iteration of

[147] Future work may wish to explore methods that directly optimize the average execution time of the passenger query order in place of our upper bound. The average execution time could be estimated and optimized using machine learning techniques. Vieira, Francis-Landau, et al. (2017) describe a general framework for such optimization. In their terminology, the specific passenger ordering would be a *strategy* for the *task* of answering the passenger query. Their method would additionally allow the ordering choice to be made dynamically by a *policy* that has access to the specific passenger query (i.e., it is not limited to just its type) and the status of the available indexing data structures.

**Algorithm 19** runtime: Infer a runtime bound for $\mathfrak{p}$ given a derived type $\overline{\mathfrak{m}}$.

```
def runtime(𝔭, m̄):
    t ← 0                                                        ▷ Accumulator
    for (y̅′ :- 𝒞′) ∈ m̄:                                          ▷ Possible driver types
        for (x ⊕= y_1 ∘ ⋯ ∘ y_K) ∈ 𝔭:          ┐
            for k ∈ [1:K]:                     │ Matching rules against driver y′
                for _ ∈ unify(y_k, y′):        ┘
                    ▷ Add the cost of driving the current rule
                    t += runtime′({y_1, …, y_K} − {y_k}, 𝒞′, vars(y_k))   ▷ Note: y′ = y_k due to unify
                                   (Passengers)
    return t
▷ This recursive function (which can be memoized for efficiency) tells us the optimal runtime for a set of subgoals (𝒫), with 𝒞 being the set of conjunctive constraints that have been accumulated and 𝒱 the set of bound variables.
def runtime′(𝒫, 𝒞, 𝒱):
    if 𝒫 = ∅: return 1                                           ▷ Base case: no subgoals remaining
    best ← ∞
    ▷ Optimize over the choice of next subgoal p from the set of remaining subgoals
    for p ∈ 𝒫:
        t ← 0                                   ▷ Accumulate total suffix runtime (given choice p)
        for ⟨𝒞′, θ⟩ ∈ m̄[p̄]:                     ▷ Lookup simple types in m̄ that match p. Note: we
                                                  require an explicit result stream here (§3.1.2.1).
            𝒞″ ← propagate(θ(𝒞 ∘ 𝒞′))
            q ← ⌈𝒞″, p̄ | 𝒱⌉                      ▷ Lookup conditional cardinality for passenger p
            if q = 0: continue                   ▷ For efficiency, avoid recursive call
            t += q · runtime′(θ(𝒫 − {p}), 𝒞″, 𝒱 ∪ vars(p))                 ▷ Recurse
        best min= t
    return 1 + best
```

seminaive (Algorithm 7) as a function of the input-size parameters rather than the total complexity of execution. To obtain the total complexity, we must adjust the bound to account for the number of times each simple type may pop, which may depend on the input-size parameters and the prioritization strategy used.[148,149]

↪ *Other limitations*: Because our runtime bounds are built on the space complexity analysis from §6.3.1, we inherit its limitations. In particular, our bounds are (piecewise) multivariate polynomials in the input-size parameters. This means, for example, that our system cannot infer an exponential runtime bound; it would instead predict an infinite running time, as it is impossible to upper bound an exponential with a polynomial. We also mention that our runtime analyzer produces an infinite upper bound whenever an infinite-cardinality delayed constraint might appear in the chart, even though our execution algorithms do not incur an infinite runtime cost when working with them. Lastly, our bounds assume that semiring operations run in constant time.

[148]Discussion of approximate fixpoints and convergence rates is provided in §4.4.

[149]Recall (§5.4) that a cyclical program can be solved more efficiently by decomposing it into strongly connected components. Future systems for reasoning about the running time of solving cyclical programs should include a static SCC decomposition analysis.

↪ *Ideas for future work*: Static analysis of cyclicity or, more generally, the strongly connected components of a program would clearly be useful for improving our running time analysis. It would also be useful to have runtime bounds for the other algorithms in § 5 (e.g., newton). Our general approach of simulating the work done in each iteration of these algorithms should translate straightforwardly. It would also be interesting to provide runtime bounds that are specialized to specific families of semirings (e.g., commutative idempotent semirings), as they may have finite-time convergence guarantees.

## 6.4 Applications

This section provides additional examples of our static analyzer.

**Example 21.** *Kuhlmann, Gómez-Rodríguez, et al. (2011) introduce an exact dynamic program for arc-eager parsing (Nivre, 2003) when the scoring function is arc-factored (Eisner, 1996). A Dyna encoding of their algorithm is given below:*

```
266 β(I,I^0,K) += word(I,K).
267 β(I,I^1,K) += word(I,K).
268 β(I,H^B,K) += β(I,H^B,J) * score(H ← H') * β(J,H'^0,K).
269 β(I,H^B,K) += β(I,H^B,J) * score(H → H') * β(J,H'^1,K).
270 goal += β(0,0^0,N) * len(N).
271 params: word; score; len.
```

*Naïvely, one could think that the arc-eager parser runs in $\mathcal{O}(n^5)$. However, a tighter analysis reveals it actually runs in $\mathcal{O}(n^3)$. To derive this, consider the following type declaration:*

```
272 params: n(N).
273 score̅(H → H') :- n(H), n(H').
274 score̅(H ← H') :- n(H), n(H').
275 word̅(I,K) :- n(I), n(K).
276 len̅(N) :- n(N).
```

*The derived types are*

```
277 β̄(I,I^0,K) :- n(I), n(K).
278 β̄(I,I^1,K) :- n(I), n(K).
279 goal̅.
```

*What is surprising about the derived types is that, although the rules suggest that* `H` *in* `β(I,H^B,K)` *may independently range over* `n`*, it turns out that it is always equal to* `I` *because only types that unify with it are* `β̄(I,I^0,K)` *and* `β̄(I,I^1,K)`*. Now, we consider the running time with the following input-size declarations:*

```
280 |n(N)| ≤ n.
```

*The inferred runtime bound is $\mathcal{O}(n^3)$.*

*Abbreviation (below) successfully drops the unnecessary variable, which improves the running time by a constant factor. The program below is isomorphic to the one that Kuhlmann, Gómez-Rodríguez, et al. (2011) discovered manually. We have dropped the* `n(I)` *constraints to make the output more readable. We have chosen abbreviated names that are similar to their original ones; we use subscripts to indicate the simple type's identifier.*

```
% Map each input type its abbreviation
score_0(H,H') += score(H → H').      } these items are flattened (nested terms are remove)
score_1(H,H') += score(H ← H').      }
word_2(I,K) += word(I,K).
len_3(N) += len(N).
% recovery rules
β(I,I^0,K) += β_4(I,K).   } these are useless recovery rules that can be dropped
β(I,I^1,K) += β_5(I,K).   }
goal += goal_6.
% abbreviated program rules
β_4(I,K) += word_2(I,K).
β_4(H,K) += β_4(H,H') * score_0(H,H') * β_5(H',K).
β_4(H,K) += β_4(H,H') * score_1(H,H') * β_4(H',K).
β_5(I,K) += word_2(I,K).
β_5(H,K) += β_5(H,H') * score_0(H,H') * β_5(H',K).
β_5(H,K) += β_5(H,H') * score_1(H,H') * β_4(H',K).
goal_6 += β_4(0,N) * len_3(N).
```

*The inferred runtime bound is unchanged by the abbreviation transformation.*

**Example 22.** *Below we give another version of CKY (Example 6) that encodes the input sentence as a list, e.g.,* `sentence([papa, ate, the caviar])`. *This version takes the encodes the weighted production rules as* `γ(X,Y,Z)` *and* `γ(X,Y)` *like before, but it does not require the sentence to be encoded with the* `word(W,I,K)` *or* `length(N)` *items. As before, the item* `goal` *sums the total weight of all complete parses.*

```
params: input; γ.  output: goal.
buffer(Ws) += sentence(Ws).     % initialize the buffer
buffer(Ws) += buffer([W|Ws]).   % shift
β(X,[W|Ws],Ws) += γ(X,W) * buffer([W|Ws]).
β(X,I,K) += γ(X,Y,Z) * β(Y,I,J) * β(Z,J,K).
β(X,I,K) += γ(X,Y) * β(Y,I,K).
goal += β(s,Ws,[]) * sentence(Ws).
```

*We specify the following input type:*

```
params: w; ws; k.
γ̄(X,Y,Z) :- k(X), k(Y), k(Z).   }
γ̄(X,Y) :- k(X), k(Y).           } same as Example 18
γ̄(X,W) :- k(X), w(W).           }
sentence̅(Ws) :- ws(Ws).
```

*Recall that* `k` *is the set of nonterminals and* `w` *is the set of words. The types for*

*the grammar rules are the same as the version of CKY that we analyzed in §6.1.*

*We define the input as type* `ws`*, which is a list of words. When we add the following propagation rules to encode that each of suffix of our list of words is also a list of words.*

```
ws([]) <== true.
ws(Xs) <== ws([X|Xs]).
```

*Lastly, we add the following propagation rules*

```
w(X) <== ws([X|Xs]).
fail <== k(X), w(X).
```

*The first rule says that the first element of a word list is a word, and the second rule says that the set of nonterminals* `k` *is disjoint from the set of words* `w`*.*

*Abstract forward chaining derives the following type for the derived items*

```
buffer(Xs) :- ws(Xs).
β(X,I,K) :- k(X), ws(I), ws(K).
goal.
```

*The runtime bound is the same as in ordinary CKY,* $\mathcal{O}(k^3\, n^3)$*, except that instead of encoding the input length as the number of sentence positions*

```
|n(I)| ≤ n
```

*we encode it as the number of suffixes in the input sentence*

```
|ws(Ws)| ≤ n.
```

*The abbreviation transformation, in this case, is not interesting; all it does is change the names of the existing items with no benefit.*

# Summary

This chapter developed several techniques for reasoning about a given program independently from specific input data. We described our system for deriving symbolic upper bounds on the program's set of non-0-valued items as a function of input data characterized by unknown, symbolic user parameters. We built on that technique to estimate the set of *useful* for answering queries about a specific set of output items. We then used the estimated set of useful items to *specialize* the program so that it computes them more efficiently by omitting useless items. We then turned our attention to space and time complexity analysis. We describe an approach for bounding the size of the set of non-0 items as a function of user-specified size parameters. We related this bound to the space complexity of agenda-based fixpoint iteration. We then gave a technique for analyzing the runtime complexity of agenda-based fixpoint iteration as a function of the input-size parameters. We will revisit the complexity bounds and the program specialization techniques in our program optimizer (§8).

# Chapter 7

# Program Transformations

In this chapter, we will develop a collection of meaning-preserving program-to-program transformations. These transformations will be used to systematically transform a given initial program into a faster program that computes the same values in §8.

Intuitively, program transformations may be regarded as *reusable tricks* for transforming one set of recursive definitions (a Dyna program) into a different (possibly faster-to-evaluate) set of recursive definitions (a different Dyna program). Since a Dyna program is essentially a system of equations that defines a valuation function (§4), we may regard program transformations much like the mathematical identities used to manipulate systems of mathematical equations (e.g., to solve them or to prove a new identity). Program transformations may be applied manually or automatically.

Program transformations—unlike execution algorithms (§5)—can operate without concrete input data.[150] Thus, many transformations correspond to steps of abstract interpretation (Cousot and Cousot, 1977) in that they simulate steps taken by some concrete solver. Thus, we introduce **input and output declarations**: we assume that the initial program in our transformation sequence declares some items as **input** and/or **output** items. The rest are considered **intermediate items**. Below is an example of CKY with input–output declarations.

```
β(X,I,K) += γ(X,Y,Z) * β(Y,I,J) * β(Z,J,K).
β(X,I,K) += γ(X,Y) * β(Y,I,K).
β(X,I,K) += γ(X,Y) * word(Y,I,K).
goal += β(s,0,N) * len(N).

input: word(X,I,K); γ(X,Y,Z); γ(X,Y); len(N).
output: goal.
```

same code

input–output declarations

To be considered *correct*, a program transformation must preserve the mapping from the valuation of the input items to the original program's valuation of the output items (§ 4). However, a program transformation can introduce, destroy, or alter intermediate items.[151]

**Outline.** In what follows, we will give an informal overview of the speedups that are achievable through program transformations—providing a more formal treatment of each transformation in a later section.

[150] Static analysis (§6.1) are another example.

[151] We will formalize notions of transformation correctness and program equivalence in §7.2.

## 7.1 Informal Introduction

This section provides an overview and general flavor of the types of transformations we will develop in this chapter. We defer formal details of each transformation to later sections. We also note that the method of §6.3 can be used to *automatically* derive the big-$\mathcal{O}$ bounds on space and time complexity in the examples below.

**Define and fold.** We start our discussion with non-recursive folding, as we expect it will be familiar to most readers.

**Example 23.** *The following program computes the total weight of length-4 paths in a graph with edge weights* `w`*:*

```
total += w(Y_1,Y_2) * w(Y_2,Y_3) * w(Y_3,Y_4) * w(Y_4,Y_5).
```

*By exploiting the distributive property, we can rewrite this program into one that reuses partial sums of products for each path-suffix length:*

```
total += rest_1(Y_1).
rest_1(Y_1) += w(Y_1,Y_2) * rest_2(Y_2).
rest_2(Y_2) += w(Y_2,Y_3) * rest_3(Y_3).
rest_3(Y_3) += w(Y_3,Y_4) * rest_4(Y_4).
rest_4(Y_4) += w(Y_4,Y_5).
```

*This program computes the same value for* `total`*, but it is significantly more efficient to evaluate than the original program. Specifically, the original program runs in time* $\mathcal{O}(N^5)$ *where* $N$ *is the number of nodes in the graph, and* *the transformed program runs in time $\mathcal{O}(N^2)$. This is an example of the definition transformation, which proposes the new* $\texttt{rest}_i$ *definitions, and the fold transformation, which grafts these definitions into the program.*

Readers may recognize that the junction tree algorithm exploits the same transformation scheme (Lauritzen and Spiegelhalter, 1988; Shafer and Shenoy, 1990) for inference in probabilistic graphical models and join-order optimization in database query planning (Chekuri and Rajaraman, 1997). Gildea (2011) discusses many uses of the junction tree algorithm in weighted deduction systems used in NLP. However, these approaches do not transform *recursive* computations—they are targeted at a fixed set of variables rather than an unbounded set as in recursive computations.[152]

**Example 24.** *Recall that CKY (Example 6) has the following expensive rule*

```
332  β(X,I,K) += γ(X,Y,Z) * β(Y,I,J) * β(Z,J,K).
```

*This rule can be sped up by summing over the variable* `Y` *separately from* `K`*:*

```
333  β(X,I,K) += tmp(I,J,X,Z) * β(Z,J,K).
334  tmp(I,J,X,Z) += γ(X,Y,Z) * β(Y,I,J).
```

*The transformed rules reduce the running time from $\mathcal{O}(K^3N^3)$ to $\mathcal{O}(K^2N^3 + K^3N^2)$ where $K$ is the number of nonterminals, and $N$ is the sentence length.*[153]

**Unfold.** The unfold transformation is essentially the inverse of the fold transformation and corresponds to inlining code. It takes as input a specific subgoal $y_k \in \mathsf{body}(\mathrm{R})$ of some rule R. The goal is to replace $y_k$ within R by its definition; however, its definition may span multiple rules R′. Therefore, we will remove R from the program and replace it with a specialized version of R for each rule R′ whose head unifies with $y_k$. These rules R′ define $y_k$—except in the special case where $y_k$ matches any input items, in which case we cannot unfold $y_k$ because its complete definition is not available. Note that unfolding can make some rules obsolete.[154]

[152] We note that GMTK (Bilmes and Zweig, 2002) provided a Junction tree optimizer that could optimize a recursive graphical model, but to our knowledge, their system did not statically optimize across recursive levels, as we do in this chapter.

[153] In NLP, this kind of transformation is often referred to as a *hook trick* (Eisner and Satta, 1999a; Huang, Zhang, et al., 2005; Gildea, 2011).

**Example 25.** *Suppose the user provided the inefficient program below, which could have been obtained by folding Example 23 in a suboptimal manner.*

```
335  goal += tmp1(X1,X4,X5).
336  tmp1(X1,X4,X5) += tmp2(X1,X4) * w(X4,X5).
337  tmp2(X1,X4) += tmp4(X1,X2) * tmp3(X2,X4).
338  tmp3(X2,X4) += w(X2,X3) * w(X3,X4).
339  tmp4(X1,X2) += w(X1,X2).
```

*While the above program is correct, its runtime is $\mathcal{O}(N^3)$, which is worse than the $\mathcal{O}(N^2)$ version we saw before. It has no variables that can be eliminated, so there are no fold actions or new definitions that can improve its runtime complexity. To improve it, we first have to* undo *the poor choices. If we unfold Example 25 multiple times, we can arrive at the version in Example 23. From there, we can refold to get the same efficient version.*

An unfold transformation will usually increase or preserve the program's runtime degree, but there are exceptions:

**Example 26.**

[154]Recall that the methods in §6.2 can identify and remove rules that cannot fire based on the declared inputs or are unused by any of the declared outputs.

```
340 a(I,K) += b(I,J) * c(J,K).
341 trace += a(L,L).
```

*Unfolding* `a(L,L)` *in #341 decreases the program's runtime from $\mathcal{O}(N^3)$ to $\mathcal{O}(N^2)$ assuming $N$ upper bounds cardinality of any of variables in the program.*

```
342 trace += b(L,J) * c(J,L).
343 ~~a(I,K) += b(I,J) * c(J,K).~~
```

*Thus, this is an example of an immediately useful unfold. The program, in this case, computes the trace of a matrix product, and the role of unfold is to* specialize *the sub-program that computes the entire matrix product* `a(I,K)` *to the site where the product is used, which only seeks its diagonal* `a(L,L)`. *Such optimizations are easy for programmers to miss.*[155]

**Transforming recursion.** In the following example, we transform an inductive definition—effectively eliminating a variable that is propagated through an (unbounded) recursion. Such a rewrite is impossible with only the tools from probabilistic graphical models and database query planning.

**Example 27.** *We begin with the following program, which computes the total weight (*`goal`*) of all paths in a graph that start in a set* `start` *and end in a set* `stop`.

```
344 goal += start(I) * path(I,K) * stop(K).
345 path(I,I).
346 path(I,K) += path(I,J) * edge(J,K).
```

*The running time of this program is $\mathcal{O}(N^3)$, where $N$ is the number of nodes in the graph. We will derive a more efficient $\mathcal{O}(N^2)$ program by replacing the intermediate* `path(I,K)` *items, which compute the total weight of all paths from* `I` *to*

[155]Many numerical libraries (e.g., Theano Development Team (2016), Abadi et al. (2016), and Smith and Gray (2018)) capture optimizations like this one. However, like the case of graphical models, these systems are limited because they do not reason about recursively defined computations.

*`K` for all pairs of nodes `I` and `K`, with `startpath(K)`, which computes the total weight of paths from any node in `start` to a node `K`.*

*Suppose we generate a new set of items that are—at least initially—defined by the following rule, which is appended to our program:*

```
347 startpath(K) += start(I) * path(I,K).
```

*Next, we fold #344 by #347:*

```
348 goal += startpath(K) * stop(K).
```

*As in Example 23, by introducing an intermediate item—in this case, `startpath`—we have managed to reduce the running time of #344 as its replacements #348 and #347 run in $\mathcal{O}(N + N^2)$ instead of $\mathcal{O}(N^3)$. Unfortunately, `startpath` still depends on `path`, which is currently the running time bottleneck for the program. Thus, we will attempt to rewrite `startpath` in terms of itself rather than `path`. Our next step is to unfold the `path` subgoal in the `startpath` rule (#347):*

```
349 startpath(K) += start(K).
350 startpath(K) += start(I) * path(I,J) * edge(J,K).
```

*Next, we recognize that `start(I) * path(I,J)` on #350 can be replaced by `startpath(J)` by folding according to its earlier definition (#347). We now have*

```
351 startpath(K) += start(K).
352 startpath(K) += startpath(J) * edge(J,K).
```

*Finally, a useless-rule elimination procedure (§6.2) detects that the `path` rules are not reachable from `goal`, which is our only declared output. Our final program is*

```
353 goal += startpath(K) * stop(K).
354 startpath(K) += start(K).
355 startpath(K) += startpath(J) * edge(J,K).
```
356 ~~`path(I,I).`~~
357 ~~`path(I,K) += path(I,J) * edge(J,K).`~~

*The running time of our improved program is $\mathcal{O}(N^2)$.*

*Notice that not all folds are safe. For example, if instead of unfolding #347, we immediately folded the definition of* `startpath` *with itself, we would have the following ill-founded definition:* `startpath(K) += startpath(K)`. *In §7.4, we will describe sufficient conditions to reject these bad folds and ensure safe folding.*

*Notice that to replace* `path` *in this example, we needed to generate a new relation* `startpath`*, which could replace it. Although* `startpath` *was* initially *defined in terms of* `path`*—through a few transformations—it became self-sufficient. Because discovering new definitions—like* `startpath`*—-often require "divine inspiration;" such definitions are affectionately called* Eureka definitions *(Pettorossi and Proietti, 1994). Fortunately, there are many common strategies for proposing plausible Eureka definitions, such as joining existing subgoals as we did. We discuss the strategies used by our system in §7.3.*

The next example follows the same general recipe as Example 27, except here we synthesize an efficient dynamic programming algorithm from an inefficient program that explicitly sums over the infinitely many paths in a graph where each path is specified as a list of its visited nodes.

**Example 28.** *The following program sums the total weight of all paths in a graph. It is written in a very direct style as a list of nodes visited along the path.*

```
358 goal += start(X) * β([X|Xs]).
359 β([X₁,X₂|Xs]) += edge(X₁,X₂) * β([X₂|Xs]).
360 β([X]) += stop(X).
```

*As it is written, this program is horribly inefficient as there might be infinitely*

*many paths, which translates into infinitely many non-0 β items. To speed this program up, we introduce the following definition, which sums the tail of the list* `Xs` *while exposing the first node* `X` *in the path:*

```
361 b(X) += β([X|Xs]).
```

*Next, we unfold #361 and obtain the following two rules that replace it:*

```
362 b(X1) += edge(X1,X2) * β([X2|Xs]).
363 b(X) += stop(X).
```

*We subsequently fold #362 by #361:*

```
364 b(X) += b(Y) * edge(X,Y).
365 b(X) += stop(X).
```

*Now,* `b` *is recursively defined in terms of itself rather than in terms of its original specification in #361 that made use of the inefficient* `β` *items. Moreover, there are finitely many non-0* `b` *items since there are finitely many nodes in the graph. Lastly, we fold #358 by #361 to graft* `b` *into the rule for* `goal`*:*

```
366 goal += start(X) * b(X).
```

*The final program is given below; we note that the* `β` *items are available, but the rules are now useless for deriving the goal, so the solver does not need to run them.*

```
367 goal += start(X) * b(X).
368 b(X) += b(Y) * edge(X,Y).
369 b(X) += stop(X).
370 β([X,Y|Xs]) += edge(X,Y) * β([Y|Xs]).
371 β([X]) += stop(X).
```

**Speculation and left-corner transformations.** The example below briefly shows a useful rearrangement of recursive equations through the speculation transformation (§7.5).

**Example 29** (Unary Cycle Factoring; adapted from Eisner and Blatz (2007))**.**

*Consider again the CKY program (Example 6):*

```
372 β(X,I,K) += γ(X,Y,Z) * β(Y,I,J) * β(Z,J,K).
373 β(X,I,K) += γ(X,Y) * β(Y,I,K).
374 β(X,I,K) += γ(X,W) * word(W,I,K).
375 goal += β(s,0,N) * len(N).
```

*This program has an inefficiency: it repeats the same unary-chain closure for all spans* ⟨I,K⟩. *Our use of speculation will refactor the program (see below) so that unary chain closures may be precomputed (independently of the sentence) and reused at each span. This implementation trick—attributed to Stolcke (1995)—is standard in probabilistic parsing.*

```
376 u(X,X’) += γ(X,Y) * u(Y,X’).
377 u(X,X’) += γ(X,X’).
378 β(X,I,K) += u(X,X’) * β’(X’,I,K).
379 β(X,I,K) += β’(X,I,K).
380 β’(X,I,K) += γ(X,Y,Z) * β(Y,I,J) * β(Z,J,K).
381 β’(X,I,K) += γ(X,W) * word(W,I,K).
382 goal += len(N) * β(s,0,N).
```

(A nearly identical program optimization can be achieved with the closely related left-corner transformation (§7.5.2).)

**Next steps.** In this section, we have provided a high-level idea of the program transformations that we develop in this chapter. In what follows, we will formalize each of the transformations and give further examples of how they are used and when they are useful. But first, we will formalize program equivalence and transformation safety.

## 7.2 Semantics-Preserving Transformations

We say that a program-to-program transformation $\mathfrak{p}^{(s)} \to \mathfrak{p}^{(t)}$ is **semantics-preserving** (or **safe**, for short) if the programs $\mathfrak{p}^{(s)}$ and $\mathfrak{p}^{(t)}$ are **semantically equivalent**, i.e., they define precisely the same valuation mapping (§4).[156] As a shorthand, we write $\mathfrak{p}^{(s)} \equiv \mathfrak{p}^{(t)}$ to assert (or stipulate) that the two programs are semantically equivalent. In other words, $\mathfrak{p}^{(s)} \equiv \mathfrak{p}^{(t)} \iff \mathfrak{p}^{(s)}[\![x]\!] = \mathfrak{p}^{(t)}[\![x]\!]$ for all $x \in \mathbb{H}$. We will also make use of the following stronger notion of semantic equivalence, which implies the former (Proposition 5).

**Definition 31.** *We say that $\mathfrak{p}^{(s)}$ and $\mathfrak{p}^{(t)}$ are **strongly equivalent** if there exists a value-preserving bijection $\underset{s \to t}{\phi} : \mathfrak{p}^{(s)}\{\cdot\} \to \mathfrak{p}^{(t)}\{\cdot\}$ where $\mathfrak{p}^{(s)}\{\cdot\}$ and $\mathfrak{p}^{(t)}\{\cdot\}$ are the set of all derivations in $\mathfrak{p}^{(s)}$ and $\mathfrak{p}^{(t)}$, resepectively.*

**Definition 32.** *The function $\phi : \mathfrak{p}^{(s)}\{\cdot\} \to \mathfrak{p}^{(t)}\{\cdot\}$ is a **value-preserving bijection** if and only if for all $\delta' \in \mathfrak{p}^{(t)}\{\cdot\}$, there exists a unique $\delta \in \mathfrak{p}^{(s)}\{\cdot\}$ such that $\phi(\delta) = \delta'$, $[\![\delta]\!] = [\![\delta']\!]$, and that $\delta$ and $\delta'$ have the same root label.*

Intuitively, for each item $x$, the function $\underset{s \to t}{\phi}$ describes the translation from derivations $\mathfrak{p}^{(s)}\{x\}$ that use the rules in $\mathfrak{p}^{(s)}$ to derivations $\mathfrak{p}^{(t)}\{x\}$ that use the rules in $\mathfrak{p}^{(t)}$. However, value preservation restricts the translation such that the value contributed by each derivation of $x$ is the same. Bijectivity ensures that no derivations are lost and no spurious derivations are created.

[156] Note that §4 does not directly discuss the semantics of partially specified programs, i.e., those with input relations. Fortunately, input relations can be reduced to built-in relations because they are simply items whose meanings are specified outside of the current program.

**Example 30.** *Consider the geometric series program. We will illustrate how systematically different sets of derivations of different programs can index a common bag of contributions to the value of a given item* x*.*

$\mathfrak{p}^{(0)} =$
```
383  x += b.
384  x += a * x.
```

*The value* x *is equal to the sum of a bag of (infinitely many) values,* $[\![\mathtt{x}]\!] = \mathtt{b} + \mathtt{a}\,\mathtt{b} + \mathtt{a}^2\,\mathtt{b} + \mathtt{a}^3\,\mathtt{b} + \mathtt{a}^4\,\mathtt{b} + \mathtt{a}^5\,\mathtt{b} + \mathtt{a}^6\,\mathtt{b} + \cdots$*. Below, we show 7 (of the infinitely many) derivations of the item* x*. We see that the values of each derivation (i.e., the product of its leaves) correspond to the sequence of summands above.*

$\mathfrak{p}^{(0)}\{\mathtt{x}\} =$

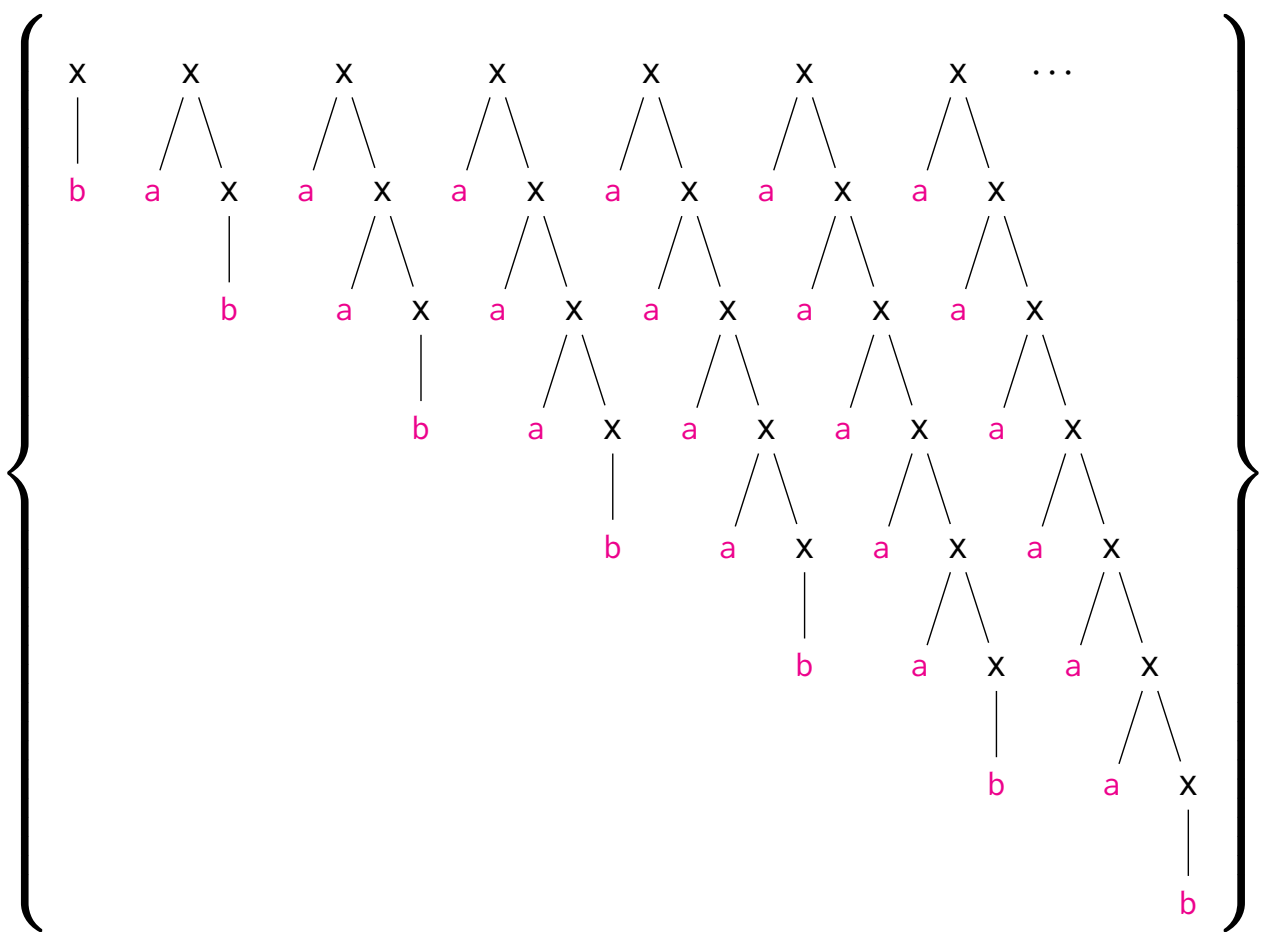

*The program* $\mathfrak{p}^{(1)}$ *below (obtained by unfolding #384) is strongly equivalent to the program above,* $\mathfrak{p}^{(1)} \overset{\text{strong}}{\equiv} \mathfrak{p}^{(0)}$.

$\mathfrak{p}^{(1)} =$
```
385  x += b.
386  x += a * b.
387  x += a * a * x.
```

*To see why we inspect the derivations of the item* x*. Below, we have ordered the derivations* $\mathfrak{p}^{(1)}\{\mathtt{x}\}$ *so that the left-to-right ordering shows the correspondence (i.e., bijection) to* $\mathfrak{p}^{(0)}\{\mathtt{x}\}$.

$\mathfrak{p}^{(1)}\{\mathtt{x}\} =$ 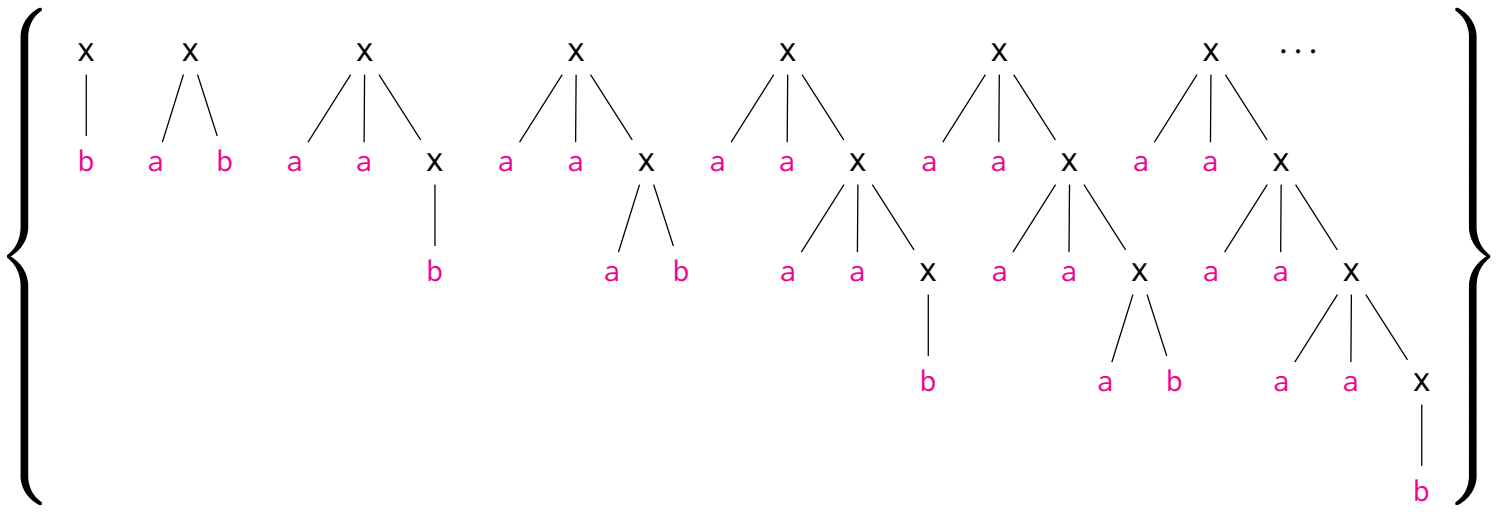

*By inspecting the leaves of the trees, we see that each of the values of the derivations can be aligned to exactly one of the original derivations (and vice versa). In other words, there is a value-preserving bijection between $\mathfrak{p}^{(0)}\{\mathtt{x}\}$ and $\mathfrak{p}^{(1)}\{\mathtt{x}\}$.*

*In the program below, we have systematically moved* `b` *up to the root of each derivation—leaving $\underline{1}$ in* `b`*'s original position.*

$\mathfrak{p}^{(2)} =$

```
388  x += x' * b.
389  x' += a * x'.
390  x' += 1.
```

$\mathfrak{p}^{(2)}\{\mathtt{x}\} =$ 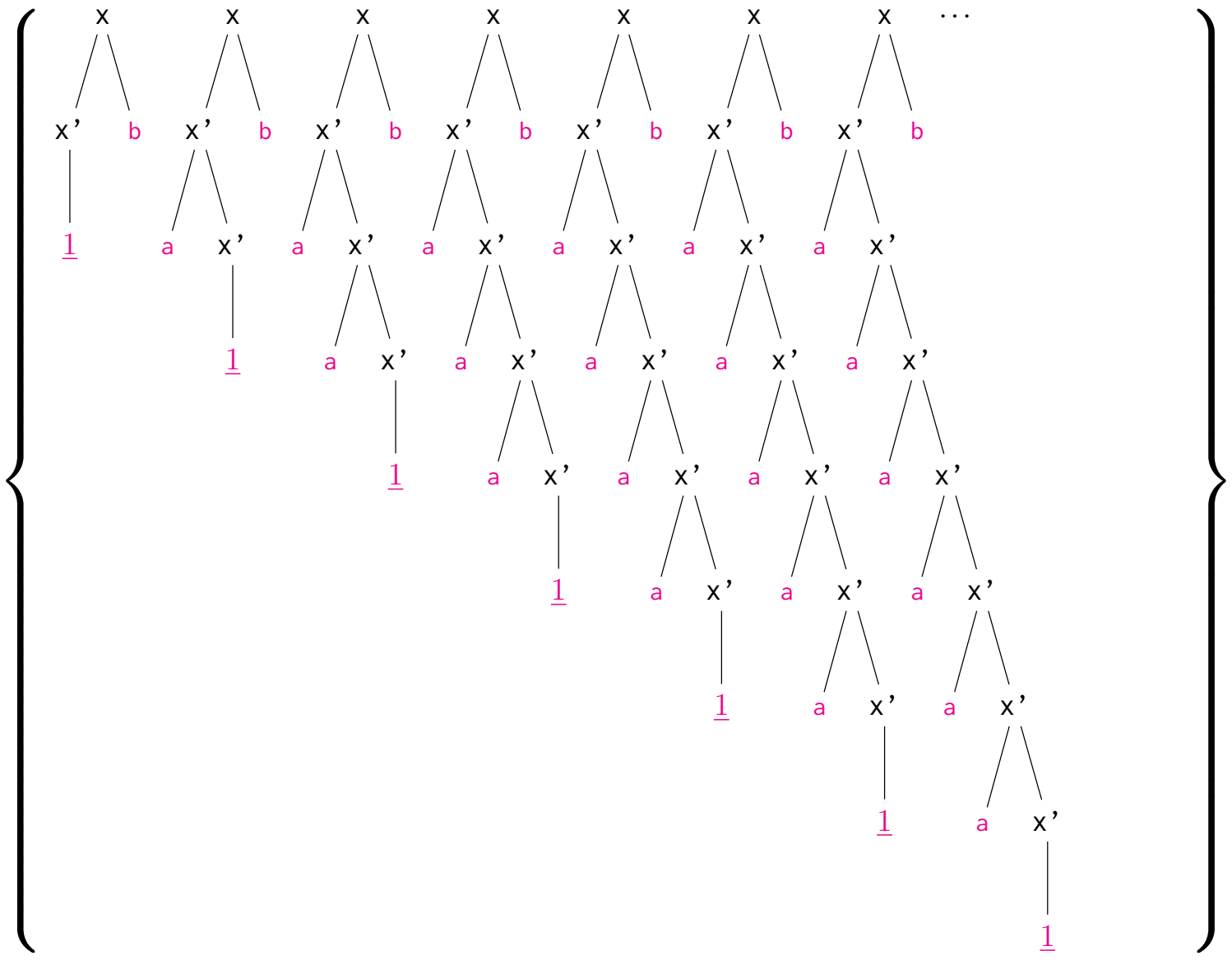

*Notice, that $\mathfrak{p}^{(2)}$ defines a new item* `x'`*, which has the value $[\![\mathtt{x'}]\!] = \underline{1} + \mathtt{a} + \mathtt{a}^2 + \cdots = \mathtt{a}^\star$. However, $\mathfrak{p}^{(0)}[\![\mathtt{x'}]\!] = \underline{0}$ because* `x'` *was not defined in $\mathfrak{p}^{(0)}$. To be semantics-preserving, we require that* `x'` *be defined in $\mathfrak{p}_0$. Since the name* `x'` *is not taken in*

*$\mathfrak{p}^{(0)}$ it is safe to add the following (unused) rules to $\mathfrak{p}_0$ that define the* `x'` *items.*

$$\mathfrak{p}^{(0')} =$$

```
x += b.
x += a * x.
x' += a * x'.
x' += 1.
```

*By adding the definitions to $\mathfrak{p}^{(0)} \not\equiv \mathfrak{p}^{(2)}$, we get $\mathfrak{p}^{(0')} \equiv \mathfrak{p}^{(2)}$. Augmenting the program with new definitions is what we call a* definition introduction transformation *(§7.3). Introducing new intermediate quantities is one of the most powerful methods at our disposal for devising a more efficient version of a given program. In general, whenever a transformation introduces a new definition, we ensure that the newly named items do not conflict with any other items in the universe of programs. Furthermore, we will add these definitions to all programs in the universe so they remain semantically equivalent.*

**Definition introduction.** We treat definition introduction (§7.3) as in Tamaki and Sato (1984). When a transformation introduces a definition, we pretend it was introduced in the original program and propagate it throughout all programs derived from it. Under this scheme, the earlier programs would be retroactively augmented with the definition even though they did not technically have it when they were created. Maintaining a global symbol table for all programs in the universe is important to ensure that a transformation does not introduce a new definition that conflicts with any prior definition's name, including inputs and builtins.

**Equivalence relation.** As the name suggests, semantic equivalence ($\equiv$) is an equivalence relation. Thus, we can represent programs found through transformations using an undirected graph where the nodes are programs, and an edge exists between programs $\mathfrak{p}$ and $\mathfrak{p}'$ if ($\mathfrak{p} \equiv \mathfrak{p}'$). However, since $\equiv$ is reflexive, symmetric, and transitive, not all edges need to be explicit. Instead, $\mathfrak{p}$ and $\mathfrak{p}'$ are equivalent if an (undirected) path exists between them. Since testing the semantic equivalence of any pair of programs is undecidable,[157] the graph of equivalent programs will be incomplete. In other words, we can only approximate it by including (directed) edges between programs that we can prove are equivalent (e.g., if a semantics-preserving transformation from $\mathfrak{p}$ to $\mathfrak{p}'$ exists, we add the edge)—this means that if an edge exists, we are certain the programs are equivalent, and if no edge, then the programs *might* be equivalent. Thus, any equivalence deduced by closure is sound.

## 7.3 Definition-Introduction Transformation

The definition-introduction transformation augments an existing program with rules defining some new set of items. We assume that the names of these items are distinct from the existing items. It is subsequently the job of the fold and unfold transformations to graft these new definitions into the program. We can often effectively replace the original definitions with new ones. Example 27 demonstrated how a new definition `startpath` could be transformed to be self-sufficient and grafted into the main program by folding it into the `goal` rule—replacing the more expensive `path` relation. In general, discovering alternative definitions can require "divine inspiration." Thus, as we mentioned earlier, such rules are affectionately called *Eureka definitions* in many works in the program transformation literature (Pettorossi and Proietti, 1994).

[157]Even testing the equivalence of a Dyna program *without variables* is undecidable. This is because Dyna programs can directly encode the language generated by a context-free grammar (CFG). Thus, determining if two Dyna programs are equivalent is at least as hard as deciding whether two context-free grammars generate the same language, which is an undecidable problem (Hopcroft et al., 2007).

The most common technique for proposing plausible Eureka definitions is the **tupling technique** (Pettorossi and Proietti, 1994).[158] Tupling is very general; it is parameterized as follows: Given a product of nonground terms $z_1 \circ \cdots \circ z_N$, and a set of variables $\{\mathtt{X}_1, \ldots, \mathtt{X}_L\}$ (which may be any subset of $\mathsf{vars}(z_1 \circ \cdots \circ z_N)$ as well as fresh variables), the new definition formed by the tupling technique is then

$$\mathtt{tmp}(\mathtt{X}_1, \ldots, \mathtt{X}_L) \mathrel{\oplus=} z_1 \circ \cdots \circ z_N$$

where `tmp` is a novel function symbol; the name `tmp` and the order of its arguments $\mathtt{X}_1, \ldots, \mathtt{X}_L$ are chosen arbitrarily. The new definitions proposed in the earlier examples of this chapter (23, 24, 27, and 28) were instances of a version of the tupling technique known as the variable-elimination technique (defined below).[159]

The **variable-elimination technique** is an instance of the tupling technique where (1) we use existing subproducts in the program, and (2) we pick the minimal subset of variables from the chosen subproduct for which the new definition can correctly replace the subproduct. More formally, for a given program $\mathfrak{p}$, the variable-elimination technique takes a rule $\mathrm{R} \in \mathfrak{p}$. Suppose that R is of the form $\boldsymbol{x} \oplus\!= \boldsymbol{\alpha} \circ \boldsymbol{\mu} \circ \boldsymbol{\beta}$ where $\boldsymbol{\alpha}$, $\boldsymbol{\mu}$, and $\boldsymbol{\beta}$ are sequence of subgoals. If $\circ$ is commutative, then we may permute R's subgoals when matching this pattern. If any subset of $\boldsymbol{\mu}$'s variables does not appear in any of $\boldsymbol{x}$, $\boldsymbol{\alpha}$, and $\boldsymbol{\beta}$, then those variables may be locally summed in a new definition—providing a more efficient rule because it has fewer variables. The set of variables $\{\texttt{X}_1, \ldots, \texttt{X}_L\} = \mathsf{vars}(\boldsymbol{\mu}) \setminus (\mathsf{vars}(\boldsymbol{\alpha}) \cup \mathsf{vars}(\boldsymbol{\beta}) \cup \mathsf{vars}(\boldsymbol{x}))$ is the minimal set of variables for the new definition to fold into this position of $\mathfrak{p}$ successfully. The remainder of the technique follows tupling.

[158] Cohen, Simmons, et al. (2011) used a combination of tupling, fold, and unfold to give a general recipe for deriving programs that compute the weighted intersection of logic programs. Using this recipe, they derived several complex programs as the product of simpler programs (e.g., a CKY is the intersection of a WCFG and a WFSA).

[159] This technique is sometimes called *loop absorption* (Pettorossi and Proietti, 1994). In NLP, this form of variable elimination is known as a *hook trick* (Eisner and Satta, 1999a; Huang, Zhang, et al., 2005; Gildea, 2011).

We have already seen a number of examples of variable elimination in §7.1. In particular, Examples 23–25, 27 and 28 use variable elimination to propose new definitions that enable a speed-up. We highlight one of them below.

**Example 31.** *We revisit Example 24 to see how it uses the variable elimination. Recall that Example 24 sped up the rule*

```
395  β(X,I,K) += γ(X,Y,Z) * β(Y,I,J) * β(Z,J,K).
```

*by transforming it into two rules where the variable* `Y` *is summed separately from* `K`*:*

```
396  β(X,I,K) += tmp(I,J,X,Z) * β(Z,J,K).
397  tmp(I,J,X,Z) += γ(X,Y,Z) * β(Y,I,J).
```

*We will show how the rule #397 is the output of variable elimination.*[160]

[160]Below we padded the body with a 1 to match the $(\boldsymbol{x} \oplus \boldsymbol{\alpha} \circ \boldsymbol{\mu} \circ \boldsymbol{\beta})$ pattern.

$$\underbrace{\texttt{β(X,I,K)}}_{x} \mathrel{\texttt{+=}} \underbrace{\underline{1}}_{\alpha} * \underbrace{\texttt{γ(X,Y,Z) * β(Y,I,J)}}_{\mu} * \underbrace{\texttt{β(Z,J,K)}}_{\beta}$$

*We named our new relation* `tmp` *and determined its arguments as follows:*

$$\{\texttt{I},\texttt{J},\texttt{X},\texttt{Z}\} = \mathsf{vars}(\texttt{γ(X,Y,Z) * β(Y,I,J)}) \setminus (\mathsf{vars}(\underline{1}) \cup \mathsf{vars}(\texttt{β(Z,J,K)}) \cup \mathsf{vars}(\texttt{β(X,I,K)}))$$

*We now see how the #397 is the output of the variable elimination technique. Note that we have to subsequently fold #395 by our new rule to get #396 to complete the example. Each definition introduction should be followed by fold to graft the new definitions into the program.*

Later, we will see that the speculation and left-corner transformations (§7.5) are also ways to generate new definitions based on an existing program.

# 7.4 Fold and Unfold Transformations

This section describes a pair of transformations called fold and unfold. This pair of transformations has a long history in unweighted logic programming[161]. They were generalized to the semiring-weighted logic programming (Eisner and Blatz, 2007). At a high level, these transformations are analogous to applying the semiring distributive property (Definition 17) in either the expanding (unfold) or factoring (fold) direction. The folding transform is essential for grafting new definitions into the existing rules of a program and rewriting those definitions to be efficient self-supporting recursive definitions (as we have seen in §7.1). We start with a basic version of the transformations (§7.4.1) and, then, develop a generalized version (§7.4.2). The generalized transformations are novel in the weighted setting. The generalized transformations themselves are largely the same as in the unweighted setting. However, we had to overhaul the correctness proofs in order to prove that these transformations preserve our sum-over-derivations semantics (§ 4)—the existing theory for the unweighted setting only considered the preservation of *at least one* derivation rather than *all* derivations (Definition 31). As noted in the unweighted setting, the generalized fold and unfold transformations are quite powerful, but they require careful technical considerations to avoid mistakes.

[161] E.g., Tamaki and Sato (1984), Kawamura and Kanamori (1990), Bossi et al. (1992), Gergatsoulis and Katzouraki (1994), Pettorossi and Proietti (1994), Roychoudhury et al. (1999).

### 7.4.1 Basic Fold and Unfold

**Unfold.** At a high level, what the unfolding transformation $\mathfrak{p}^{(s)} \xrightarrow{\text{unfold}} \mathfrak{p}^{(t)}$ does is to replace a given rule $r \in \mathfrak{p}^{(s)}$ with versions of that rule $\mathcal{S}$ that correspond to expanding a specific subgoal against the rules that define that subgoal's value. The result of the fold is a new program $\mathfrak{p}^{(t)} = \mathfrak{p}^{(s)} \setminus \{r\} \cup \mathcal{S}$.

**Example 32.** *Below, we unfold the first subgoal (*q*) in #398:*

```
398  p += q * t.
399
400  q += r.
401  q += s.
```

$\xrightarrow{\text{unfold}(\#398,1)}$

```
402  p += r * t.
403  p += s * t.
404  q += r.
405  q += s.
```

*Because there are two rules #400 and #401 that define* q*, we get two variants (#402 and #403) of #398 in the new program. In this example* $r = \#398$ *and* $\mathcal{S} = \{\#402, \#403\}$ *Below, we show the derivations for* p *and* q*.*

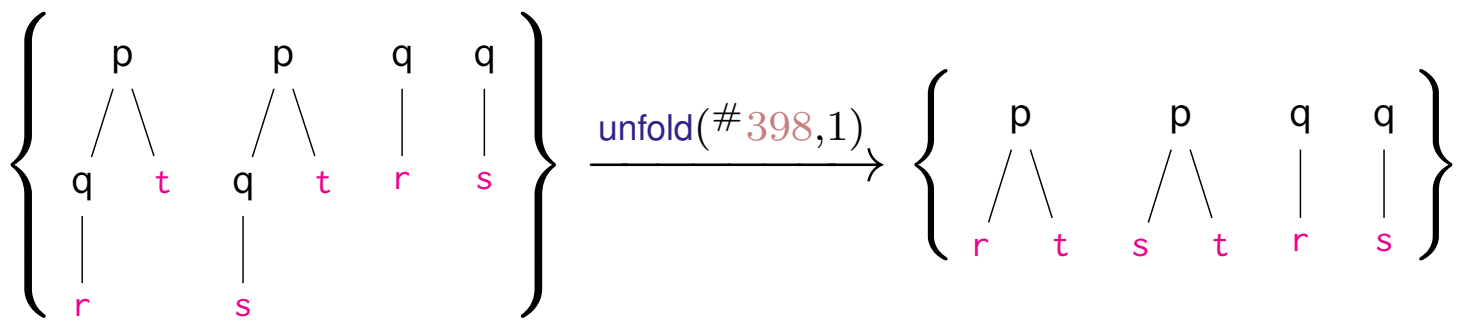

*Notice how #398 is replaced in the first derivation with #402 and how #398 is replaced in the first derivation with #403. The two derivations of* q *are unchanged because none of the rules that appear in them were altered.*

**Fold.** Folding is the inverse of unfolding. At a high level, a folding transformation $\mathfrak{p}^{(s)} \xrightarrow{\text{fold}} \mathfrak{p}^{(t)}$ replaces a collection of rules $\mathcal{S} \subseteq \mathfrak{p}^{(s)}$ with a single rule $r$. In other words, the result of the fold is a new program $\mathfrak{p}^{(t)} = \mathfrak{p}^{(s)} \setminus \mathcal{S} \cup \{r\}$.

**Example 33.** *Continuing Example 32, we use* fold *to invert the* unfold *transforma-*

*tion from Example 32. We simultaneously fold #406 by #408, and #407 by #409.*

```
406  p += r * t.                  410  p += q * t.
407  p += s * t.        fold      411
408  q += r.            --->      412  q += r.
409  q += s.                       413  q += s.
```

*This replaces the set of rules* $\mathcal{S} = \{{}^{\#}406, {}^{\#}407\}$ *by the single rule* $r = {}^{\#}410$*. To check that we folded correctly, verify that unfolding #410 gives the program on the left (this is known as the reversibility test).*

**Definition 33.** *The* ***basic unfold transformation*** *takes* $\langle \mathfrak{p}^{(s)}, r, j \rangle$ *where* $\mathfrak{p}^{(s)}$ *is a program,* $r$ *is a rule (*$r \in \mathfrak{p}^{(s)}$*), and* $j$ *is an index of* $r$*'s body (*$0 \le j \le |\,\mathsf{body}(r)\,|$*). And produces a program* $\mathfrak{p}^{(t)}$*, which is given by the following procedure:*

---

**Algorithm 20** unfold: Apply the basic unfold transformation (Definition 33).

---

**def** unfold($\mathfrak{p}^{(s)}, r, j$):

$\quad \left(x \mathrel{\oplus=} y_{[:j)} \circ y_j \circ y_{(j:]}\right) \leftarrow r$

$\quad \mathfrak{p}^{(t)} \leftarrow \mathfrak{p}^{(s)} \setminus \wr r \wr$ $\quad\quad$ ▷ *New program with rule r removed*

$\quad$ ▷ *Add new rules by expanding subgoal* $y_j$ *in rule r with its defining rules in* $\mathfrak{p}^{(s)}$

$\quad \mathfrak{p}^{(t)}[x] \mathrel{\oplus=} y_{[:j)} \circ \mathfrak{p}^{(s)}[y_j] \circ y_{(j:]}$

$\quad$ **return** $\mathfrak{p}^{(t)}$

---

*Let* $\mathcal{S}$ *denote the set of rules added by the transformation. There is a one-to-one alignment* $\underset{t\to s}{\sigma} : \mathcal{S} \to \mathfrak{p}^{(s)}$*; each rule in* $\mathcal{S}$ *maps to a distinct rule in* $\mathfrak{p}^{(s)}$ *that gave rise to it. The inverse of this alignment is denoted* $\underset{s\to t}{\sigma} : \mathfrak{p}^{(s)} \to \mathcal{S}$*.*[162]

**Definition 34.** *The* ***basic fold transformation*** *takes* $\langle \mathfrak{p}^{(s)}, \mathcal{S}, r \rangle$ *where* $\mathfrak{p}^{(s)}$ *is a program,* $\mathcal{S} \subseteq \mathfrak{p}^{(s)}$*, and* $r$ *is a new rule. And, produces a program* $\mathfrak{p}^{(t)} =$

[162] The following behaviors are inherited from $\mathfrak{p}[\cdot]$. If subgoal $j$ is a constant or input, then the unfold is a no-op. If subgoal $j$ is a built-in relation, then it may evaluate if it is sufficiently instantiated. Recall from §3.1.2.3 that expanding a rule can lead to summing over a free variable. That case is handled here thanks to the result-stream machinery developed in §3.1.2.1.

$(\mathfrak{p}^{(s)} \smallsetminus \mathcal{S}) \cup \{r\}$. *However, the fold is valid if and only if it passes the* ***reversibility test***: $\exists j : \mathfrak{p}^{(s)} = \mathsf{unfold}(\mathfrak{p}^{(t)}, r, j)$.[163]

**Algorithm 21** fold: Apply the basic fold transformation (Definition 34).

```
def fold(p^(s), S, r):
    p^(t) ← (p^(s) ∖ S) ∪ {r}
    assert ∃j : p^(s) = unfold(p^(t), r, j)        ▷ Reversibility test
    return p^(t)
```

Notice that the new rule $r$ must be carefully synthesized for this transformation method to be useful and pass the reversibility test.[164] The reversibility test was originally studied by Maher (1987).

**Theorem 4.** *Suppose* $\mathfrak{p}^{(s)} \xrightarrow{\mathsf{unfold}} \mathfrak{p}^{(t)}$ *according to Definition 33, then* $\mathfrak{p}^{(s)} \overset{\text{strong}}{\equiv} \mathfrak{p}^{(t)}$.

**Theorem 5.** *Suppose* $\mathfrak{p}^{(s)} \xrightarrow{\mathsf{fold}} \mathfrak{p}^{(t)}$ *according to Definition 34, then* $\mathfrak{p}^{(s)} \overset{\text{strong}}{\equiv} \mathfrak{p}^{(t)}$.

Proofs of correctness in the unweighted setting are available in many works (e.g., Maher (1987)). However, the weighted setting (first proposed in Eisner and Blatz (2007)) lacks a formal proof of correctness for fold and unfold; we fill this gap in the literature in §B.2. Furthermore, our theoretical results strengthen Eisner and Blatz (2007)'s (informal) correctness claims by demonstrating *strong* equivalence of fold and unfold, rather than (weak) equivalence.

### 7.4.2 Generalized Fold and Unfold

In this section, we generalize the basic unfold transformation (Definition 33) to allow the rules used to expand the subgoal to come from a (possibly different) program $\mathfrak{p}^{(d)} \overset{\text{strong}}{\equiv} \mathfrak{p}^{(s)}$.[165] We call $\mathfrak{p}^{(d)}$ the **auxiliary definitions**. As we shall see in Example 34, auxiliary definitions give us additional flexibility because we can fruitfully mix and match semantically equivalent versions of a given program.

[163] For this test, syntactic equality is sufficient.

[164] For a given $r$, we can search over $\mathcal{S} \subseteq \mathfrak{p}^{(s)}$ to see if any of those folds pass the reversibility test.

**Example 34.** *Consider unfolding the geometric series $\mathfrak{p}^{(0)}$ to get $\mathfrak{p}^{(1)}$:*

$\mathfrak{p}^{(0)} =$

```
414  x += b.
415  x += x * a.
```

$\xrightarrow{\text{unfold}(\#415,1)}$ $\mathfrak{p}^{(1)} =$

```
416  x += b.
417  x += b * a.
418  x += x * a * a.  % two a's
```

*So far, so good. What if we unfold again? Frustratingly,* $\text{unfold}(\mathfrak{p}^{(1)}, \#418, 1)$ *gives*

```
419  x += b.
420  x += b * a.
421  x += b * a * a.
422  x += b * a * a * a.
423  x += x * a * a * a * a.   % four a's
```

*which has unrolled the recurrence 4 levels instead of 3 levels.*[166] *Using generalized unfold,* $\text{unfold}(\mathfrak{p}^{(1)}, \#418, 1, \mathfrak{p}^{(0)})$ *with* $\mathfrak{p}^{(0)}$ *as auxiliary definitions, gives*

```
424  x += b.
425  x += b * a.
426  x += b * a * a.
427  x += x * a * a * a.   % three a's
```

*which unrolls the recurrence to 3 levels.*[167]

**Generalized fold.** We also consider a generalized fold transformation, which relaxes the reversibility test in basic fold (Definition 34) to allow unfolding against auxiliary definitions in its reversibility test. Specifically, the generalized fold transformation modifies the reversibility test in Definition 34 from $\exists j : \mathfrak{p}^{(s)} = \text{unfold}(\mathfrak{p}^{(t)}, r, j, \mathfrak{p}^{(t)})$ to $\exists j : \mathfrak{p}^{(s)} = \text{unfold}(\mathfrak{p}^{(t)}, r, j, \mathfrak{p}^{(d)})$. This added flexibility enhances the power of our transformation system: Example 27 and Example 28 would not be possible with only basic folds. We revisit Example 27 below.

[165] Future work may want to explore other families of equivalence beyond strongly equivalent as strong equivalence is clearly not *necessary* for correctness.

[166] Unfolding the bottom rule once more would unroll the recurrence up to 8 levels.

[167] Unfolding the bottom rule once more (against $\mathfrak{p}^{(0)}$) would unroll the recurrence up to 4 levels.

**Example 35.** *Recall Example 27, which computes the total weight (*`goal`*) of all paths in a graph that start in a set* `start` *and end in a set* `stop`*. Suppose we start with* $\mathfrak{p}^{(0)}$ *and unfold it to* $\mathfrak{p}^{(1)}$*:*

$\mathfrak{p}^{(0)} =$

```
428  goal += startpath(K) * stop(K).
429  startpath(K) += start(I) * path(I,K).
430  path(I,I) += 1.
431  path(I,K) += path(I,J) * edge(J,K).
```

$\xrightarrow{\text{unfold}(\#429,2)} \mathfrak{p}^{(1)} =$

```
432  goal += startpath(K) * stop(K).
433  startpath(K) += start(K).
434  startpath(K) += start(I) * path(I,J) * edge(J,K).
435  path(I,I) += 1.
436  path(I,K) += path(I,J) * edge(J,K).
```

*Next, we use generalized folding to rewrite* `startpath` *more efficiently in terms of itself rather than* `path`*. We recognize that* $\mathcal{S} = \{\#434\}$ *can be replaced by*

$$r = (\texttt{startpath(K) += startpath(J) * edge(J,K)})$$

*This transformation folds #434 by #429. Since #429 is in* $\mathfrak{p}^{(0)}$*, this transformation is a generalized fold. Specifically, the transformation* $\text{fold}^{0}_{\{\#434\}\to r}$ *gives*

$\mathfrak{p}^{(2)} =$

```
437  goal += startpath(K) * stop(K).
438  startpath(K) += start(K).
439  startpath(K) += startpath(J) * edge(J,K).
440  path(I,I) += 1.
441  path(I,K) += path(I,J) * edge(J,K).
```

*We deactivate the* ~~`path`~~ *rules as they are no longer used by* `goal`*. The running time of* $\mathfrak{p}^{(2)}$ *is* $\mathcal{O}(N^2)$ *where* $N$ *is the number of nodes in the graph, which is considerably better than the* $\mathcal{O}(N^3)$ *running time of the initial program* $\mathfrak{p}^{(0)}$*.*

**Warning!** The generalized fold and unfold transformations require some technical restrictions to be semantics-preserving. The definitions above are only suffi-cient to guarantee **partial correctness**—meaning that we *can* lose derivations, but we *cannot* gain spurious derivations. To give a quick example, consider Example 35. If instead of unfolding #429, we immediately folded #429 by itself, we would arrive at the ill-founded definition: `startpath(K) += startpath(K)`. In the remainder of this section, we will develop sufficient conditions to ensure specific fold and unfold transformations are semantics-preserving.

#### 7.4.2.1 Formal Definitions

**Generalized Unfold.** Below, we generalize the basic unfold transformation (Definition 33) to allow auxiliary definitions ($\mathfrak{p}^{(d)}$, below) to be used.[168]

**Definition 35.** *The* ***generalized unfold transformation*** *takes* $\langle \mathfrak{p}^{(s)}, r, j, \mathfrak{p}^{(d)} \rangle$ *where* $\mathfrak{p}^{(s)}$ *is a program,* $r$ *is a rule* ($r \in \mathfrak{p}^{(s)}$)*,* $j$ *is an index of* $r$*'s body* ($0 \le j \le |\,\mathsf{body}(r)\,|$)*, and* $\mathfrak{p}^{(d)}$ *is a program* ($\mathfrak{p}^{(d)} \overset{\text{strong}}{\equiv} \mathfrak{p}^{(s)}$)*. And produces a program* $\mathfrak{p}^{(t)}$*, which is given by the following procedure.*

**Algorithm 22** unfold: Apply the generalized unfold transformation (Definition 35).

1. ▷ *The pseudocode is similar to basic unfold (Algorithm 20). The difference is that we have introduced an additional argument* $\mathfrak{p}^{(d)}$*, and we use* $\mathfrak{p}^{(d)}[y_j]$ *instead of* $\mathfrak{p}^{(s)}[y_j]$ *on line 22.6.*
2. **def** unfold($\mathfrak{p}^{(s)}, r, j, \mathfrak{p}^{(d)}$):
3. $\left(\boldsymbol{x} \oplus\!= \boldsymbol{y}_{[:j)} \circ \boldsymbol{y}_j \circ \boldsymbol{y}_{(j:]}\right) \leftarrow r$
4. $\mathfrak{p}^{(t)} \leftarrow \mathfrak{p}^{(s)} \setminus \{\!\!\{ r \}\!\!\}$ ▷ *New program with rule* $r$ *removed*
5. ▷ *Add new rules where subgoal* $y_j$ *in rule* $r$ *with its defining rules in* $\mathfrak{p}^{(d)}$
6. $\mathfrak{p}^{(t)}[\boldsymbol{x}] \oplus\!= \boldsymbol{y}_{[:j)} \circ \mathfrak{p}^{(d)}[\boldsymbol{y}_j] \circ \boldsymbol{y}_{(j:]}$
7. **return** $\mathfrak{p}^{(t)}$

[168] Pettorossi and Proietti (1994) have generalized unfold, but Roychoudhury et al. (1999) did not. Thus, it was missing in Roychoudhury et al. (1999)'s safety framework, which we have based ours on, and we will add it in §7.4.2.4.

*Let $\mathcal{S}$ denote the set of rules the transformation adds. There is a one-to-one alignment $\underset{t\to d}{\sigma} : \mathcal{S} \to \mathfrak{p}^{(s)}$; each rule in $\mathcal{S}$ maps to a distinct rule in $\mathfrak{p}^{(d)}$ that gave rise to it. The inverse of this alignment is $\underset{d\to t}{\sigma} : \mathfrak{p}^{(d)} \to \mathcal{S}$.*

Unlike basic unfolds (Definition 33), not all generalized unfolds are safe (cf. Example 36); sufficient conditions are presented in Theorems 1 and 2. Notice when $\mathfrak{p}^{(d)} = \mathfrak{p}^{(s)}$, we recover basic unfold. We often write $\mathfrak{p}^{(s)} \xrightarrow{\mathsf{unfold}^{d,j}_{r\to\mathcal{S}}} \mathfrak{p}^{(t)}$ to describe a specific unfolding relationship between $\mathfrak{p}^{(s)}$ and $\mathfrak{p}^{(t)}$. This provides a convenient shorthand to reference $\mathcal{S}$ and $r$, which are functions of the tuple $\langle \mathfrak{p}^{(s)}, r, j, \mathfrak{p}^{(d)} \rangle$. When we omit the definition index $d$, e.g., $\mathfrak{p}^{(s)} \xrightarrow{\mathsf{unfold}^{j}_{r\to\mathcal{S}}} \mathfrak{p}^{(t)}$, we take $d = s$ as a default value for the auxiliary definition argument.

**Generalized Fold.** Below, we generalize the basic fold transformation (Definition 34) to allow auxiliary definitions to be used in its reversibility test.

**Definition 36.** *The **generalized fold transformation** works as follows: Given $\langle \mathfrak{p}^{(s)}, r, \mathcal{S}, \mathfrak{p}^{(d)} \rangle$ where $r$ is a rule that replaces a subset of rules $\mathcal{S} \subseteq \mathfrak{p}^{(s)}$ and $\mathfrak{p}^{(d)} \overset{\text{strong}}{\equiv} \mathfrak{p}^{(s)}$ are auxiliary definitions. The result is a program $\mathfrak{p}^{(t)} = (\mathfrak{p}^{(s)} \smallsetminus \mathcal{S}) \cup \{r\}$. However, $\mathfrak{p}^{(t)}$ must satisfy the **generalized reversibility test**: $\exists j \colon \mathfrak{p}^{(s)} = \mathsf{unfold}(\mathfrak{p}^{(t)}, r, j, \mathfrak{p}^{(d)})$.*

**Algorithm 23** fold: Apply the generalized fold transformation (Definition 36).

**def** fold($\mathfrak{p}^{(s)}, r, \mathcal{S}, \mathfrak{p}^{(d)}$):
  $\mathfrak{p}^{(t)} \leftarrow (\mathfrak{p}^{(s)} \smallsetminus \mathcal{S}) \cup \{r\}$  ▷ *Replace rules $\mathcal{S}$ by the rule $r$*
  **assert** $\exists j \colon \mathfrak{p}^{(s)} = \mathsf{unfold}(\mathfrak{p}^{(t)}, r, j, \mathfrak{p}^{(d)})$  ▷ *Generalized reversibility test*
  **return** $\mathfrak{p}^{(t)}$

Unlike basic folds (Definition 34), not all generalized folds are safe (cf. Example 25); sufficient conditions are presented in Theorems 1 and 3. Similar to our notation for unfold, we often write $\mathfrak{p}^{(s)} \xrightarrow{\mathsf{fold}^{d,j}_{\mathcal{S}\to r}} \mathfrak{p}^{(t)}$. Notice that taking $\mathfrak{p}^{(d)} = \mathfrak{p}^{(s)}$ does *not* recover the basic fold. Specifically, the generalized reversibility condition would not be the same because the basic fold would unfold with respect to $\mathfrak{p}^{(t)}$, and the generalized fold would unfold with respect to $\mathfrak{p}^{(d)}$, which equals $\mathfrak{p}^{(s)}$ rather than $\mathfrak{p}^{(t)}$.

#### 7.4.2.2 Challenges

Unfortunately, not all generalized unfolds and folds are semantics-preserving. We provide examples of each in this section.

**Example 36** (Unsafe generalized unfold)**.** *We start with*

$$\mathfrak{p}^{(s)} = \begin{cases} {}^{442}\ \texttt{p += q.} \\ {}^{443}\ \texttt{q += r.} \\ {}^{444}\ \texttt{r += 1.} \end{cases} \qquad \mathfrak{p}^{(d)} = \begin{cases} {}^{445}\ \texttt{p += 1.} \\ {}^{446}\ \texttt{q += r.} \\ {}^{447}\ \texttt{r += p.} \end{cases}$$

*Clearly,* $\mathfrak{p}^{(s)} \equiv \mathfrak{p}^{(t)}$ *as they both have* $[\![\mathtt{p}]\!] = [\![\mathtt{q}]\!] = [\![\mathtt{r}]\!] = 1$. *Furthermore,* $\mathfrak{p}^{(s)} \overset{\text{strong}}{\equiv} \mathfrak{p}^{(t)}$ *as we can see that there is a value-preserving bijection between*

$$\mathfrak{p}^{(s)}\{\cdot\} = \left\{ \begin{array}{c} \mathtt{p} \\ | \\ \mathtt{q} \\ | \\ \mathtt{r} \\ | \\ \mathtt{1} \end{array} \quad \begin{array}{c} \mathtt{q} \\ | \\ \mathtt{r} \\ | \\ \mathtt{1} \\ \\ \\ \\ \\ \end{array} \quad \begin{array}{c} \mathtt{r} \\ | \\ \mathtt{1} \\ \\ \\ \\ \\ \end{array} \right\} \qquad \textit{and} \qquad \mathfrak{p}^{(d)}\{\cdot\} = \left\{ \begin{array}{c} \mathtt{p} \\ | \\ \mathtt{1} \\ \\ \\ \\ \\ \end{array} \quad \begin{array}{c} \mathtt{q} \\ | \\ \mathtt{r} \\ | \\ \mathtt{p} \\ | \\ \mathtt{1} \end{array} \quad \begin{array}{c} \mathtt{r} \\ | \\ \mathtt{p} \\ | \\ \mathtt{1} \\ \\ \\ \end{array} \right\}$$

*However, unfolding #443 against* $\mathfrak{p}^{(d)}$ *gives the following incorrect program:*

$$\mathfrak{p}^{(t)} = \begin{cases} {}^{448}\ \texttt{p += q.} \\ {}^{449}\ \texttt{q += p.} \\ {}^{450}\ \texttt{r += 1.} \end{cases} \qquad \mathfrak{p}^{(t)}[\![\cdot]\!] = \begin{cases} {}^{451}\ \texttt{p += 0.} \\ {}^{452}\ \texttt{q += 0.} \\ {}^{453}\ \texttt{r += 1.} \end{cases} \qquad \mathfrak{p}^{(t)}\{\cdot\} = \left\{ \begin{array}{c} \mathtt{r} \\ | \\ \mathtt{1} \end{array} \right\}$$

*Inspecting the rules, we see an ill-founded cycle between #448 and #449 that causes its valuation to differ and derivation set to shrink in size.*

**Example 37** (Unsafe generalized fold)**.** *We start with*[169]

$$\mathfrak{p}^{(0)} = \begin{cases} {\scriptstyle 454} & \texttt{p += q.} \\ {\scriptstyle 455} & \texttt{q += 1.} \end{cases} \qquad \mathfrak{p}^{(0)}\{\cdot\} = \left\{ \begin{array}{cc} \texttt{q} & \texttt{p} \\ | & | \\ \texttt{1} & \texttt{q} \\ & | \\ & \texttt{1} \end{array} \right\}$$

*Since* $[\![\texttt{p}]\!] = [\![\texttt{q}]\!] = 1$*, we might try to replace all instances of* q *on the right-hand side with* p*. So we define the following new program* $\mathfrak{p}^{(*)}$ *by folding with* $\mathcal{S} = \{{}^{\#}454\}$, $r = (\texttt{p += p})$, $j = 0$*, and definitions* $\mathfrak{p}^{(d)} = \mathfrak{p}^{(0)}$*. The output is the following (partially correct) program.*

$$\mathfrak{p}^{(*)} = \begin{cases} {\scriptstyle 456} & \texttt{p += p.} \\ {\scriptstyle 457} & \texttt{q += 1.} \end{cases} \qquad \mathfrak{p}^{(*)}\{\cdot\} = \left\{ \begin{array}{c} \texttt{q} \\ | \\ \texttt{1} \end{array} \right\}$$

*Although, the generalized reversibility condition is satisfied (*$\mathsf{unfold}(\mathfrak{p}^{(0)}, {}^{\#}456, 1, \mathfrak{p}^{(0)}) = \mathfrak{p}^{(0)}$*), this transformation has lost the derivation of* p*. At a high level, what has happened is that we have created an ill-founded definition for* p*; this buggy definition is cyclical without any base cases.*

#### 7.4.2.3 What Causes Partial Correctness?

To better understand the fold and unfold transformations, we will construct their derivation mappings. Recall from Definition 31 that the derivation mapping is a function $\underset{s\to t}{\phi}$ that maps derivations from $\mathfrak{p}^{(s)}\{\cdot\}$ into derivations from $\mathfrak{p}^{(t)}\{\cdot\}$. We require that $\underset{s\to t}{\phi}$ is a value-preserving bijection from the transformation $\mathfrak{p}^{(s)} \longrightarrow \mathfrak{p}^{(t)}$ to be strongly semantics-preserving. What is interesting about the generalized fold/unfold case is the auxiliary definitions' role in construct-ing these mappings. Roughly speaking, the derivation mappings $\underset{s\to t}{\phi}$ will pivot through $\mathfrak{p}^{(d)}\{\cdot\}$ using $\underset{s\to d}{\phi}$ and $\underset{d\to s}{\phi}$. The challenge of pivoting is that it can result in infinite recursion, corresponding to (possibly) losing a derivation (i.e., partial correctness). We will use what we learn from the derivation mapping's failure modes to develop sufficient conditions for when the transform is safe (i.e., when the mapping is a value-preserving bijection).

[169]This is a Dyna adaptation of Example 4.1 from Pettorossi and Proietti (1994).

We direct the reader to Algorithm 24 and Algorithm 25, which provide pseudocode for the derivation mappings of the generalized unfold and fold transformations. We wish to emphasize two things: (1) the derivation mappings are never run as part of the transformations—they are just used in our analysis of the transformations, and (2) neither Algorithm 24 or Algorithm 25 are guaranteed to define a value-preserving bijection—in fact, they are not even guaranteed to be functions, as they do not terminate for all inputs.

To see why the transformations go awry, we will study how the derivation mapping chokes on the following example:

**Example 38.** *Continuing with the unsafe fold example (Example 37). Recall:*

$$
\begin{aligned}
r &= (\texttt{p += p}) \\
\mathcal{S} &= \{\texttt{p += q}\} \\
\mathfrak{p}^{(d)} &= \mathfrak{p}^{(s)}
\end{aligned}
\qquad
\mathfrak{p}^{(s)}\{\cdot\} = \left\{ \begin{array}{cc} \texttt{q} & \texttt{p} \\ | & | \\ \texttt{1} & \texttt{q} \\ & | \\ & \texttt{1} \end{array} \right\}
\qquad
\boldsymbol{\delta}^{(s)} = \left( \begin{array}{c} \texttt{p} \\ | \\ \texttt{q} \\ | \\ \texttt{1} \end{array} \right)
$$

*And we saw that the derivation $\boldsymbol{\delta}^{(s)}$ is lost in this transformation. Consider the evaluation of $\underset{s\to t}{\phi}\left(\boldsymbol{\delta}^{(s)}\right)$ through Algorithm 25. Since the top rule of $\boldsymbol{\delta}^{(s)}$ is*

$\mathbf{S}^{(\ell)} = (\texttt{p += q})$ *and* $\mathbf{S}^{(\ell)} \in \mathcal{S}$*, we fall in the bottom case (line 25.7).*

$$\underset{s\to t}{\phi}\left(\begin{array}{c}\mathsf{p}\\ |\\ \mathsf{q}\\ |\\ 1\end{array}\right) = \begin{array}{c}\mathsf{p}\\ |\\ \underset{s\to t}{\phi}\left(u^{-1}\left(\begin{array}{c}\mathsf{q}\\ |\\ 1\end{array}\right)\right)\end{array}$$

*The invocation of* $u^{-1}$ *returns*[170]

$$\left(\begin{array}{c}\mathsf{p}\\ |\\ \mathsf{q}\\ |\\ 1\end{array}\right) = u^{-1}\left(\begin{array}{c}\mathsf{q}\\ |\\ 1\end{array}\right)$$

*Now, when we return to line 25.7, we observe the problem:*

$$\underset{s\to t}{\phi}\left(\begin{array}{c}\mathsf{p}\\ |\\ \mathsf{q}\\ |\\ 1\end{array}\right) = \begin{array}{c}\mathsf{p}\\ |\\ \underset{s\to t}{\phi}\left(u^{-1}\left(\begin{array}{c}\mathsf{q}\\ |\\ 1\end{array}\right)\right)\end{array} = \begin{array}{c}\mathsf{p}\\ |\\ \underset{s\to t}{\phi}\left(\begin{array}{c}\mathsf{p}\\ |\\ \mathsf{q}\\ |\\ 1\end{array}\right)\end{array} = \begin{array}{c}\mathsf{p}\\ |\\ \mathsf{p}\\ |\\ \mathsf{p}\\ |\\ \mathsf{p}\\ \vdots\end{array}$$

*This recurrence never terminates as* $\underset{s\to t}{\phi}$ *immediately calls itself with the same argument. It tries to build an infinitely tall tree and never succeeds because it does not hit base cases. Thus, the derivation* $\boldsymbol{\delta}^{(s)}$ *is lost by the derivation mapping—meaning that our procedure for* $\underset{s\to t}{\phi}$ *cannot find any corresponding* $\boldsymbol{\delta}^{(t)}$ *in* $\mathfrak{p}^{(t)}\{\cdot\}$*. This means that we have failed to define a value-preserving bijection (* $\underset{s\to t}{\phi}$ *is not even a* function*). Therefore, we cannot promise safety for this instance of generalized fold, nor should we!*

[170]To see why, we trace the steps of $u^{-1}$. Since $\mathfrak{p}^{(d)} = \mathfrak{p}^{(s)}$ the $\underset{s\to t}{\phi}$ (and $\underset{t\to s}{\phi}$) transformations are no-ops. The folder rule $\mathbf{D}^{(\ell)} = (\texttt{p += q})$ (#454). So when we reconstruct, we get $y_j = \mathsf{p}$. We also have the following alignment: $\underset{d\to t}{\sigma}(\mathbf{D}^{(\ell)}) = (\texttt{p += q})$.

The lesson learned from this example is that for the derivation maps (Algorithm 24 and Algorithm 25) to be correct, we need to ensure that cyclical recursion—like we just saw in Example 38—will not occur. We will use the nontermination of $\underset{s\to t}{\phi}$ as a diagnostic for when generalized fold/unfold goes awry. When $\underset{s\to t}{\phi}(\delta)$ (Algorithm 24 and Algorithm 25) fails to terminate on some $\delta \in \mathfrak{p}^{(s)}\{\cdot\}$, the problematic $\delta$ corresponds to a derivation that might be lost. Thus, to ensure correctness, we need to reject transformations for which $\underset{s\to t}{\phi}$ might not always terminate. More formally, we have the following theorems, which are proven in §B.2:

**Theorem 2.** *Given* $\mathfrak{p}^{(s)} \xrightarrow{\text{unfold}^{d,j}_{r\to\mathcal{S}}} \mathfrak{p}^{(t)}$ *and* $\mathfrak{p}^{(s)} \overset{\text{strong}}{\equiv} \mathfrak{p}^{(d)}$. *Suppose the function* $\underset{s\to t}{\phi}\left(\delta^{(s)}\right)$, *as specified in Algorithm 24, terminates for all* $\delta^{(s)} \in \mathfrak{p}^{(s)}\{\cdot\}$. *Then,* $\underset{s\to t}{\phi}$ *is a value-preserving bijection between* $\mathfrak{p}^{(s)}\{\cdot\}$ *and* $\mathfrak{p}^{(t)}\{\cdot\}$. *Furthermore,* $\mathfrak{p}^{(s)} \overset{\text{strong}}{\equiv} \mathfrak{p}^{(t)}$.

**Theorem 3.** *Given* $\mathfrak{p}^{(s)} \xrightarrow{\text{fold}^{d,j}_{\mathcal{S}\to r}} \mathfrak{p}^{(t)}$ *and* $\mathfrak{p}^{(s)} \overset{\text{strong}}{\equiv} \mathfrak{p}^{(d)}$. *Suppose the function* $\underset{s\to t}{\phi}\left(\delta^{(s)}\right)$, *as specified in Algorithm 25, terminates for all* $\delta^{(s)} \in \mathfrak{p}^{(s)}\{\cdot\}$. *Then,* $\underset{s\to t}{\phi}$ *is a value-preserving bijection between* $\mathfrak{p}^{(s)}\{\cdot\}$ *and* $\mathfrak{p}^{(t)}\{\cdot\}$. *Furthermore,* $\mathfrak{p}^{(s)} \overset{\text{strong}}{\equiv} \mathfrak{p}^{(t)}$.

The termination condition manifests in our proof of correctness (§B.2); termination ensures that a well-founded ordering exists for our inductive proofs that establish value-preserving bijectivity. Verifying that these derivation mappings always terminate can be difficult. We describe a relatively simple system in §7.4.2.4 for ensuring termination that is based on static analysis.

**Algorithm 24** Derivation mapping for $\mathfrak{p}^{(s)} \xrightarrow{\text{unfold}^{d,j}_{r\to S}} \mathfrak{p}^{(t)}$.

▷ *The pseudocode assumes that rules are distinct. This can be arranged by marking each rule with an identifier that is checked by equality tests. When a rule is copied from one program to another, its identifier is unchanged. When a rule is created, it is assigned a unique identifier.*

▷ *We use derivation mappings* $\underset{s\to d}{\phi}$ *and* $\underset{d\to s}{\phi}$ *that establish* $\mathfrak{p}^{(s)} \overset{\text{strong}}{\equiv} \mathfrak{p}^{(d)}$. *They must be fixed for all* $\boldsymbol{\delta}^{(s)}$.

▷ *In the code below,* $\boldsymbol{\alpha}$, $\boldsymbol{\mu}$, *and* $\boldsymbol{\beta}$ *are sequences.*

**def** $\underset{s\to t}{\phi}\left(\boldsymbol{\delta}^{(s)}\right)$:

  **case** constant or builtin: **return** $\boldsymbol{\delta}^{(s)}$

  **case** $\overset{y_1}{\triangle} \circ \cdots \circ \overset{y_K}{\triangle}$: **return** $\underset{s\to t}{\phi}\left(\overset{y_1}{\triangle}\right) \circ \cdots \circ \underset{s\to t}{\phi}\left(\overset{y_K}{\triangle}\right)$

  **case** $\left(\begin{array}{c} x \\ | \\ \overset{\boldsymbol{\mu}}{\triangle} \end{array}\right)$ **and** its top rule $\in \mathfrak{p}^{(t)}$: **return** $\begin{array}{c} x \\ | \\ \underset{s\to t}{\phi}\left(\overset{\boldsymbol{\mu}}{\triangle}\right) \end{array}$

  **case** $\left(\begin{array}{ccc} & x & \\ \swarrow & \downarrow & \searrow \\ \overset{\boldsymbol{\alpha}}{\triangle} & \overset{y_j}{\triangle} & \overset{\boldsymbol{\beta}}{\triangle} \end{array}\right)$ **and** its top rule $\notin \mathfrak{p}^{(t)}$:

    **return** $\begin{array}{ccc} & x & \\ \swarrow & \downarrow & \searrow \\ \underset{s\to t}{\phi}\left(\overset{\boldsymbol{\alpha}}{\triangle}\right) & \underset{s\to t}{\phi}\left(u\left(\overset{y_j}{\triangle}\right)\right) & \underset{s\to t}{\phi}\left(\overset{\boldsymbol{\beta}}{\triangle}\right) \end{array}$ ▷ *The top rule of the transformed derivation is given by the* $2^{nd}$ *element of the pair returned from* $u$ *(notation abuse). The call to* $u$ *replaces the* $y_j$ *subtree with a value-equivalent derivation in* $\mathfrak{p}^{(s)}\{\cdot\}$ *(the* $1^{st}$ *element of the pair), which is then recursively transformed.*

▷ *Pivoting function: Returns pair of a value-equivalent derivation and the new top-level rule in* $\mathfrak{p}^{(t)}$. *See Lemma 2 for a formal statement and proof.*

**def** $u\left(\overset{y_j}{\triangle}\right)$: ▷ *Note: This function references* $\langle s, t, d\rangle$ *from the calling context.*

  $\boldsymbol{\delta}_1^{(s)} \leftarrow \overset{y_j}{\triangle}$; $\boldsymbol{\delta}_1^{(d)} \leftarrow \underset{s\to d}{\phi}\left(\boldsymbol{\delta}_1^{(s)}\right)$ ▷ *Convert subtree into* $\mathfrak{p}^{(d)}\{\cdot\}$.

  $\left(\begin{array}{c} y_j \\ | \\ \overset{\boldsymbol{\mu}}{\triangle} \end{array}\right) \leftarrow \boldsymbol{\delta}_1^{(d)}$; $\boldsymbol{\delta}_2^{(d)} \leftarrow \overset{\boldsymbol{\mu}}{\triangle}$ ▷ *Remove the intermediate item* $y_j$

  $\mathbf{D}^{(\ell)} \leftarrow$ top rule of $\boldsymbol{\delta}_1^{(d)}$

  $\mathbf{T}^{(\ell)} \leftarrow \underset{d\to t}{\sigma}\left(\mathbf{D}^{(\ell)}\right)$ ▷ *Use the alignment to retrieve the new top rule*

  $\boldsymbol{\delta}_2^{(s)} \leftarrow \underset{d\to s}{\phi}\left(\boldsymbol{\delta}_2^{(d)}\right)$ ▷ *Convert back into* $\mathfrak{p}^{(s)}\{\cdot\}$

  **return** $\left\langle \boldsymbol{\delta}_2^{(s)}, \mathbf{T}^{(\ell)} \right\rangle$

**Algorithm 25** Derivation mapping for $\mathfrak{p}^{(s)} \xrightarrow{\text{fold}^{d,j}_{\mathcal{S}\to r}} \mathfrak{p}^{(t)}$.

▷ *The comments at the top of Algorithm 24 apply here as well.*

**def** $\underset{s\to t}{\phi}\left(\boldsymbol{\delta}^{(s)}\right)$:

  **case** constant or builtin: **return** $\boldsymbol{\delta}^{(s)}$

  **case** $\overset{y_1}{\triangle} \circ \cdots \circ \overset{y_K}{\triangle}$: **return** $\underset{s\to t}{\phi}\left(\overset{y_1}{\triangle}\right) \circ \cdots \circ \underset{s\to t}{\phi}\left(\overset{y_K}{\triangle}\right)$

  **case** $\left(\begin{array}{c} x \\ | \\ \overset{\mu}{\triangle} \end{array}\right)$ **and** its top rule $\in \mathfrak{p}^{(t)}$: **return** $\begin{array}{c} x \\ | \\ \underset{s\to t}{\phi}\left(\overset{\mu}{\triangle}\right) \end{array}$

  **case** $\left(\begin{array}{ccc} & x & \\ \overset{\alpha}{\triangle} & \overset{\mu}{\triangle} & \overset{\beta}{\triangle} \end{array}\right)$ **and** its top rule $\notin \mathfrak{p}^{(t)}$:

    **return** $\begin{array}{ccc} & x & \\ \underset{s\to t}{\phi}\left(\overset{\alpha}{\triangle}\right) & \underset{s\to t}{\phi}\left(u^{-1}\left(\overset{\mu}{\triangle}\right)\right) & \underset{s\to t}{\phi}\left(\overset{\beta}{\triangle}\right) \end{array}$ ▷ *The rule passed to $u^{-1}$ is $\mathbf{S}^{(\ell)}$ the top rule of $\boldsymbol{\delta}^{(s)}$ (abusing notation), and the alignment ($\sigma$) used in $u^{-1}$ is from the* fold *transformation's generalized reversibility test.*

▷ *Inverse of the pivoting functions $\boldsymbol{u}$. See Lemma 3 for formal guarantees.*

**def** $u^{-1}\left(\boldsymbol{\delta}_2^{(s)}, \mathbf{T}^{(\ell)}\right)$: ▷ *Note: With apologies, the argument name $\mathbf{T}^{(\ell)}$ was chosen to match those of $\boldsymbol{u}$. (It is not a mistake that we pass in a rule from $\mathfrak{p}^{(s)}$ for a variable named $\mathbf{T}^{(\ell)}$ when we call it from the derivation mapping for fold.)*

  $\boldsymbol{\delta}_2^{(d)} \leftarrow \underset{s\to d}{\phi}\left(\boldsymbol{\delta}_2^{(s)}\right)$

  $\mathbf{D}^{(\ell)} \leftarrow \underset{t\to d}{\sigma}\left(\mathbf{T}^{(\ell)}\right)$

  $\boldsymbol{y}_j \leftarrow \text{reconstruct}\left(\mathbf{T}^{(\ell)}, \mathbf{D}^{(\ell)}\right)$

  $\boldsymbol{\delta}_1^{(d)} \leftarrow \left(\begin{array}{c} y_j \\ | \\ \overset{\mu}{\triangle} \end{array}\right)$ ▷ *The rule added is an instantiation of $\mathbf{D}^{(\ell)}$*

  $\boldsymbol{\delta}_1^{(s)} \leftarrow \underset{d\to s}{\phi}\left(\boldsymbol{\delta}_1^{(d)}\right)$

  **return** $\boldsymbol{\delta}_1^{(s)}$

▷ *Reconstruct the intermediate item that would have been erased by $\mathfrak{p}^{(s)} \xrightarrow{\text{unfold}^{d,j}_{r\to\mathcal{S}}} \mathfrak{p}^{(t)}$.*

**def** reconstruct$(\mathbf{T}^{(\ell)}, \mathbf{D}^{(\ell)})$:

  $(x' \oplus\!= \boldsymbol{\alpha}' \circ \boldsymbol{y}' \circ \boldsymbol{\beta}') \leftarrow r$

  $(y'' \oplus\!= \boldsymbol{\mu}'') \leftarrow \mathbf{D}^{(\ell)}$

  $\boldsymbol{\theta} \leftarrow \text{cover}\left((x' \oplus\!= \boldsymbol{\alpha}' \circ \boldsymbol{\mu}'' \circ \boldsymbol{\beta}'), \mathbf{T}^{(\ell)}\right)$ ▷ *Solve $\mathbf{T}^{(\ell)} = \boldsymbol{\theta}((x' \oplus\!= \boldsymbol{\alpha}' \circ \boldsymbol{\mu}'' \circ \boldsymbol{\beta}'))$*

  **return** $\boldsymbol{\theta}(y')$

#### 7.4.2.4 Practical Sufficient Conditions for Correctness

This section will identify a practical way to test for the impractical sufficient conditions we identified in the theorems of §7.4.2.3. What will result is a system that can guarantee termination of $\underset{s\to t}{\phi}(\boldsymbol{\delta})$ for all derivations $\boldsymbol{\delta}$ (where $\underset{s\to t}{\phi}$ is either Algorithm 24 or Algorithm 25). We will use a simple abstract interpretation technique to ensure that each recursive call made in $\underset{s\to t}{\phi}$ descends in a certain well-founded order—ensuring that the recursive calls eventually hit a base case.

Our strategy for avoiding unsafe fold and unfold transformations is adapted from Roychoudhury et al. (1999). At a high level, the strategy keeps track of the provenance of each rule (i.e., how each rule was derived using previous transformations). The rule provenance data will give rise to additional constraints on folding and unfolding actions between rules of different programs to prevent circular definition cycles. These are merely sufficient conditions but general enough to cover most use cases. A condition worth highlighting is that $\mathfrak{p}^{(s)}$ and $\mathfrak{p}^{(d)}$ must be related through a series of transformations; they cannot be unrelated programs.

We start with an initial program $\mathfrak{p}^{(0)}$. The initial program's derivations each have a finite size we can use as a well-founded ordering for induction. As we derive new programs from the initial program through transformations, we can use a common *relative* size with respect to $\mathfrak{p}^{(0)}$ to perform induction on derivations from the later programs. The relative size enables the comparison of derivations across programs (such as the derivations resulting from pivoting in $u$ and $u^{-1}$). However, we need to ensure that the relative-size measure remains accurate and well-founded in the new program after each fold or unfold.

**Modeling Decisions.** To make static analysis tractable, we make the following modeling decisions:

1 For any generalized fold or unfold, the programs $\mathfrak{p}^{(s)}$, $\mathfrak{p}^{(d)}$, and $\mathfrak{p}^{(t)}$ are related through a series of fold or unfold transformations to an initial program $\mathfrak{p}^{(0)}$.

2 The relative size of each derivation will be (approximately) preserved by each transformation. We will define a function $\Psi$, such that $\Psi\left(\boldsymbol{\delta}^{(s)}\right) = \Psi\left(\boldsymbol{\delta}^{(t)}\right)$ for all $\left\langle \boldsymbol{\delta}^{(s)}, \boldsymbol{\delta}^{(t)} \right\rangle$ in the bijection between $\mathfrak{p}^{(s)}\{\cdot\}$ and $\mathfrak{p}^{(t)}\{\cdot\}$.

3 The relative size of a derivation $\boldsymbol{\delta} \in \mathfrak{p}\{\cdot\}$ is an additive function $\Psi$ that is parameterized by values associated with each of the program's rules.

We now make 3 more precise; we define a derivation's measure.

**Definition 37.** *Suppose a given program* $\mathfrak{p}$ *assigns a* ***rule measure*** $\psi(\mathrm{R})$ *to each* $\mathrm{R} \in \mathfrak{p}$.[171] *The* ***derivation measure*** $\Psi(\boldsymbol{\delta})$ *for* $\boldsymbol{\delta} \in \mathfrak{p}\{\cdot\}$ *is*

$$\Psi(\boldsymbol{\delta}) \stackrel{\text{def}}{=} \begin{cases} 0 & \textbf{\textit{if}}\ \boldsymbol{\delta} \text{ is a constant or builtin} \\ \Psi(\boldsymbol{\delta}_1) + \cdots + \Psi(\boldsymbol{\delta}_K) & \textbf{\textit{if}}\ \boldsymbol{\delta} \text{ is a product } \boldsymbol{\delta}_1 \circ \cdots \circ \boldsymbol{\delta}_K \\ \psi(\mathrm{R}) + \Psi(\mathsf{body}(\boldsymbol{\delta})) & \textbf{\textit{if}}\ \boldsymbol{\delta} \text{ is a derivation with top rule } \mathrm{R} \end{cases} \tag{7.1}$$

We will carefully choose the rule measure parameters (3) so that 2 holds. After illustrating the following example, we will give the precise recipe.

[171]We assume, for simplicity, that rule measures are integer-valued. However, it is possible to generalize to certain well-ordered commutative groups (Roychoudhury et al., 1999, Def. 1).

**Example 39.** *Consider the following* unfold *with rule measures provided to the left of each rule:*

```
458  γ0: p += q * t.
459
460  γ1: q += r.
461  γ2: q += s.
```

$\xrightarrow{\text{unfold}}$

```
462  γ0 + γ1: p += r * t.
463  γ0 + γ2: p += s * t.
464  γ1: q += r.
465  γ2: q += s.
```

*We have associated the variables* $\gamma_0$*,* $\gamma_1$*, and* $\gamma_2$ *with the 3 rules of the program on the left. The rule measures on the right were chosen to preserve the derivation measure (2). Consider the derivations of 2 derivations for* p *in each program:*

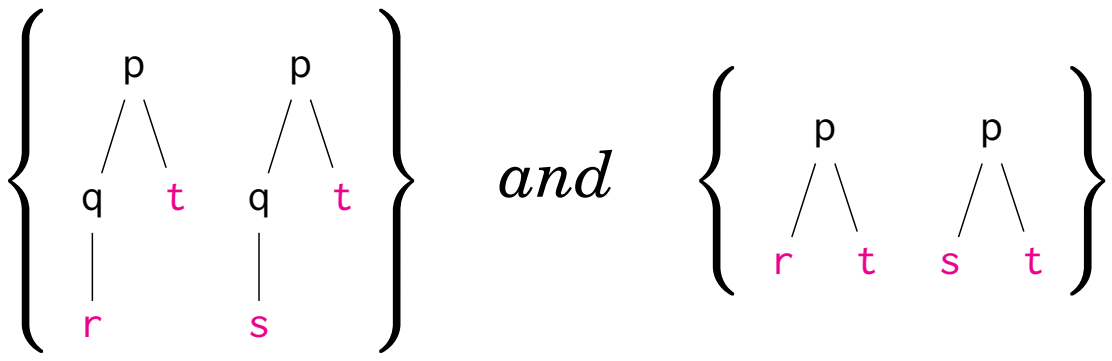

*We verify that the derivation measures are preserved:*

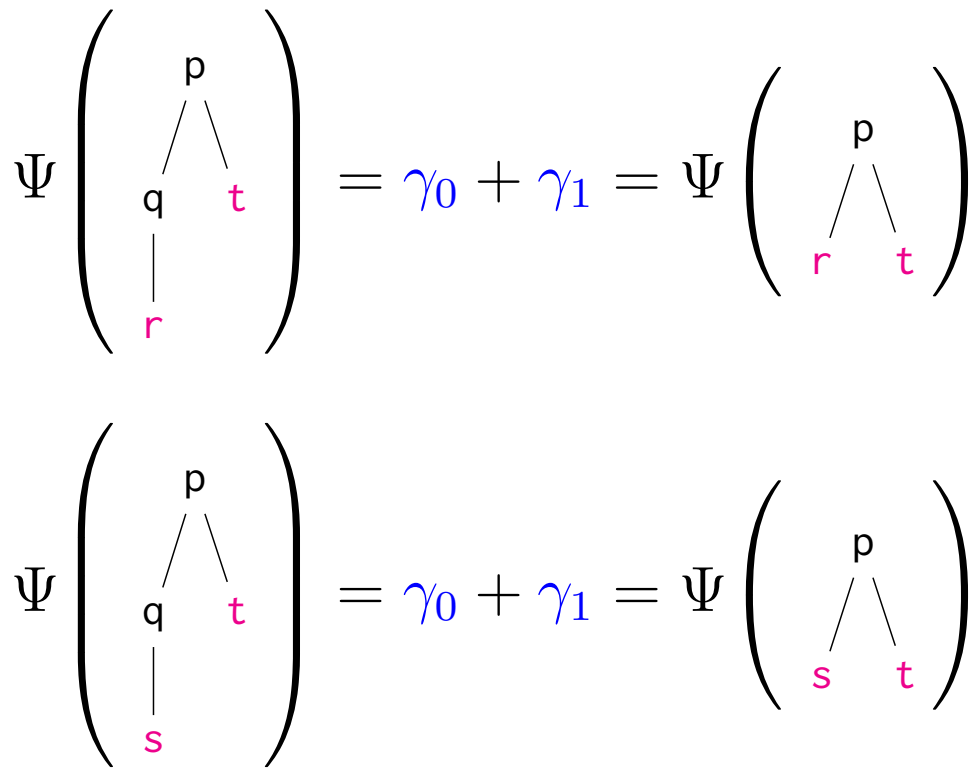

*Intuitively, the scheme works because the rules arising from the* unfold *count for* two *rules each time they appear in a derivation on the right.*

*We now consider* fold*, which works in the opposite direction. Here the measure for each rule appears to its left.*

```
466  γ0: q += r.
467  γ1: q += s.
468  γ2: p += r * t.
469  γ3: p += s * t.
```

$\xrightarrow{\text{fold}}$

```
470  γ0: q += r
471  γ1: q += s
472
473  {γ3 − γ1, γ2 − γ0}: p += q * t.
```

*The measure of* #473 *is interesting: we are uncertain about whether it should be*

$\gamma_3 - \gamma_1$ *or* $\gamma_2 - \gamma_0$. *So, we represent the measure of the new rule using an* uncertainty set*:* $\{\gamma_3 - \gamma_1, \gamma_2 - \gamma_0\}$. *We do this because we have lost the information about which of the two rules was folded. Now, we can conservatively estimate the size as an uncertainty set for each pair of derivations in the bijection between*

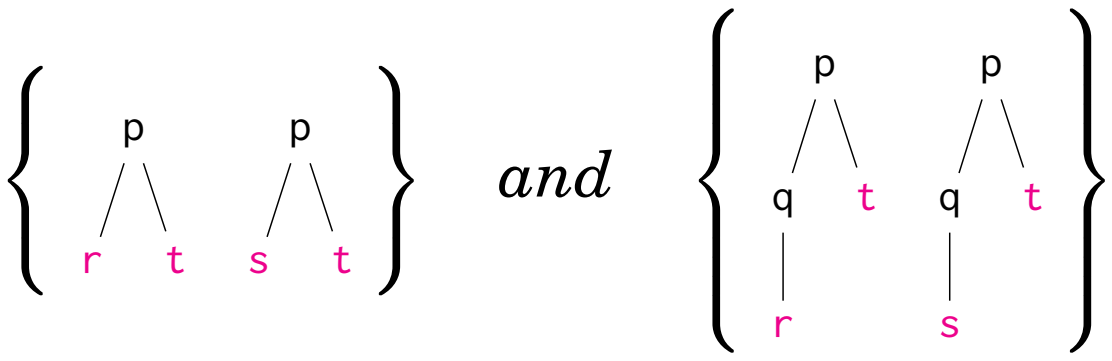

*We verify that the derivation measures are preserved (i.e., they are contained within the uncertainty sets):*

$\Psi\Big($ 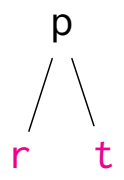 $\Big) = \{\gamma_2\} \subseteq \{\gamma_3 - \gamma_1 + \gamma_0, \gamma_2\} = \{\gamma_3 - \gamma_1, \gamma_2 - \gamma_0\} + \{\gamma_0\} = \Psi\Big($ 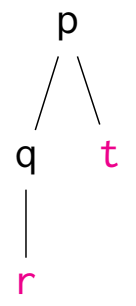 $\Big)$

$\Psi\Big($ 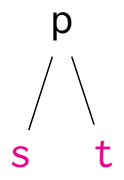 $\Big) = \{\gamma_3\} \subseteq \{\gamma_3, \gamma_2 - \gamma_0 + \gamma_1\} = \{\gamma_3 - \gamma_1, \gamma_2 - \gamma_0\} + \{\gamma_1\} = \Psi\Big($ 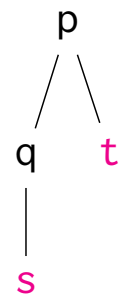 $\Big)$

We make 1 more precise by defining a transformation sequence:

**Definition 38.** *A **transformation sequence*** $\mathfrak{p}^{(0)}, \ldots, \mathfrak{p}^{(T)}$ *is defined recursively. Base case* ($T = 0$)*: We allow* $\mathfrak{p}^{(0)}$ *to be any program. We call* $\mathfrak{p}^{(0)}$ *the **reference program**. Recursive case* ($T \geq 1$)*: Suppose* $\mathfrak{p}^{(0)}, \ldots, \mathfrak{p}^{(T-1)}$ *is a transformation sequence, then* $\mathfrak{p}^{(0)}, \ldots, \mathfrak{p}^{(T-1)}, \mathfrak{p}^{(T)}$ *is a transformation sequence as long as* $\mathfrak{p}^{(T)} = \mathsf{fold}^{d,j}_{\mathcal{S}\to r}(\mathfrak{p}^{(s)})$ *or* $\mathfrak{p}^{(T)} = \mathsf{unfold}^{d,j}_{r\to\mathcal{S}}(\mathfrak{p}^{(s)})$ *where* $\mathfrak{p}^{(s)}, \mathfrak{p}^{(d)}$ *are earlier programs in the sequence, the new transformation is well-formed* ($\langle \mathcal{S}, j, r\rangle$ *are valid arguments), and the new transformation is safe.*

**Definition 39** (Rule measure updates)**.** *For each program in a transformation sequence $\mathfrak{p}^{(t)}$, we define a rule measure $\psi_t(\mathbf{R})$ for each of its rules $\mathbf{R} \in \mathfrak{p}^{(t)}$. The value of the rule measure will generally be an uncertainty set. The rule measure is defined inductively.*

*Base case: For each rule $\mathbf{R} \in \mathfrak{p}^{(0)}$, we take its measure $\psi_0(\mathbf{R}) > 0$ as an input parameter.*[172]

*Recursive case: Suppose the last program in our sequence is $\mathfrak{p}^{(t)}$. Then it must have been derived from one of the following cases:*

- *Case $\mathfrak{p}^{(t)} = \mathsf{unfold}^{d,j}_{r \to \mathcal{S}}(\mathfrak{p}^{(s)})$: Here, we have an alignment between each new rule $\mathbf{T}^{(\ell)} \in \mathfrak{p}^{(t)}$ and some $\mathbf{D}^{(\ell)} \in \mathfrak{p}^{(d)}$. That alignment is $\{\langle \mathbf{D}^{(\ell)}, \mathbf{T}^{(\ell)} \rangle\}_{\ell=1}^{L} = \underset{d \to t}{\sigma}$. The rule measures for each new rule $\mathbf{T}^{(\ell)}$ is then*

$$\psi_t(\mathbf{T}^{(\ell)}) = \psi_s(r) + \psi_d(\mathbf{D}^{(\ell)}) \quad \textit{for } \ell \in [1{:}L]$$

- *Case $\mathfrak{p}^{(t)} = \mathsf{fold}^{d,j}_{\mathcal{S} \to r}(\mathfrak{p}^{(s)})$: Let $r$ be the new rule that results from folding $\mathcal{S} = \{\mathbf{S}^{(\ell)}\}_{\ell=1}^{L}$ by the definitions $\{\mathbf{D}^{(\ell)}\}_{\ell=1}^{L}$. Then, $r$ has the following rule measure*[173,174]

$$\psi_t(r) = \bigcup_{\ell=1}^{L} \left(\psi_s(\mathbf{S}^{(\ell)}) - \psi_d(\mathbf{D}^{(\ell)})\right)$$

[172] The choice of $\psi_0(\mathbf{R}) = 1$ for all $\mathbf{R}$ is fine; however, a clever choice may make certain transformations available and others not. We will use a symbolic system to infer a value for the initial measure that designates a transformation as safe if such a value exists.

[173] The alignment $\ell$ that we are using here is given by $\sigma$ of the alignment of the unfold transformation that is used in the generalized reversibility test.

[174] Notice that when we fold *multiple* rules $|\mathcal{S}| > 1$, we might introduce approximation since multiple uncertainty sets are merged into $\psi_t(r)$.

*In both recursive cases, any unchanged rule* $\mathbf{R} \in \left(\mathfrak{p}^{(t)} \cap \mathfrak{p}^{(s)}\right)$ *copies its* $\psi$ *from* $\mathfrak{p}^{(s)}$. *The rule measures for rules in* $\mathfrak{p}^{(s)}$ *and* $\mathfrak{p}^{(d)}$ *are defined inductively.*

**Using relative size to ensure termination.** We will devise a simple test that ensures each recursive call of $\underset{s \to t}{\phi}$ *descends* (i.e., decreases the relative size). We will consider the fold and unfold cases separately.

Consider the case of fold. In this case, we will have an alignment of the form

$$\left\{\left\langle \mathbf{S}^{(\ell)}, \mathbf{D}^{(\ell)} \right\rangle\right\}_{\ell=1}^{L} = \left\{\left\langle \left(\boldsymbol{x}^{(\ell)} \oplus\!\!= \boldsymbol{\alpha}^{(\ell)} \circ \boldsymbol{\mu}^{(\ell)} \circ \boldsymbol{\beta}^{(\ell)}\right), \left(\boldsymbol{y}'^{(\ell)} \oplus\!\!= \boldsymbol{\mu}'^{(\ell)}\right) \right\rangle\right\}_{\ell=1}^{L}$$

We now inspect Algorithm 25's recursive calls. Since the other cases are straightforward, we will focus on line 25.7. Below, we drop the $\ell$ annotations to avoid notational clutter. In this case, we make the recursive call

$$\underset{s \to t}{\phi}\left(\boldsymbol{\delta}^{(s)}\right) = \underset{s \to t}{\phi}\left(\begin{array}{c} \boldsymbol{x} \\ \swarrow \;\downarrow\; \searrow \\ \overset{\boldsymbol{\alpha}}{\triangle} \;\; \overset{\boldsymbol{\mu}}{\triangle} \;\; \overset{\boldsymbol{\beta}}{\triangle} \end{array}\right) = \overbrace{\underset{s \to t}{\phi}\left(\overset{\boldsymbol{\alpha}}{\triangle}\right) \underset{s \to t}{\phi}\left(u^{-1}\left(\overset{\boldsymbol{\mu}}{\triangle}\right)\right) \underset{s \to t}{\phi}\left(\overset{\boldsymbol{\beta}}{\triangle}\right)}^{\boldsymbol{x}}$$

The tricky thing about these recursive calls is that we need to make sure that the recursive call to $\underset{s \to t}{\phi}\left(u^{-1}\left(\overset{\boldsymbol{\mu}}{\triangle}\right)\right)$ will not end up calling $\underset{s \to t}{\phi}\left(\boldsymbol{\delta}^{(s)}\right)$ again. The way that we will operationalize this quest is to use the derivation measure that we developed earlier in this section. Assuming that each derivation has equal measure ( 2 ) means that we can focus our attention on proving that $\Psi\left(\boldsymbol{\delta}^{(s)}\right) > \Psi\left(u^{-1}\left(\overset{\boldsymbol{\mu}}{\triangle}\right)\right)$ or, equivalently, $\Psi\left(\boldsymbol{\delta}^{(s)}\right) - \Psi\left(u^{-1}\left(\overset{\boldsymbol{\mu}}{\triangle}\right)\right) > 0$. If we can do that, there is no way that the $\underset{s \to t}{\phi}$ program could recurse forever: each step decreases by some nonzero positive integer amount, and the derivation measure is $\geq 0$. It is straightforward to verify that the sibling calls, $\underset{s \to t}{\phi}\left(\overset{\boldsymbol{\alpha}}{\triangle}\right)$

and $\underset{s \to t}{\phi}\left(\overset{\boldsymbol{\beta}}{\triangle}\right)$, are descending because they are subtrees of $\boldsymbol{\delta}^{(s)}$.

We will now derive the test. First, we expand the derivation measure of $\boldsymbol{\delta}^{(s)}$ by Definition 37:

$$\Psi\left(\boldsymbol{\delta}^{(s)}\right) = \Psi\left(\begin{array}{c} \boldsymbol{x} \\ \overset{\boldsymbol{\alpha}}{\triangle} \;\; \overset{\boldsymbol{\mu}}{\triangle} \;\; \overset{\boldsymbol{\beta}}{\triangle} \end{array}\right) = \psi_s(\mathbf{S}) + \Psi\left(\overset{\boldsymbol{\alpha}}{\triangle}\right) + \Psi\left(\overset{\boldsymbol{\mu}}{\triangle}\right) + \Psi\left(\overset{\boldsymbol{\beta}}{\triangle}\right)$$

Recall from Algorithm 25 that

$$u^{-1}\left(\overset{\boldsymbol{\mu}}{\triangle}\right) = \underset{d \to s}{\phi}\left(\begin{array}{c} y \\ | \\ \underset{s \to d}{\phi}\left(\overset{\boldsymbol{\mu}}{\triangle}\right) \end{array}\right)$$

where the rule $\mathbf{D} = (y \oplus\!= \boldsymbol{\mu})$ is an instantiation of a rule in $\mathfrak{p}^{(d)}$. Since this operation is measure-preserving (by assumption), we know that the measure of the new derivation is $\Psi\left(u^{-1}\left(\overset{\boldsymbol{\mu}}{\triangle}\right)\right) = \psi_d(\mathbf{D}) + \Psi\left(\overset{\boldsymbol{\mu}}{\triangle}\right)$.

Now, we can put these pieces together:

$$\begin{aligned}
\Psi\left(\boldsymbol{\delta}^{(s)}\right) &= \psi_s(\mathbf{S}) + \Psi\left(\overset{\boldsymbol{\alpha}}{\triangle}\right) + \Psi\left(\overset{\boldsymbol{\mu}}{\triangle}\right) + \Psi\left(\overset{\boldsymbol{\beta}}{\triangle}\right) \\
&= \psi_s(\mathbf{S}) + \Psi\left(\overset{\boldsymbol{\alpha}}{\triangle}\right) + \left(\Psi\left(u^{-1}\left(\overset{\boldsymbol{\mu}}{\triangle}\right)\right) - \psi_d(\mathbf{D})\right) + \Psi\left(\overset{\boldsymbol{\beta}}{\triangle}\right) \\
\Psi\left(\boldsymbol{\delta}^{(s)}\right) - \Psi\left(u^{-1}\left(\overset{\boldsymbol{\mu}}{\triangle}\right)\right) &= \psi_s(\mathbf{S}) + \Psi\left(\overset{\boldsymbol{\alpha}}{\triangle}\right) - \psi_d(\mathbf{D}) + \Psi\left(\overset{\boldsymbol{\beta}}{\triangle}\right)
\end{aligned}$$

Thus, we have descent if the following holds for all possible siblings $\overset{\boldsymbol{\alpha}}{\triangle}$ and $\overset{\boldsymbol{\beta}}{\triangle}$

and all possible measures in the uncertainty set:[175,176]

$$(\psi_s(\mathbf{S}) - \psi_d(\mathbf{D})) + \Psi\left(\overset{\alpha}{\triangle}\right) + \Psi\left(\overset{\beta}{\triangle}\right) > 0$$

In practice, we replace $\Psi\left(\overset{\alpha}{\triangle}\right)$ and $\Psi\left(\overset{\beta}{\triangle}\right)$ with estimates $\widehat{\Psi}(\boldsymbol{\alpha})$ and $\widehat{\Psi}(\boldsymbol{\beta})$ that only depend on the top of the derivation.[177] These estimates must satisfy

$$\widehat{\Psi}(\boldsymbol{x}) \leq \min_{\boldsymbol{\delta} \in \mathfrak{p}^{(s)}\{x\}} \Psi(\boldsymbol{\delta}) \quad \text{for all } \boldsymbol{x} \in \mathbb{H}$$

Notice that measure preservation means that we can use any program's derivation set for the base of the minimization. Thus, $\widehat{\Psi}$ can be estimated from the initial program $\mathfrak{p}^{(0)}$, which has the additional benefit that its derivation measure has the highest fidelity.[178] We describe possible estimators in Definition 40.

For the case of the unfold transformation, we arrive through a similar argument at the following condition:[179]

$$\psi_s(\mathbf{S}) + \psi_d(\mathbf{D}) + \Psi\left(\overset{\alpha}{\triangle}\right) + \Psi\left(\overset{\beta}{\triangle}\right) > 0$$

Notice that $\psi_s(\mathbf{S}) + \psi_d(\mathbf{D})$ need not be positive.

Lastly, we must ensure the entire provenance chain of $\mathfrak{p}^{(t)}$ is safe. This means that as we crawl up the sequence of prior fold and unfold transformations, we need to check that each is safe and they derive from a common $\mathfrak{p}^{(0)}$.

---

[175] When $x$ is an uncertainty set, the stipulation $x > 0$ is shorthand for $\forall x_i \in x : x_i > 0$.

[176] This condition is equivalent to that of Roychoudhury et al. (1999). However, we have arrived at it through different arguments.

[177] Notice that our choice to use the relative size of a derivation rather than relative height pays off here because the siblings in the derivation can provide some slack for the condition.

[178] Its uncertainty sets ought to be singletons.

[179] This condition was missing in Roychoudhury et al. (1999) since they had not considered generalized unfold.

**Definition 40.** *The following are possible estimators for the* ***minimum derivation measure*** *of an item.*

- *The simplest estimator is* $\widehat{\Psi}(\cdot) = 0$.
- *A simple estimator is*

$$\widehat{\Psi}(x) \overset{\text{def}}{=} \begin{cases} 0 & \textbf{\textit{if}}\ x \text{ is a constant, builtin, or input} \\ \widehat{\Psi}(x_1) + \cdots + \widehat{\Psi}(x_K) & \textbf{\textit{if}}\ x \text{ is a product } x_1 \circ \cdots \circ x_K \\ 1 & \textbf{\textit{otherwise}} \end{cases}$$

- *A more precise* $\widehat{\Psi}$ *can be expressed through the following system of linear inequalities that are a sufficient characterization of the set of feasible estimators:*[180]

$$\begin{cases} \widehat{\Psi}(\cdot) \geq 0 \\ \textbf{\textit{for}}\ x \text{ in constants, builtins, or inputs: } \widehat{\Psi}(x) \leq 0 \\ \textbf{\textit{for}}\ \mathbf{R} \in \mathfrak{p}^{(0)}: \\ \quad (x \mathrel{\oplus=} y_1 \circ \cdots \circ y_K) \leftarrow \mathbf{R} \\ \quad \widehat{\Psi}(x) \leq \psi_0(\mathbf{R}) + \widehat{\Psi}(y_1) + \cdots + \widehat{\Psi}(y_K) \end{cases}$$

  *For* $\mathbf{R} \in \mathfrak{p}^{(0)}$, $\psi_0(\mathbf{R})$ *should be a singleton set; thus, we can treat it like a scalar in these inequalities.*

We summarize the complete system in Algorithm 26.

**Theorem 1** (Correctness of Algorithm 26)**.** *Let* $\mathfrak{p}^{(0)}, \mathfrak{p}^{(1)}, \ldots, \mathfrak{p}^{(t)}$ *be a (measure-safe) transformation sequence. Then, we can assess the safety of a candidate* fold *or* unfold *transformation,* $\mathfrak{p}^{(t)} \to \mathfrak{p}^{(?)}$*, using* measure_safe$(\mathfrak{p}^{(?)})$*. If it is true, then the transformation is safe.*

[180]For efficiency, we might coarsen the program, e.g., by a maximum depth, as we did in §6.1.

*Proof Sketch.* The measure safety test (Algorithm 26) ensures that the recursive procedures for derivation mappings (Algorithm 24 and Algorithm 25) always terminate. The method measure_safe$\left(\mathfrak{p}^{(?)}\right)$ will only return `true` when the recursive calls provably descend safely under the candidate fold or unfold. The argument for why this is the case was detailed earlier in this section. Therefore, the candidate transformation must be safe by Theorem 2 and Theorem 3. ∎

**Summary.** This section has provided an effective safety test for the weighted setting based on prior work by Roychoudhury et al. (1999) for the unweighted setting. We have found that this system provides excellent recall when measure_safe leverages a constraint solver.[181] Not requiring an initial rule measure specification makes the overall system much more practical.

[181] We have found the complicated stratified counters in Roychoudhury et al. (1999) to be unnecessary thanks to the constraint solver.

**Algorithm 26** Fold–unfold transformation sequence safety test. We assume an initial reference program $\mathfrak{p}^{(0)}$ with rule measure $\psi_0(\cdot)$ as input.

$\widehat{\Psi}(\cdot) \leftarrow$ {one of the estimators in Definition 40}

**def** $\psi_t(\mathbf{R})$:
  **case** $\mathfrak{p}^{(t)} = \mathfrak{p}^{(0)}$: **return** $\psi_0(\mathbf{R})$
  **case** $\mathfrak{p}^{(t)} = \mathsf{fold}^{d,j}_{\mathcal{S}\to r}(\mathfrak{p}^{(s)})$ with alignment $\sigma$:
    **if** $\mathrm{R} = r$:
      **return** $\bigcup_{\langle \mathbf{S},\mathbf{D}\rangle \in \sigma} (\psi_s(\mathbf{S}) - \psi_d(\mathbf{D}))$
    **else**
      **return** $\psi_s(\mathbf{R})$ ▷ *Copy the unchanged rule*
  **case** $\mathfrak{p}^{(t)} = \mathsf{unfold}^{d,j}_{r\to\mathcal{S}}(\mathfrak{p}^{(s)})$ with alignment $\sigma$:
    **if** $\mathrm{R} \in \underset{t\to d}{\sigma}$:
      **return** $\psi_s(r) + \psi_d\left(\underset{t\to d}{\sigma}(\mathbf{R})\right)$
    **else**
      **return** $\psi_s(\mathbf{R})$ ▷ *Copy the unchanged rules*

**def** $\varphi(t)$:
  **case** $\mathfrak{p}^{(t)} = \mathfrak{p}^{(0)}$: **return** `true`
  **case** $\mathfrak{p}^{(t)} = \mathsf{fold}^{d,j}_{\mathcal{S}\to r}(\mathfrak{p}^{(s)})$ with alignment $\sigma$:
    ▷ *In the code below, $\mathbf{S}$ is of the form $(\boldsymbol{x} \oplus\!= \boldsymbol{\alpha} \circ \boldsymbol{\mu} \circ \boldsymbol{\beta})$ where $\boldsymbol{\alpha}$, $\boldsymbol{\mu}$, and $\boldsymbol{\beta}$ are determined by how $\mathbf{S}$ was folded.*
    **return** $\varphi(s) \wedge \varphi(d) \wedge \bigwedge_{\langle \mathbf{S},\mathbf{D}\rangle \in \sigma} \left(\psi_s(\mathbf{S}) - \psi_d(\mathbf{D}) + \widehat{\Psi}(\boldsymbol{\alpha}) + \widehat{\Psi}(\boldsymbol{\beta}) > 0\right)$
  **case** $\mathfrak{p}^{(t)} = \mathsf{unfold}^{d,j}_{r\to\mathcal{S}}(\mathfrak{p}^{(s)})$ with alignment $\sigma$:
    $(\boldsymbol{x} \oplus\!= \boldsymbol{\alpha} \circ \boldsymbol{y}_j \circ \boldsymbol{\beta}) \leftarrow r$ ▷ *Here $\boldsymbol{\alpha}$, $y_j$, and $\boldsymbol{\beta}$ are determined by how $r$ was unfolded.*
    **return** $\varphi(s) \wedge \varphi(d) \wedge \bigwedge_{\langle \mathbf{D},\mathbf{T}\rangle \in \sigma} \left(\psi_s(\mathbf{S}) + \psi_d(\mathbf{D}) + \widehat{\Psi}(\boldsymbol{\alpha}) + \widehat{\Psi}(\boldsymbol{\beta}) > 0\right)$

**def** $\mathsf{measure_safe}(\mathfrak{p}^{(t)})$:
  ▷ *For best results, use a constraint solver to find values for $\{\gamma_{\mathbf{R}}\}_{\mathbf{R}\in\mathfrak{p}^{(0)}}$ that satisfy the proposition $\varphi(t)$. Recall that each $\gamma_{\mathbf{R}}$ must be $> 0$ and an integer. Our implementation uses the popular Z3 solver (Moura and Bjørner, 2008), but we note that any integer linear solver would suffice.*
  ▷ *(For faster results, we can check the feasibility of $\varphi(t)$ using $\gamma_{\mathbf{R}} = 1$ for all $\mathbf{R} \in \mathfrak{p}^{(0)}$.)*
  ▷ *(It is always ok to return `false`, e.g., if we want to set a timeout on the SMT solver.)*
  **return** `true` **if** $\varphi(t)$ is satisfiable **else** `false`

## 7.5 Hoisting Transformations

This section gives two related transformations: the speculation transformation (Eisner and Blatz, 2007) and the left-corner transformations (Rosenkrantz and Lewis, 1970; Johnson, 1998; Johnson and Roark, 2000). Our left-corner transformation generalizes the well-known left-corner transformation for context-free grammars to Dyna programs.[182] We find an interesting connection between the left-corner and speculation transformations: both transformations "hoist" a subgoal up the tree, but the left-corner transformation additionally *transposes* the structure of each derivation—replacing left recursion with right recursion.

### 7.5.1 Speculation Transformation

The speculation transformation was first introduced in Eisner and Blatz (2007). Our presentation will differ from theirs because we will describe it in terms of how it transforms derivations.

**Example 40.** *Consider again the CKY program (Example 6).*[183]

```
474  p(X,I,K) += p(Y,I,J) * p(Z,J,K) * r(X,Y,Z).
475  p(X,I,K) += p(Y,I,K) * r(X,Y).
476  p(X,I,K) += w(Y,I,K) * r(X,Y).
```

*This program has an inefficiency: it repeats the same unary-chain closure for all spans* ⟨I,K⟩. *Our use of speculation will refactor the program so that unary chain closures may be precomputed (independently of the sentence) and reused at*

[182] The left-corner transformation eliminates left recursion in context-free grammars, which makes them amenable to specialized top-down parsing algorithms.
[183] The following example is adapted from Eisner and Blatz (2007).

*each span. This trick—attributed to Stolcke (1995)—is standard in probabilistic parsing. The final program is as follows:*

```
477 u(X,X') += u(Y,X') * r(X,Y).
478 u(X,X') += r(X,X').
479 p(X,I,K) += p'(X',I,K) * u(X,X').
480 p(X,I,K) += p'(X,I,K).
481 p'(X,I,K) += p(Y,I,J) * p(Z,J,K) * r(X,Y,Z).
482 p'(X,I,K) += w(Y,I,K) * r(X,Y).
```

*But how were we able to arrive at it? Below is a derivation of the original program that contains a long chain of #475 all with* `I=3, K=5` *( highlighted ):*

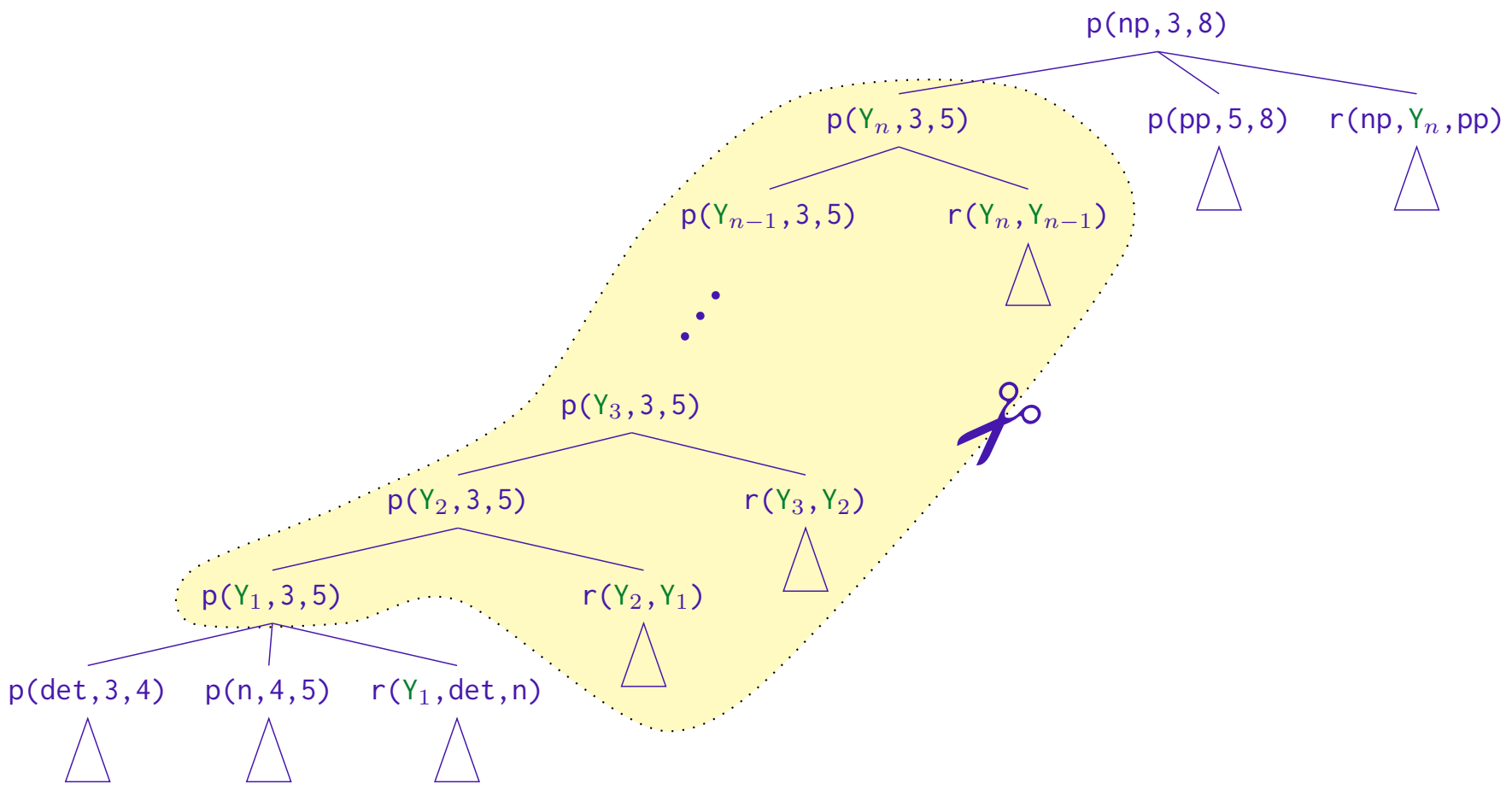

*Our transformed program will use a slashed item*[184] `p(Y`$_n$`,3,5)/p(Y`$_1$`,3,5)` *that computes the total weight of the intermediate chains—specifically, those that start at* `p(Y`$_1$`,3,5)`*, end at* `p(Y`$_n$`,3,5)`*, and go through a subset of rules (in this case {#475}).*[185] *These different chains will correspond to different derivations of the slashed item. We can think of the new item as abstracting away the chain from the specific starting point* `p(Y`$_1$`,3,5)`*. We can commit to a specific starting point of the form* `p(Y`$_1$`,3,5)` *later (i.e., higher up in the derivation tree) using a* recovery rule*, such as*

[184] There is nothing special about the use of / here—it is not a special operator; it is just a name for the new item that is meant to *suggest* that `p(Y`$_n$`,3,5) = p(Y`$_1$`,3,5) * p(Y`$_n$`,3,5)/p(Y`$_1$`,3,5)` as it would if / meant division under real arithmetic.

[185] The specific subset of rules is a parameter of the transformation.

p($Y_n$,3,5) += $\widetilde{\text{p}}$($Y_1$,3,5) * p($Y_n$,3,5)/p($Y_1$,3,5). *For technical reasons, we rename the bottom element of the chain apart from its original by marking it with* $\widetilde{\cdot}$ *(in this case,* $\widetilde{\text{p}}$($Y_1$,3,5)*). Our transformed derivation looks as follows.*

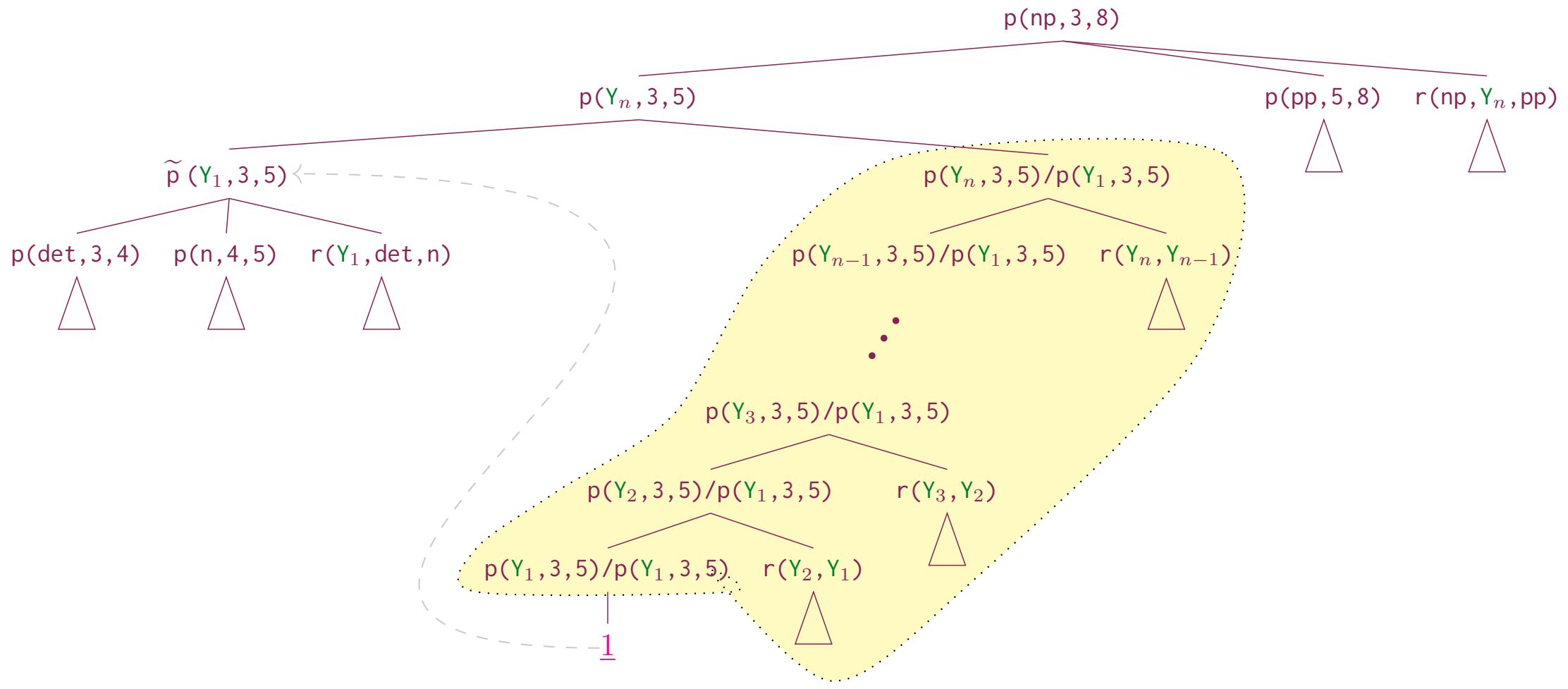

*In the diagram, we can see that the chain has been stashed into a derivation of the slashed item. That subderivation turns out not to depend on the specific values for the span* $\langle 3, 5\rangle$*—meaning that it can be reused by different spans and leads to a runtime improvement after we apply abbreviation (§6.2.3). The subderivations represented as generic triangles* △ *will be transformed recursively.*

*In the speculation notation, our transformed program is*

```
p(X,I,K) / p(X',I',K') += p(Y,I,K) / p(X',I',K') * r(X,Y).
p(X',I',K') / p(X',I',K').
p(X,I,K) += p̃(X',I,K) * p(X,I,K) / p(X',I',K').
p̃(X,I,K) += p(Y,I,J) * p(Z,J,K) * r(X,Y,Z).
p̃(X,I,K) += w(Y,I,K) * r(X,Y).
```

*Program specialization (§6.2) recognizes that* I=I' *and* K=K'*. Additionally, it recognizes that* I*,* I'*,* K*, and* K' *are always free in the slashed items so they can be dropped. So after specialization, we have the final program from earlier.*

The speculation transformation is parameterized by $\boldsymbol{\sigma} \subseteq \mathfrak{p}$ and a nonground term $z$. Let $\delta \in \mathfrak{p}\{\cdot\}$ be an arbitrary derivation tree of a program $\mathfrak{p}$. We may represent $\delta$ schematically as a derivation of the form on the left. The speculation transform will transform $\delta$ into a derivation of the form on the right.

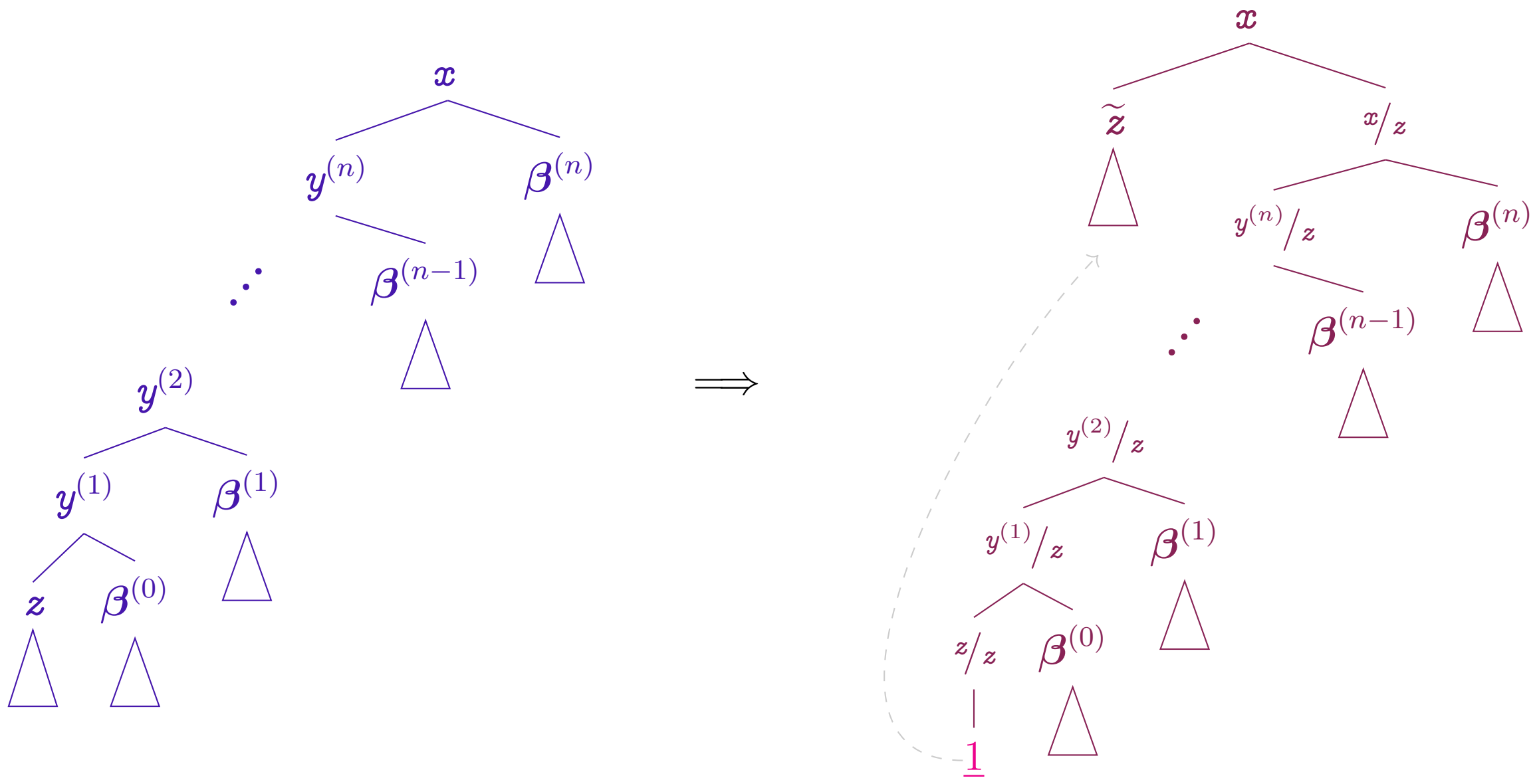

The schematic above exposes a **spine** of the derivation: the (maximal length) path $x \rightsquigarrow z$ along the left edge of the derivation that uses only some designated subset of rules $\boldsymbol{\sigma}$ and ends at an item matching the nonground term $z$.[186] As we can see with the derivation on the right, the speculation transformation has *dragged up* (hoisted) the subderivation rooted at $z$ leaving behind a **trace** ${}^{x}\!/_{z} \rightsquigarrow {}^{z}\!/_{z}$ along the spine $x \rightsquigarrow z$ with a $\underline{1}$ in the original position of the hoisted subderivation. We call each ${}^{\cdot}\!/_{\cdot}$ item a **slashed item**. The new derivation still has root $x$, but now $z$ attaches at the top with a derivation of the $x$ item without $z$ at the base of its spine. Each subtree rooted at $\boldsymbol{\beta}^{(i)}$ remains a derivation of the same $\boldsymbol{\beta}^{(i)}$ items, but the internals of each subtree may change because the transformation is applied recursively to it.

Recall from Example 40 that the marking $\widetilde{z}$ on the $z$ item means that the item is the bottom of a chain. We will use this marking for an additional purpose. In $\mathfrak{p}$, it might also be possible to derive $x$ in other ways using the rules in $\boldsymbol{\sigma}$ such that the base of the spine does not match $z$. We want to keep these derivations intact, so we create a route marked with $\widetilde{\cdot}$ that allows the derivations built with rules in $\boldsymbol{\sigma}$ that do not have $z$ along their spine to continue contributing values to $x$.

[186]Thus, if $z$ appears multiple times, we take the bottom-most occurrence.

These marked items will contribute to the value of $x$ using a rule ($x \oplus\!= \widetilde{x}$), but notice that we need to ensure that $x \notin z$ because the $\widetilde{x}$ is supposed to sum up the derivations of $x$ that either do not have $z$ on their spine—either because (a) they do not go through $\boldsymbol{\sigma}$, or (b) they do go through $\boldsymbol{\sigma}$, but they do not end in a $z$ item. If $x$ is a $z$ item, it reaches $z$ with a zero-length spine. To drop these unwarranted contributions, Eisner and Blatz (2007) add a constraint that $x \notin z$.

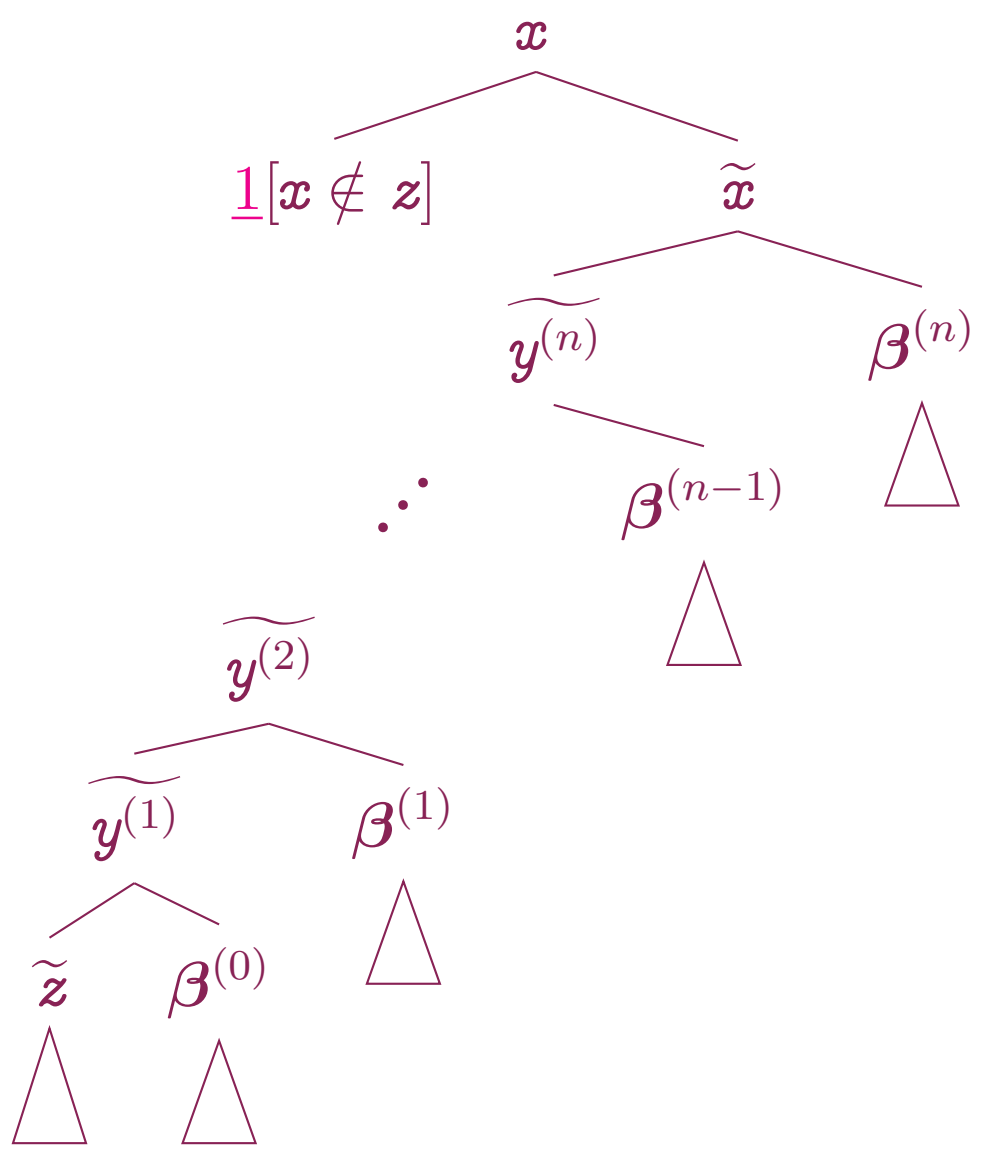

The last case we need to consider is when multiple $z$ items might appear along the spine. If we do not distinguish this case, we could overcount, as there are many places where the $\widetilde{z}$ item will try to attach itself. We use the solution proposed by Eisner and Blatz (2007): hoist the bottom-most $z$ item as high as it can go along the spine (i.e., breaking the ambiguity in where to attach it by attaching at the highest point along the spine).

**Definition 41.** *Given a program* $\mathfrak{p}$*, the* ***speculation transformation*** $\mathfrak{p}' = \text{speculation}(\mathfrak{p}, \boldsymbol{z}, \boldsymbol{\sigma})$ *is parameterized by a nonground term* $\boldsymbol{z}$ *representing a set of items to slash out, and* $\boldsymbol{\sigma}$ *is a set of rule indices to slash from. For simplicity, the transformation given below requires that (1)* $\mathfrak{p}$ *does not have inputs,*[187] *(2) the variables in* $\boldsymbol{z}$ *cannot appear in* $\mathfrak{p}$*, (3) for each* $i \in \boldsymbol{\sigma}$*,* $(\boldsymbol{x} \mathrel{\oplus=} \boldsymbol{y}_1 \circ \cdots \circ \boldsymbol{y}_K) = \mathfrak{p}_i$*: (a)* $K > 0$*, (b)* $\boldsymbol{y}_1$ *is not a constant or builtin,*[188] *(c)* $\boldsymbol{y}_1 \subseteq \boldsymbol{z}$ *or* $\boldsymbol{y}_1 \cap \boldsymbol{z} = \emptyset$*.*[189]

**Algorithm 27** speculation: Apply the speculation transformation (Definition 41).

1. **def** $\text{speculation}(\mathfrak{p}, \boldsymbol{z}, \boldsymbol{\sigma})$:
2. $\quad \mathfrak{p}' \leftarrow \{\!\!\{\}\!\!\}$
3. $\quad$ **for** $i = 1 \ldots |\,\mathfrak{p}\,|$:
4. $\quad\quad$ **if** $i \notin \boldsymbol{\sigma}$:
5. $\quad\quad\quad (\boldsymbol{x} \mathrel{\oplus=} \boldsymbol{\mu}) \leftarrow \mathfrak{p}_i$
6. $\quad\quad\quad \mathfrak{p}'[\widetilde{\boldsymbol{x}}] \mathrel{\oplus=} \boldsymbol{\mu}$
7. $\quad\quad$ **else**
8. $\quad\quad\quad (\boldsymbol{x} \mathrel{\oplus=} \boldsymbol{y} \circ \boldsymbol{\beta}) \leftarrow \mathfrak{p}_i$
9. $\quad\quad\quad \mathfrak{p}'[\boldsymbol{x}/\boldsymbol{z}] \mathrel{\oplus=} \boldsymbol{y}/\boldsymbol{z} \circ \boldsymbol{\beta}$
10. $\quad\quad\quad$ **if** $\boldsymbol{y} \cap \boldsymbol{z} = \emptyset$:
11. $\quad\quad\quad\quad \mathfrak{p}'[\widetilde{\boldsymbol{x}}] \mathrel{\oplus=} \widetilde{\boldsymbol{y}} \circ \boldsymbol{\beta}$
12. $\quad \triangleright$ *Base case for slashed items*
13. $\quad \mathfrak{p}'[\boldsymbol{z}/\boldsymbol{z}] \mathrel{\oplus=} \underline{1}$
14. $\quad \triangleright$ *Rules that recover the original items in* $\mathfrak{p}$
15. $\quad \mathfrak{p}'[\mathsf{X}] \mathrel{\oplus=} \widetilde{\boldsymbol{z}} \circ \mathsf{X}/\boldsymbol{z}$
16. $\quad \mathfrak{p}'[\mathsf{X}] \mathrel{\oplus=} \text{not-matches}_{\boldsymbol{z}}(\mathsf{X}) \circ \widetilde{\mathsf{X}}$
17. $\quad$ **return** $\mathfrak{p}'$

[187] Annoyingly, when the program has inputs, the transformation given will rename the inputs (wrapping them in $\widetilde{\cdot}$)—breaking the existing program API. It is straightforward to relax this assumption when the names of the inputs are disjoint from items that are defined by the program: (1) override $\widetilde{\boldsymbol{x}}$ not to mark $\boldsymbol{x}$ when it is an input relation, and (2) constrain line 27.16 so that it will not produce the problematic rule $\boldsymbol{x} \mathrel{\oplus=} \text{not-matches}_{\boldsymbol{z}}(\boldsymbol{x}) \circ \boldsymbol{x}$.

[188] To support slashing-out constants and built-in relations, we can wrap them in temporary items, and then eliminate them (by unfolding) once the speculation transformation is complete.

[189] It is sometimes useful to partition $\mathfrak{p}_i$ into three more specialized rules using boolean side conditions: $(\boldsymbol{y}_1 \cap \boldsymbol{z})$, $(\boldsymbol{y}_1 \setminus \boldsymbol{z})$, and $(\boldsymbol{z} \setminus \boldsymbol{y}_1)$.

**Algorithm 28** not-matches: Implementation of the not-matches relation.

▷ *This method implements* $\text{not-matches}_z(\text{Y}) \stackrel{\text{def}}{=} \underline{1}[\text{Y} \notin z]$*, a built-in relation that checks if* Y *is not an element of the set of terms denoted by the nonground term* $z$*. Importantly, the variables in* $z$ *are part of the* name *of the builtin, so they are not summed over when the builtin is used in a rule. This method should be included in the library of built-in relations. It is implemented as a result stream to be compatible with §3.1.2.2.*

```
def not-matches_z(y):
    if unify(z, y) = FAIL: yield 1                 ▷ certain, satisfied: z and y cannot match
    else if cover(z, y) ≠ FAIL: return             ▷ certain, fail: z ⊇ y means that y must be in z
    else yield not-matches_z(y)                    ▷ uncertain, delay: y could still become an instance of z
```

*The transformation will generate novel function symbols to denote the* $\cdot/\cdot$ *and* $\tilde{\cdot}$ *items. These names must be disjoint from the names of the items in* $\mathfrak{p}$*. The pseudocode utilizes* $\text{not-matches}_z(\text{Y}) \stackrel{\text{def}}{=} \underline{1}[\text{Y} \notin z]$*, a built-in relation that checks if* Y *is not an element of the set of terms denoted by the nonground term* $z$*. Importantly, the variables in* $z$ *should not be summed over—they are meant to be local to the builtin.*[190] *For completeness, we include pseudocode for* not-matches *in Algorithm 28. The transformation should often be followed by* abbreviate *(§6.2.3) to refine its output, as it often contains useless rules and variables.*

**Extensions.** We also note that a straightforward adaptation to Definition 41 speculates over the rightmost subgoal instead of the leftmost subgoal. Our implementation provides both leftmost and rightmost options. When the semiring is commutative, we can permute the subgoals (as a pre-transformation) to make any subgoal the first subgoal in its rule. Even without commutativity, some subgoals may commute with one another—in particular, built-in constraints can be permuted because they multiply a $\underline{1}$ or $\underline{0}$, which can be freely moved around thanks to the semiring properties. Lastly, we note that many of the $\tilde{\cdot}$ items created by the speculation transformation are unnecessary; thus, it is often worth following speculation with unfolding and useless-rule elimination to remove some unnecessary intermediate $\tilde{\cdot}$ items. An easy choice is to avoid creating $\tilde{\cdot}$ rules for any item that is not in $\{z\} \cup \{y \mid (x \mathrel{\oplus=} y \circ \boldsymbol{\beta}) = \mathfrak{p}_i, i \in \boldsymbol{\sigma}\}$.[191]

[190] The strategy of using a built-in relation is a workaround that avoids a mixed semiring program as was done in Eisner and Blatz (2007)'s presentation.

**Correctness.** We believe that the speculation transformation is strongly semantics-preserving. The $/\cdot$ and $\tilde{\cdot}$ items should be regarded as new definitions (§7.3), as they are not defined in $\mathfrak{p}$. We have provided a derivation mapping for the transformation in §B.3. However, we have not provided a complete proof of correctness.

## 7.5.2 Left-Corner Transformation

We extend the left-corner transformation (Rosenkrantz and Lewis, 1970; Johnson, 1998) to Dyna programs. To have fine-grained control over how the transformation is applied, we build on the selective left-corner transformation (Johnson and Roark, 2000), which permits the transformation to apply to a subset of rules. The left-corner transformation is closely related to speculation: both transformations hoist up the subderivations, but the left-corner transformation will, additionally, switch left recursion to right recursion (i.e., it will transpose the spine of the derivation). Below is a schematic that builds on the earlier discussion of speculation.

[191] This set of items is useful for specializing the recovery rules on line 27.15 and line 27.16.

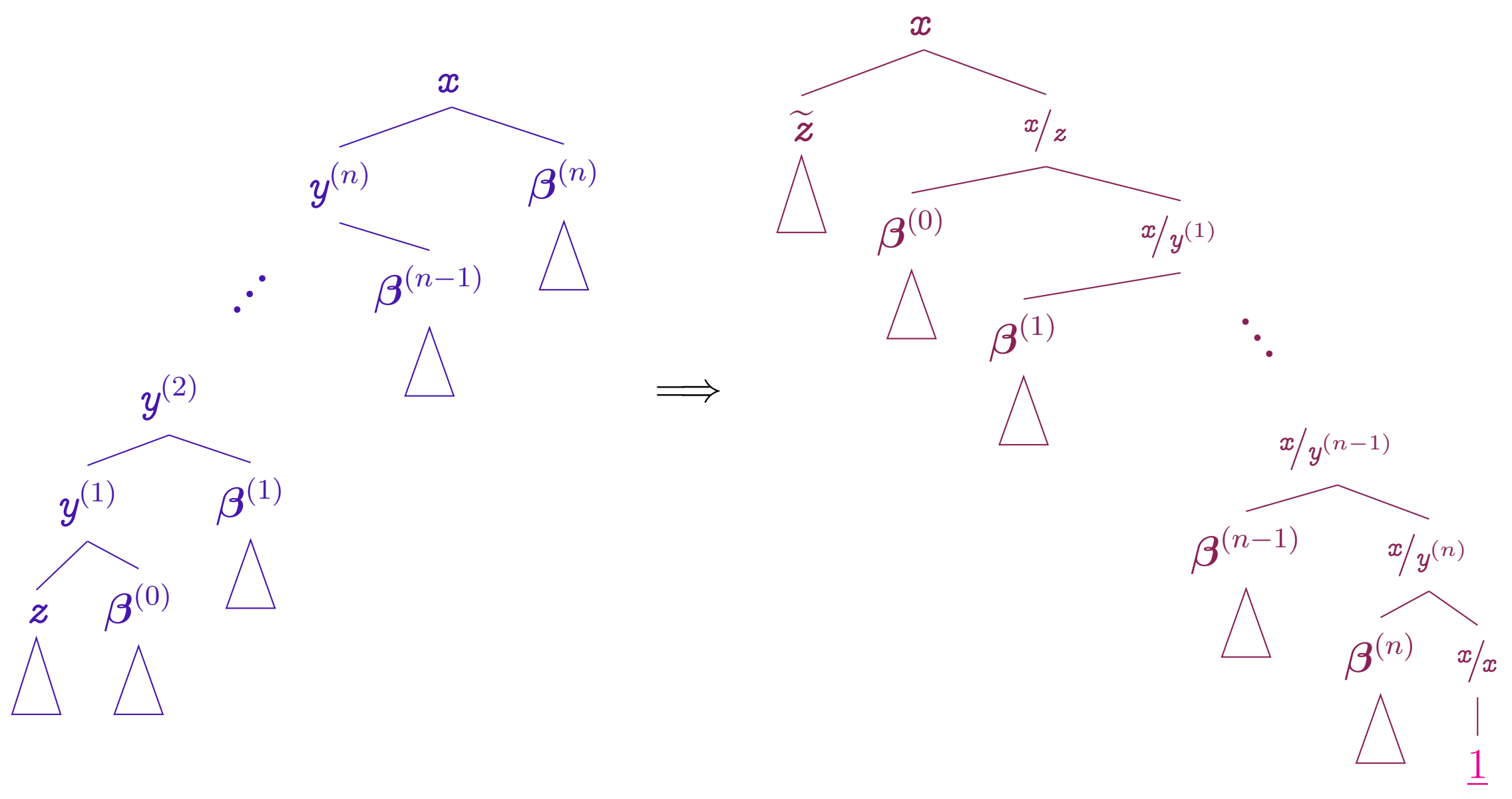

In contrast with speculation, the slashed items in the trace of the transformed derivation are named differently; rather than naming them by the item $z$ that is hoisted up, they are named by the destination $y$. This is how the transformation manages to transpose the derivation tree. This effect can be beneficial when the program defines many items irrelevant to a set of output items.[192] Our version of the left-corner transformation does not constrain the base of the spine, so there is no need to create the recursive rules of $\widetilde{\cdot}$ items. We maintain the $\widetilde{\cdot}$ items to distinguish items at the base of the spine from the top.

**Definition 42.** *Given a program* $\mathfrak{p}$*,* ***left-corner transformation*** $\mathfrak{p}' = \mathsf{leftcorner}(\mathfrak{p}, \boldsymbol{\sigma})$ *is parameterized by a set of rule indices* $\boldsymbol{\sigma}$*. For simplicity, the transformation be-*

[192]This is analogous to forward-mode vs. backward-mode automatic differentiation (Griewank and Walther, 2008). Specifically, speculation is to left-corner as forward-mode is to backward-mode. The differentiation methods build items of the form $\frac{\partial x}{\partial z}$, much like hoisting methods build items of the form $x/z$. However, they build different subsets: backward-mode and left-corner target the numerators, and forward-mode and speculation target the denominators. It is well-known in the automatic differentiation setting, that backward-mode is more efficient when the number of numerators needed is less than the number of denominators needed, and forward-mode is more efficient in the opposite regime.

*low requires that (1) $\mathfrak{p}$ does not have inputs,*[193] *(2) for each $i \in \boldsymbol{\sigma}$, $(\boldsymbol{x} \mathrel{\oplus}= y_1 \circ \cdots \circ y_K) = \mathfrak{p}_i$: (a) $K > 0$, (b) $y_1$ is not a constant, builtin, or input.*[194]

**Algorithm 29** leftcorner: Apply the left-corner transformation (Definition 42).

```
1.  def leftcorner(𝔭, σ):
2.      𝔭′ ← ⦃⦄
3.      for i = 1 … |𝔭|:
4.          if i ∉ σ:
5.              (x ⊕= μ) ← 𝔭_i
6.              𝔭′[x̃] ⊕= μ
7.          else
8.              (x ⊕= y ∘ β) ← 𝔭_i
9.              𝔭′[X/y] ⊕= β ∘ X/x
10.     ▷ Base case of slashed items
11.     𝔭′[X/X] ⊕= 1
12.     ▷ Rules that recover the original items in 𝔭
13.     𝔭′[X] ⊕= Z̃ ∘ X/Z
14.     𝔭′[X] ⊕= X̃
15.     return 𝔭′
```

*The comments in the bottom paragraph of Definition 41 apply here.*

**Extensions.** The same extensions for the case of speculation apply here: the right-corner transformation that hoists the rightmost subgoals, and permuting commutative subgoals.

**Correctness.** We believe that the left-corner transformation is strongly semantics-preserving. It is sufficiently similar to the grammar transformation that we do not foresee any issues. However, we have not formally checked all of the details.

[193]See footnote 187 for a workaround.
[194]See footnote 188 for a workaround.

### 7.5.3 Examples

**Example 41.** *What happens if we apply the left-corner transformation to Example 40? The transformed program is almost identical: both produce precisely the same items. However, as expected, the left-corner transformation will transpose the recursion: the slashed items are the same but defined using a right-recursive rule instead of a left-recursive rule. Specifically, the new program is the same as #483–487, except that we replace #483 (repeated below)*

```
488 p(X,I,K) / p(X',I',K') += p(Y,I,K) / p(X',I',K') * r(X,Y).
```

*with the rule*

```
489 p(X',I',K') / p(Y,I,K) += r(X,Y) * p(X',I',K') / p(X,I,K).
```

**Example 42.** *This example explores the connection between speculation and the variable-elimination strategy we saw in §7.3. Consider the following single-rule program, which may be regarded as the product of three matrices:*

```
490 a(I,L) += b(I,J) * c(J,K) * d(K,L).
```

*Factoring the last two subgoals into an intermediate item* `tmp(J,L)`*:*

```
491 a(I,L) += b(I,J) * tmp(J,L).
492 tmp(J,L) += c(J,K) * d(K,L).
```

*The new program is more efficient because a rule's maximum number of variables is reduced. We can mimic this transformation by speculation with* $\sigma = \{\#490\}$ *and* $z =$ `b(I',J')`*:*

```
493 a(I,L) / b(I',J') += b(I,J) / b(I',J') * c(J,K) * d(K,L).
494 a(I,L) += a(I,L) / b(I',J') * b(I',J').
495 b(I',J') / b(I',J').
```

*Next, we eliminate #495 by unfolding:*

```
496 a(I,L) += a(I,L) / b(I,J) * b(I,J).
497 a(I,L) / b(I,J) += c(J,K) * d(K,L).
```

*Notice that the second rule always has* `I` *free. Thus, speculation's items* `a(I,L) / b(I,J)` *are effectively just a naming scheme for the* `tmp(J,L)` *items, a step that we can automate with the abbreviation transformation (§6.2.3).*

**Example 43.** *Suppose we have a program such as*

```
498 goal += x(I) * f(I).
499 goal += x(I) * g(J) * z.
```

*and we want to factor it into the following form:*

```
500 goal += x(I) * (???).
```

*We can do so by speculation with* $z =$ `x(I)` *and* $\boldsymbol{\sigma} = \{\#498, \#499\}$*:*

```
501 goal += x(I') * goal/x(I').
502 goal/x(I') += x(I)/x(I') * f(I).
503 goal/x(I') += x(I)/x(I') * g(J).
504 x(I')/x(I').
```

*Next, we unfold the* `x(I)/x(I')` *subgoals:*

```
505 goal += x(I') * goal/x(I').  % unchanged
506 goal/x(I') += f(I').
507 goal/x(I') += g(J) * z.
```

*Notice that unfolding binds* `I` *to* `I`$'$ *in #506, and #507 is non-range-restricted (i.e., has a free variable in its head). We see that* `goal/x(I')` *is a clever scheme for naming the contributions from both of these rules. This program has the form that we were after.*

**Example 44.** *A split bilexical grammar is one where the head word must combine with all its right children before any of its left children. The naive algorithm, given below, runs in* $\mathcal{O}(n^5)$*. Eisner and Satta (1999a) were able to speed it up to* $\mathcal{O}(n^3)$*. Eisner and Blatz (2007) described the speedup in terms of speculation. To* *simplify the example, we assume an unlabeled grammar.*

```
508 r(I,I,K) += word(I,K).                           % no right children yet
509 r(P,I,K) += r(P,I,J) * arc(P → C) * l(C,J,K).  % add right child
510
511 l(P,I,K) += r(P,I,K).                            % no left children yet
512 l(P,I,K) += l(C,I,J) * arc(C ← P) * l(P,J,K).  % add left child
513
514 goal += l(P,0,N) * len(N).
```

*The crux of the speedup is to transform left constituents so they can be built without a specific right constituent. This work can be done independently of the right constituent, and the slashed items we will derive in the transformed program below do not instantiate* `K`*—leading to an improved runtime. Below is the output of speculation from the right with* $\boldsymbol{z} =$ `r(H',I',K')` *and* $\boldsymbol{\sigma} = \{\#511, \#512\}$ *after omitting the useless* $\widetilde{\texttt{goal}}$ *item and useless-rule elimination:*

```
515 r̃(I,I,K) += word(I,K).
516 r̃(P,I,K) += r(P,I,J) * arc(P → C) * l(C,J,K).
517
518 l(P,I,K) / r(H',I',K') += r(P,I,K) / r(H',I',K').
519 l(P,I,K) / r(H',I',K') += l(C,I,J) * arc(C ← P) * l(P,J,K) / r(H',I',K').
520
521 goal += += l(H,0,N) * len(N).
522
523 % base case for slashed items
524 (r(H',I',K') / r(H',I',K')).
525
526 % recovery rules
527 r(P,I,K) += r̃(H',I',K') * r(P,I,K) / r(H',I',K').
528 l(P,I,K) += r̃(H',I',K') * l(P,I,K) / r(H',I',K').
```

*Next, we eliminate all uses of the original* `r` *and* `l` *items using unfold; and we run program specialization, which exposes variables that are equated:*

```
529 r̃(I,I,K) += word(I,K).
530 r̃(I,I,K) += r̃(I,I,J) * arc(I → C) * r̃(C,C,K) * l(C,J,K) / r(C,C,K).
531 l(I',I',K') / r(I',I',K').
532 l(I',I,K') / r(I',I',K') += r̃(C,C,J) * l(C,I,J) / r(C,C,J)
533     * arc(C ← I') * l(I',J,K') / r(I',I',K').
534 goal += r̃(H,H,N) * l(H,0,N) / r(H,H,N) * len(N).
```

*We see that the first two arguments of* `r̃(H,H,N)` *are always equal; so we can*

*abbreviate* $\widetilde{\texttt{r}}$(H,H,N) ⟼ rtrap(H,N).

```
535 rtrap(I,K) += word(I,K).
536 rtrap(I,K) += rtrap(I,J) * arc(I → C) * rtrap(C,K) * l(C,J,K) / r(C,C,K).
537 l(I',I',K') / r(I',I',K').
538 l(I',I,K') / r(I',I',K') += rtrap(C,J) * l(C,I,J) / r(C,C,J)
539     * arc(C ← I') * l(I',J,K') / r(I',I',K').
540 goal += rtrap(H,N) * l(H,0,N) / r(H,H,N) * len(N).
```

*We chose this name because it corresponds to Eisner and Satta (1999a)'s notion of a "right trapezoid;" we will derive their "left trapezoid" next. First, we observe the redundancy in* l(I',I,K') / r(I',I',K')*, which has three variables equated. We can abbreviate* l(I',I,K') / r(I',I',K') ⟼ lr(I',I,K').

```
541 rtrap(I,K) += word(I,K).
542 rtrap(I,K) += rtrap(I,J) * arc(I → C) * rtrap(C,K) * lr(C,J,K).
543 lr(I',I',K').
544 lr(I',I,K') += rtrap(C,J) * lr(C,I,J) * arc(C ← I') * lr(I',J,K').
545 goal += lr(H,0,N) * rtrap(H,N) * len(N).
```

*Next, we observe that* K' *in* lr(I',I,K') *is always free, so we drop it by abbreviating* lr(I',I,K') ⟼ ltrap(I',I).

```
546 rtrap(I,K) += word(I,K).
547 rtrap(I,K) += rtrap(I,J) * arc(I → C) * rtrap(C,K) * ltrap(C,J).
548 ltrap(I',I').
549 ltrap(I',I) += rtrap(C,J) * ltrap(C,I) * arc(C ← I') * ltrap(I',J).
550 goal += ltrap(H,0) * rtrap(H,N) * len(N).
```

*Finally, we apply variable elimination to derive the following new items:*

```
551 tmp(C,I) += ltrap(I,J) * rtrap(C,J).
```

*Eagerly folding this new definition into the program gives us our final* $\mathcal{O}(n^3)$ *time algorithm:*

```
552 rtrap(I,K) += word(I,K)
553 rtrap(I,K) += tmp(I,C) * arc(I → C) * rtrap(C,K).
554 ltrap(I',I').
555 ltrap(I',I) += ltrap(C,I) * arc(C ← I') * tmp(C,I').
556 tmp(C,I) += ltrap(I,J) * rtrap(C,J).
557 goal += ltrap(H,0) * rtrap(H,N) * len(N).
```

**Example 45.** *Consider the following left-recursive program for computing the total weight of all paths in a graph.*

```
558 startpath(X) += start(X).
559 startpath(Y) += startpath(X) * edge(X,Y).
560 goal += startpath(X) * stop(X).
```

*If we apply the left-corner transformation to it, we get the following right-recursive program:*

```
561 (goal / start(X)) += goal / startpath(X).
562 (goal / startpath(X)) += edge(X,Y) * goal / startpath(Y).
563 (goal / startpath(X)) += stop(X).
564 goal += start(X) * goal / start(X).
```

*Next, we unfold* `goal / start(X)` *in the bottom rule and abbreviate* `(goal / startpath(X))` $\longmapsto$ `stoppath(X)`*:*

```
565 stoppath(X) += edge(X,Y) * stoppath(Y).
566 stoppath(X) += stop(X).
567 goal += start(X) * stoppath(X).
```

*This program computes the same value for* `goal`*, but it does so using the total weight of paths* starting *at a given node instead of the total weight of all paths* ending *at a given node. Both are well-known methods for computing the total weight of all paths. Here, we have presented a mechanical technique for deriving one from the other. (Note that if we apply the same exercise to this last program using the right-corner transformation, we will arrive at the original program.)*

**Example 46.** *Recall Example 28. That program (repeated below) sums the total weight of all paths in a graph where the paths are written in a very direct style as a list of nodes visited along the path.*

```
568 goal += start(X₁) * β([X₁|Xs]).
569 β([X₁,X₂|Xs]) += edge(X₁,X₂) * β([X₂|Xs]).
570 β([X]) += stop(X).
```

*We apply the right-corner transformation to this program with* $\boldsymbol{\sigma}$ *equal to all the rules. We get the following after useless-rule elimination:*

```
(Z / β([X1|Xs])) += start(X1) * Z / goal.
(Z / β([X2|Xs])) += edge(X1,X2) * Z / β([X1,X2|Xs]).
(Z / stop(X)) += Z / β([X]).
(Z / Z).
goal += stop(X1) * goal / stop(X1).
```

*After specialization:*

```
(goal / path([X1|Xs])) += start(X1) * goal / goal.
(goal / path([X2|Xs])) += edge(X1,X2) * goal / path([X1,X2|Xs]).
(goal / stop(X)) += goal / path([X]).
(goal / goal).
goal += stop(X1) * goal / stop(X1).
```

*When we run this program, the tail of the list* `Xs` *in* `goal/path([X2|Xs])` *is never instantiated. Thus, we can abbreviate* `goal/path([X|Xs])` $\mapsto$ `starpath(X)`*, and we recover the efficient algorithm from Example 27.*

## 7.6 Discussion

**Implementation.** We have implemented and tested these transformations. They are available in our system: `https://github.com/timvieira/dyna-pi`. Using these transformations in the context of an automated optimizer (§8) has stress-tested them since buggy transformations often lead to fast programs (e.g., the empty program), and search probes the possible arguments to the transformations.

**Eisner and Blatz (2007).** The original paper on transforming semiring-weighted logic programs is Eisner and Blatz (2007). They extended the fold/unfold transformations (Tamaki and Sato, 1984), and the magic sets transformations (Ramakrishnan, 1991) from the unweighted setting to the weighted setting and proposed a novel speculation transformation.

↪ **Single vs. Multiple Semirings.** Eisner and Blatz (2007) allowed programs to use more than one semiring to combine values. They supported this case by transforming semiring program fragments; our work should also be compatible with semiring fragments. As we discussed in §4, our semantics are based on sums over derivations, and Eisner and Blatz (2007)'s are based on the weaker notion of fixpoints. The sum-over-derivations semantics is deeply tied to the distributive property of the semiring. Mixing semirings means that property is not always available.[195]

↪ **Bringing Generalized Fold and Unfold to the Weighted Setting.** Eisner and Blatz (2007)'s extensions of the fold and unfold transformations to the weighted setting were limited to *basic* folds. Consequently, their exploration of what fold and unfold can accomplish was severely limited. In particular, they never discovered self-supporting recursive definitions like `startpath` (Example 27). We generalized Eisner and Blatz (2007)'s treatment of fold/unfold to allow fold and unfold with respect to auxiliary programs, such as prior versions of the program in a sequence of transformations. The generalization to auxiliary definitions was previously observed to be powerful in the unweighted setting (e.g., Tamaki and Sato (1984), Roychoudhury et al. (1999), Bossi et al. (1992), and Pettorossi and Proietti (1994)). Examples like Example 27 and Example 28 are impossible with only basic fold. We give examples of transformed programs going from non-terminating to terminating and exponential to polynomial time. Unlike the limited examples presented in Eisner and Blatz (2007) (and Gildea (2011)), which only employed simple, superficial "hook tricks," our approach led to the discovery of self-supporting recursive definitions. However, some additional care must be taken to prevent ill-founded circular definitions. We characterized the ill-founded definition problem in §7.4.2.2 as the nontermination of a derivation mapping. We proposed a practical system for avoiding ill-founded definitions in §7.4.2.4 that builds on Roychoudhury et al. (1999).

[195] A notable exception is when the mixed-semiring program is *stratified*—meaning that we can partition its non-$\underline{0}$ items into strongly connected components (as we did in §5.4) such that each component uses a single semiring.

↪ **Speculation as a Hoisting Transformation.** We deepen the discussion of the speculation transformation by characterizing it as a structural change to the derivations of a program. We also connected it to the better-known left-corner grammar transformations (Rosenkrantz and Lewis, 1970; Johnson, 1998; Johnson and Roark, 2000), which we generalized to Dyna programs (§7.5.2).

↪ **Correctness Proofs.** Eisner and Blatz (2007) never formally proved the correctness of (basic) folding or unfolding; we have filled this gap in the semiring-weighted logic programming literature.

**Contributions to the Fold and Unfold Literature.** We analyze the fold–unfold transformation's ability to preserve the entire set of derivations for each item. Prior work[196] focused on the preservation of *at least one* derivation

[196] E.g., Tamaki and Sato (1984), Maher (1987), Kawamura and Kanamori (1990), Bossi et al. (1992), Roychoudhury et al. (1999), and Pettorossi and Proietti (1994).

for each item, which is insufficient for our semiring-weighted setting.

↪ **Extending Roychoudhury et al. (1999).** Our practical fold/unfold safety test (§7.4.2.4) builds directly on Roychoudhury et al. (1999) for unweighted logic programming. We proved that their technique extends directly to weighted logic programming. However, we sought a different notion of correctness: strong equivalence. As a result, we provide an alternative lens on why their technique is correct. We have also structured our presentation (and proofs) to disentangle the structure of the transformation from the static analysis technique used to check its safety. Additionally, §7.4.2.4 makes their approach's static analysis modeling assumptions more explicit than in other work. One minor contribution is that we have extended their method to support the generalized unfold transformation, which was straightforward but missing in their work. Lastly, as a practical matter, we proposed using constraint solvers to eliminate the need to specify an initial measure, which improves our ability to approve certain transformations that would otherwise be rejected.

**A Notable Omission: The Magic Sets Transformation.** Eisner and Blatz (2007) gave a weighted generalization of the magic sets transformation (Ramakrishnan, 1991), a well-known transformation to the Datalog community as it is a simple method for partially specializing a program to some specific queries. An exciting application of the magic sets transformation is that it can recover the top-down filtering part of Earley's algorithm when applied to CKY (Minnen, 1996; Eisner and Blatz, 2007). Unfortunately, we do not support Eisner and Blatz (2007)'s version of magic sets because the transformed program mixes semirings. Support for mixed semiring programs is interesting future work. However, a different perspective on the magic sets transformation is that it is a recipe for prioritization of agenda-based evaluation (§5.1.3). It is possible to derive tighter heuristics than the simple boolean heuristic given by magic sets. Felzenszwalb and McAllester (2007) and Prieditis (1993) show how to use the outside context of a *relaxed* program to prioritize the fine-grained inside pass. The relaxed programs can be automatically discovered and optimized by searching over which constraints and variables to drop. The values that it computes for prioritization can be more expressive than booleans. Moreover, in the case of superior semirings, we can even automatically derive admissible heuristics for $A^*$-like decoding. In future work, we would like to explore this direction.

# Summary

We presented several program transformations that can be used to optimize a given program. We have many examples where the transformations dramatically improve the running time. In § 8, we will use search to automatically find a transformation sequence that minimizes the program's running time within a search budget.

# Chapter 8

# Searching For Faster Dynamic Programs

This chapter develops a tool for automating the discovery of faster programs. At a high level, our approach is straightforward: given an initial, correct program, we will search for a sequence of meaning-preserving, source-to-source transformations (§7) that minimizes a given cost function, such as the worst-case runtime bounds of §6.3 or the average execution time of the program on a given input workload. We will discuss the benefits and drawbacks of several cost functions. We will describe how we structure our transformation sequence search space. We conclude with experiments demonstrating that our proof-of-concept system is capable of replicating the examples given throughout the thesis as well as many of the missed speed-ups found in the literature (outlined in Example 1).

# 8.1 The Graph Search Problem

We consider an abstract **graph search problem**, $\langle \mathcal{S}, s_0, \mathcal{T}, \textsf{transitions}, \textsf{cost} \rangle$, where $\mathcal{S}$ is a state space, $\textsf{transitions} : \mathcal{S} \to 2^{\mathcal{S}}$ is a transition function, $s_0 \in \mathcal{S}$ is an initial state, $\mathcal{T} \subseteq \mathcal{S}$ is a set of terminal states, and $\textsf{cost} : \mathcal{T} \to \mathbb{R}_{\geq 0}$ is a cost function on the terminal states.[197] The search algorithm will treat the cost and transitions functions as black boxes. The goal of the search problem is to find the (reachable) terminal state in the graph that has the lowest cost, $s^* = \min_{s \in \mathcal{T}} \textsf{cost}(s)$.

Our program optimization problem (depicted in Figure 8.1) can be straightforwardly mapped into this notation. Our states $s \in \mathcal{S}$ are programs.[198] The initial state $s_0$ is the initial program. Each transition is the application of a semantics-preserving program transformation (§ 7). This means every state reachable from $s_0$ in this graph is a semantically equivalent program. In our setting, every state is terminal. In §8.1.1, we will describe the search algorithm that we use to traverse this graph. In §8.1.2, we will discuss the details of how we structured the search space used in our experiments (§8.3). In §8.2, we will discuss the cost function.

[197] In principle, the cost function does not have to be real-valued; it can be any mapping from states to a totally ordered set.

[198] To use the generalized fold/unfold transformations, the state must be augmented with the metadata required to check the measure-based safety conditions of §7.4.2.4.

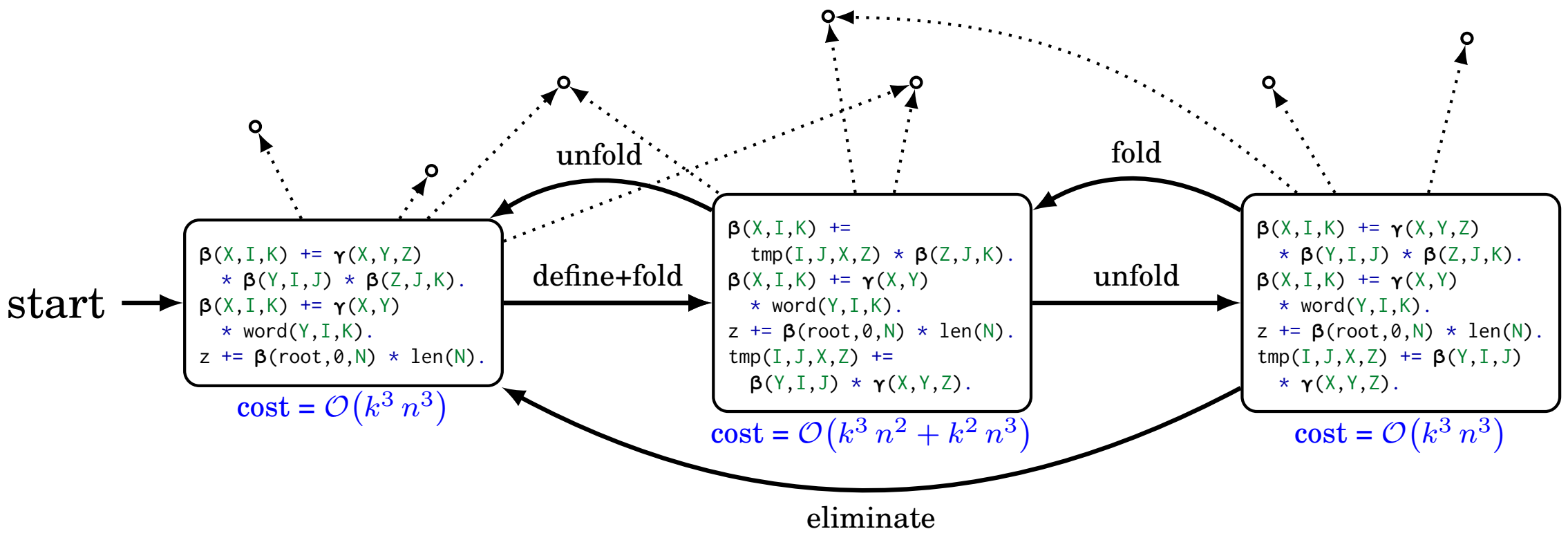

**Figure 8.1:** Depiction of the program optimization graph search problem. The program used in this figure is CKY (Example 6). Nodes are Dyna programs (§ 2). The node pointed to by "start" indicates the user's program. Edges are program transformations (§ 7). Costs are derived by program analysis (§ 6). Only a tiny subset of the nodes and edges in the search graph is shown. The dotted unlabeled outgoing edges represent additional transformations we did not elaborate on in the diagram to reduce clutter.

### 8.1.1 Prioritized Search Algorithm

Our search algorithm is based on best-first search, which is often used to solve difficult combinatorial optimization problems (Russell and Norvig, 2020). Pseudocode is given in Algorithm 30. The basic idea of the best-first search algorithm is to maintain a set of candidate states $\mathcal{Q}$ to explore next and to expand the state that is estimated to be the most promising based on a search policy $\pi$ that selects among the candidates. Once a candidate is selected, its successor states are enumerated according to transitions and added to the candidate set $\mathcal{Q}$ (if they have not previously been added).

Our overall search algorithm $\text{search}_\pi$ is a kind of *anytime* algorithm (Zilberstein, 1996) that traverses the graph until the search budget is exceeded. While $\text{search}_\pi$ traverses the graph, it keeps track of the least cost terminal state it has found so far. If the search budget is reached before the algorithm has explored

**Algorithm 30** search: Algorithm for a given graph search problem $\langle \mathcal{S}, \mathcal{A}, s_0, \mathcal{T}, \text{transitions}, \text{cost} \rangle$. Requires a search policy $\pi$ to select which candidate state to expand. We also require that a search budget is specified, such as an upper limit on the total execution time spent searching.

1. **def** $\text{search}_\pi(s_0)$:
2. $\quad \mathcal{Q} \leftarrow \{s_0\}$ ▷ *Set of candidate states to explore next*
3. $\quad \text{visited} \leftarrow \{s_0\}$ ▷ *Set of states that have already been visited*
4. $\quad$ **while** $|\,\mathcal{Q}\,| > 0$ **and** the search budget has not been exceeded:
5. $\quad\quad s \leftarrow \pi(\mathcal{Q})$ ▷ *Pick any state among the candidates*
6. $\quad\quad \mathcal{Q}.\text{remove}(s)$
7. $\quad\quad$ **if** $s \in \mathcal{T}$:
8. $\quad\quad\quad$ evaluate $\text{cost}(s)$
9. $\quad\quad$ **for** $s' \leftarrow \text{transitions}(s)$:
10. $\quad\quad\quad$ **if** $s' \notin \text{visited}$: ▷ *Add $s'$ as a candidate, if we have not already done so in the past*
11. $\quad\quad\quad\quad \mathcal{Q}.\text{add}(s')$
12. $\quad\quad\quad\quad \text{visited.add}(s')$

$\quad$ **return** the lowest cost state evaluated

the entire graph, it will return the best solution found up until that point. This makes it an *anytime* algorithm because it can provide a solution even if it has not fully explored the search space (i.e., before $|\,\mathcal{Q}\,| = 0$). The search policy $\pi$ governs the order in which the states of the graph are explored. Abstractly, $\pi$ takes the set of candidate states $\mathcal{Q} \subseteq \mathcal{S}$ and returns an element from $\mathcal{Q}$, possibly randomly. It can encode a variety of search orders (e.g., depth-first search, breadth-first search, greedy search, and A*). In practice, we use a priority queue to implement $\mathcal{Q}$ and $\pi$, and we compute the priority of a given state when we add the state to $\mathcal{Q}$ (line 30.11).

This algorithm can guarantee that the best reachable state is eventually found if the search budget is sufficiently large and $\pi$ does not *starve* any candidate state.[199] A simple policy that guarantees this is the breadth-first policy, which enumerates states in first-in-first-out order (FIFO); or, equivalently, by the length of the path by which it was first discovered. This strategy does not have good anytime performance. On the other hand, the greedy strategy of always selecting the cheapest state often has good anytime performance, but it may starve states. Therefore, we use a hybrid strategy that interleaves greedy and FIFO choices to get the benefits of both.

[199]Starving means that the state says on the priority queue indefinitely.

### 8.1.2 Search Space Details

The most basic version of the search graph would say that all valid transformations from §7 are available at all times.[200] However, this would result in an astronomical number of states to explore, and it would ignore some useful problem structure. This section provides the details of how our search space is structured.

**Eureka definitions.** For proposing Eureka definitions (§7.3), our automated system considers variable-elimination actions.[201] To enumerate the available variable-elimination definitions, we consider each rule one at a time. For a given rule R, the variable $v \in \mathsf{vars}(\mathbf{R})$ can be usefully eliminated if it does not appear in *all* of the subgoals in the rule's body and it does not appear in the head of the rule.[202] Thus, to enumerate all variable-elimination definitions, we

[200]In the sections of § 7 describing each transformation, we discussed the conditions for the transformations to be valid.

[201]In principle, other Eureka definitions §7.3 be straightforwardly added within our search framework as additional transformations.

[202]More formally, the set of variables that can be eliminated is $\mathsf{elim}(\mathbf{R}) \overset{\text{def}}{=} \{v \mid \mathsf{factors}(\mathbf{R}, v) \neq \mathsf{body}(\mathbf{R}), v \notin \mathsf{head}(\mathbf{R})\}$ where $\mathsf{factors}(\mathbf{R}, v) \overset{\text{def}}{=} \{y \mid y \in \mathsf{body}(\mathbf{R}), v \in \mathsf{vars}(y)\}$. If any rule $\mathbf{R}$ of the program $\mathfrak{p}$ contains variables that can be eliminated ($|\mathsf{elim}(\mathbf{R})| > 0$), then eliminating any variable $v \in \mathsf{elim}(\mathbf{R})$ by folding $\mathsf{factors}(\mathbf{R}, v)$ out of $\mathbf{R}$ into a new rule reduces that $\mathbf{R}$'s runtime bound, which may reduce (and never increases) the runtime bound of the program. When

identify each variable that can be eliminated and propose a new definition using the recipe detailed in §7.3.

**Unfold.** Recall that the generalized unfold transformation (Definition 35) is parameterized by a source program $\mathfrak{p}^{(s)}$, rule index $r \in \mathfrak{p}^{(s)}$, a subgoal index of $r$, and auxiliary definitions $\mathfrak{p}^{(d)}$. Under this parameterization, it is straightforward to enumerate all instances of unfold: any rule and any non-input subgoal in that rule are candidates for unfolding. The choice of $\mathfrak{p}^{(d)}$ is discussed later. Generalized unfolds must pass the measure-based safety test.

**Fold.** Recall that the generalized fold transformation (Definition 36) is parameterized by a source program $\mathfrak{p}^{(s)}$, auxiliary definitions $\mathfrak{p}^{(d)}$, a new rule $r$, and a set of rules $\mathcal{S} \subseteq \mathfrak{p}^{(s)}$ that will be replaced.

- We generate candidate rules $r$ as follows: for each pair $\langle \mathbf{S}, \mathbf{D} \rangle \in \mathfrak{p}^{(s)} \times \mathfrak{p}^{(d)}$, we attempt to replace subproducts of $\mathbf{S}$'s body by $\mathbf{D}$.[203] From there, we obtain a set of $r$ candidates and filter out those that do not satisfy the reversibility condition of the fold transformation.

- The set $\mathcal{S}$ is determined by $r$ and the position $j$ that we unfold in $r$'s body in the generalized reversibility test. In the generalized reversibility test, we have a generalized unfold transformation. That unfold transformation adds a set of rules, and that set of rules is $\mathcal{S}$.

more than one variable can be eliminated ($|\mathsf{elim}(\mathbf{R})| > 1$), the order in which the variables are eliminated will affect the eventual runtime bound. Unfortunately, finding an optimal sequence of variable-elimination steps is NP-hard by reduction from variable-elimination ordering in probabilistic graphical models (Eisner and Blatz, 2007; Gildea, 2011).

[203] The relevant algorithms for enumerating subproducts are given in §C.2. Once the rule bodies are aligned, we replace the matched subproduct with the head of $\mathbf{D}$. We will subsequently check the validity of this replacement using the reversibility condition.

- Generalized folds must pass the measure-based safety test.

**Auxiliary definitions used for generalized fold/unfold.** For efficiency, we limit the auxiliary definitions considered by generalized folds and unfold transformations to the original definitions of each item.[204] We have found our choice sufficient for our benchmark problems.

**Reducing the number of safety checks.** We found it more efficient overall to check the fold/unfold safety conditions after the transformed program is selected by $\pi$. This results in significantly fewer calls to the constraint solver. This way, we only have to check the safety conditions of the *most promising* transformations rather than *every* transformation. Technically, this means that a state is only terminal (i.e., a member of $\mathcal{T}$) after it passes the safety test. Also, we do not continue transforming states deemed unsafe by the test.

**Hoisting transformations.** Recall the hoisting transformations (i.e., speculation and left-corner) are parameterized by a source program $\mathfrak{p}^{(s)}$ and a subset of rules $\boldsymbol{\sigma} \subseteq \mathfrak{p}^{(s)}$. For speculation, we have an additional parameter $\boldsymbol{z} \in \mathbb{H}$ that describes the type of the item. In the non-commutative case, the choice of left vs. right is also a parameter. In the commutative case, we consider moving any of the subgoals in each of the rules in $\boldsymbol{\sigma}$ to be the leftmost subgoal as an additional parameter. Unfortunately, the hoisting transformations are difficult to search over because there are far too many of them available: the number of valid hoisting transformations scales exponentially with the number of rules—and it is even worse if the program's semiring is commutative! This may be tolerable for a small set of programs, each with only a few rules, but incorporating it into search will result in far too many candidate states. We have included methods for enumerating these transformations in the software library, but they must be used sparingly, or else the search space grows too large.

[204] In principle, any program visited in the entire search space can be used as an auxiliary definition. However, that yields far too many options, and these options do not appear to increase the set of states that are reachable in practice.

**Program specialization.** Recall that the abbreviation transformation (§6.2) can speedup a given program, but may do so by eliminating useless items (i.e., making their values $\underline{0}$). However, our transformation framework (§ 7) assumes all programs are (strongly) equivalent; thus, we may only apply abbreviation as a post-processing step. The cost function, on the other hand, may wish to account for the cost of the abbreviated program. However, running abbreviation often may slow down search.

**Rule to-do list.** The unfold and Eureka generation methods we consider are centered around a specific rule in the current version of the transformed program. Applying transformations to rules $\mathrm{R}, \mathrm{R}' \in \mathfrak{p}^{(s)}$ in either order will get the same result if neither transform makes the other one impossible. Thus, we consider transformations in a canonical order. Each state will now consist of a program together with a **to-do list** of rules that can still be transformed. The possible actions at that state consist of removing the top rule R from the list (declining to transform R) or applying the next transformation. Applying such a transform may delete and/or add program rules, which are correspondingly deleted from the list and/or added at the bottom. (If the list is empty, no more actions are possible.) This design reduces the branching factor by an amount proportional to the number of rules. Of course, a potential downside is that it makes the search tree deeper by a factor proportional to the number of rules.

## 8.2 Cost Functions

There are two broad families of cost functions: empirical and analytical. Our experiments will only use analytical bounds; we will explain our position here.

**Empirical cost functions.** The most direct cost function to consider optimizing is the empirical execution time of running a given program on a specific workload of representative inputs (e.g., running a transformed parser on actual sentences from the Penn Treebank (Marcus et al., 1993)) with a specific solver configuration (e.g., prioritized fixpoint iteration with SCC-based prioritization) on a specific computer (e.g., your laptop).

**Analytical cost functions.**[205] Analytical runtime bounds, such as worst-case runtime bounds, are the method of choice for the theoretical comparison of algorithms in computer science. Indeed it is how we have compared algorithms throughout this dissertation. In our setting, worst-case, analytical running time bounds can be derived by the approach that we gave in §6.3, which automates McAllester (2002)'s well-known meta-theorem.[206] We will additionally provide a cheaper-to-compute proxy later in this section.

[205] Although we do not explore it in this work, we briefly mention that it is possible to derive average-case runtime estimates from statistical models of the input data. Such models are used in database query optimizers (Deshpande et al., 2007; Chekuri and Rajaraman, 1997; Leis et al., 2017).

[206] This theorem is used throughout natural language processing to analyze algorithms, e.g., Gildea (2011), Nederhof and Satta (2011), Gilroy et al. (2017), Melamed (2003), Kuhlmann (2013), Nederhof and Sánchez-Sáez (2011), Büchse et al. (2011), Lopez (2009), and Eisner and Blatz (2007).

**Direct execution can be expensive.** Empirical cost functions can be very expensive to evaluate—especially when the initial program is inefficient. For example, an initial program could have an infinite running time, but its optimized version might be finite. Similarly, evaluating the execution time of an $\mathcal{O}(n^{1000})$ program, unsurprisingly, requires $\mathcal{O}(n^{1000})$ time. To mitigate the issue of some programs running very long on certain inputs, it is common to impose a time limit on the execution of any given program. We also have the challenge that the runtime measurements must be averaged over many runs to estimate the expected running time precisely. Often a deterministic proxy measurement is preferable. In our setting, the observed number of total prefix firings or the program's rules (§6.3) strongly correlates with the execution time of fixpoint iteration algorithms.

**Overfitting the workload.** The direct execution approach requires us to specify particular inputs to the program; this has the downside in that we can *overfit* to these inputs. We may also overfit to specific details of how the program was executed (e.g., hardware-specific details). Optimizing analytical measures, or other proxies for execution time, can lead to *underfitting*.

**Practical efficiency vs. lack of fidelity.** Practical efficiency can easily outweigh the lack of fidelity. Consider evaluating the cost of an $\mathcal{O}(n^{1000})$ program: evaluating the wallclock time takes $\mathcal{O}(n^{1000})$ time, whereas assessing the runtime complexity does not depend on $n$ in this way. Optimizing the analytical runtime can provide a crucial speedup over direct execution, enabling a more exhaustive search in practice. Additionally, it sidesteps the need to specify a specific workload. Of course, the efficacy of comparing analytical bounds is highly dependent on the fidelity of the analysis used to derive the bound: comparing very loose bounds can lead to incorrect conclusions about which algorithm is more efficient.[207] The assumptions behind the bounds are also important, e.g., the tightness of the input-type specification.

**Solver configuration.** We are considering the optimization of programs, not solver configurations. However, we want to highlight that the solver configuration, including the choice of the solver itself, is an important choice when developing an efficient, end-to-end algorithm for a given problem.[208]

- The choice of solver from § 5 is important: prioritized fixpoint iteration and Newton's method are good choices, but they have their pros/cons that depend on the program and its input data. Additionally, each of the solvers has its configuration options. For example, the prioritized fixpoint iteration algo-rithm's execution time will depend on the choice of the prioritization function. The choice of convergence threshold (if applicable) affects the running time for cyclical programs using the SCC-based approach. Newton's method has overhead that fixpoint iteration does not have, but it is quicker at solving cyclical programs.

[207] We caution that comparing upper bounds does not lead to sound conclusions about which system is better. In principle, given $f_{lo} \leq f \leq f_{hi}$ and $g_{lo} \leq g \leq g_{hi}$ we can only correctly conclude that $f$ is definitely smaller than $g$ iff $f_{hi} \leq g_{lo}$. In other words, the correct conclusion requires both an upper and lower bound. Unfortunately, we do not have a method for deriving meaningful lower bounds on running time. Nonetheless, the comparison can be accurate if the runtime bounds are tight and characterize the typical behavior.

[208] Many high-performance computing systems are tuned by optimizing execution time. Some well-known examples are numerical libraries, such as ATLAS (Whaley, Petitet, et al., 2001), PhiPAC (Bilmes, Asanović, et al., 1996), FFTW (Frigo and Johnson, 1998), and SPIRAL (`http://www.spiral.net`), where execution time is used to adapt low-level implementation details to specific hardware. These details include loop {order, tiling, unrolling} and whether to use instruction-level parallelism. One of the primary reasons these libraries are so popular is that performance improvements on one computer do not necessarily generalize to other architectures or all workloads; that is why they tune the implementation for specific hardware. A portable performance research community has formed around tuning software to be as fast as possible on a given computing platform. We also mention the configuration of solvers for computationally hard problems. Hutter (2009) used black-box optimization techniques to tune the parameters of solver heuristics for mixed-integer programming solvers and satisfiability solvers to optimize efficiency for specific problem workloads. This approach won several competitions, demonstrating the effectiveness of tuning solvers to a specific workload.

- For any given program and solver, many low-level details affect the execution time. For example, memory layouts (e.g., row-order or column-order layout of a dense array in memory), sparse vs. dense representations of relations (e.g., hash tables vs. arrays), and indexes on relations (including sorted order) can have a dramatic effect on the running time in practice. However, they will not manifest in the analytical runtime bounds.

Fortunately, improvements to the program are generally orthogonal and complementary to the lower-level solver configurations. Our perspective is that we should get the high-level algorithm right—i.e., find an algorithm with a good big-$\mathcal{O}$ running time—before optimizing the lower-level choices. From there, we can investigate low-level tricks, such as memory layouts, that improve the constant factors and average execution time.

### 8.2.1 Fine-grained Runtime Bounds

In §6.3, we devised an automated method for assessing a given program's worst-case time complexity as a function of user-defined input-size parameters.[209]

[209]The validity of this bound was discussed in §6.3, but we remind the reader that the bound assumes that the program is acyclic for the bound to be valid. Additionally, the bound assumes

**Comparing bounds.** The analytical bounds provided by the method need to be compared. The most obvious approach is to evaluate the runtime expressions on specific values for the input-size parameters. For runtime bounds in a single variable, it is possible to determine the faster of two runtime bounds in the sense of big-$\mathcal{O}$ using a computer algebra system, such as SymPy (Meurer et al., 2017) or Mathematica (Wolfram, 2003).[210] Comparing multi-variable runtime expressions in this way is trickier. One option is for the user to define the input-size parameters as a function of a single parameter $t$ (e.g., $k = \log(t)$, $n = t$), and reduce to the single variable case.

**Efficiency.** The method from §6.3 can be slow for a few reasons:

- For some programs, abstract forward chaining may need to do a lot of work. At one extreme, a program could have no input data; thus, abstract interpretation will essentially solve the relaxed, booleanized version of the program. Thus, similar to concrete execution, abstract forward chaining may also require limits on its execution, as discussed in (§6.1.2.5).

- Overhead: Constraint propagation systems can be slow if the constraint propagation rules are very productive (produce a lot of consequences). Additionally, the method from §6.3 optimizes subgoal-ordering decisions, which can be inefficient for rules with long bodies.[211]

that prioritized fixpoint iteration is used with SCC-based prioritization (§5.1.4).

[210] These computer algebra systems can compare the asymptotic behavior of two functions $f(n)$ and $g(n)$ in the sense of big-$\mathcal{O}$: $f \leq g \iff f = \mathcal{O}(g) \iff \limsup_{n\to\infty} \frac{f(n)}{g(n)} < \infty$. If the runtime bound is a simple polynomial, this is just a comparison of the largest degree terms. However, the runtime bounds of §6.3 may include $\min$ in addition to summation and multiplication, which makes the comparison more detailed; hence we suggest the computer algebra system.

[211] A suboptimal subgoal ordering could be used to speed things up.

These inefficiencies motivate the coarser-grained analytical bound that we describe in §8.2.2.

### 8.2.2 Coarse-grained Runtime Bounds

In this section, we will describe a very fast cost function based on a folk theorem from the Datalog community with a long history of use. The folk theorem says that, under certain conditions (discussed later), the running time of forward chaining execution of a given program is, at worst, linear in the number of ways to instantiate its rules, i.e., bind the variables to constants.[212] A relatively simple bound on rule instantiations is available if we can establish that each variable in the program can be bound in at most $\eta$ different ways. Given some other conditions discussed at the end of this section, the number of ways to instantiate a rule with $k$ variables is bounded by $\mathcal{O}(\eta^k)$. Program $\mathfrak{p}$'s total runtime is $\mathcal{O}(\eta^{\mathsf{degree}(\mathfrak{p})})$ where $\mathsf{degree}(\mathfrak{p})$ is the maximum number of variables in any rule of $\mathfrak{p}$. We can, therefore, use degree as a cost function to minimize during search, and it is very efficient to compute and easy to implement.

As an example, consider CKY. The degree bound says that the program below runs in $\mathcal{O}(\eta^6)$, as the first rule must sum over 6 variables (X, Y, Z, I, J, K).

```
600 β(X,I,K) += γ(X,Y,Z) * β(Y,I,J) * β(Z,J,K).
601 β(X,I,K) += γ(X,Y) * β(Y,I,J).
602 β(X,I,K) += γ(X,Y) * word(Y,I,K).
```

This is a coarse-grained analysis: the runtime is usually given more specifically as $\mathcal{O}(n^3\,k^3)$ where $n$ is the number of sentence positions, and $k$ is the number of grammar symbols. (This finer-grained bound can be achieved using type analysis §6.3: the intuition is that the variables I, J, K can each be bound in $\mathcal{O}(n)$ ways while X, Y, Z can each be bound in $\mathcal{O}(k)$ ways.) However, the simpler analysis $\mathcal{O}(\eta^6)$ gives us a single exponent to reduce, namely 6. Below is a sped-up variant, which sums over the variable Y separately from K:

```
603  β(X,I,K) += tmp(I,J,X,Z) * β(Z,J,K).
604  tmp(I,J,X,Z) += γ(X,Y,Z) * β(Y,I,J).
605  β(X,I,K) += γ(X,Y) * β(Y,I,J).
606  β(X,I,K) += γ(X,Y) * word(Y,I,K).
```

[212]This is not to be confused with the number of non-$\underline{0}$ instantiations of the rule.

This variant is more efficient as its running time is $\mathcal{O}(\eta^5)$. It is also more efficient under the finer-grained analysis, $\mathcal{O}(k^2\, n^3 + k^3\, n^2)$.

For programs with multiple rules, we can refine the degree analysis to measure the degree of each rule in the program rather than summarize it by the maximum degree. Then, we can say that a program is faster if its list of rule degrees—-sorted in descending order—is lexicographically smaller than that of the other program. This has the benefit of reducing lower-order terms in the running time, not just the highest-order term.

**Correctness of the bound.** The degree analysis of a Dyna program only leads to a *valid* $\mathcal{O}$-expression under some conditions, which we will now discuss.

1. The degree bound assumes the *grounded* program to be acyclic. Cycles slow down forward chaining because we must iterate to a fixpoint (§5). Generally, the number of iterations required to reach a fixpoint is data-dependent.

2. The degree bound assumes that all of the relations in the program are finite. This prevents the user from encoding infinite sets, such as the Peano integers.

Although technically, the bound is not *wrong*; it is just that $\eta$ is infinite.

3. The degree bound also assumes constant-time semiring operations.

Recall from §6.3 that the fine-grained runtime bound makes assumptions (1) and (3). However, the fine-grained bound is generally tighter as it can account for certain kinds of sparsity, and its bounds do not assume all variables have the same domain size. The fine-grained system can *check* if (2) is true by type inference, so it is not *assumed*.

**Looseness.** The degree analysis might be loose for many reasons. The upper bounds derived using our methodology assume that relations are dense and that all variables have a single type. We note that the degree analysis becomes tighter after performing program specialization (§6.2). For example, if two variables in the program are always equal, we should not count them as separate, but the degree bound would. Fortunately, specialization can equate these variables, which improves the degree bound.[213] Similarly, if a variable is always bound to a finite set of values (or it is always uninstantiated), it should not count at all in the degree; specialization transforms the program not to have such variables. Despite being a crude approximation, minimizing the degree of a program often works well. We will see in §8.3 that simply optimizing degree is sufficient to recover several asymptotic speedups noted in the NLP literature (see §1).

[213]To speed up search, our experiments only run specialization on the final program.

## 8.3 Experiments

We aim to devise a system for automatically improving typical dynamic programming problems—especially those encountered in natural language processing applications. To evaluate whether we achieved this goal, we devised a set of benchmark programs to see how well our proposed approach works. Our tests include most of the missed speedups that we mention in §1 as well as many of the examples that we described throughout the dissertation; the specific set of programs is provided in §E.

**Experiment details.**

- Results are summaried in Table 8.1.
- The name of each benchmark is hyperlinked to a section in §E that contains further information about the benchmark problem, including the Dyna code of the initial program and any notes about where the program appears in the literature. For each of the benchmark problems, we have manually specified its optimal degree. We assume a commutative semiring in our experiments.
- We show for each benchmark the initial, found, and optimal degrees as well as search time: the time (in seconds) that it took search to find the optimal solution (if one was found). Cells marked with "—" indicate that the optimal-degree solution was not found within the search-time budget of 1 minute.
- A check mark (✔) signifies that the benchmark problem was solved to optimal-

ity with respect to the degree cost function.

- For the search policy $\pi$, we use a hybrid greedy–FIFO strategy (suggested in §8.1.1) where we alternate between greedy and FIFO actions.[214] The greedy policy is prioritized by degree.

**Discussion.**

- These tests are by no means exhaustive, but they show that our proof-of-concept system can automate a variety of speedups. We see that in all but two cases, search is able to find the optimal degree program within a few seconds. We believe that we could significantly improve this already quick time by implementing each of the program transformations more efficiently. For example, in the implementation used for these experiments, there is a lot of unnecessary copying, which can be expensive as programs become larger. Improving the efficiency of the program transformations would improve our ability to search the state space more quickly and exhaustively.

- The benchmarks {path-list, even-odd-peano, explicit-pda} are unlike the others in that each of their initial versions has an infinite-size variable domain (i.e., $\eta = \infty$). However, they can each be optimized so that they have finite domains. Despite the degree bound being unaware of the change in $\eta$, the programs found by search for path-list and even-odd-peano are optimal.

[214] We did not optimize the ratio of greedy and FIFO actions since it could lead to overfitting this small set of benchmark problems. We did find that the purely FIFO policy increased average search time significantly.

| name | initial degree | found degree | optimal degree | search time (sec) |
|---|---|---|---|---|
| arc-eager | 6 | ✔ 3 | 3 | 1.375 |
| bad-chain-05 | 3 | ✔ 2 | 2 | 0.398 |
| bad-chain-10 | 3 | ✔ 2 | 2 | 1.289 |
| bar-hillel | 10 | ✔ 8 | 8 | 0.013 |
| bilexical-labeled | 8 | ✔ 7 | 7 | 0.014 |
| bilexical-unlabeled | 5 | ✔ 4 | 4 | 0.049 |
| chain-05 | 6 | ✔ 2 | 2 | 0.016 |
| chain-10 | 11 | ✔ 2 | 2 | 0.090 |
| chain-expect | 3 | ✔ 3 | 3 | 0.003 |
| cky+grammar | 6 | ✔ 3 | 3 | 1.424 |
| cky3 | 6 | ✔ 5 | 5 | 0.003 |
| cky4 | 8 | ✔ 6 | 6 | 0.012 |
| edit | 6 | ✔ 4 | 4 | 0.008 |
| even-odd-peano | 1 | ✔ 0 | 0 | 5.812 |
| explicit-pda | 8 | 6 | 5 | — |
| hmm | 5 | ✔ 4 | 4 | 0.003 |
| itg | 9 | ✔ 8 | 8 | 0.011 |
| path-list | 3 | ✔ 2 | 2 | 1.164 |
| path-start | 3 | ✔ 2 | 2 | 8.707 |
| semi-markov | 4 | ✔ 3 | 3 | 0.002 |
| split-head-EB | 5 | 4 | 3 | — |
| split-head-J | 4 | ✔ 3 | 3 | 0.780 |
| tensor-decomp-parser | 6 | ✔ 4 | 4 | 1.916 |

**Table 8.1:** Experimental demonstration of our search-based program optimizer on the benchmarks described in §E.

- We also highlight that our system was unable to solve split-head-EB benchmark to optimality. We did manage to speed it up ($\mathcal{O}(\eta^5) \to \mathcal{O}(\eta^4)$). However, we are unable to achieve the optimal $\mathcal{O}(\eta^3)$ algorithm because we do not have hoisting transformations in our search space (discussed in §8.1.2). Unfortunately, we have not yet found an efficient strategy for searching over hoisting transformations. However, we have verified that manually initializing the transformation search with one round of speculation transformations on the initial program is sufficient for our algorithm to recover the optimal degree algorithm. The issue is that making these transformations available in other states causes the search space to grow too much.

- The explicit-pda benchmark appears to be out of reach for our space of program transformations. We describe how to optimize it in §E.12. The missing trick appears to be a generalization of speculation.

# Summary

In this chapter, we developed a tool for automatically discovering faster programs. We do so by searching for a transformation sequence that minimizes a given cost function, which we cast as a kind of graph search. We gave a simple anytime algorithm for performing program optimization on this graph. We discussed the benefits and drawbacks of using various cost functions for our optimization problem. We concluded with an experimental demonstration of how our prototype system can optimize several benchmark problems drawn from the literature on overlooked speedups (Example 1). In all but two of our benchmarks, search finds the fast algorithm in a matter of seconds. However, there is room for improvement. A particular issue we ran into experimentally was the difficulty of searching over hoisting transformations because too many of them were available. We also showed an example that is out of reach for our current space of transformations—suggesting that future work will want to broaden the available transformations. Nonetheless, we are pleased to see that so many of the speedups found in the literature can be rapidly found automatically. Thus, we believe this tool already provides significant value as these previously overlooked speedups and similar ones will no longer be missed when using our tool.

# Chapter 9

# Related and Future Work

## 9.1 Related Work

**Dyna 1.** We have drawn heavily on Eisner, Goldlust, et al. (2005) and Eisner and Blatz (2007). We have discussed these connections extensively throughout this dissertation.

**Dyna 2.** The Dyna 2 language (Eisner and Filardo, 2011) is closely related to the Dyna 1 language that we have discussed in this work, as Dyna 1 programs are contained in the set of Dyna 2 programs. We have provided a short section in §D describing the language and possible extensions of our work.

**Algorithm = Logic + Control.** The decomposition of an algorithm (for parsing or otherwise) into a logical specification and controller is due to Kowalski (1979), who wrote that "Algorithm = Logic + Control." Later, in the context of parsing algorithms, Shieber et al. (1995) wrote:

> “Although the deduction systems do not specify detailed control structure, the control information needed to turn them into full-fledged parsers is uniform, and can therefore be given by a single deduction engine that performs sound and complete bottom-up interpretation of the rules of inference.”

The logical specification of algorithms was further studied in many works, such as McAllester (2002), Ganzinger and McAllester (2002). However, it was truly espoused by the natural language parsing community after the seminal papers of Pereira and Warren (1983) and Shieber et al. (1995). Outside the NLP community, the programming languages community widely uses logical deduction notation to specify operational semantics (Plotkin, 2004) and program analyses (Whaley, Avots, et al., 2005; Ullman, 1988; Dawson et al., 1996; Reps, 1993; Esparza et al., 2010).

**Parsing as Deduction.** Pereira and Warren (1983) wrote a seminal paper to the NLP and parsing community, which cast parsing as logical deduction. Later, Shieber et al. (1995) established deduction systems as a valuable framework for concisely specifying parsing algorithms.

Goodman (1998) modernized the deduction framework to accommodate what was at that time a budding interest in *statistical* parsing. He showed that many quantities of interest, such as the total probability, the most-likely parse, and the set of most-likely parses, could be computed using the same weighted inference algorithm where differences are encapsulated into a choice of semiring weights.[215] This was an important discovery, and it aligns with work in graphical models that was going on around the same time (Aji and McEliece, 2006). At this point, the need for a programming language for specifying *weighted* deductive system was evident. Thus, the Dyna notation emerged (Eisner, Goldlust, et al., 2005; Eisner, Goldlust, et al., 2004).

Deduction systems have been used to express several formalisms:

- Earley parsing (Pereira and Warren, 1983)
- projective dependency parsing (Eisner and Satta, 1999b)
- nonprojective dependency parsing (Cohen, Gómez-Rodríguez, et al., 2011)
- tree adjoining grammars (Alonso et al., 1999)
- combinatory categorical grammars (Kuhlmann and Satta, 2014)
- linear context-free rewriting systems (Burden and Ljunglöf, 2005)
- machine translation models (Melamed et al., 2004; Lopez, 2009)

We also mention Sikkel (1997), Gómez-Rodríguez (2010), and Goodman (1998), which include extensive examples of deduction systems related to syntactic analyses in NLP.

**Deduction System Notation vs. Dyna Notation.** The difference between semiring-weighted deduction systems and Dyna is largely superficial.

[215] We also mention Teitelbaum (1973) who showed that the CKY algorithm could be extended to work in any semiring. Goodman (1998) acknowledges this paper as laying the foundation for most of the work on semiring-weighted deduction.

Consider a Dyna rule of the form:

$$x \mathrel{\oplus}= y_1 \circ \cdots \circ y_K$$

It is rendered in deduction systems notation as

$$\frac{y_1 \quad \cdots \quad y_K}{x}$$

Deduction notation is often used in research papers and books. However, it is not a user-friendly programming language.[216] Additionally, the Dyna notation is more explicit about how item values are combined.[217]

**Deductive Databases and Logic Programming.** The Dyna notation is a natural generalization of Prolog (and Datalog), which allows items to have non-boolean values. Green, Karvounarakis, et al. (2007)—independently from Eisner, Goldlust, et al. (2005)—discovered the beauty and utility of attaching semiring values to Datalog rules. Their work, however, focused on data provenance[218] rather than dynamic programming. We have built on the logic programming literature in many ways: our semantics, execution, and source-to-source transformations have roots in the logic programming literature. These connections are discussed throughout the dissertation.

**Einsum Notation.** Dyna programs are similar to `einsum` notation, which is widely used in scientific computing libraries toolkits, such as `numpy` (Harris

[216]Gómez-Rodríguez et al. (2009) built a system for interpreting unweighted rules written in notation that mimics the deduction systems notation but in plain text.

[217]In Dyna 2 (discussed in §D), item values can be combined in arbitrary ways (i.e., by more than the two semiring operations), which would not be possible in the deduction notation.

[218]Data provenance is a record trail that traces why certain database items exist. In effect, data provenance provides a derivation that an item belongs in the database.

et al., 2020), as a concise notation for tensor products.

**Example 47.** *Below are example pairs of* `einsum` *code (in a* `% comment` *above) and its equivalent Dyna code.*

```
% out_0 = einsum('i->', a)
out_0 += a(I).
% out_1 = einsum('i,i->i', a, b)
out_1(I) += a(I) * b(I).
% out_2 = einsum('i,j->ij', a, b)
out_2(I,J) += a(I) * b(J).
% out_3 = einsum('ii->', c)
out_3 += c(I,I).
% out_4 = einsum('ij,ji->ij', c, d)
out_4(I,J) += c(I,J) * d(J,I).
% out_5 = einsum('ij,kl->ijkl', c, d)
out_5(I,J,K,L) += c(I,J) * d(K,L).
% out_6 = einsum('ij,jk->ik', c, d)
out_6(I,K) += c(I,J) * d(J,K).
```

Dyna notation extends `einsum` in many ways—the most important being that it supports interlocking and recursive equations. We hope that someday Dyna—or some variant thereof—will supplant `einsum` as the default domain-specific language for tensor manipulation.

**Context-Free Grammars.** It is interesting to relate Dyna programs to context-free grammars (CFGs). The mapping is as follows

(i) terminal $\Longrightarrow$ value

(ii) nonterminal $\Longrightarrow$ ground item

(iii) x $\to$ y z $\Longrightarrow$ `x += y * z`

(iv) parse tree $\Longrightarrow$ derivation

(v) the yield of a parse tree $\Longrightarrow$ the value of a derivation

(vi) the language generated (by a nonterminal) $\Longrightarrow$ the total value of all derivations (of that item)

The primary difference is that Dyna does not restrict the set of terminals and nonterminals to finite sets. Anecdotally, exploring the connection between Dyna programs and CFGs is how we discovered the left-corner transformation (§7.5.2) and the Newton solver (§5.2). Lastly, we note that our program transformations may be useful as grammar transformations, and our search method may be effective at reducing grammar size.

## 9.2 Future Work

**Beyond semirings.** In future work, exploring the extensions of the Dyna 1 language would be interesting.

- We discuss extension to Dyna 2 in §D.
- We also mention that a relaxation of semirings that allows nonassociative product operations (e.g., Gildea (2020)) might be interesting to explore. This would allow the item values to be trees rather than strings of values. With some attention to detail, we believe many of our results will likely extend to this more general framework, but we leave that exploration for future work.

**Improving efficiency.** Efficiency can be improved in many ways that we have not explored in this dissertation. We highlight a few below.

- Code generation and compilation are crucial to efficiently running Dyna programs. Eisner, Goldlust, et al. (2005) provided a compiler for Dyna, but it is no longer actively developed. We hope to provide a similar compiler someday.

- There are optimizations for certain semirings that can lead to more efficient execution strategies. For example, superior semirings are amenable to A*-like execution algorithms (Huang, 2008; Felzenszwalb and McAllester, 2007; Knuth, 1977). We want to explore these in future work.

- Prioritization function optimization for use in the prioritized fixpoint iteration algorithm (§5.1.3) could improve our execution times considerably as we will not require an SCC-decomposition pass. It may be possible to synthesize prioritization functions using SMT as in Bagnara et al. (2010) and Podelski and Rybalchenko (2004), which was applied to imperative programs. Synthesizing an effective prioritization heuristic could greatly improve our ability to synthesize efficient code.

- Data structure optimization can dramatically improve our ability to store and retrieve values at the innermost loops of our execution algorithms. The following data structure choices can have a dramatic effect on the execution time in practice: memory layouts (e.g., row-order or column-order layout of a dense array in memory), sparse vs. dense representations of relations (e.g., hash tables vs. arrays), and indexes on relations (including sorted order).[219]

- It would also be interesting to explore the memory-efficient mixed-chaining algorithms of (Filardo and Eisner, 2012; Filardo, 2017).

[219] Vieira, Francis-Landau, et al. (2017) describe a reinforcement-learning technique for dynamically favoring faster data structures during the course of execution. It uses a policy that dynamically adjusts how much data is stored in each of the competing data structures without duplicating data.

**Improving transformation search.** We provided a proof-of-concept search algorithm, which is not very sophisticated. The following are some ideas for how it might be improved:

- Decomposing the set of programs explored into subtasks that may depend on one another enables a kind of dynamic programming that should improve search efficiency because each subtask can be optimized recursively and combined into programs that efficiently compute the required output tasks.

- Vieira, Francis-Landau, et al. (2017) proposed dynamic optimization which adapts the solver to specific workloads *as it executes* using reinforcement learning rather than ahead of time as we do in §8. They also propose decomposition into subtasks to speed up learning.

**Broadening the set of program transformations.** There are still many reusable tricks employed by algorithms designers that have not yet been systematized into program transformations, and other works are exploring the efficient compilation of combinatorial algorithms. One notable recent work is PERPL (Chiang, McDonald, et al., 2023), a probabilistic programming language that can efficiently compile away certain kinds of "linear" data structures, such as stacks and lists. We want to port over their compilation technique as a source-to-source transformation for Dyna programs. It would allow us to recover useful speed-ups relevant to NLP algorithms (e.g., push-down automata and linear indexed grammars). In principle, the efficient compilation strategy used in PERPL can transform the following inefficient program for parsing with a

push-down automaton

```
621 pda([],0).
622 pda([X|Xs],K) += pda(Xs,J) * rewrite(X,Y) * word(Y,J,K).
623 pda([X|Xs],K) += pda([Z,Y|Xs],K) * rewrite(X,Y,Z).
624 goal += length(N) * pda([s],N).
625
626 input: rewrite(_,_,_); rewrite(_,_); length(_); word(_,_,_).
627 output: goal.
```

into an efficient algorithm that is equivalent to CKY.

```
628 p(I,X,K) += rewrite(X,Y) * word(Y,I,K).
629 p(I,X,K) += pda(I,Y,J) * p(J,Z,K) * rewrite(X,Y,Z).
630 goal += length(N) * p(0,s,N).
```

We discuss this example further in §E.12.

# Chapter 10

# Conclusion

We presented Dyna as a notation for dynamic programming (§ 2). We showed in Example 1 and Example 2 that NLP authors have traditionally overlooked speedups. We argued that this demonstrates a need for tools to facilitate the analysis and development of NLP algorithms.

We studied Dyna's foundations: semantics (§ 4) and execution (§ 5). We provided a new sum-over-derivations semantics. We discovered a new execution algorithm, Newton, and we generalized Eisner, Goldlust, et al. (2005)'s forward-chaining algorithm to support non-range-restricted programs. We identified and explored several rich families of source-to-source transformations that systematize the speedup tricks found across many papers for deriving more efficient algorithms, including the generalized fold/unfold, speculation, and left-corner transformations.

We provided a search-based system (§ 8) for finding a sequence of meaning-preserving source-to-source transformations to a given initial program that minimizes its worst-case running time. Our program analyzer (§6) can automate much of the algorithmic analysis found in NLP papers, including deriving precise big-$\mathcal{O}$ bounds on runtime and space complexity. We showed that our system recovers many interesting oversights published in the NLP literature, including oversights in the analysis (Example 2) and overlooked speedups (Example 1).

We have released a freely available prototype implementation `https://github.com/timvieira/dyna-pi`. We have found that the utility of our system is that it aids in discovering the right (i.e., efficient) high-level algorithm. In other words, our tool is most helpful to algorithms developers and the authors of NLP systems in finding the right recurrence—similar to a theorem-proving assistant. Our implementation does not yet perform efficient code generation as Eisner, Goldlust, et al. (2005). The implementation includes several useful development utilities, such as visualization of derivations, that we have not discussed here.

We believe that this dissertation represents a significant step toward automating the development of efficient dynamic programs, especially those found in natural language processing applications.

# Appendix A

# Inference Algorithms

This section shows how our algorithms can solve certain programs with infinitely many non-0 items.

**Example 48.** *Below, we have three versions of the geometric series program. On the left, we have the basic geometric series. In the middle, we have threaded through a useless variable* `I`*. On the right, we add an inequality constraint on* `I`*. Beneath each program variant, we show their respective charts on iterations 1, 2, 3, and the final iteration.*

*Basic:*

$$\mathfrak{p} = \begin{cases} \texttt{x += 1.} \\ \texttt{x += 0.5 * x.} \end{cases}$$

$$\begin{aligned} \mathfrak{m}^{(1)} &= \{\texttt{x += 1}\} \\ \mathfrak{m}^{(2)} &= \{\texttt{x += 1.5}\} \\ \mathfrak{m}^{(3)} &= \{\texttt{x += 1.75}\} \\ &\vdots \\ \mathfrak{m}^{(\star)} &= \{\texttt{x += 2}\} \end{aligned}$$

*Uninstantiated:*

$$\mathfrak{p} = \begin{cases} \texttt{x(I) += 1.} \\ \texttt{x(I) += 0.5 * x(I).} \end{cases}$$

$$\begin{aligned} \mathfrak{m}^{(1)} &= \{\texttt{x(I) += 1}\} \\ \mathfrak{m}^{(2)} &= \{\texttt{x(I) += 1.5}\} \\ \mathfrak{m}^{(3)} &= \{\texttt{x(I) += 1.75}\} \\ &\vdots \\ \mathfrak{m}^{(\star)} &= \{\texttt{x(I) += 2}\} \end{aligned}$$

*Delayed constraint:*

$$\mathfrak{p} = \begin{cases} \texttt{x(I) += (I > 10).} \\ \texttt{x(I) += 0.5 * x(I).} \end{cases}$$

$$\begin{aligned} \mathfrak{m}^{(1)} &= \{\texttt{x(I) += 1.0 * (I > 10)}\} \\ \mathfrak{m}^{(2)} &= \{\texttt{x(I) += 1.5 * (I > 10)}\} \\ \mathfrak{m}^{(3)} &= \{\texttt{x(I) += 1.75 * (I > 10)}\} \\ &\vdots \\ \mathfrak{m}^{(\star)} &= \{\texttt{x(I) += 2 * (I > 10)}\} \end{aligned}$$

**Example 49** (Constraint Accumulation)**.** *Our method does not, however, solve*

$$\mathfrak{p} = \begin{cases} \texttt{x(I) += (I > 0).} \\ \texttt{x(I) += 0.5 * (I > J) * x(J).} \end{cases}$$

*After five iterations, we see that the constraints are just accumulating,*

$$\mathfrak{m}^{(5)} = \begin{cases} \texttt{x(I) += (I > 0).} \\ \texttt{x(I) += 0.5 * (I > J) * (J > 0).} \\ \texttt{x(I) += 0.25 * (I > J) * (J > J}_1\texttt{) * (J}_1\texttt{ > 0).} \\ \texttt{x(I) += 0.125 * (I > J) * (J > J}_1\texttt{) * (J}_1\texttt{ > J}_2\texttt{) * (J}_2\texttt{ > 0).} \\ \texttt{x(I) += 0.0625 * (I > J) * (J > J}_1\texttt{) * (J}_1\texttt{ > J}_2\texttt{) * (J}_2\texttt{ > J}_3\texttt{) * (J}_3\texttt{ > 0).} \end{cases}$$

*To solve this program, we need to simplify the rule bodies. Specifically, we need a method that counts the number of satisfying assignments to the delayed constraints and projects any constraints on the variables in the head. For example, to simplify the bottom rule, we need an efficient mechanism for representing the solution to*

```
644 multiplicity(I) += (I > J) * (J > J₁) * (J₁ > J₂) * (J₂ > J₃) * (J₃ > 0)
```

*Efficiently representing and reasoning about the solutions to the multiplicity problems encountered during fixpoint iteration in a general way is challenging. It is possible that SMT solvers (e.g., Moura and Bjørner (2008)) may be helpful for certain constraint domains like arithmetic inequalities. We think this is an interesting area for future work.*

**Example 50.** *Consider the program for computing the reflexive, transitive closure of a graph in the min-plus semiring:*

```
645 path(I,I) min= 0
646 path(I,K) min= path(I,J) + edge(J,K).
```

*Notice that the top rule is not range-restricted, i.e., it has a free variable in its head. As discussed in §5.1.3, this kind of free variable will cause problems for*

*Eisner, Goldlust, et al. (2005)'s forward chaining algorithm.*

*Below, we show naïve fixpoint iteration on the following input data:*

```
647 edge(a,b) min= 1.
648 edge(b,c) min= 1.
649 edge(c,a) min= 1.
```

*The chart of each iteration is shown below:*

$\mathfrak{m}^{(1)} =$
```
edge(a,b) min= 1.
edge(b,c) min= 1.
edge(c,a) min= 1.
path(I,I) min= 0.
```

$\mathfrak{m}^{(2)} =$
```
edge(a,b) min= 1.
edge(b,c) min= 1.
edge(c,a) min= 1.
path(I,I) min= 0.
path(a,b) min= 1.
path(b,c) min= 1.
path(c,a) min= 1.
```

$\mathfrak{m}^{(3)} =$
```
edge(a,b) min= 1.
edge(b,c) min= 1.
edge(c,a) min= 1.
path(I,I) min= 0.
path(a,b) min= 1.
path(b,c) min= 1.
path(c,a) min= 1.
path(a,c) min= 2.
path(b,a) min= 2.
path(c,b) min= 2.
```

$\mathfrak{m}^{(4)} =$
```
edge(a,b) min= 1.
edge(b,c) min= 1.
edge(c,a) min= 1.
path(I,I) min= 0.
path(a,b) min= 1.
path(b,c) min= 1.
path(c,a) min= 1.
path(a,c) min= 2.
path(b,a) min= 2.
path(c,b) min= 2.
path(a,a) min= 3.
path(b,b) min= 3.
path(c,c) min= 3.
```

*We see that fixpoint iteration converges after 4 iterations. The thing to notice about these charts is that they contain a nonground rule* `path(I,I) min= 0`*. What that means is when we query for the value of* `path(a,a)`*, for example, two rules in* $\mathfrak{m}^{(4)}$ *contribute to its value giving* $\min(0,3) = 0$.

# Appendix B

# Program Transformations

## B.1 Strong Equivalence

**Proposition 5.** *If* $\mathfrak{p}^{(s)} \overset{\text{strong}}{\equiv} \mathfrak{p}^{(t)}$, *then* $\mathfrak{p}^{(s)} \equiv \mathfrak{p}^{(t)}$.

*Proof.* Fix an arbitrary item $\boldsymbol{x} \in \mathbb{H}$.

$$\mathfrak{p}^{(s)}[\![\boldsymbol{x}]\!] = \bigoplus_{\boldsymbol{\delta}^{(s)} \in \mathfrak{p}^{(s)}\{\boldsymbol{x}\}} \mathfrak{p}^{(s)}\left[\!\!\left[\boldsymbol{\delta}^{(s)}\right]\!\!\right] \qquad \text{[Definition 22]} \tag{B.1}$$

$$= \bigoplus_{\boldsymbol{\delta}^{(s)} \in \mathfrak{p}^{(s)}\{\boldsymbol{x}\}} \mathfrak{p}^{(t)}\left[\!\!\left[\underset{s \to t}{\phi}\left(\boldsymbol{\delta}^{(s)}\right)\right]\!\!\right] \qquad \text{[value preserving]} \tag{B.2}$$

$$= \bigoplus_{\boldsymbol{\delta}^{(t)} \in \mathfrak{p}^{(t)}\{\boldsymbol{x}\}} \mathfrak{p}^{(t)}\left[\!\!\left[\boldsymbol{\delta}^{(t)}\right]\!\!\right] \qquad \text{[bijection]} \tag{B.3}$$

$$= \mathfrak{p}^{(t)}[\![\boldsymbol{x}]\!] \qquad \text{[Definition 22]} \tag{B.4}$$

■

## B.2 Correctness of the Fold and Unfold Transformations

This appendix provides several results that characterize the correctness of the $\{\text{basic}, \text{generalized}\} \times \{\text{unfold}, \text{fold}\}$ transformations. We have chosen to give the most detailed correctness proofs for the case of generalized unfold since the proof of correctness for the generalized fold transformation follows the same argument. Furthermore, the correctness proofs for basic unfold and fold transformations (Theorem 4) and Theorem 5) are corollaries of the generalized unfold proof. The correctness proof for the generalized unfold transformation is given by two lemmas, Lemma 1 and Lemma 4, that are assembled into the main theorem: Theorem 2. Lemma 1 provides generalize conditions under which the derivation mapping $\underset{s\to t}{\phi}$ (provided in Algorithm 24) is a value-preserving mapping. Then, Lemma 4 demonstrates that $\underset{t\to s}{\phi}$ (provided in Algorithm 32) is the inverse of $\underset{s\to t}{\phi}$. Analogous results are provided for the generalized fold transformation. We remind the reader that the conditions provided here are not easy to verify (§7.4.2.3). A set of practical sufficient conditions are provided in §7.4.2.4.

**Overview.** An overview of the results is given below.

| | unfold | fold |
|---|---|---|
| Derivation mapping ( $\underset{s\to t}{\phi}$ ) | | |
| ↪ Algorithm | Algorithm 24 | Algorithm 25 |
| ↪ $\underset{s\to t}{\phi}$ is a value-preserving mapping | Lemma 1 | Lemma 5 |
| Inverse derivation mapping ( $\underset{t\to s}{\phi}$ ) | | |
| ↪ Algorithm | Algorithm 31 | Algorithm 32 |
| ↪ $\underset{t\to s}{\phi}$ is the inverse of $\underset{s\to t}{\phi}$ | Lemma 4 | Lemma 6 |
| Semantics-preserving | | |
| ↪ basic | Theorem 4 | Theorem 5 |
| ↪ generalized (impractical) | Theorem 2 | Theorem 3 |
| ↪ generalized (practical) | Theorem 1 | |

**Additional definitions.** Some of our proofs make use of the following ordering relation on derivations:

**Definition 43.** *We define the* ***subtree-ordering relation****. Let* $\delta, \delta' \in \mathfrak{p}\{\cdot\}$ *be derivations (of possibly different items). We say that* $\delta' \underset{\textit{subtree}}{\prec} \delta$ *if and only if* $\delta'$ *is a (strict) subtree of* $\delta$*.*

**Lemma 1.** *Suppose*

A₁ $\mathfrak{p}^{(s)} \xrightarrow{\text{unfold}^{d,j}_{r \to \mathcal{S}}} \mathfrak{p}^{(t)}$

A₂ $\mathfrak{p}^{(s)} \overset{\text{strong}}{\equiv} \mathfrak{p}^{(d)}$

A₃ *the method* $\underset{s \to t}{\phi} : \mathfrak{p}^{(s)}\{\cdot\} \to \mathfrak{p}^{(t)}\{\cdot\}$ *specified in Algorithm 24 always terminates.*

*Then,* $\underset{s \to t}{\phi}$ *satisfies the following,* $\forall \boldsymbol{\delta}^{(s)} \in \mathfrak{p}^{(s)}\{\cdot\}$*:*

C₁ $\underset{s \to t}{\phi}\left(\boldsymbol{\delta}^{(s)}\right)$ *exists*

C₂ $\underset{s \to t}{\phi}\left(\boldsymbol{\delta}^{(s)}\right) \in \mathfrak{p}^{(t)}\{\boldsymbol{x}\}$

C₃ $\left[\!\left[\underset{s \to t}{\phi}\left(\boldsymbol{\delta}^{(s)}\right)\right]\!\right] = \left[\!\left[\boldsymbol{\delta}^{(s)}\right]\!\right]$.

*where* $\boldsymbol{x}$ *is the root label of* $\boldsymbol{\delta}^{(s)}$.

*Proof.* Our correctness argument uses induction on the ordering $\underset{\text{steps}}{\prec}$ over $\mathfrak{p}^{(s)}\{\cdot\}$.

↪ Since $\underset{s \to t}{\phi}$ terminates by A₃ and it is a side-effect-free deterministic procedure, there exists a finite number of steps per call to $\underset{s \to t}{\phi}$. Let $\text{steps}\left\{\underset{s \to t}{\phi}(\boldsymbol{\delta})\right\}$ be the number of steps when $\underset{s \to t}{\phi}$ is run on $\boldsymbol{\delta}$. The ordering relation $\underset{\text{steps}}{\prec}$ is defined as follows: $\boldsymbol{\delta}' \underset{\text{steps}}{\prec} \boldsymbol{\delta} \iff \text{steps}\left\{\underset{s \to t}{\phi}(\boldsymbol{\delta}')\right\} < \text{steps}\left\{\underset{s \to t}{\phi}(\boldsymbol{\delta})\right\}$. We may assume that every function call has a positive overhead cost so that no subroutine call can have zero cost; this ensures that $\underset{\text{steps}}{\prec}$ is irreflexive. Clearly, if $\underset{s \to t}{\phi}(\boldsymbol{\delta})$ recursively calls $\underset{s \to t}{\phi}(\boldsymbol{\delta}')$ (i.e., $\underset{s \to t}{\phi}(\boldsymbol{\delta}')$ is a subroutine of $\underset{s \to t}{\phi}(\boldsymbol{\delta})$), and A₃ holds, then $\boldsymbol{\delta}' \underset{\text{steps}}{\prec} \boldsymbol{\delta}$. Furthermore, $\underset{\text{steps}}{\prec}$ is a well-founded ordering on $\underset{s\to t}{\phi}(\cdot)$ because there are no infinite descending chains of comparison. Thus, $\underset{\text{steps}}{\prec}$ is safe for use in an inductive proof using the principle of Noetherian induction (Cohn, 1969).

Choose $\boldsymbol{\delta}^{(s)} \in \mathfrak{p}^{(s)}\{\cdot\}$ arbitrarily.

Inductive hypothesis:

IH $\forall \boldsymbol{\delta}' \underset{\text{steps}}{\prec} \boldsymbol{\delta}^{(s)} : \underset{s\to t}{\phi}(\boldsymbol{\delta}')$ exists, $\underset{s\to t}{\phi}(\boldsymbol{\delta}') \in \mathfrak{p}^{(t)}\{\boldsymbol{x}\}$, and $[\![\boldsymbol{\delta}']\!] = \left[\!\!\left[\underset{s\to t}{\phi}(\boldsymbol{\delta}')\right]\!\!\right]$.

Base case: **($\boldsymbol{\delta}^{(s)}$ is a constant or builtin)**: This case is straightforward.

C$_1$: $\underset{s\to t}{\phi}\left(\boldsymbol{\delta}^{(s)}\right) \overset{\text{def}}{=} \boldsymbol{\delta}^{(s)}$

C$_2$: $\underset{s\to t}{\phi}\left(\boldsymbol{\delta}^{(s)}\right) \in \mathfrak{p}^{(t)}\{\cdot\}$ because constants and builtins are, by Definition 23, always included in the set of derivations.

C$_3$: $\left[\!\!\left[\underset{s\to t}{\phi}\left(\boldsymbol{\delta}^{(s)}\right)\right]\!\!\right] = \left[\!\!\left[\boldsymbol{\delta}^{(s)}\right]\!\!\right]$ by definition of the valuation map (Definition 22).

Inductive case: There are three cases to consider:

Case 1 **($\boldsymbol{\delta}^{(s)}$ is a product** $\overset{y_1}{\triangle} \circ \cdots \circ \overset{y_K}{\triangle}$): This case is straightforward. Notice $\overset{y_1}{\triangle}, \ldots, \overset{y_K}{\triangle} \underset{\text{steps}}{\prec} \boldsymbol{\delta}^{(s)}$ because they are subroutines.

C$_1$: $\underset{s\to t}{\phi}\left(\overset{y_1}{\triangle} \circ \cdots \circ \overset{y_K}{\triangle}\right) \overset{\text{def}}{=} \underset{s\to t}{\phi}\left(\overset{y_1}{\triangle}\right) \circ \cdots \circ \underset{s\to t}{\phi}\left(\overset{y_K}{\triangle}\right)$

C$_2$:

$$\overset{y_1\circ\cdots\circ y_K}{\triangle} \in \mathfrak{p}^{(s)}\{y_1 \circ \cdots \circ y_K\} \implies \overset{y_1}{\triangle} \in \mathfrak{p}^{(s)}\{y_1\}, \ldots, \overset{y_K}{\triangle} \in \mathfrak{p}^{(s)}\{y_K\}$$

$$\overset{\text{IH}}{\implies} \underset{s\to t}{\phi}\left(\overset{y_1}{\triangle}\right) \in \mathfrak{p}^{(t)}\{y_1\}, \ldots, \underset{s\to t}{\phi}\left(\overset{y_K}{\triangle}\right) \in \mathfrak{p}^{(t)}\{y_K\}$$

$$\implies \underset{s\to t}{\phi}\left(\overset{y_1}{\triangle} \circ \cdots \circ \overset{y_K}{\triangle}\right) \in \mathfrak{p}^{(t)}\{y_1 \circ \cdots \circ y_K\}$$

$$\boxed{\mathsf{C}_3}: \left[\!\left[ \underset{s\to t}{\phi}\left( \overset{y_1}{\triangle} \circ \cdots \circ \overset{y_K}{\triangle} \right) \right]\!\right] = \left[\!\left[ \underset{s\to t}{\phi}\left( \overset{y_1}{\triangle} \right) \circ \cdots \circ \underset{s\to t}{\phi}\left( \overset{y_K}{\triangle} \right) \right]\!\right] \overset{\boxed{\mathsf{IH}}}{=} \left[\!\left[ \overset{y_1}{\triangle} \circ \cdots \circ \overset{y_K}{\triangle} \right]\!\right]$$

**Case 2** **(top rule of $\delta^{(s)}$ is in $\mathfrak{p}^{(t)}$):** This case is straightforward. Observe that $\overset{\mu}{\triangle} \underset{\text{steps}}{\prec} \delta^{(s)}$ because it is a subroutine.

$$\boxed{\mathsf{C}_1}: \underset{s\to t}{\phi}\left( \begin{array}{c} x \\ | \\ \overset{\mu}{\triangle} \end{array} \right) \overset{\text{def}}{=} \begin{array}{c} x \\ | \\ \underset{s\to t}{\phi}\left( \overset{\mu}{\triangle} \right) \end{array}$$

$\boxed{\mathsf{C}_2}$: Since, the top rule of $\delta^{(t)} = \underset{s\to t}{\phi}\left(\delta^{(s)}\right)$ is a specialization of a rule in $\mathfrak{p}^{(t)}$ and its subtrees are in $\mathfrak{p}^{(t)}\{\cdot\}$ by $\boxed{\mathsf{IH}}$, then $\delta^{(t)} \in \mathfrak{p}^{(t)}\{x\}$.

$$\boxed{\mathsf{C}_3}: \left[\!\left[ \underset{s\to t}{\phi}\left( \begin{array}{c} x \\ | \\ \overset{\mu}{\triangle} \end{array} \right) \right]\!\right] = \left[\!\left[ \begin{array}{c} x \\ | \\ \underset{s\to t}{\phi}\left( \overset{\mu}{\triangle} \right) \end{array} \right]\!\right] \overset{\boxed{\mathsf{IH}}}{=} \left[\!\left[ \begin{array}{c} x \\ | \\ \overset{\mu}{\triangle} \end{array} \right]\!\right] = \left[\!\left[ \delta^{(s)} \right]\!\right]$$

**Case 3** **(top rule of $\delta^{(s)}$ is not in $\mathfrak{p}^{(t)}$):** By Definition 35, the derivation $\delta^{(s)}$ must have the form:

$$\delta^{(s)} = \begin{array}{c} x \\ \diagup \quad | \quad \diagdown \\ \overset{\alpha}{\triangle} \quad \overset{y_j}{\triangle} \quad \overset{\beta}{\triangle} \end{array}$$

We now unfold the derivation $\overset{y_j}{\triangle} : \left\langle \overset{\mu}{\triangle}, \mathbf{T}^{(\ell)} \right\rangle = u\left( \overset{y_j}{\triangle} \right)$. This step is required for the (distinct) replacement rule $\mathbf{T}^{(\ell)} \in \mathcal{S}$ to fire in the transformed derivation (below). Observe that $\overset{\alpha}{\triangle}, \overset{\mu}{\triangle}, \overset{\beta}{\triangle} \underset{\text{steps}}{\prec} \delta^{(s)}$ because they are subroutines.

$$\boxed{\mathsf{C}_1}: \underset{s\to t}{\phi}\left( \begin{array}{c} x \\ \diagup \quad | \quad \diagdown \\ \overset{\alpha}{\triangle} \quad \overset{y_j}{\triangle} \quad \overset{\beta}{\triangle} \end{array} \right) \overset{\text{def}}{=} \begin{array}{c} x \\ \diagup \qquad | \qquad \diagdown \\ \underset{s\to t}{\phi}\left( \overset{\alpha}{\triangle} \right) \quad \underset{s\to t}{\phi}\left( \overset{\mu}{\triangle} \right) \quad \underset{s\to t}{\phi}\left( \overset{\beta}{\triangle} \right) \end{array}$$

C$_2$: Lemma 2 guarantees that the remapped tree $\overset{\mu}{\triangle} \in \mathfrak{p}^{(s)}\{\boldsymbol{\mu}\}$. Furthermore, the replacement rule $\mathbf{T}^{(\ell)} \in \mathcal{S}$ is a (distinct) rule in $\mathfrak{p}^{(t)}$ of the form $\boldsymbol{x} \oplus\!= \boldsymbol{\alpha} \circ \boldsymbol{\mu} \circ \boldsymbol{\beta}$. Now, since $\underset{s\to t}{\phi}\left(\overset{\alpha}{\triangle}\right), \underset{s\to t}{\phi}\left(\overset{\mu}{\triangle}\right), \underset{s\to t}{\phi}\left(\overset{\beta}{\triangle}\right) \in \mathfrak{p}^{(t)}\{\boldsymbol{x}\}$ by IH, it follows that $\boldsymbol{\delta}^{(t)} \in \mathfrak{p}^{(t)}\{\boldsymbol{x}\}$.

C$_3$:

$$
\begin{aligned}
&\left[\!\left[ \underset{s\to t}{\phi}\left( \begin{array}{c} \boldsymbol{x} \\ \overset{\boldsymbol{\alpha}}{\triangle} \quad \overset{y_j}{\triangle} \quad \overset{\boldsymbol{\beta}}{\triangle} \end{array} \right) \right]\!\right] \\
&= \left[\!\left[ \underset{s\to t}{\phi}\left(\overset{\alpha}{\triangle}\right) \right]\!\right] \circ \left[\!\left[ \underset{s\to t}{\phi}\left(u\left(\overset{y_j}{\triangle}\right)\right) \right]\!\right] \circ \left[\!\left[ \underset{s\to t}{\phi}\left(\overset{\beta}{\triangle}\right) \right]\!\right] && \text{[by C}_1\text{]} \\
&= \left[\!\left[ \overset{\alpha}{\triangle} \right]\!\right] \circ \left[\!\left[ u\left(\overset{y_j}{\triangle}\right) \right]\!\right] \circ \left[\!\left[ \overset{\beta}{\triangle} \right]\!\right] && \text{[by IH]} \\
&= \left[\!\left[ \overset{\alpha}{\triangle} \right]\!\right] \circ \left[\!\left[ \overset{y_j}{\triangle} \right]\!\right] \circ \left[\!\left[ \overset{\beta}{\triangle} \right]\!\right] && \text{[by Lemma 2]} \\
&= \left[\!\left[ \boldsymbol{\delta}^{(s)} \right]\!\right]
\end{aligned}
$$

■

**Lemma 2.** *Suppose $u$ is called by Algorithm 24 with $\mathfrak{p}^{(s)} \xrightarrow{\mathsf{unfold}^{d,j}_{r \to \mathcal{S}}} \mathfrak{p}^{(t)}$ and*

$$\delta^{(s)} =$$

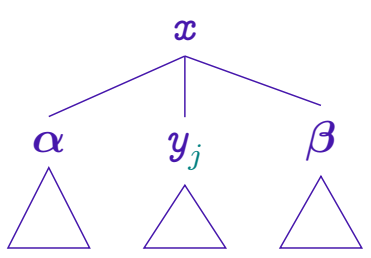

*where the top rule of $\delta^{(s)}$ is the rule to be unfolded $r$. Then, $\left\langle \overset{\mu}{\triangle}, \mathbf{T}^{(\ell)} \right\rangle \leftarrow u\left(\overset{y_j}{\triangle}\right)$ has the following properties:*

- C₁ $\overset{\mu}{\triangle} \in \mathfrak{p}^{(s)}\{\cdot\}$
- C₂ $\left[\!\left[ \overset{\mu}{\triangle} \right]\!\right] = \left[\!\left[ \overset{y_j}{\triangle} \right]\!\right]$
- C₃ $\mathbf{T}^{(\ell)} \in \mathfrak{p}^{(t)}$ *and* $\mathbf{T}^{(\ell)}$ *has the form* $x \mathrel{\oplus}= \alpha \circ \mu \circ \beta$

*Proof sketch.* Conclusions C₁ and C₂ are obvious consequences of $\mathfrak{p}^{(s)} \overset{\text{strong}}{\equiv} \mathfrak{p}^{(d)}$ (assumed by the calling context).

C₃: By the construction of the unfold transformation (Definition 35), we know that $r$ has the form $x \mathrel{\oplus}= \alpha \circ y_j \circ \beta$. By inspecting Definition 35, we can see that the rule $\mathbf{T}^{(\ell)}$ must be of the form $x \mathrel{\oplus}= \alpha \circ \mu \circ \beta$ for some $\mu$. The transformation replaces the item $y_j$ with the various bodies of the defining rules from $\mathfrak{p}^{(d)}$. Furthermore, when we know the definition rule $\mathbf{D}^{(\ell)}$ that was used, we know which rule in $\mathfrak{p}^{(t)}$ replaced $r$. Specifically, $\mathbf{T}^{(\ell)} = \underset{d \to t}{\sigma}(\mathbf{D}^{(\ell)})$ using the one-to-one alignment $\underset{d \to t}{\sigma}$ (described in Definition 35). Any variable bindings have already been propagated when unfold created $\mathbf{T}^{(\ell)}$.

■

**Lemma 3.** *The method $u^{-1}$ correctly implements the inverse of $u$.*

*Proof (Sketch).* Our argument is that $u^{-1}$ inverts $u$ by working backwards from the output of u and inverting each line one-at-a-time.[220]

Suppose the calling context has $\boldsymbol{\delta}^{(s)}$ with top rule $\mathbf{T}^{(\ell)} \in \mathcal{S}$. Thus, $\mathbf{T}^{(\ell)}$ has the form $\boldsymbol{x} \oplus\!= \boldsymbol{\alpha} \circ \boldsymbol{\mu} \circ \boldsymbol{\beta}$ and the unfolded rule $r$ has the form $\boldsymbol{x} \oplus\!= \boldsymbol{\alpha} \circ \boldsymbol{y}_j \circ \boldsymbol{\beta}$.

The only tricky part of inverting $u$ is that we have to reconstruct the intermediate item $\boldsymbol{y}_j$ that u erased on line 24.13. Fortunately, there is just enough information in the transformation to reconstruct it from the identity of $\mathbf{T}^{(\ell)}$. The generalized unfold transformation (Definition 35) is careful to ensure this is possible by tracking the alignment $\underset{d \to t}{\sigma}$.

The way that line 25.15 manages to reconstruct the erased item is as follows: We look up the distinct unfolder rule $\mathbf{D}^{(\ell)} \in \mathfrak{p}^{(d)}$ that was used to create $\mathbf{T}^{(\ell)}$ during the generalized unfold transformation (Definition 35). We have $\mathbf{D}^{(\ell)} = \underset{t \to d}{\sigma}(\mathbf{T}^{(\ell)})$. Furthermore, $\mathbf{T}^{(\ell)}$ must be of the form $\boldsymbol{y}' \oplus\!= \boldsymbol{\mu}'$. Now, an intermediate subtree must exist (that unfold erased on line 24.13), built from an instantiation of the rule $\mathbf{D}^{(\ell)}$. To build this node, we determine the substitution that makes the folder's body $\boldsymbol{\mu}'\,\boldsymbol{\theta} = \boldsymbol{\mu}$. We can determine a unique substitution $\boldsymbol{\theta} = \mathsf{cover}(\boldsymbol{\mu}', \boldsymbol{\mu})$. We can now determine the name $\boldsymbol{y}_j$ of the missing intermediate item $\boldsymbol{y}_j = \boldsymbol{\theta}(\boldsymbol{y}')$.

The remaining operations are correct by the assumption that $\mathfrak{p}^{(s)} \overset{\text{strong}}{\equiv} \mathfrak{p}^{(d)}$. ■

[220]Recall that if $h(x) = f(g(x))$, then $h^{-1}(x) = g^{-1}(f^{-1}(x))$ (assuming these inverses exist).

**Algorithm 31** Inverse derivation mapping for $\mathfrak{p}^{(s)} \xrightarrow{\text{unfold}^{d,j}_{r\to\mathcal{S}}} \mathfrak{p}^{(t)}$.

▷ *The comments at the top of Algorithm 24 apply here as well.*

**def** $\underset{t\to s}{\phi}\left(\delta^{(t)}\right)$: ▷ *Inverse derivation mapping*

  **case** constant or builtin: **return** $\delta^{(t)}$

  **case** $\overset{y_1}{\triangle} \circ \cdots \circ \overset{y_K}{\triangle}$ : **return** $\underset{t\to s}{\phi}\left(\overset{y_1}{\triangle}\right) \circ \cdots \circ \underset{t\to s}{\phi}\left(\overset{y_K}{\triangle}\right)$

  **case** $\left(x \,\big|\, \overset{\mu}{\triangle}\right)$ and its top rule $\in \mathfrak{p}^{(s)}$: **return** $x \,\big|\, \underset{t\to s}{\phi}\left(\overset{\mu}{\triangle}\right)$

  **case** $\left(x \to \overset{\alpha}{\triangle}\ \overset{\mu}{\triangle}\ \overset{\beta}{\triangle}\right)$ and its top rule $\notin \mathfrak{p}^{(s)}$: ▷ *Top rule must be in $\mathcal{S}$*

    **return** $x \to \underset{t\to s}{\phi}\left(\overset{\alpha}{\triangle}\right)\ \ u^{-1}\left(\underset{t\to s}{\phi}\left(\overset{\mu}{\triangle}\right)\right)\ \ \underset{t\to s}{\phi}\left(\overset{\beta}{\triangle}\right)$ ▷ *The top rule of the transformed derivation is $r$*

**Algorithm 32** Inverse derivation mapping for $\mathfrak{p}^{(s)} \xrightarrow{\text{fold}^{d,j}_{\mathcal{S}\to r}} \mathfrak{p}^{(t)}$.

▷ *The comments at the top of Algorithm 24 apply here as well.*

**def** $\underset{t\to s}{\phi}\left(\delta^{(t)}\right)$: ▷ *Inverse derivation mapping*

  **case** constant or builtin: **return** $\delta^{(t)}$

  **case** $\overset{y_1}{\triangle} \circ \cdots \circ \overset{y_K}{\triangle}$ : **return** $\underset{t\to s}{\phi}\left(\overset{y_1}{\triangle}\right) \circ \cdots \circ \underset{t\to s}{\phi}\left(\overset{y_K}{\triangle}\right)$

  **case** $\left(x \,\big|\, \overset{\mu}{\triangle}\right)$ and its top rule $\in \mathfrak{p}^{(s)}$: **return** $x \,\big|\, \underset{t\to s}{\phi}\left(\overset{\mu}{\triangle}\right)$

  **case** $\left(x \to \overset{\alpha}{\triangle}\ \overset{y_j}{\triangle}\ \overset{\beta}{\triangle}\right)$ and its top rule $\notin \mathfrak{p}^{(s)}$: ▷ *Top rule must be $r$*

    **return** $x \to \underset{t\to s}{\phi}\left(\overset{\alpha}{\triangle}\right)\ \ u\left(\underset{t\to s}{\phi}\left(\overset{y_j}{\triangle}\right)\right)\ \ \underset{t\to s}{\phi}\left(\overset{\beta}{\triangle}\right)$ ▷ *The top rule of the transformed derivation is given by the call to $u$*

**Lemma 4.** *Suppose*

A1 $\mathfrak{p}^{(s)} \xrightarrow{\text{unfold}^{d,j}_{r \to \mathcal{S}}} \mathfrak{p}^{(t)}$

A2 $\mathfrak{p}^{(s)} \overset{\text{strong}}{\equiv} \mathfrak{p}^{(d)}$

A3 $\underset{s \to t}{\phi} : \mathfrak{p}^{(s)}\{\cdot\} \to \mathfrak{p}^{(t)}\{\cdot\}$ *is the partial*[221] *function specified in Algorithm 24.*

*Then,* $\forall \boldsymbol{\delta}^{(t)} \in \mathfrak{p}^{(t)}\{\cdot\} : \exists! \, \boldsymbol{\delta}^{(s)} \in \mathfrak{p}^{(s)}\{\cdot\} : \underset{s \to t}{\phi}\left(\boldsymbol{\delta}^{(s)}\right) = \boldsymbol{\delta}^{(t)}$. *Furthermore, the function* $\underset{t \to s}{\phi}$ *specified in Algorithm 31 computes* $\boldsymbol{\delta}^{(s)} = \underset{t \to s}{\phi}\left(\boldsymbol{\delta}^{(t)}\right)$.

*Proof.* Our proof uses induction on $\underset{\text{subtree}}{\prec}$ over $\mathfrak{p}^{(t)}\{\cdot\}$.[222] Fix an arbitrary $\boldsymbol{\delta}^{(t)} \in \mathfrak{p}^{(t)}\{\cdot\}$.

Inductive hypothesis:

IH $\forall \boldsymbol{\delta}_1^{(t)} \underset{\text{subtree}}{\prec} \boldsymbol{\delta}^{(t)} : \exists! \boldsymbol{\delta}_1^{(s)} \in \mathfrak{p}^{(s)}\{\cdot\} : \underset{s \to t}{\phi}(\boldsymbol{\delta}_1^{(s)}) = \boldsymbol{\delta}_1^{(t)}$. Furthermore, $\boldsymbol{\delta}_1^{(s)} = \underset{t \to s}{\phi}\left(\boldsymbol{\delta}_1^{(t)}\right)$.

Base case: ($\boldsymbol{\delta}^{(t)}$ **is a constant or builtin**): This case is straightforward.

We seek to identify all $\boldsymbol{\delta}^{(s)}$ that satisfy the following equation:

$$\boldsymbol{\delta}^{(t)} = \underset{s \to t}{\phi}\left(\boldsymbol{\delta}^{(s)}\right)$$

It is clear from the implementation of $\underset{s \to t}{\phi}$ (see line 24.5) that $\boldsymbol{\delta}^{(t)}$ has this form is if and only if

$$\boldsymbol{\delta}^{(t)} = \boldsymbol{\delta}^{(s)}$$

[221]Here, we take nontermination on a given input to mean that the function is undefined on it.
[222]The fact that we call pull off the induction on an ordering ($\underset{\text{subtree}}{\prec}$) that is guaranteed to be well-founded without additional assumptions is why $\underset{t \to s}{\phi}$ is a total function and why $\underset{s \to t}{\phi}$ is only guaranteed to be a partial function. The induction scheme used in the proof of Lemma 1 required us to stipulate that $\underset{s \to t}{\phi}$ terminates to perform induction in the number of steps performed by Algorithm 24.

This is clearly a unique solution. Furthermore, $\boldsymbol{\delta}^{(s)} = \underset{t \to s}{\phi}\left(\boldsymbol{\delta}^{(t)}\right) = \boldsymbol{\delta}^{(t)}$.

Inductive case: As one might expect, the construction of $\underset{t \to s}{\phi}$ essentially inverts each step of the construction of $\underset{s \to t}{\phi}$. Thus, there are three analogous cases to consider as in Lemma 1:

Case 1 **($\boldsymbol{\delta}^{(t)}$ is a product $\overset{y_1}{\triangle} \circ \cdots \circ \overset{y_K}{\triangle}$)**: This case is straightforward.

We seek to identify all $\boldsymbol{\delta}^{(s)}$ that satisfy the following equation:

$$\boldsymbol{\delta}^{(t)} = \overset{y_1}{\triangle} \circ \cdots \circ \overset{y_K}{\triangle} = \underset{s \to t}{\phi}\left(\boldsymbol{\delta}^{(s)}\right)$$

It is clear from the implementation of $\underset{s \to t}{\phi}$ (see line 24.6) that $\boldsymbol{\delta}^{(t)}$ has this form is if and only if

$$\overset{y_1}{\triangle} \circ \cdots \circ \overset{y_K}{\triangle} = \underset{s \to t}{\phi}\left(\overset{y_1}{\triangle}\right) \circ \cdots \circ \underset{s \to t}{\phi}\left(\overset{y_K}{\triangle}\right)$$

Now, by IH, we have that for each $k \in [1{:}K] : \exists!\, \boldsymbol{\delta}_k^{(s)} : \overset{y_k}{\triangle} = \underset{s \to t}{\phi}\left(\boldsymbol{\delta}_k^{(s)}\right)$; and, furthermore, that $\boldsymbol{\delta}_k^{(s)} = \underset{t \to s}{\phi}\left(\overset{y_k}{\triangle}\right)$. Therefore, the equation $\boldsymbol{\delta}^{(t)} = \underset{s \to t}{\phi}\left(\boldsymbol{\delta}^{(s)}\right)$ has exactly one solution for $\boldsymbol{\delta}^{(s)}$. Furthermore,

$$\boldsymbol{\delta}^{(s)} = \boldsymbol{\delta}_1^{(s)} \circ \cdots \circ \boldsymbol{\delta}_K^{(s)} = \underset{t \to s}{\phi}\left(\overset{y_1}{\triangle}\right) \circ \cdots \circ \underset{t \to s}{\phi}\left(\overset{y_K}{\triangle}\right) = \underset{t \to s}{\phi}\left(\boldsymbol{\delta}^{(t)}\right)$$

Case 2 **(top rule of $\boldsymbol{\delta}^{(t)} \in \mathfrak{p}^{(s)}$)**: This case is straightforward.

We seek to identify all $\boldsymbol{\delta}^{(s)}$ that satisfy the following equation:

$$\boldsymbol{\delta}^{(t)} = \begin{array}{c} x \\ | \\ \overset{\mu}{\triangle} \end{array} = \underset{s \to t}{\phi}\left(\boldsymbol{\delta}^{(s)}\right)$$

It is clear from the implementation of $\underset{s\to t}{\phi}$ (see line 24.7) that $\boldsymbol{\delta}^{(t)}$ has this form is if and only if

$$\begin{array}{c} x \\ | \\ \overset{\mu}{\triangle} \end{array} = \begin{array}{c} x \\ | \\ \underset{s\to t}{\phi}\left(\overset{\mu}{\triangle}\right) \end{array}$$

Now, by IH, we have that $\exists! \boldsymbol{\delta}_1^{(s)} : \overset{\mu}{\triangle} = \underset{s\to t}{\phi}\left(\boldsymbol{\delta}_1^{(s)}\right)$; and, furthermore, that $\boldsymbol{\delta}_1^{(s)} = \underset{t\to s}{\phi}\left(\overset{\mu}{\triangle}\right)$. From line 24.7, it is clear that the top rule must be the same in $\boldsymbol{\delta}^{(s)}$ and $\boldsymbol{\delta}^{(t)}$; therefore, the equation $\boldsymbol{\delta}^{(t)} = \underset{s\to t}{\phi}\left(\boldsymbol{\delta}^{(s)}\right)$ has exactly one solution for $\boldsymbol{\delta}^{(s)}$. Furthermore,

$$\boldsymbol{\delta}^{(s)} = \underset{t\to s}{\phi}\left(\begin{array}{c} x \\ | \\ \overset{\mu}{\triangle} \end{array}\right) \overset{\text{def}}{=} \begin{array}{c} x \\ | \\ \underset{t\to s}{\phi}\left(\overset{\mu}{\triangle}\right) \end{array}$$

Case 3 **(top rule of $\boldsymbol{\delta}^{(t)} \notin \mathfrak{p}^{(s)}$):**

We seek to identify all $\boldsymbol{\delta}^{(s)}$ that satisfy the following equation:

$$\boldsymbol{\delta}^{(t)} = \begin{array}{ccc} & x & \\ \diagup & | & \diagdown \\ \overset{\alpha}{\triangle} & \overset{\mu}{\triangle} & \overset{\beta}{\triangle} \end{array} = \underset{s\to t}{\phi}\left(\boldsymbol{\delta}^{(s)}\right)$$

It is clear from the implementation of $\underset{s\to t}{\phi}$ (see line 24.9) that $\boldsymbol{\delta}^{(t)}$ has this form is if and only if

$$\begin{array}{ccc} & x & \\ \diagup & | & \diagdown \\ \overset{\alpha}{\triangle} & \overset{\mu}{\triangle} & \overset{\beta}{\triangle} \end{array} = \begin{array}{ccc} & x & \\ \diagup & | & \diagdown \\ \underset{s\to t}{\phi}\left(\overset{\alpha}{\triangle}\right) & \underset{s\to t}{\phi}\left(u\left(\overset{y_j}{\triangle}\right)\right) & \underset{s\to t}{\phi}\left(\overset{\beta}{\triangle}\right) \end{array}$$

By $\boxed{\text{IH}}$, we have unique solutions to the following equations:

$$\overset{\boldsymbol{\alpha}}{\triangle} = \underset{s\to t}{\phi}\left(\boldsymbol{\delta}_0^{(s)}\right) \quad ; \quad \overset{\boldsymbol{\mu}}{\triangle} = \underset{s\to t}{\phi}\left(\boldsymbol{\delta}_1^{(s)}\right) \quad ; \quad \overset{\boldsymbol{\beta}}{\triangle} = \underset{s\to t}{\phi}\left(\boldsymbol{\delta}_2^{(s)}\right)$$

Furthermore, those solutions are given by $\underset{t\to s}{\phi}$ :

$$\underset{t\to s}{\phi}\left(\overset{\boldsymbol{\alpha}}{\triangle}\right) = \boldsymbol{\delta}_0^{(s)} \quad ; \quad \underset{t\to s}{\phi}\left(\overset{\boldsymbol{\mu}}{\triangle}\right) = \boldsymbol{\delta}_1^{(s)} \quad ; \quad \underset{t\to s}{\phi}\left(\overset{\boldsymbol{\beta}}{\triangle}\right) = \boldsymbol{\delta}_2^{(s)}$$

Next, we solve $\boldsymbol{\delta}_1^{(s)} = u\left(\boldsymbol{\delta}_3^{(s)}\right)$. According to Lemma 3, this equation has a unique solution that is computed by $\boldsymbol{\delta}_3^{(s)} = u^{-1}\left(\boldsymbol{\delta}_1^{(s)}\right)$ with line 25.7. Lastly, we have that the unique solution $\boldsymbol{\delta}^{(s)}$ is given by

$$\boldsymbol{\delta}^{(s)} = \underset{t\to s}{\phi}\left(\begin{array}{ccc} & \boldsymbol{x} & \\ \overset{\boldsymbol{\alpha}}{\triangle} & \overset{\boldsymbol{\mu}}{\triangle} & \overset{\boldsymbol{\beta}}{\triangle} \end{array}\right) \overset{\text{def}}{=} \begin{array}{ccc} & \boldsymbol{x} & \\ \underset{t\to s}{\phi}\left(\overset{\boldsymbol{\alpha}}{\triangle}\right) & u^{-1}\left(\underset{t\to s}{\phi}\left(\overset{\boldsymbol{\mu}}{\triangle}\right)\right) & \underset{t\to s}{\phi}\left(\overset{\boldsymbol{\beta}}{\triangle}\right) \end{array}$$

■

**Theorem 2.** *Given* $\mathfrak{p}^{(s)} \xrightarrow{\mathsf{unfold}^{d,j}_{r\to\mathcal{S}}} \mathfrak{p}^{(t)}$ *and* $\mathfrak{p}^{(s)} \overset{\text{strong}}{\equiv} \mathfrak{p}^{(d)}$. *Suppose the function* $\underset{s\to t}{\phi}\left(\boldsymbol{\delta}^{(s)}\right)$*, as specified in Algorithm 24, terminates for all* $\boldsymbol{\delta}^{(s)} \in \mathfrak{p}^{(s)}\{\cdot\}$. *Then,* $\underset{s\to t}{\phi}$ *is a value-preserving bijection between* $\mathfrak{p}^{(s)}\{\cdot\}$ *and* $\mathfrak{p}^{(t)}\{\cdot\}$. *Furthermore,* $\mathfrak{p}^{(s)} \overset{\text{strong}}{\equiv} \mathfrak{p}^{(t)}$.

*Proof.* Under the above conditions, it follows from Lemma 1 and Lemma 4 that $\underset{s\to t}{\phi}$ is a value-preserving bijection between $\mathfrak{p}^{(s)}\{\cdot\}$ and $\mathfrak{p}^{(t)}\{\cdot\}$. Such a bijection ensures that $\mathfrak{p}^{(s)} \overset{\text{strong}}{\equiv} \mathfrak{p}^{(t)}$ by Definition 31. ∎

**Lemma 5.** *Suppose*

A₁ $\mathfrak{p}^{(s)} \xrightarrow{\textsf{fold}^{d,j}_{\mathcal{S}\to r}} \mathfrak{p}^{(t)}$

A₂ $\mathfrak{p}^{(s)} \overset{\text{strong}}{\equiv} \mathfrak{p}^{(d)}$

A₃ *the method* $\underset{s\to t}{\phi} : \mathfrak{p}^{(s)}\{\cdot\} \to \mathfrak{p}^{(t)}\{\cdot\}$ *specified in Algorithm 25 always terminates.*

*Then,* $\underset{s\to t}{\phi}$ *satisfies the following,* $\forall \boldsymbol{\delta}^{(s)} \in \mathfrak{p}^{(s)}\{\cdot\}$:

C₁ $\underset{s\to t}{\phi}\left(\boldsymbol{\delta}^{(s)}\right)$ *exists*

C₂ $\underset{s\to t}{\phi}\left(\boldsymbol{\delta}^{(s)}\right) \in \mathfrak{p}^{(t)}\{x\}$

C₃ $\left[\!\!\left[\underset{s\to t}{\phi}\left(\boldsymbol{\delta}^{(s)}\right)\right]\!\!\right] = \left[\!\!\left[\boldsymbol{\delta}^{(s)}\right]\!\!\right]$.

*where* $x$ *is the root label of* $\boldsymbol{\delta}^{(s)}$.

*Proof.* Our correctness argument uses induction on the ordering $\underset{\textsf{steps}}{\prec}$ over $\mathfrak{p}^{(s)}\{\cdot\}$ that we defined in the proof of Lemma 1.

Choose $\boldsymbol{\delta}^{(s)} \in \mathfrak{p}^{(s)}\{\cdot\}$ arbitrarily.

Inductive hypothesis:

IH $\forall \boldsymbol{\delta}' \underset{\textsf{steps}}{\prec} \boldsymbol{\delta}^{(s)} : \underset{s\to t}{\phi}(\boldsymbol{\delta}')$ exists, $\underset{s\to t}{\phi}(\boldsymbol{\delta}') \in \mathfrak{p}^{(t)}\{x\}$, and $[\![\boldsymbol{\delta}']\!] = \left[\!\!\left[\underset{s\to t}{\phi}(\boldsymbol{\delta}')\right]\!\!\right]$.

Base case: (**$\boldsymbol{\delta}^{(s)}$ is a constant or builtin**): This case is exactly the same as in the proof of Lemma 1.

Inductive case: There are three cases to consider:

Case 1 (**$\boldsymbol{\delta}^{(s)}$ is a product** $\overset{y_1}{\triangle} \circ \cdots \circ \overset{y_K}{\triangle}$): This case is exactly the same as in the proof of Lemma 1.

Case 2 **(top rule of $\boldsymbol{\delta}^{(s)}$ is in $\mathfrak{p}^{(t)}$)**: This case is exactly the same as in the proof of Lemma 1.

Case 3 **(top rule of $\boldsymbol{\delta}^{(s)}$ is not in $\mathfrak{p}^{(t)}$)**: The derivation $\boldsymbol{\delta}^{(s)}$ must have the following form with top rule $\mathbf{S}^{(\ell)} \in \mathcal{S}$:

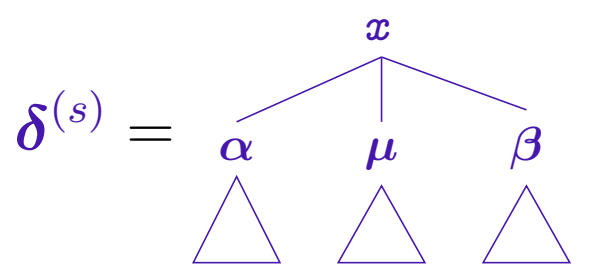

We now fold the derivation $\overset{\mu}{\triangle} : u^{-1}\left(\overset{\mu}{\triangle}, \mathbf{S}^{(\ell)}\right) = \overset{y_j}{\triangle}$. Observe that $\overset{\alpha}{\triangle}, \overset{y_j}{\triangle}, \overset{\beta}{\triangle} \prec_{\text{steps}} \boldsymbol{\delta}^{(s)}$ because they are subroutines.

$C_1$:

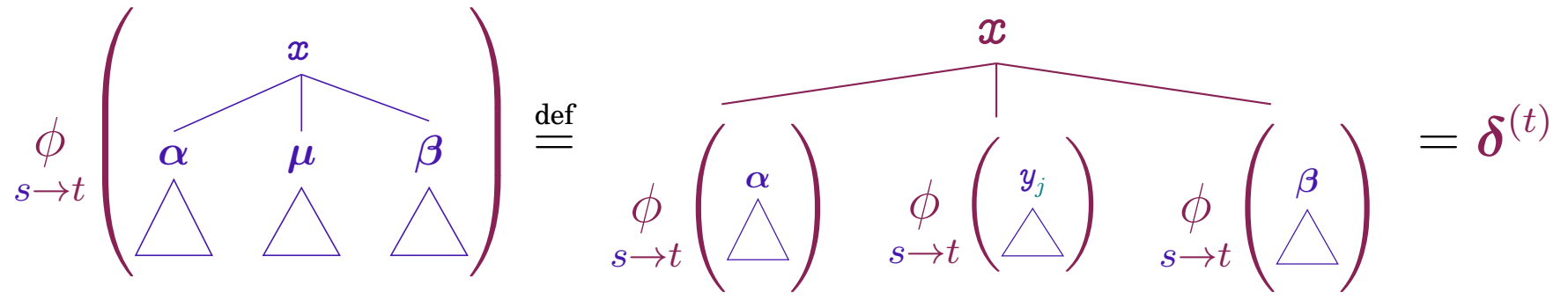

$C_2$: Lemma 3 guarantees that the remapped tree $\overset{y_j}{\triangle} \in \mathfrak{p}^{(s)}\{y_j\}$. Furthermore, the folder rule $r$ is a (distinct) rule in $\mathfrak{p}^{(t)}$ of the form $\boldsymbol{x} \oplus\!= \boldsymbol{\alpha} \circ \boldsymbol{y}_j \circ \boldsymbol{\beta}$. Now, since $\underset{s\to t}{\phi}\left(\overset{\alpha}{\triangle}\right), \underset{s\to t}{\phi}\left(\overset{y_j}{\triangle}\right), \underset{s\to t}{\phi}\left(\overset{\beta}{\triangle}\right) \in \mathfrak{p}^{(t)}\{\boldsymbol{x}\}$ by IH, it follows that $\boldsymbol{\delta}^{(t)} \in \mathfrak{p}^{(t)}\{\boldsymbol{x}\}$.

$\boxed{C_3}$:

$$\left[\!\!\left[ \underset{s\to t}{\phi}\left( \begin{array}{ccc} & \boldsymbol{x} & \\ \overset{\boldsymbol{\alpha}}{\triangle} & \overset{\boldsymbol{\mu}}{\triangle} & \overset{\boldsymbol{\beta}}{\triangle} \end{array} \right) \right]\!\!\right]$$

$$= \left[\!\!\left[ \underset{s\to t}{\phi}\left( \overset{\boldsymbol{\alpha}}{\triangle} \right) \right]\!\!\right] \circ \left[\!\!\left[ \underset{s\to t}{\phi}\left( u^{-1}\left( \overset{\boldsymbol{\mu}}{\triangle} \right) \right) \right]\!\!\right] \circ \left[\!\!\left[ \underset{s\to t}{\phi}\left( \overset{\boldsymbol{\beta}}{\triangle} \right) \right]\!\!\right] \qquad \text{[by } \boxed{C_1}\text{]}$$

$$= \left[\!\!\left[ \overset{\boldsymbol{\alpha}}{\triangle} \right]\!\!\right] \circ \left[\!\!\left[ u^{-1}\left( \overset{\boldsymbol{\mu}}{\triangle} \right) \right]\!\!\right] \circ \left[\!\!\left[ \overset{\boldsymbol{\beta}}{\triangle} \right]\!\!\right] \qquad \text{[by } \boxed{\text{IH}}\text{]}$$

$$= \left[\!\!\left[ \overset{\boldsymbol{\alpha}}{\triangle} \right]\!\!\right] \circ \left[\!\!\left[ \overset{y_j}{\triangle} \right]\!\!\right] \circ \left[\!\!\left[ \overset{\boldsymbol{\beta}}{\triangle} \right]\!\!\right] \qquad \text{[by Lemma 3]}$$

$$= \left[\!\!\left[ \boldsymbol{\delta}^{(s)} \right]\!\!\right]$$

■

**Lemma 6.** *Suppose*

[A₁] $\mathfrak{p}^{(s)} \xrightarrow{\mathsf{fold}^{d,j}_{\mathcal{S}\to r}} \mathfrak{p}^{(t)}$

[A₂] $\mathfrak{p}^{(s)} \overset{\text{strong}}{\equiv} \mathfrak{p}^{(d)}$

[A₃] $\underset{s\to t}{\phi} : \mathfrak{p}^{(s)}\{\cdot\} \to \mathfrak{p}^{(t)}\{\cdot\}$ *is the partial function specified in Algorithm 25.*

*Then,* $\forall \boldsymbol{\delta}^{(t)} \in \mathfrak{p}^{(t)}\{\cdot\} : \exists!\, \boldsymbol{\delta}^{(s)} \in \mathfrak{p}^{(s)}\{\cdot\} : \underset{s\to t}{\phi}\left(\boldsymbol{\delta}^{(s)}\right) = \boldsymbol{\delta}^{(t)}$. *Furthermore, the function* $\underset{t\to s}{\phi}$ *specified in Algorithm 32 computes* $\boldsymbol{\delta}^{(s)} = \underset{t\to s}{\phi}\left(\boldsymbol{\delta}^{(t)}\right)$.

*Proof.* Our proof uses induction on $\underset{\text{subtree}}{\prec}$ over $\mathfrak{p}^{(t)}\{\cdot\}$. All but one of the cases in this proof are identical to Lemma 4. Fix an arbitrary $\boldsymbol{\delta}^{(t)} \in \mathfrak{p}^{(t)}\{\cdot\}$.

Inductive hypothesis:

[IH] $\forall \boldsymbol{\delta}_1^{(t)} \underset{\text{subtree}}{\prec} \boldsymbol{\delta}^{(t)} : \exists!\, \boldsymbol{\delta}_1^{(s)} \in \mathfrak{p}^{(s)}\{\cdot\} : \underset{s\to t}{\phi}(\boldsymbol{\delta}_1^{(s)}) = \boldsymbol{\delta}_1^{(t)}$. Furthermore, $\boldsymbol{\delta}_1^{(s)} = \underset{t\to s}{\phi}\left(\boldsymbol{\delta}_1^{(t)}\right)$.

Base case: **($\boldsymbol{\delta}^{(t)}$ is a constant or builtin)**: Same as the proof of Lemma 4.

Inductive case: As one might expect, the construction of $\underset{t\to s}{\phi}$ essentially inverts each step of the construction of $\underset{s\to t}{\phi}$. Thus, there are three analogous cases to consider as in Lemma 5:

[Case 1] **($\boldsymbol{\delta}^{(t)}$ is a product** $\overset{y_1}{\triangle} \circ \cdots \circ \overset{y_K}{\triangle}$**)**: Same as the proof of Lemma 4.

[Case 2] **(top rule of $\boldsymbol{\delta}^{(t)} \in \mathfrak{p}^{(s)}$)**: Same as the proof of Lemma 4.

[Case 3] **(top rule of $\boldsymbol{\delta}^{(t)} \notin \mathfrak{p}^{(s)}$)**:

We seek to identify all $\boldsymbol{\delta}^{(s)}$ that satisfy the following equation:

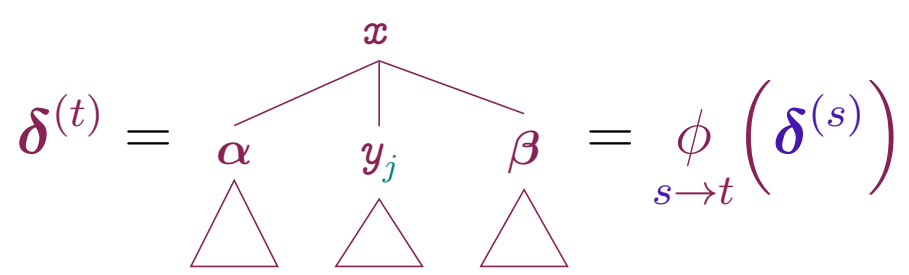

It is clear from the implementation of $\underset{s\to t}{\phi}$ (see line 25.7) that $\boldsymbol{\delta}^{(t)}$ has this form is if and only if

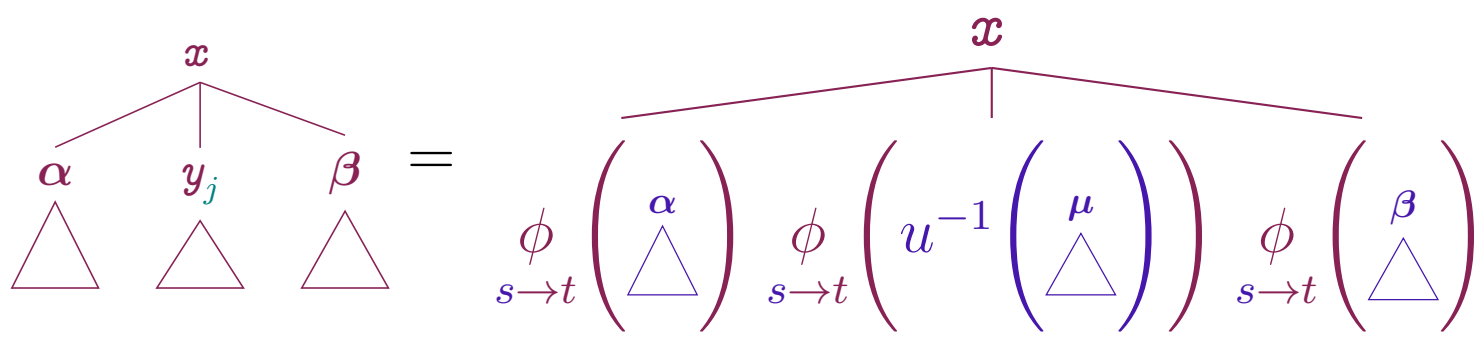

By IH, we have unique solutions to the following equations:

$$\overset{\alpha}{\triangle} = \underset{s\to t}{\phi}\left(\boldsymbol{\delta}_0^{(s)}\right) \quad ; \quad \overset{y_j}{\triangle} = \underset{s\to t}{\phi}\left(\boldsymbol{\delta}_1^{(s)}\right) \quad ; \quad \overset{\beta}{\triangle} = \underset{s\to t}{\phi}\left(\boldsymbol{\delta}_2^{(s)}\right)$$

Furthermore, those solutions are given by $\underset{t\to s}{\phi}$ :

$$\underset{t\to s}{\phi}\left(\overset{\alpha}{\triangle}\right) = \boldsymbol{\delta}_0^{(s)} \quad ; \quad \underset{t\to s}{\phi}\left(\overset{y_j}{\triangle}\right) = \boldsymbol{\delta}_1^{(s)} \quad ; \quad \underset{t\to s}{\phi}\left(\overset{\beta}{\triangle}\right) = \boldsymbol{\delta}_2^{(s)}$$

Next, we solve $\boldsymbol{\delta}_1^{(s)} = u^{-1}\left(\boldsymbol{\delta}_3^{(s)}\right)$. According to Lemma 3, this equation has a unique solution that is computed by $\boldsymbol{\delta}_3^{(s)} = u^{-1}\left(\boldsymbol{\delta}_1^{(s)}\right)$ with line 25.7. Lastly, we have that the unique solution $\boldsymbol{\delta}^{(s)}$ is given by

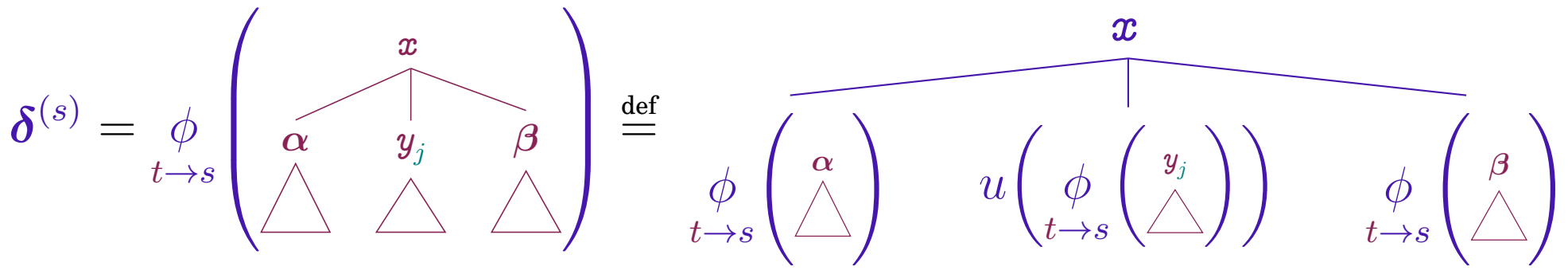

■

**Theorem 3.** *Given* $\mathfrak{p}^{(s)} \xrightarrow{\text{fold}\, \mathcal{S}^{d,j}_{\to r}} \mathfrak{p}^{(t)}$ *and* $\mathfrak{p}^{(s)} \overset{\text{strong}}{\equiv} \mathfrak{p}^{(d)}$. *Suppose the function* $\underset{s\to t}{\phi}\left(\boldsymbol{\delta}^{(s)}\right)$, *as specified in Algorithm 25, terminates for all* $\boldsymbol{\delta}^{(s)} \in \mathfrak{p}^{(s)}\{\cdot\}$. *Then,* $\underset{s\to t}{\phi}$ *is a value-preserving bijection between* $\mathfrak{p}^{(s)}\{\cdot\}$ *and* $\mathfrak{p}^{(t)}\{\cdot\}$. *Furthermore,* $\mathfrak{p}^{(s)} \overset{\text{strong}}{\equiv} \mathfrak{p}^{(t)}$.

*Proof.* Under the above conditions, it follows from Lemma 5 and Lemma 6 that $\underset{s\to t}{\phi}$ is a value-preserving bijection between $\mathfrak{p}^{(s)}\{\cdot\}$ and $\mathfrak{p}^{(t)}\{\cdot\}$. Such a bijection ensures that $\mathfrak{p}^{(s)} \overset{\text{strong}}{\equiv} \mathfrak{p}^{(t)}$ by Definition 31. ■

**Theorem 4.** *Suppose* $\mathfrak{p}^{(s)} \xrightarrow{\textsf{unfold}} \mathfrak{p}^{(t)}$ *according to Definition 33, then* $\mathfrak{p}^{(s)} \overset{\text{strong}}{\equiv} \mathfrak{p}^{(t)}$.

*Proof.* Any basic unfold transformation is an instance of a generalized unfold transformation with $\mathfrak{p}^{(d)} = \mathfrak{p}^{(s)}$. Thus, it inherits the correctness proof for the generalized unfold transformation (Theorem 2). We can strengthen the claim: *all* basic unfold transformations are semantics-preserving because we can drop Theorem 2's stipulation that the derivation mapping (Algorithm 24) terminates as it is always true for basic unfold transformations. To see why, we observe that the call to $u$ always returns a (strict) subtree of the $\boldsymbol{\delta}^{(s)}$, and subtrees are a well-founded order. Thus, $\underset{s \to t}{\phi}$ is guaranteed to terminate for any basic unfold transformation. ■

**Theorem 5.** *Suppose* $\mathfrak{p}^{(s)} \xrightarrow{\textsf{fold}} \mathfrak{p}^{(t)}$ *according to Definition 34, then* $\mathfrak{p}^{(s)} \overset{\text{strong}}{\equiv} \mathfrak{p}^{(t)}$.

*Proof.* Theorem 4 establishes a symmetrical result, which we can leverage here. Definition 34 requires reversibility: $\mathfrak{p}^{(t)} \xrightarrow{\textsf{unfold}^{t,j}_{r \to \mathcal{S}}} \mathfrak{p}^{(s)}$. Since $\mathfrak{p}^{(t)} \xrightarrow{\textsf{unfold}^{t,j}_{r \to \mathcal{S}}} \mathfrak{p}^{(s)}$ implies $\mathfrak{p}^{(t)} \overset{\text{strong}}{\equiv} \mathfrak{p}^{(s)}$ (by Theorem 4) and semantic equivalence is symmetric, it follows that $\textsf{fold}^{s,j}_{\mathcal{S} \to r}$ is also semantics-preserving because *any* transformation from $\mathfrak{p}^{(s)}$ to $\mathfrak{p}^{(t)}$ (or vice versa) would be semantics-preserving under the condition that a strongly equivalent transformation relates $\mathfrak{p}^{(t)}$ and $\mathfrak{p}^{(s)}$. ■

## B.3 Speculation

We provide a derivation mapping as an (incomplete) justification for the correctness of the speculation transformation (Definition 41). To complete the proof of correctness, we would need to prove that this derivation mapping is a bijective value-preserving mapping (as we did in our proofs of correctness for fold and unfold in §B.2). This should be a fairly straightforward argument to make by structural induction (on the subtree relation); however, we do not provide such an argument.

**Algorithm 33** Derivation mapping for $\mathfrak{p}^{(s)} \xrightarrow{\text{speculation}_{z,\sigma}} \mathfrak{p}^{(t)}$.

**def** $\underset{s\to t}{\phi}\left(\boldsymbol{\delta}^{(s)}\right)$:

  **case** $\boldsymbol{\delta}^{(s)}$ is a builtin or constant: **return** $\boldsymbol{\delta}^{(s)}$

  **case** $\overset{y_1}{\triangle} \circ \cdots \circ \overset{y_K}{\triangle}$: **return** $\underset{s\to t}{\phi}\left(\overset{y_1}{\triangle}\right) \circ \cdots \circ \underset{s\to t}{\phi}\left(\overset{y_K}{\triangle}\right)$

  **case** $\left(\begin{array}{c} x \\ | \\ \overset{\mu}{\triangle} \end{array}\right)$ **and** the top rule $\notin \boldsymbol{\sigma}$: ▷ *Bottom of the spine*

    **return** 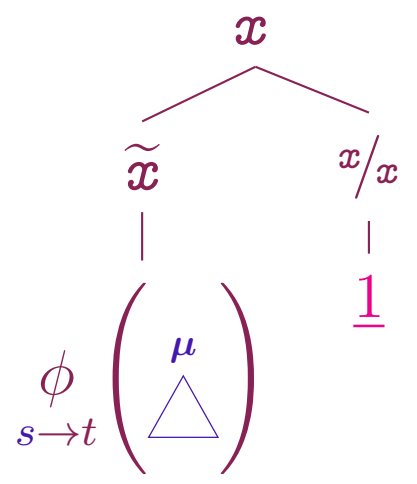 **if** $x \subseteq z$ **else** 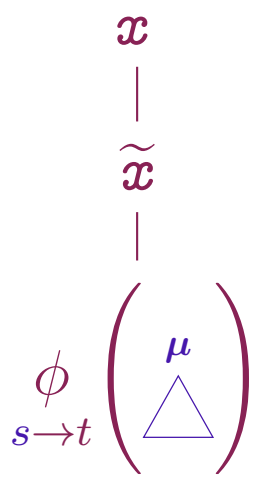

  **case** 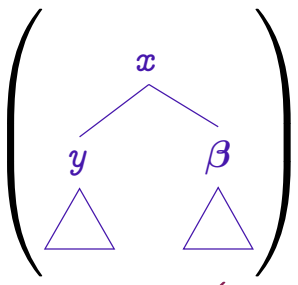 **and** top rule $\in \boldsymbol{\sigma}$:

    **match** $\underset{s\to t}{\phi}\left(\overset{y}{\triangle}\right)$:

      **case** 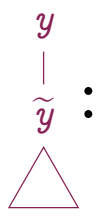 : **return** 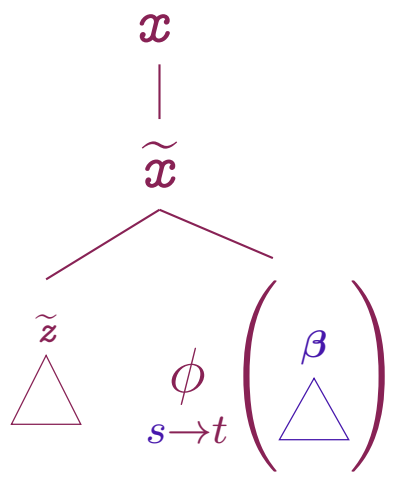

      **case** 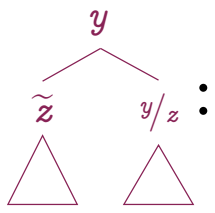 : **return** 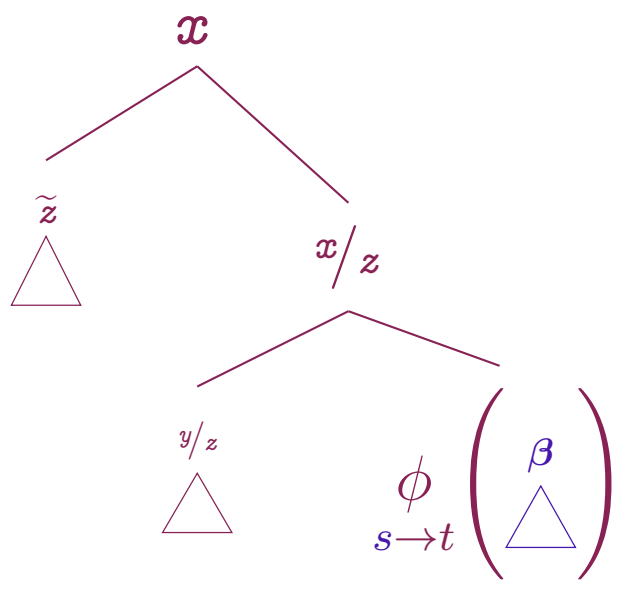

# Appendix C

# Pattern Matching Algorithms

## C.1 Unification

In this section, we describe algorithms for unification, which we briefly described in §3.1.1.[223] Much has been written about unification (Herbrand, 1930; Robinson, 1965; Knight, 1989; Norvig, 1991; Martelli and Montanari, 1982) so we will not talk too much about the algorithm—we have primarily included it for completeness. Our implementation of explicit unification `unify` is given in Algorithm 34; it closely follows the pseudocode from the textbook of Russell and Norvig (2020). Overall, the algorithm is a relatively straightforward recursive procedure that simultaneously traverses a pair of terms from the top down, acquiring the minimal set of variable bindings needed to make the terms equal (or failing if equality is impossible). Algorithm 35 provides a "one-sided" unification algorithm (cover) which solves unification problems of the form $\boldsymbol{\theta}(x) = y$. This algorithm can be used to determine if the set of terms denoted by $x$ is a superset of those denoted by $y$: $(x \supseteq y) \iff (\text{cover}(x, y) \neq \text{FAIL})$. In the pseudocode, $\text{deref}^*(x, \boldsymbol{\theta})$ recursively applies $\boldsymbol{\theta}(x)$ until $x \notin \boldsymbol{\theta}$.

[223] The implementation in this section uses explicit substitution maps, whereas in §3.1.2.1 the map is represented implicitly using a mutable data structure.

**Algorithm 34** unify: Unification algorithm. Solves any equation of the form $\boldsymbol{\theta}(x) = \boldsymbol{\theta}(y)$ for $\boldsymbol{\theta}$ given possibly nonground terms $x$ and $y$.

```
def unify(x, y, θ = {}):
  if θ = FAIL: return FAIL
  else if x = y: return θ
  else if x is a variable:
    if x ∈ θ: return unify(θ[x], y, θ)
    else if y ∈ θ: return unify(x, θ[y], θ)
    else if x occurs in y after substitution θ: return FAIL
    else
      θ[x] ← y                                    ▷ Mutates θ
      return θ
  else if y is a variable: return unify(y, x, θ)
  else if x and y are terms with the same name and arity:
    for i ∈ 1 … |x|: θ ← unify(x_i, y_i, θ)
    return θ
  else
    return FAIL
```

**Algorithm 35** cover: One-sided unification algorithm. Solves any equation of the form $\boldsymbol{\theta}(x) = y$ for $\boldsymbol{\theta}$ given possibly nonground terms $x$ and $y$.

```
def cover(x, y, θ = {}):
  x ← deref*(x, θ); y ← deref*(y, θ)
  if x = y: return θ
  else if x is a variable:
    if x ∈ θ:
      if θ[x] = y: return θ
    else if x does not occur in θ(y):
      θ[x] ← y                                    ▷ Mutates θ
      return θ
    return FAIL
  else if x and y are terms with the same name and arity:
    for i ∈ 1 … |x|: θ ← cover(x_i, y_i, θ)
  return θ
```

# C.2 Sub-Product Enumeration

## C.2.1 Commutative Sub-Products

Algorithm 36 is a procedure to enumerate alignments between a collection of subgoals $x_1 \circ \cdots \circ x_N$ to another collection of subgoals $y_1 \circ \cdots \circ y_M$ assuming a $\circ$ is commutative. The main usage is during folding where $x_1 \circ \cdots \circ x_N$ is the right-hand side of a folder rule and $y_1 \circ \cdots \circ y_M$ is the right-hand side of a rule that is folded.

For example,

- covers_product(`f(X) * f(Y)`, `f(A) * f(B) * f(C)`) has six alignments:

$$\begin{gathered}
\langle\{\mathtt{X} \mapsto \mathtt{A}, \mathtt{Y} \mapsto \mathtt{B}\}, \{0 \mapsto 0, 1 \mapsto 1\}\rangle \\
\langle\{\mathtt{X} \mapsto \mathtt{A}, \mathtt{Y} \mapsto \mathtt{C}\}, \{0 \mapsto 0, 1 \mapsto 2\}\rangle \\
\langle\{\mathtt{X} \mapsto \mathtt{B}, \mathtt{Y} \mapsto \mathtt{A}\}, \{0 \mapsto 1, 1 \mapsto 0\}\rangle \\
\langle\{\mathtt{X} \mapsto \mathtt{B}, \mathtt{Y} \mapsto \mathtt{C}\}, \{0 \mapsto 1, 1 \mapsto 2\}\rangle \\
\langle\{\mathtt{X} \mapsto \mathtt{C}, \mathtt{Y} \mapsto \mathtt{A}\}, \{0 \mapsto 2, 1 \mapsto 0\}\rangle \\
\langle\{\mathtt{X} \mapsto \mathtt{C}, \mathtt{Y} \mapsto \mathtt{B}\}, \{0 \mapsto 2, 1 \mapsto 1\}\rangle
\end{gathered}$$

- covers_product(`f(X,Y), g(Y,Z)`, `f(I,J), g(J,I)`) has one alignment:

$$\langle\{\mathtt{X} \mapsto \mathtt{I}, \mathtt{Y} \mapsto \mathtt{J}, \mathtt{Z} \mapsto \mathtt{K}\}, \{0 \mapsto 0, 1 \mapsto 1\}\rangle$$

Each iterate is a pair of a substitution map $\boldsymbol{\theta}$ and index-to-index alignment $\boldsymbol{\sigma}$. Substitutions maps $\boldsymbol{\theta}$ and alignment $\boldsymbol{\sigma}$ are such that

1. Each alignment $\boldsymbol{\sigma}$ is a function $[1{:}N] \to [1{:}M]$; in other words, from the indices of $x$ to the indices of $y$.

2. $\boldsymbol{\theta}(x_i) = y_{\boldsymbol{\sigma}(i)}$ for $i \in [1{:}N]$; like in cover, $\boldsymbol{\theta}$ may only bind variables in $x$.

**Algorithm 36** covers_product: Lazily enumerate commutative matches from a collection of subgoals $x_1 \circ \cdots \circ x_N$ to another collection of subgoals $y_1 \circ \cdots \circ y_M$. Each iterate is a pair of a substitution mapping and an index-to-index alignment.

**def** covers_product($x_1 \circ \cdots \circ x_N$, $y_1 \circ \cdots \circ y_M$, $\boldsymbol{\theta}_0 = \{\}$):
  **def** $f(i, \boldsymbol{\beta}, \boldsymbol{\theta})$:
    **if** $\boldsymbol{\theta} = \text{FAIL}$: **return** ▷ *Failure*
    **else if** $|x_{[i:]}| > |y_{\boldsymbol{\beta}}|$: **return** ▷ *Failure: too many subgoals*
    **else if** $|x_{[i:]}| = 0$: **yield** $\langle \boldsymbol{\theta}, \{\} \rangle$ ▷ *Success: reached the end of x's*
    **else**
      **for** $j \in \boldsymbol{\beta}$:
        **for** $\langle \boldsymbol{\theta}', \boldsymbol{\sigma}' \rangle \in f(i+1, \boldsymbol{\beta} \setminus \{j\}$, cover($x_i$, $y_j$, copy($\boldsymbol{\theta}$))):
          **yield** $\langle \boldsymbol{\theta}', \boldsymbol{\sigma}' \cup (i \to j) \rangle$
  **yield from** $f(1, \{1, \ldots, M\}, \boldsymbol{\theta}_0)$

## C.2.2 Noncommutative Sub-Products

Algorithm 37 is a procedure to enumerate subproducts; it will find all the subproducts $x_1 \circ \cdots \circ x_N$ contained in $y_1 \circ \cdots \circ y_M$. Unlike the commutative case, no reordering or gaps in the match are allowed.

**Algorithm 37** covers_product: Lazily enumerate non-commutative matches from a collection of subgoals $x_1 \circ \cdots \circ x_N$ to another collection of subgoals $y_1 \circ \cdots \circ y_M$. Each iterate is a pair of a substitution mapping and the start position of the match.

**def** covers_product($x_1 \circ \cdots \circ x_N$, $y_1 \circ \cdots \circ y_M$, $\boldsymbol{\theta} = \{\}$):
  **for** $j \in [1{:}M-N]$:
    $\boldsymbol{\theta}' \leftarrow$ cover($x$, $y_{[j+i:j+N]}$, copy($\boldsymbol{\theta}$))
    **if** $\boldsymbol{\theta}' \neq \text{FAIL}$: **yield** $\langle \boldsymbol{\theta}', j \rangle$

# Appendix D

# Dyna 2

We give a brief overview of the Dyna 2 language and thoughts about how to extend this dissertation's results to it.

**Dyna 2** (Eisner and Filardo, 2011) extends Dyna 1 (cf. § 2) by allowing values to be combined in more than the two ways that a single semiring allows. Thus, Dyna 2 program can be used as a concise notation for specifying general computation graphs[224] via rules that are still written in a Prolog-like notation (as with Dyna 1). Each rule continues to be a template whose *variables* may be instantiated with arbitrary terms. The rule instantiations collectively define the *contributions* to the *values* of many *items*. The contributions are *aggregated* as in Dyna 1 with an *aggregation operator* (*aggregator*, for short). Dyna programs support general mathematical expressions (not just semiring operations). A feed-forward neural network and its training objective can be defined via

[224]Other notations and libraries for defining computation graphs include Theano (Theano Development Team, 2016), PyTorch (Paszke et al., 2017), TensorFlow (Abadi et al., 2016), JAX (Bradbury et al., 2018), and Named Tensor Notation (Chiang, Rush, et al., 2021).

```
684 σ(X) = 1/(1+exp(-X)).         % define sigmoid function
685 out(J) = σ(in(J)).            % apply sigmoid function
686 in(J) += out(I) * edge(I,J).     % vector-matrix product
687 loss += (out(J) - target(J))².  % L2 loss}
```

The structure and weights of the network are specified by defining the values of items of the form `edge(I,J)`. This could be done by listing one rule per edge, but it can also be done systematically by writing edge-defining rules in terms of structured node names, where these names will instantiate `I`, `J` above. For example,

```
688 edge(input(X,Y),hidden(X+DX,Y+DY)) = weight(DX,DY).
689 weight(DX,DY) := random(*,-1,1) for DX:-4..4, DY:-4..4.
```

defines a convolutional layer with initially random weights[225] and a $9 \times 9$ convolution filter. For example, the node named `hidden(10,10)` has a connection from the node named `input(8,11)` with weight `weight(-2,1)`. The input to the network is specified at runtime by updating the values of `out(input(...))`; these items otherwise have no value as the input nodes have no incoming edges.

The Dyna solver seeks a *fixed point* in which all items are consistent, i.e., each item's value matches its definition from other values. A cyclic program may not have a unique fixed point. The solver is permitted to choose arbitrarily if there is more than one. In general, the solver may fail to terminate because no fixed point exists or because it cannot discover one. Sometimes the solver terminates only at numerical convergence, as for the geometric series sum

```
690 a += 1.
691 a += a/2.
```

Or this program that implements the Babylonian method for numerically

[225]Each expression `random(*,-1,1)` names a distinct random variate, since the special argument `*` generates a symbol ("gensym") that is different in each instantiation of the rule. Thus, the last line defines 81 distinct weights.

approximating square root: $y = \sqrt{x}$:

```
692 y := 1.              % arbitrary initial value ≠ 0
693 y := 1/2 * (x/y + y).
694 x := 2.
```

Here the := aggregator is the "last rule wins" aggregator. The top rule specifies the initialization for fixpoint iteration.

Eisner and Filardo (2011) give a more complete introduction to Dyna 2 with more complex examples, including message passing in variational inference and graphical models, constraint propagation in backtracking search, backpropagation, and Markov chain Monte Carlo.

# Extending this Dissertation to Dyna 2

**Execution.** The naive, semi-naive, and prioritized fixpoint iterations algorithms in § 5 have analogs in the Dyna 2 setting. They have been studied in Filardo (2017) and to some extent in Eisner and Blatz (2007).[226] In settings where there is a semiring subprogram, any of our execution algorithms could be used to solve it. That said, the Dyna 2 framework appears to be moving in the direction of term-rewriting systems as the basis of its execution system (Francis-Landau et al., 2020).

**Type Analysis.** Our type analyzer §6.1 can be extended to Dyna 2. The main difference is that we would need to prove a type for the *value* that an item may have since it is no longer homogeneous. In other words, we would reason about possible value–item *pairs*. In principle, this should not be a difficult extension. Our approach to space and time complexity analysis method (§6.3) should also work in the case of Dyna 2, as the complexity of executing a Dyna 2 program is also proportional to the cost of instantiating its rules. Our program specialization methods (§6.2) should also work with little modification.

[226] Eisner and Blatz (2007) described forward chaining for mixed semirings, which covers most but not all Dyna 2 programs.

**Program Transformations.** Although Eisner and Blatz (2007) predate Dyna 2, they had some foresight in their approach. Essentially, they devised semiring transformations that could be applied in mixed-semiring programs. Their trick was simply to be careful to check that there is a *locally* distributive property in a program fragment. Essentially our transformations should still work, but we just need to check for local distributivity. We also note that Dyna 2 programs will require other kinds of transformations that pull from a wider range of mathematical rewrites. Such transformations are used in other computation graph libraries.

**Program Optimization by Transform Search.** Our transformation search approach is directly applicable to a transformation search space tailored to Dyna 2 programs.

# Appendix E

# Programs Used for Experiments

## E.1 Chains (chain-05, chain-10, bad-chain-05, bad-chain-10)

Chain structures are common in NLP as they model the interaction between adjacent words or labels in a sequence (e.g., Lafferty et al. (2001)).

The chain-N programs are adaptations of Example 3,

```
goal += w(X₁,X₂) * w(X₂,X₃) * w(X₃,X₄) * w(X₄,X₅).
```

which is for the specific case of length 4, to have length $N$. The original program's degree is $N$. The optimal degree for chains is 2 regardless of $N$.

We also have the following examples of bad variable-elimination orderings

§7.3. Both of the examples below can be sped up to degree 2.

**bad-chain-05**:

```
goal += gen6(X5) * tmp(X1,X₄,X₅).
tmp(X₁,X₄,X₅) += gen2(X₁,X₄) * w(X₄, X₅).
gen2(X₁,X₄) += gen5(X₁,X₂) * gen4(X₂, X₄).
gen4(X₂,X₄) += w(X₂, X₃) * w(X₃, X₄).
gen5(X₁,X₂) += w(X₁, X₂).
gen6(X₅) += w(X₅, X₆).
```

**bad-chain-10**:

```
gen56(X₄, X₆) += w(X₄, X₅) * w(X₅, X₆).
gen57(X₂) += w(X₁, X₂).
gen59(X₄, X₇) += gen56(X₄, X₆) * w(X₆, X₇).
goal += gen124(X₃) * gen122(X₃).
gen124(X₃) += gen123(X₄) * w(X₃, X₄).
gen123(X₄) += gen125(X₁₀) * gen121(X₁₀, X₄).
gen125(X₁₀) += w(X₁₀, X₁₁).
gen122(X₃) += gen57(X₂) * w(X₂, X₃).
gen121(X₁₀, X₄) += gen119(X₄, X₈) * gen120(X₁₀, X₈).
gen120(X₁₀, X₈) += w(X8, X9) * w(X₉, X₁₀).
gen119(X₄, X₈) += gen59(X4, X7) * w(X₇, X₈).
```

# E.2 Projective Dependency Parsing

This program performs a version of CKY (§E.3) where the non-terminals have been lexically annotated with their head word. In contrast to CKY's $\mathcal{O}(n^3)$ runtime, the original presentation of this algorithm ran in $\mathcal{O}(n^5)$ (Collins, 1996). In subsequent work, Eisner and Satta (1999a) gave a faster version of the algorithm that runs in $\mathcal{O}(n^4)$.

## E.2.1 Bilexical Unlabeled (bilexical-unlabeled)

```
phrase(I,H,K) += phrase(I,H,J) * phrase(J,H',K) * score(H ← H').
phrase(I,H',K) += phrase(I,H,J) * phrase(J,H',K) * score(H → H').
phrase(I,I,K) += word(I,_,K).
```

```
4 goal(N) += phrase(0,0,N).
5
6 inputs: word(_,_,_); score(_ → _); score(_ ← _).
7 outputs: goal(_).
```

Degree: 5. Optimal: 4.

### E.2.2 Bilexical Labeled (bilexical-labeled)

Extends bilexical-unlabeled with labels (i.e., grammar relations).

```
1 phrase(I,H,X,K)  += phrase(I,H,Y,J) * phrase(J,H',Z,K) * rewrite(H,Y, H',Z, H,X).
2 phrase(I,H',X,K) += phrase(I,H,Y,J) * phrase(J,H',Z,K) * rewrite(H,Y, H',Z, H',X).
3 phrase(I,I,X,K) += rewrite(X,W) * word(I,W,K).
4 goal(N) += phrase(0,0,s,N).
5
6 inputs: word(_,_,_); rewrite(_,_,_,_,_,_); rewrite(_,_).
7 outputs: goal(_).
```

Degree: 8. Optimal: 7.

### E.2.3 Arc Eager (arc-eager)

This benchmark is based on Example 21; please refer to that discussion.

```
1 β(I,I^0,K) += word(I,K).
2 β(I,I^1,K) += word(I,K).
3 β(I,H^B,K) += β(I,H^B,J) * score(H ← H') * β(J,H'^0,K).
4 β(I,H^B,K) += β(I,H^B,J) * score(H → H') * β(J,H'^1,K).
5 goal(N) += β(0,0^0,N).
6
7 inputs: word(_,_); score(_ → _); score(_ ← _).
8 outputs: goal(_).
```

Degree: 6, Optimal: 3.

## E.2.4 Split Bilexical Dependency Parsing

A split bilexical grammar is one where the headword must combine with all its right children before any of its left children. Johnson (2007) and Eisner and Satta (1999a) each gave initial, inefficient algorithms that they show how to speed up to $\mathcal{O}(n^3)$. Their initial algorithms are different, so, we have provided them as separate benchmarks.

### E.2.4.1 Johnson (2007)'s Version (split-head-J)

The following program is the initial, inefficient algorithm of Johnson (2007).

```
1  goal(N) += x(0,V,N).
2
3  % words are duplicated, as in Johnson (2007).
4  l(I,K) += word(left,I,K).
5  r(I,K) += word(right,I,K).
6
7  % l(I, K) is a span with K as the head
8  l(I,K) += x(I,V,J) * score(V ← K) * l(J,K).
9
10 % r(I, K) is a span with I as the head
11 r(I,K) += r(I,J) * score(I → V) * x(J,V,K).
12
13 % J is the head of l(I, J) and J is the head of r(J, K)
14 x(I,J,K) += l(I,J) * r(J,K).
15
16 input: word(_,_,_); score(_ → _); score(_ ← _).
17 output: goal(_).
```

Degree: 4, Optimal: 3.

### E.2.4.2 Eisner and Blatz (2007)'s Version (split-head-EB)

The following program is the initial, inefficient algorithm of Eisner and Blatz (2007), which we discussed in Example 44.

```
1  r(I,I,K) += word(I,K).                     % no right children yet
```

```
2  r(P,I,K) += r(P,I,J) * arc(P → C) * l(C,J,K).  % add right child
3
4  l(P,I,K) += r(P,I,K).                           % no left children yet
5  l(P,I,K) += l(C,I,J) * arc(C ← P) * l(P,J,K).  % add left child
6
7  goal(N) += l(H,0,N).
8
9  inputs: arc(C ← P); arc(P → C); word(_, _).
10 outputs: goal(N).
```

Degree: 5, Optimal: 3.

## E.3 Constituency Parsing

The following programs are variants of CKY (Cocke and Schwartz, 1970; Younger, 1967; Kasami, 1965; Lange and Leiß, 2009; Eisner and Blatz, 2007; Tomita, 1985; Tomita, 1991; Johnson, 1989; Baker, 1979; Jelinek, 1985). We briefly discussed CKY in Example 6 of the main text.

### E.3.1 CKY (cky3)

```
1 β(X,I,K) += γ(X,Y,Z) * β(Y,I,J) * phrase(Z,J,K).
2 β(X,I,K) += γ(X,Y) * β(Y,I,K).
3 β(X,I,K) += γ(X,Y) * word(Y,I,K).
4 z += β(root, 0, N) * len(N).
5 input: word(W,I,K); len(N); γ(X,Y,Z); γ(X,Y).
6 output: z.
```

Degree: 6, Optimal: 5.

### E.3.2 CKY with 4-ary Productions (cky4)

The following program implements Tomita (1985)'s algorithm where the grammar can have a production rule of length up to 4. It is a simple modification to CKY4. We just add the following rule and input declaration.

```
7  phrase(X,I₁,I₄) += γ(X,Y₁,Y₂,Y₃) * phrase(Y₁,I₁,I₂) * phrase(Y₂,I₂,I₃)
8      * phrase(Y₃,I₃,I₄).
9  input: γ(X,Y₁,Y₂,Y₃)
```

Degree: 7, Optimal: 6.

### E.3.3 CKY with a Fixed Grammar (cky+grammar)

For CKY+grammar, we use CKY3, but we remove the input declaration for $\gamma$ and declare the following grammar for the program to specialize to.

```
10 γ("S", "NP", "VP") += 1.0.
11 γ("NP", "Det", "N") += 1.0.
12 γ("NP", "NP", "PP") += 1.0.
13 γ("VP", "V", "NP") += 1.0.
14 γ("VP", "V") += 1.0.
15 γ("VP", "VP", "PP") += 1.0.
16 γ("PP", "P", "NP") += 1.0.
17 γ("<.>", ".") += 1.0.
18 γ("NP", "Papa") += 1.0.
19 γ("N", "caviar") += 1.0.
20 γ("N", "spoon") += 1.0.
21 γ("N", "fork") += 1.0.
22 γ("N", "telescope") += 1.0.
23 γ("N", "boy") += 1.0.
24 γ("N", "girl") += 1.0.
25 γ("N", "baby") += 1.0.
26 γ("N", "man") += 1.0.
27 γ("N", "woman") += 1.0.
28 γ("N", "moon") += 1.0.
29 γ("N", "cat") += 1.0.
30 γ("V", "ate") += 1.0.
31 γ("V", "saw") += 1.0.
32 γ("V", "fed") += 1.0.
33 γ("V", "said") += 1.0.
34 γ("V", "jumped") += 1.0.
35 γ("P", "with") += 1.0.
36 γ("P", "over") += 1.0.
37 γ("P", "under") += 1.0.
38 γ("P", "above") += 1.0.
39 γ("P", "below") += 1.0.
40 γ("P", "on") += 1.0.
41 γ("P", "in") += 1.0.
```

Degree: 6, Optimal: 3.

### E.3.4 Tensor Decomposition Parser (tensor-decomp-parser)

The following program is a variation on CKY where the grammar (i.e., the `rewrite` relation) is factored: `rewrite(X,Y,Z) += λ(R) * u(R, X) * v(R, Y) * w(R, Z)`. This approach was proposed by Cohen, Satta, et al. (2013) whereby the grammar was approximated using low-rank tensor decomposition and subsequently parsed with the algorithm below. However, the algorithm below can be sped up by exploiting the structure of the `rewrite` relation.

```
1 rewrite(X,Y,Z) += λ(R) * u(R,X) * v(R,Y) * w(R,Z).
2 β(I,K,X) += β(I,J,Y) * β(J,K,Z) * rewrite(X,Y,Z).
3 β(I,K,X) += word(I,K,X).
4 goal += β(0,N,s) * length(N).
5
6 inputs:  λ(R); u(R,X); v(R,Y); w(R,Z); word(I,K,X); length(_).
7 outputs: goal.
```

Degree: 6, Optimal: 4.

## E.4 Inversion Transduction Grammars (itg)

Inversion transduction grammars were introduced by Wu (1996), who gave an $\mathcal{O}(n^7)$ algorithm, which was later sped up to $\mathcal{O}(n^6)$ by Huang, Zhang, et al. (2005). The model simultaneously parses a pair of related sentences—typically a target language sentence and a source language sentence, as in machine translation. The model allows for a restricted syntactic reordering of phrases called inversion.

```
1 β(A,I₁,K₁,I₂,K₂) += word₁(X,I₁,K₁) * word₂(X',I₂,K₂) * transduce(A,X,X').
2 β(A,I₁,K₁,I₂,K₂) += β(B,I₁,J₁,I₂,J₂) * β(C,J₁,K₁,J₂,K₂) * rewrites(A,B,C).
3 β(A,I₁,K₁,I₂,K₂) += β(B,J₁,K₁,I₂,J₂) * β(C,I₁,J₁,J₂,K₂) * rewritesinv(A,B,C).
4 goal += β(s,0,N₁,0,N₂) * length(N₁,N₂).
5
6 inputs: word₁(_,_,_); word₂(_,_,_); transduce(_,_,_); rewrites(_,_,_);
7     rewritesinv(_,_,_); length(_,_).
8 output: goal.
```

Degree: 9, Optimal: 8.

## E.5 Edit Distance (edit)

The following program implements a weighted generalized monotonic alignment between two sequences word and word'. It is essentially the well-known Levenstein distance (Levenshtein, 1966).

```
1 align(0,0) min= 1.
2 align(J,J') min= align(I,I') * word(W,I,J) * word'(W',I',J') * score(W,W').
3 align(I,J') min= align(I,I')               * word'(W',I',J') * score(ε,W').
4 align(J,I') min= align(I,I') * word(W,I,J)                   * score(W,ε).
5 goal min= align(N,N') * len(N) * len'(N').
6 input: word(_,_,_); word'(_,_,_); score(_,_); len(_); len'(_).
7 output: goal.
```

Degree 6, Optimal: 4.

## E.6 Bar-Hillel Construction (bar-hillel)

The following program implements a parser for the (weighted) intersection of a context-free parser and a bigram model on the part-of-speed sequences. It is essentially Bar-Hillel et al. (1961)'s construction of a context-free language that accepts the intersection of a regular language and a context-free language.

```
goal += β(0,_,root,_,N) * len(N).
β(I,A,X,D,K) += β(I,A,Y,B,J) * β(J,C,Z,D,K) * γ(X,Y,Z) * bigram(B,C).
β(I,X,X,X,K) += tag(X,W) * word(W,I,K).
input: len(_); word(_,_,_); bigram(_,_); γ(_,_,_); tag(_,_).
output: goal.
```

Degree 10, Optimal: 8.

## E.7 Expectations under a Linear-Chain Conditional Random Field (chain-expect)

This example implements the inside-outside speedup (Li and Eisner, 2009) for computing the expectation of an additively decomposable function $f : \mathtt{S} \times \mathtt{S}' \to \mathbb{R}^d$ over randomly drawn sequences from a weighted graph (e.g., a conditional random field (Lafferty et al., 2001)). The graph is specified as a collection of weights `w`, as well as `start` and `end` nodes. The relations $\alpha$ and $\beta$ implement the forward-backward algorithm (discussed in Rabiner (1989)), and `z` is the normalization constant of the distribution. The expectation of the $i^{\text{th}}$ dimension of `f` is `fbar(I)/z`.

```
% forward algorithm
α(S) += start(S).
α(S') += α(S) * w(S,S').
% backward algorithm
β(S) += end(S).
β(S) += w(S,S') * β(S').
% normalization constant
z += α(S) * end(S).
% unnormalized expected value via inside-outside speedup
fbar(R) += α(S) * w(S,S') * β(S') * r(S,S',R).
input: w(_,_); r(_,_,_); start(_); end(_).
```

```
12 output: fbar(_). z.
```

Degree 3, Optimal: 3.

## E.8 Intersection of Even and Odd Peano Numbers (even-odd-peano)

The following program counts the number of Peano numbers that are both even and odd. No such numbers exist, but as the program is written, fixpoint iteration will enumerate the even and odd numbers forever. Thus, its running time is infinite. However, it is possible to transform this program into a version that runs in constant time using fold/unfold and define; the optimal program is the empty program.

```
1  % even peano numbers
2  even(0).
3  even(s(s(X))) += even(X).
4
5  % odd peano numbers
6  odd(X) += even(s(X)).
7
8  % intersection
9  evenodd(X) += even(X) * odd(X).
10
11 % size of the intersection
12 goal += evenodd(X).
13
14 output: goal.
```

Degree: 1, Optimal: 0.

## E.9 Hidden Markov Models (hmm)

Hidden Markov models (HMMs) are the generative and locally normalized analog of CRFs, which are discussed in §E.7. Rabiner (1989) provides a classic tutorial.

```
1 v(0,start) += 1.
2 v(T',Y') += v(T,Y) * emission(Y,X) * transition(Y,Y') * obs(T,X,T').
3 goal += v(N,stop) * len(N).
4 input: obs(_,_,_); len(_); emission(_,_); transition(_,_).
5 output: goal.
```

Degree 5, Optimal: 4.

## E.10 Semi-Markov Model (semi-markov)

The semi-Markov model (Sarawagi and Cohen, 2004) generalizes a Markov model to score spans rather than individual words. In terms of runtime, one can perform inference in a Markov model in $\mathcal{O}(n)$ time (omitting dependence on the number of tags). In contrast, inference in a semi-Markov model takes $\mathcal{O}(n^2)$.

```
1 β(start, 0) += 1.
2 β(Y, J) += β(X, I) * transition(X, Y) * chunk(Y, I, J).
3 goal += β(_, N) * len(N).
4 input: transition(_,_); chunk(_, _, _); len(_).
5 output: goal.
```

Degree 4, Optimal: 3.

## E.11 Shortest Path in a Graph

### E.11.1 Explicit Path (path-list)

The code below was previously discussed in Example 28. This example is interesting because its initial running time is infinite, but it can be sped up to $\mathcal{O}(n^2)$ where $n$ is the number of nodes in the graph. This example requires either generalized fold/unfold or the left-corner transformation (Example 46)

```
1 β([X,Y|Xs]) += edge(X,Y) * β([Y|Xs]).
2 β([X]) += stop(X).
3 goal += start(X) * β([X|Xs]).
4
5 outputs: goal.
6 inputs: start(_); edge(_,_); stop(_).
```

### E.11.2 Start Path (path-start)

The code below was previously discussed in Example 27. Here we speed up an initial $\mathcal{O}(n^3)$ program to $\mathcal{O}(n^2)$ where $n$ is the number of nodes in the graph. This example requires generalized fold/unfold.

```
1 path(I,I).
2 path(I,K) += path(I,J) * edge(J,K).
3
4 goal += start(I) * path(I,K) * stop(K).
5
6 outputs: goal.
7 inputs: start(_); edge(_,_); stop(_).
```

# E.12 Pushdown Automaton with an Explicit Stack (explicit-pda)

The program below can, in principle, be converted into CKY. However, doing so requires program transformations that appear to go beyond those in our arsenal.

```
1 pda([],0).                                                % initial stack is empty
2 pda([X|Xs],K) += pda(Xs,J) * rewrite(X,Y) * word(Y,J,K).                    % shift
3 pda([X|Xs],K) += pda([Z,Y|Xs],K) * rewrite(X,Y,Z).                         % reduce
4 goal += pda([s], N) * len(N).                % parse of the entire input labeled s
5
6 inputs: rewrite(_,_); rewrite(_,_,_); word(_,_,_); len(_).
7 outputs: goal.‘
```

The following program was manually specified using an extension of speculation, which permits a CCG-like composition rule.[227] Unfortunately, we have not been able to fully formalize the correctness conditions for this extended transformation as of the time of writing. We can, however, use our fold/unfold transformations to prove that the program below is equivalent to the original.

Call the original program $\mathfrak{p}^{(0)}$ and the program below $\mathfrak{p}^{(1)}$.

```
1 goal += len(N) * pda([s],N).
2 pda(Ys,I)/pda(Ys,I).   % base case
3 pda([X|Xs],K)/pda(Xs,I) += pda(Xs,J)/pda(Xs,I) * rewrite(X,Y) * word(Y,J,K).
4 pda([X|Xs],K)/pda(Xs,I) +=
5     pda([Z,Y|Xs],K)/pda([Y|Xs],J) * pda([Y|Xs],J)/pda(Xs,I) * rewrite(X,Y,Z).
6 % recovery rules
7 pda([X|Ys],K) += (pda([X|Ys],K)/pda(Ys,I)) * pda(Ys,I).
8 pda([],0).
```

**Recovering CKY.** First, let us see how the program converts into CKY. If we unfold the goal rule (twice), we arrive at a program that no longer depends

[227] It appears as if our hoisting transformations are incapable of this transformation because they only have a CCG-like *application* rule.

on the pda relation:

```
pda(Ys,I) / pda(Ys,I).
pda([X|Xs],K)/pda(Xs,I) += pda(Xs,J)/pda(Xs,I) * rewrite(X,Y) * word(Y,J,K).
pda([X|Xs],K)/pda(Xs,I) +=
    pda([Z,Y|Xs],K)/pda([Y|Xs],J) * pda([Y|Xs],J)/pda(Xs,I) * rewrite(X,Y,Z).
goal += len(N) * pda([s],N)/pda([],0).
```

This program does not instantiate the Xs. We could run this program as it is, and it will have the $\mathcal{O}(k^3\,n^3)$ running time of CKY. We can also use our program specialization §6.2 transformation to arrive at the following program, which is visually more similar to CKY. Notice that the argument ordering to the new relation p is chosen arbitrarily.

```
p(X,K,I) += rewrite(X,Y) * word(Y,I,K).
p(X,K,I) += p(Z,K,J) * p(Y,J,I) * rewrite(X,Y,Z).
goal += len(N) * p(s,N,0).
```

It is easy to see that this program is equivalent to CKY.

**Correctness by fold/unfold.** Now, we investigate the correctness of our manually specified program. The key is to show that the recovery rules can be transformed into the original definition $\mathfrak{p}^{(0)}$. If we can do that using semantics-preserving transformations, we would have a proof that $\mathfrak{p}^{(0)}$ and $\mathfrak{p}^{(1)}$ are equivalent.

```
pda([X|Ys],K) += (pda([X|Ys],K)/pda(Ys,I)) * pda(Ys,I).
pda([],0).
```

Can be transformed into the original rules:

```
pda([],0).
pda([X|Xs],K) += pda(Xs,J) * rewrite(X,Y) * word(Y,J,K).
pda([X|Xs],K) += pda([Y2,Y1|Xs],K) * rewrite(X,Y1,Y2).
```

Unfolding the recursive rule twice (using $\mathfrak{p}^{(1)}$ as auxiliary definitions):

```
pda([],0).
pda([X|Ys],K) += pda([Z,Y|Ys],K) / pda([Y|Ys],J) * pda([Y|Ys],J) / pda(Ys,I) *
```

```
      rewrite(X,Y,Z) * pda(Ys,I).
3  pda([X|Ys],K) += rewrite(X,Y) * word(Y,I,K) * pda(Ys,I).
```

Folding this program twice (using $\mathfrak{p}^{(1)}$ as auxiliary definitions) us what we were after.

```
1  pda([],0).
2  pda([X|Ys],K) += pda([Z,Y|Ys],K) * rewrite(X,Y,Z).
3  pda([X|Ys],K) += rewrite(X,Y) * word(Y,I,K) * pda(Ys,I).
```

Since we used semantics-preserving rules to arrive at this definition, it follows that $\mathfrak{p}^{(0)} \equiv \mathfrak{p}^{(1)}$. We used our Dyna library to work through this example, meaning we did not have to manually perform the transformations or check their safety conditions; all that was handled seamlessly by the library. We used our search algorithm to enumerate the valid folds but chose among them manually.

# Index